\documentclass[a4paper,11pt]{book}

\usepackage[latin1]{inputenc}
\usepackage[american]{babel}

\usepackage{euscript}
\usepackage{amssymb}
\usepackage{amsfonts}
\usepackage{amsbsy}
\usepackage{amsmath}
\usepackage{epsfig}
\usepackage{hyperref}

 \usepackage{a4wide}
 \usepackage{graphics}
 \usepackage{subfigure}
 \usepackage{fancyhdr}

\newcommand{\bee}{\begin{equation}}
\newcommand{\ee}{\end{equation}}
\newcommand{\bea}{\begin{eqnarray}}
\newcommand{\eea}{\end{eqnarray}}
\newcommand{\baa}{\begin{array}}
\newcommand{\eaa}{\end{array}}

\newcommand{\MM}{\mathcal{M}}
\newcommand{\NN}{\mathcal{N}}
\newcommand{\FF}{\mathcal{F}}
\newcommand{\KK}{\mathcal{K}}
\newcommand{\RR}{\mathcal{R}}
\newcommand{\Ss}{\mathcal{S}}
\newcommand{\PP}{\mathcal{P}}
\newcommand{\OO}{\mathcal{O}}
\newcommand{\Rbb}{\mathbb{R}}
\newcommand{\Zbb}{\mathbb{Z}}
\newcommand{\Cbb}{\mathbb{C}}
\newcommand{\Obb}{\mathbb{O}}

\newcommand{\nn}{\nonumber}

\newcommand{\re}{\mbox{Re}}
\newcommand{\im}{\mbox{Im}}

\newcommand{\D}{{\bf \not}  D}

\def\bj{\bar{j}}

\def\bz{\bar{z}}
\def\bF{\bar{F}}

\begin{document}

\begin{flushright}

%RUNHETC-2000-44
%26th May 1998
\end{flushright}
\begin{flushright}
%{\sf \today}
\end{flushright}
\begin{center}
\huge{ \sc Phenomenology from the Landscape of String Vacua}\\
\vskip 2cm

\Large{\sc Roberto Valandro$^a$ %\ft{valandro@sissa.it}
}
\\
\vskip 15mm \normalsize
$^a$  ITP, Universit\"at Heidelberg \\ Philosophenweg 19 -- D69120 Heidelberg -- GERMANY\\
\smallskip
{\it r.valandro@thphys.uni-heidelberg.de}

\end{center}
%\renewcommand{\abstractname}{\sc Abstract}
%\begin{Abstract}
\vskip 15mm
\begin{center}
{\bf {\sc Abstract}}
\end{center}
%\bigskip
%\normalsize

This article is the author's PhD thesis. After a review of string vacua obtained through compactification (with and without fluxes), it presents and describes various aspects of the Landscape of string vacua. At first it gives an introduction and an overview of the statistical study of the set of four dimensional string vacua, giving the detailed study of one corner of this set ($G_2$-holonomy compactifications of M-theory). Then it presents the ten dimensional approach to string vacua, concentrating on the ten dimensional description of the Type IIA flux vacua. Finally it gives two examples of models having some interesting and characteristic phenomenological features, and that belong to two different corners of the Landscape: warped compactifications of Type IIB String Theory and M-theory compactifications on $G_2$-holonomy manifolds.

\newpage

\normalsize

\begin{titlepage}
 \vspace{3cm}
\center{\huge{\textsf{Roberto Valandro}}}
%\begin{flushright}
%  \huge{\textsl{Padova,  07 marzo 2003.}}
%\end{flushright}

\vspace{2cm}
\center{\LARGE{\textsl{PhD Thesis:}}}

\vspace{1cm}
% \center{\Huge{\textbf{Phenomenology from the }}}
% \vspace{5mm}
\center{\Huge{\textbf{Phenomenology}}}
\vspace{5mm}
\center{\Huge{\textbf{from the }}}
\vspace{5mm}
\center{\Huge{\textbf{Landscape of String Vacua}}}

\vspace{3cm}

\Large{Supervisor: \textsf{Prof. Bobby Samir Acharya}}

\vspace{5cm}
\center{\LARGE{\textsl{September 2007}}}

\vspace{1cm}

\center{\Large{\it International School for Advanced Studies \\ (SISSA/ISAS)}}

\end{titlepage}

{\tiny .}

\newpage

{\tiny .}

\vspace{3cm}

\begin{center}
\LARGE{\textsc{Acknowledgments}}
\end{center}

\vspace{10mm}

First of all I would like to thank my supervisor Prof. Bobby Samir Acharya, for his important and continuous support, expert and helpful advises, frequent encouragement and for the numerous discussions we had in his office.

I would like to thank Frederik Denef for the stimulating and fruitful collaboration, very important in the first period of my PhD.

Furthermore I thank Francesco Benini for the very productive collaboration and for interesting, frequent and illuminating discussions. 

I thank Giuseppe Milanesi, for the extremely fertile and insightful discussions about physics and beyond that we had throughout our PhD in our office at SISSA. He has been the best office-mate I could have ever had.

I thank also the other PhD students at SISSA, in particular Alberto Salvio, Giuliano Panico, Alessio Provenza, Stefano Cremonesi and Davide Forcella for several interesting and useful discussions. 

I must thank all my friends who have continuously shown me their appreciation thus helping me not to lose my self-confidence.
Moreover I want to thank the Gan Ainm Irish Dancers, because our fantastic activity allowed me to find new energy to spend in my PhD work. In particular I want to thank my dear Tati, who made me meet this dancing group and who in the last months has given me a peace of mind that turned out to be really fruitful in writing this thesis.

I thank my dear brother, for his wisdom, in spite of his young age.

Finally, the most important acknowledgment is for my parents, for their invaluable support throughout my entire education and life.

% Un grazie va anche a tutti i miei amici che con la loro continua dimostrazione di
% stima mi hanno aiutato a non perdere la fiducia in me stesso.

\tableofcontents

\chapter{Introduction}

String theory has long held the promise to provide us with a complete and final description of the laws of physics in our
universe. The early times of String Theory were characterized by the discovery that in its massless spectrum there is a spin-2 state with couplings similar to those of General Relativity. It was also clear that String Theory could provide Yang-Mills bosons as well. The introduction of supersymmetry allowed also massless fermions and eliminated the tachyon from the spectrum. Thus String Theory became soon a good candidate for a unifying theory of all the four interactions: electromagnetic, strong, weak and gravitational.

If we expect String Theory to describe our world, it should be possible to deduce from it the other theories that have been experimentally tested.
At present the first three interactions are described by the Standard Model (SM) of particle physics at a very high experimental precision, while the gravity is very well described by General Relativity (GR). 
Unfortunately these theories seem to be incompatible from a theoretical point of view, in the sense that neither of them allows to naturally adapt the other. It is at this point that String Theory should come, since it includes both Yang-Mills theories and gravity.

The SM is a quantum gauge theory with gauge group $SU(3)_c\times SU(2)_L\times U(1)_Y$, with three generations of fermions and one scalar, the Higgs, responsible for the fermions and gauge bosons masses. The SM has been tested to a very high precision. Experimentally, the only missing ingredient is the scalar Higgs particle. Despite its great success, it is not completely satisfactory from a theoretical point of view, for many reasons, such as the large number of free parameters, the large hierarchy between the electroweak scale and the Plack scale as well as the already mentioned missing unification with gravity. 

Various extentions of the SM have been proposed after its birth. A natural one is provided by supersymmetry, a symmetry that relates bosons and fermions. Supersymmetry predicts a superpartner for all known particles. However, so far these new particles have not been detected in the accelerator experiments. Hence supersymmetry must be broken at the electroweak scale. The supersymmetric extensions of the SM solves some problems mentioned above. In particular supersymmetry protects scalar masses from large quantum corrections, giving a solution to the hierarchy problem: the Higgs mass remains of the order of the electroweak scale, also in a theory with a large cutoff.

Another extention of the SM is given by the Grand Unified Theories (GUT's). The idea that characterizes them is that the SM gauge group is a proper subgroup of a larger simple group, with only one coupling constant. It is broken to the SM gauge group at the so called $GUT$ scale. Actually, if one includes supersymmetry into the SM and makes the three couplings run, they meet each other at one point corresponding to the energy $M_{GUT}\simeq 10^{16}GeV$. This unifies the electromagnetic, weak and strong interactions in one quantum field theory. The gravitational interaction is not included. 

Gravity is very well described by GR. It is a theory very different from the Quantum Field Theory(QFT) describing particle physics. GR is a classical theory that is hard to quantize due to its ultra-violet(UV) divergences. It actually works very well at large distances, where the quantum effects are negligible.

As we have said, these two theories seem to be incompatible. This is a problem when one wants to describe phenomena in regimes where both theories have to be applied. Early time cosmology or physics of black holes are two such examples. In order to approach this question, one should have a theory that combine the SM and GR. String Theory is a good candidate to be such unifying theory. 

The path from String Theory to the SM or GR is however not so simple. At present this program is far to be completed. Still one of the most important issues to address is how to relate String Theory to the observables in the low energy physics world. This is the main task of the branch of the theory known as {\it String Phenomenology}, {\it i.e.} to reproduce all the characteristic features of the SM: non-Abelian gauge group, chiral fermions, hierarchical Yukawa couplings, hierarchy between the electroweak scale $M_W$ and the Plack scale $M_p$. In particular String Theory should provide a framework for computing all couplings of the SM and give an explanation of the supersymmetry breaking at low energies (since spacetime supersymmetry is automatically built into String Theory). Finally, one of the main problems of string phenomenology is the translation between the low energy effective string action and the data that will be collected at LHC, starting hopefully in fall 2008.

There are two possible approaches to these problems. The first one is the {\it top-down} approach, which starts from the fundamental theory and tries to deduce from it all low energy observables. The second one is the {\it bottom-up} approach, which tries to build consistent string models that contain as many SM features as possible. The works presented in this thesis belong both to the first and to the second directions.

\

It is time to say what is String Theory\footnote{An introduction to this subject can be found in \cite{Polchinski:1998rq,Green:1987sp}}. Its characterizing feature, that distinguishes it from a QFT, is that its fundamental blocks are not particles, but one dimensional objects: the {\it strings}. There can be {\it open strings} and {\it closed strings} and the two different topologies give different spectra. The characteristic length of the strings is $\ell_s = \sqrt{2\pi \alpha '}$, where $\alpha'$ is the Regge slope. It is the only input parameter of the theory.

The fundamental string can appear in various vibrational modes which at low energies are identified with different particles. The states of minimal energy are massless, while the other has masses of the order $n/\sqrt{\alpha'}$ (with $n\in\mathbb{Z}$). 
The extended nature of the strings becomes apparent close to the string scale $m_s\sim 1/\sqrt{\alpha'}$.
Hence the point particle limit is given by $\alpha'\rightarrow 0$. In this limit, only the massless modes survive, while the massive ones are integrated out. The massless string spectrum naturally includes a mode corresponding to the graviton, providing a renormalizable  quantum theory of gravity around a given background. It avoids the UV divergences of graviton scattering in quantum field theory because of the extended nature of the strings, whose minimal length regularizes the amplitudes.

String Theory is a strongly constrained theory. Superstring Theories require spacetime supersymmetry and predict a ten dimensional spacetime at weak coupling. There are just five consistent ten dimensional String Theories: Type IIA, Type IIB, Type I, Heterotic $E_8\times E_8$ and Heterotic $SO(32)$. Exploring several kinds of duality symmetries, it is conjectured that all these string theories can be unified into the so called M-theory \cite{Polchinski:1998rq,Green:1987sp}, that lives in eleven dimensions. Together with eleven dimensional supergravity, the five ten dimensional string theories are seen as limit of this more fundamental theory.

At this stage String/M-theory is a ten(eleven) dimensional theory, while both the SM and GR are defined on a four dimensional spacetime.
One approach to reduce String/M-theory from ten(eleven) to four spacetime dimensions is the so called {\it compactification}. It consists in studying the theory on a geometric background of the form $M^{3,1}\times X$. $M^{3,1}$ is identified with our spacetime, while the manifold $X$ is chosen to be {\it small} and {\it compact}, such that the six(seven) additional dimensions are not detectable in experiments. 

The process of compactification introduces a high amount of ambiguity, as String/M-theory allows many different choices of $X$.
To get the effective four dimensional theory, one should integrate out the massive string states\cite{Polchinski:1998rq}, together with the massive Kaluza-Klein (KK)\cite{Kaluza:1921tu,Klein:1926tv} modes appearing in the process of compactification.
The structure of the obtained four dimensional theory strongly depends on the chosen internal manifold $X$. The properties of $X$ determine the amount of preserved supersymmetry and the surviving gauge group of the lower dimensional effective theory. Usually one requires $X$ to preserve some supercharges, both for phenomenological reasons and because String/M-theory on supersymmetric background is under much better control than on non-supersymmetric ones. This requirement is actually translated into a geometric condition on the compact manifold: it must have reduced holonomy. In particular in many cases this implies the internal manifold to be a Calabi-Yau (CY), {\it i.e} a six dimensional compact manifold with $SU(3)$ holonomy. After compactification and reduction to the four dimensional theory, one would like at least to obtain a realistic spectrum. But here one encounters
one of the main problems in compactification: the presence of {\it moduli}. These are parameters that label the continuous degeneracy of consistent background and can generically take arbitrary values. In four dimensions, they 
%Parmeters like the size and the shape of the compact manifold 
appear as massless neutral scalar fields. These scalars are not present in our world and one should find a mechanism to generate a potential for them, in such a way that they acquire a mass and are not dynamical in the low energy action. 
Moreover, the low energy masses and coupling constants are functions of the moduli. Thus, for example, if one wants to solve the hierarchy problem between the electroweak scale $M_W$ and the Plank scale, one has to fix the moduli and generate the hierarchy simultaneously, since $M_W$ depends on the moduli.

In order to introduce a potential that stabilizes the moduli, one should add some more ingredients to the compactification. One of them, largely studied in the last years, is the introduction of non-zero fluxes threading nontrivial cycles of the compact manifold. Each of the limits of M-theory mentioned above has certain p-form gauge fields, which are sourced by elementary branes. Background values for their field strength can actually stabilize the moduli. This is because, their contribution to the total energy will depend on the moduli controlling the size of the cycles that the fluxes are threading. If the generated potential is sufficiently general, minimizing it will stabilize the moduli to fixed values. Some beautiful recent reviews on flux compactifications are \cite{Grana:2005jc,Douglas:2006es,Blumenhagen:2006ci}.

The fluxes are subject to a Dirac-like quantization condition. Hence they take discrete values, that add to the other discrete parameters parametrizing the compactification data, such as for instance the brane charges. The four dimensional effective moduli potential depends on these discrete data. Varying them we get an ensemble of effective four dimensional potentials. Minimizing each of them gives a set of vacua. Putting all together, one gets an huge number of lower dimensional string groundstates (vacua). The set of all these four dimensional constructions is called {\it the Landscape}.
The extremely large number of distinct string vacua gives rise to the question if String Theory is actually a predictive theory or not. In fact, each point in the Landscape corresponds to a possible universe with different particle physics and cosmology. 
Another question is if among these vacua there is at least one that describes our world.

\

One fruitful approach to these problems was suggested by M. Douglas and collaborators \cite{Douglas:2003um,Denef:2004ze,Denef:2004cf}. It consists in investigating the statistical properties of the string Landscape. For example one can determine by statistical methods what is the fraction of vacua with good phenomenological properties. It would also be interesting to discover a statistical correlation between the distribution of two physical quantities, because it could be characteristic of string theory vacua \cite{Douglas:2004zg}. In addition, it was argued that the Landscape could give the possibility to address the hierarchy problems in physics, especially that concerning the smallness of the cosmological constant \cite{Bousso:2000xa}: the tiny observed values $\Lambda \simeq 10^{-120}M_p^4$ could be explained if the number of vacua was of order of $10^{120}$. Finally one could merge the statistical approach with the anthropic principle. In particular one could analyze the impact of environmental constraints on the distributions of the four dimensional couplings \cite{ArkaniHamed:2005yv}, to see if for example the considered ensembles are ``friendly neighborhoods'' of the Landscape, {\it i.e.} with peaked distribution of the dimensionless physical couplings, but uniform distributions for dimesionfull quantities such as the cosmological constant or the supersymmetry breaking scale. This gives a certain degree of predictivity as explained in \cite{ArkaniHamed:2005yv}.

The statistics due to the closed string fluxes provides estimates for the frequencies of cosmological parameters like the cosmological constant. Of course, for making contact with elementary particle physics and the SM, one has also to include the statistics of the open string sector in Type II theories. A general study of D-brane statistics was initiated in \cite{Blumenhagen:2004xx}, where for the ensemble of intersecting branes on certain toroidal orientifolds, the statistical distribution of various gauge theoretic quantities was studied, like the rank of the gauge group and the number of generations. This branch of the statistical approach to the String Landscape has been carried on in \cite{Gmeiner:2005vz,Gmeiner:2005nh,Gmeiner:2006vb} and in \cite{Douglas:2006xy}.

\

The moduli stabilization by fluxes occurs within the effective supergravity approach. %, which is limited to large volumes of the internal space. 
We take the ten dimensional effective action of String Theory, that is valid only at large volume (to neglect the $\alpha'$ corrections) and small string coupling. We extract from this the four dimensional effective action compactifying around a particular background and integrating out all but a finite number of fields. Then minimizing the resulting moduli potential we get a pletora of vacua. 
These vacua are found within some approximations and so constitute only a limited corner of the full Landscape of string vacua. In principle, it could be that our world resides outside this corner.

Moreover there are consistent string constructions on backgrounds that are not geometric \cite{Hull:2004in}, in the sense that the metric of the compact manifold is not globally defined; these nongeometric vacua were discovered through a series of T-duality applied on geometric background \cite{Kachru:2002sk} and the resulting potential has been studied in \cite{Shelton:2005cf}. 

\

Thus far we have described the four dimensional approach to the Landscape, {\it i.e.} one reduces the ten dimensional theory to lower dimensions and studies the resulting four dimensional effective action. Another approach consists in studying the solutions of the string or ten dimensional supergravity equations of motion. This is a complementary approach, because it allows to make contact with the fundamental theory from which one starts to extract the real world. The simplest way of proceding consists in finding the supersymmetric solutions of the higher dimensional theory. This is essentially because the supersymmetry equations are more simple to solve than the full set of equations of motion. Before introducing fluxes, the supersymmetric solutions consist of a compact manifold with reduced holonomy; let us say for concreteness that it should be a CY. The solutions have some continuous parameters, the moduli, that will become massless scalar fields in the effective four dimensional theory. 
If we turned on background value for the p-form field strength, the compact manifold is no more Ricci-flat and it cannot be a CY. Moreover it may happen that there are no moduli of the compact manifold, because the supersymmetry equations could fix them in the case of non-vanishing fluxes. This is how the moduli fixing occurs in ten dimensions. Clearly there must be a relation between the fixed values when they can be found by both the approaches. This is the case of a special class of Type IIB solutions\cite{Giddings:2001yu}, that we will review in the first part of this work: solving the ten dimensional equations and the four dimensional ones give the same results. 

In the last years much effort has been spent in studying the ten dimensional supersymmetry equations in presence of fluxes. In particular, the backreaction of fluxes has been considered \cite{Gurrieri:2002wz,Gauntlett:2003cy}, contrary to the earlier approaches to flux compactifications. 
In that cases, in deriving the four dimensional effective theory the ten dimensional action was reduced around a background consisting of a CY with non-zero fluxes along non-trivial cycles. But, as we have said, this background is not a solution of the ten dimensional equations of motion. This is just an approximation, that turns out to be valid when the energy scale of the fluxes is much lower the KK scale; it is realized in the limit of large volume of the compactification scale, that is also required to neglect $\alpha'$ corrections.

To classify the full solutions of the ten dimensional supersymmetry equations, the new formalism of {\it generalized geometry} \cite{Hitchin:2004ut,Gualtieri:2003dx} has been introduced \cite{Grana:2004bg,Grana:2005sn}. It is very useful because it allows to give a unifying mathematical descriptions of all internal manifolds arising in supersymmetric flux backgrounds.
Using this formalism, a four dimensional approach has also been recently initiated, to give four dimensional description that includes also the backreaction of fluxes on the geometry and possibly also the nongeometric fluxes\cite{Grana:2006kf,Micu:2007rd,Koerber:2007xk,Cassani:2007pq}.

\

All we have said so far is related to the top-down approach, {\it i.e.} starting from the fundamental theory, one extracts a four dimensional effective theory, trying to understand if it can or cannot describe our world. As we have just seen, following this way one finds a pletora of possible four dimensional worlds arising from String Theory. A statistical study of this Landscape can give some indications which region one should concentrate on to hopefully find realistic vacua. This, among other things, allows to describe different setups in String Theory that lead to the same kind of physics as the SM. Within each setup, one should construct explicit examples with low energy physics as close as possible to the SM. In doing this one is driven by the realistic features one wants to realize: we know the answer and use only string ingredients that can give results compatible with that. This is the bottom-up approach. It has this name because one starts from the phenomenological features that he wants to realize using objects of the fundamental theory. These phenomenological features can be properties of the SM itself, or can be properties of some extension of it, such as Minimal Supersymmetric Standard Model (MSSM) or warped five dimensional inspired by the Randall-Sundrum model \cite{Randall:1999ee,Randall:1999vf}.

The setups which we concentrated on in this work are Type IIB with fluxes and M-theory compactification on $G_2$ holonomy manifolds. 

As we have seen, fluxes backreacts on the geometry driving the internal manifold far from special holonomy. The most studied set of flux vacua is a special class of solutions of Type IIB. Fluxes backreact on the geometry just giving a compact manifold that is conformally CY, {\it i.e.} its metric is a CY metric multiplied by a function; in particular it is not Ricci-flat. Moreover the ten dimensional spacetime is not a product of two space, but the four dimensional metric is multiplied by a function depending on the compact coordinates \cite{Giddings:2001yu}. This is the so called {\it warp factor}. Its phenomenological importance is made clear in the famous Randall-Sundrum \cite{Randall:1999ee} paper where they studied a five dimensional non-factorisable metric. They found that warping generates a natural exponential hierarchy of four dimensional scales. This mechanism works also in warped string theory compactifications. In \cite{Giddings:2001yu}, it was found in the context of Type IIB that fluxes generate a warp factor depending on the moduli. Moreover, they can fix the moduli in such a way to generate a exponential hierarchy of scales. This provides a solution of the hierarchy problem in String Theory. In this setup, one can try to construct string models that realize the features of the phenomenological five dimensional models extensively studied and refined during the last years\cite{Randall:1999ee,Randall:1999vf,Gherghetta:2006ha}.%Davoudiasl:1999tf,Chang:1999nh,Gherghetta:2000qt,Grossman:1999ra,Huber:2000ie,Agashe:2003zs,Agashe:2004rs,Huber:2000fh,Csaki:2002gy,Hewett:2002fe,Pomarol:2000hp,Randall:2001gb,Goldberger:2002cz,Agashe:2002bx,Contino:2002kc,}.

Another setup rich for model building is the set of $G_2$ holonomy vacua (for a complete review, see \cite{Acharya:2004qe}). These are compactifications of the eleven dimensional supergravity, that is believed to be the low energy limit of M-theory. In order to get a four dimensional description, one has to compactify on a seven dimensional manifold. The requirement of supersymmetry, in absence of fluxes, implies it to have holonomy group $G_2$ (it is the analog of CY in six dimensions). Compactifications on smooth $G_2$ manifolds give only abelian gauge fields and neutral fermions. To get a realistic spectrum, the internal manifold must be singular. In particular, non-Abelian gauge fields live on a three dimensional locus of orbifold singularities of the compact manifold \cite{Acharya:2000gb}, while chiral fermions are localized on pointlike conical singularities \cite{Acharya:2001gy}. The low energy theory is a seven dimensional super Yang-Mills theory with four dimensional chiral multiplets. The localization of fermions allow to have for example exponentially suppressed Yukawa couplings, and, as we will review in the latest chapter, suppressed proton decay rate\cite{Friedmann:2002ty,Acharya:2005ks}.

%\vskip 3cm
%\newpage
\section*{Summary of the Thesis}

This thesis focuses on the various aspects of String Phenomenology described in this introduction. It is structured as follows.

\

In the first part we will give a review of string compactification and of the resulting four dimensional effective theories.

In the {\it chapter 2}, we will start discussing how CY compactifications arise in String Theory from requiring four dimensional supersymmetry. Then we will introduce the concept of {\it moduli} of the CY solutions. We will see that they give rise to four dimensional neutral massless scalars, whose vev's the physical couplings depend on. They are incompatible with experiments and must be fixed to some value, getting a large mass. We will explain how this is realized by the introduction of flux background. 

In {\it chapter 3} we will concentrate on the effective four dimensional description of String/M-theory flux vacua, in the approximation in which the backreaction of fluxes on the geometry is neglected. We will review three ensemble of vacua and we will see how fluxes stabilize the moduli of the compactification manifold. Firstly the will study the Type IIB flux vacua. We will present both the ten dimensional and the four dimensional description \cite{Giddings:2001yu}, and we will finally focus on how hierarchy scales arise in this context. Then we will briefly present the four dimensional description of Type IIA flux vacua given in \cite{DeWolfe:2005uu}. Finally we will give a detailed review on the M-theory vacua. We will firstly describe compactification of M-theory on smooth and singular $G_2$ holonomy manifolds and finally we will introduce fluxes and explain how they stabilize the geometric moduli.

\

In the second and larger part of the thesis we will describe the results of our work. 

In {\it chapter 4} we will present a short review of the Statistical program outlined above. We will see what are the main motivations for a statistical study of the string Landscape and what are the basic techniques. We will introduce the result obtained in the ensemble of Type IIB flux vacua, since it is the first ensemble of string vacua where the statistical technique were applied. Then we will present the results obtained in our work \cite{Acharya:2005ez}. Fist we will give a brief review of the Freund-Rubin statistics, and then we will describe in more details the results obtained studying the $G_2$ holonomy ensemble. We will give the results of the statistical study for general $G_2$ holonomy vacua, and then we will concentrate on a particular class of models in which the computations can be done explicitly. We will so verify that fluxes stabilize all the geometrical moduli both in supersymmetric and in non-supersymmetric vacua. Finally we will give a comparison between our results and what one obtains in the Type IIB case.

In {\it chapter 5} we will pass to the ten dimensional approach to string vacua. We will illustrate what is the effect of fluxes on the geometry, implied by the requirement of four dimensional supersymmetry. We will introduce and use the formalism of $G$-structures. Then we will present the results of \cite{Acharya:2006ne}. We will give the ten dimensional description of the Type IIA CY flux vacua, studied in \cite{DeWolfe:2005uu} with a four dimensional approach. We will study the modification of the equations, given by the introduction of an orientifold plane and we will stabilize the moduli. We will see that in the so called "smeared" approximation, we are able to get the same results that \cite{DeWolfe:2005uu} get in the CY with fluxes approximation.

In {\it chapter 6} we will study one important aspect of flux compactification of Type IIB theory. In one class of solutions of Type IIB equations, the backreaction of the fluxes on the geometry leads to a non-factorisable ten dimensional metric. The four dimensional metric is multiplied by the warp factor, a function of extradimensional coordinates. This is reminiscent of what happens in five dimensional models inspired by the seminal Randall-Sundrum paper \cite{Randall:1999ee}. We will give a brief review of such models and then we will illustrate how they can be realized in Type IIB String Theory. We will describe the setup we have constructed in this context in \cite{Acharya:2006mx}. In particular we will see how fermion localization and Yukawa hierarchy can be realized through an instanton background on a D7-brane. At the end we will compare our results with those obtained in five dimensional models.

Finally in {\it chapter 7} we will focus on the realization of GUT theories in M-theory compactifications on $G_2$ manifolds. We will firstly give a short review of four dimensional GUT theories, concentrating on their more dramatic prediction: the decay of proton. Then we will study realizations of GUT in theories with gauge fields propagating in extradimensions, but fermions localized in the bulk. This is actually what happens in M-theory compactifications on singular $G_2$ manifolds. Then we will introduce the results of our work \cite{Acharya:2005ks}. We will see how the proton decay rate can be highly suppressed in some decay channels, due to a mechanism characteristic of these M-theory like realizations.

%\vskip 15mm

\

This thesis is based on the following papers:

\begin{description}
\item {\cite{Acharya:2005ez}}
  B. S. Acharya, F. Denef and R. Valandro,
  ``Statistics of M theory vacua,''
  JHEP {\bf 0506} (2005) 056
  [arXiv:hep-th/0502060].
 % \vskip 1cm
\item {\cite{Acharya:2005ks}}
  B. S. Acharya and R. Valandro,
  ``Supressing Proton Decay in Theories with Extra Dimensions,''
  JHEP {\bf 0608} (2006) 038
  [arXiv:hep-th/0512144].
 % \vskip 1cm
\item {\cite{Acharya:2006ne}}
  B. S. Acharya, F.Benini and R. Valandro,
  `` Fixing Moduli in Exact Type IIA Flux Vacua''
  JHEP {\bf 0702} (2007) 018
  [arXiv:hep-th/0607223].
 % \vskip 1cm 
\item {\cite{Acharya:2006mx}}
  B. S. Acharya, F.Benini and R. Valandro,
  `` Warped Models in String Theories''
  $[$arXiv:hep-th/0612192$]$.
 % \vskip 1cm
\end{description}

\part{String Compactifications and Effective 4D Actions}

\chapter[String Compactifications]{String Compactifications and Moduli Stabilization}\label{SC}

\section{String Compactifications}

At present, String/M-theory is formulated in six weakly coupled limits. There are five superstring theories in ten dimensional spacetime, called Type IIA, Type IIB, Heterotic $E_8\times E_8$, Heterotic $SO(32)$ and Type I, and an eleven dimensional limit, usually called M-theory. These all are unsatisfactory from a phenomenological point of view for (at least) one main reason:
%the gauge group is not the Standard Model one and 
the number of spacetime dimensions is greater than four.

The standard way of solving this problem is what is called {\it Compactification}: one assigns the extra-dimensions to an invisible sector, by choosing them to be small and compact and not detectable in present experiments. To preserve four dimensional Poincar\'e invariance, the ten(eleven) dimensional metric is assumed to be a (possibly warped) product of a four dimensional spacetime with a six(seven) dimensional space $X$:%. The most general metric preserving foru dimensional Poincar\'e invaritance takes the form:
\bee\label{SCds}
  ds^2= e^{-2A(y)} \eta_{\mu\nu} dx^\mu dx^\nu  + g_{mn} dy^m dy^n \:.
\ee
$\eta_{\mu\nu}$ is the usual four dimensional Minkowski metric, while $g_{mn}$ is the metric on the compact internal subspace. $e^{-2A(y)}$ is the so called ``warp factor", {\it i.e.} a $y$-dependent function in front of the four dimensional metric.
In what follows, we will consider the case in which it is equal to $1$.

\

%SUSY N=1 -> from Douglas introduction
Thus far we have only required $X$ to be compact and of sufficiently small size. Another important constraint on $X$ comes from requiring to have an $\NN=1$ supersymmetric effective theory at low energy. There are many reasons to focus on this kind of compactifications.

The best reason is that supersymmetry suggests natural extensions of the Standard Model (SM) such as the Minimal Supersymmetric Standard Model (MSSM) and non Minimal Supersymmetric Standard Model (nMSSM) with additional fields. These models can solve the hierarchy problem, can explain the gauge coupling unification, can contain a dark matter candidate and have many other attractive features. 
All this is only suggestive, because these models have other problems, such as reproducing precision electroweak measurements.  
However these reasons have been enough to concentrate on $\NN=1$ models in string compactifications for twenty years. 
Another reason is the calculational power that supersymmetry provides, since String Theory on supersymmetric backgrounds is under  a much better control than on non-supersymmetric ones.

The requirement of $\NN=1$ supersymmetry at compactification scale constraints the compactification manifold $X$. We consider the Heterotic case as an illustrative example. At low energy it is described by a ten dimensional $\NN=1$ supergravity with a Yang-Mills sector. 
We want a background that leaves some supersymmetry unbroken. The condition for this is that the variations of the Fermi fields are zero. In particular the variation of the gravitino is
\bea
 \delta \psi_\mu &=& \nabla_\mu \epsilon \\\label{SCdeltapsi}
 \delta \psi_m &=& \left(\partial_m+\frac{1}{4}\omega_{mnp}\Gamma^{np}-\frac{1}{8} H_{mnp} \Gamma^{np} \right)\epsilon %\\
% \delta \chi &=& \left( \Gamma^m\partial_m\Phi-\frac{1}{12}\Gamma^{mnp} H_{mnp} \right) \epsilon \\
% \delta \lambda &=& F_{mn} \Gamma^{mn} \epsilon
\eea
%These are the variations of the gravitino, dilatino and gaugino respectively. 

The spinor $\epsilon$ is the ten dimensional supersymmetry parameter; it is in the ${\bf 16}$ spinorial representation of $SO(1,9)$. Under the decomposition $SO(1,9)\rightarrow SO(1,3)\times SO(6)$, the  ${\bf 16}$ decomposes as ${\bf 16} \rightarrow ({\bf 2,4}) \oplus ({\bf \bar{2},\bar{4}})$. So one can write $\epsilon_{\alpha\beta}=\sum_k u^{(k)}_\alpha \eta^{(k)}_\beta$, where $u^{(k)}$ are arbitrary four dimensional spinors, while $\eta^{(k)}$ are the solutions of $\delta\psi_m=0$. The number of solutions $\eta^{(k)}$ gives the number of four dimensional supersymmetries.

The condition that these variations vanish for some spinor $\eta(y)$ can be solved to obtain conditions on the background fields.
What we want to stress here is that the variation \eqref{SCdeltapsi}, in the case of null $H$ field, implies that there exists a six dimensional spinor satisfying
\bee\label{SCnablazeta}
 \nabla_m \eta =0
\ee
{\it i.e.} $\eta$ is a covariantly constant spinor on the internal space. This condition implies that the holonomy group $SO(6)$ must be reduced to a subgroup, as the ${\bf 4}$ spinorial representation of $SO(6)$ must contain the singlet representation of the reduced holonomy group. Since under $SO(6) \rightarrow SU(3)$ we have the splitting ${\bf 4} \rightarrow {\bf 3} \oplus {\bf 1}$, in order to have $\NN=1$ in four dimension the holonomy group must be $SU(3)$. This implies the compactification manifold $X$ to be a {\it Calabi-Yau}(CY)\label{4DSusyReduced}.

The reduction of the holonomy group is the requirement that leaves some supersymmetry unbroken also for the compactifications of the other corners of string/M-theory. The first studied were Heterotic compactifications, because the Type II theories seemed to lead to $\NN=2$ supersymmetry in four dimensions, while M-theory compactifications on smooth seven dimensional manifolds cannot lead to non-Abelian gauge fields and chiral fermions. As we will see, these problems have been recently solved. The $E_8\times E_8$ Heterotic compactifications were the first studied as they provide a natural GUT setup, contrary to $SO(32)$ Heterotic and Type I theories. 

\

The lower dimensional theory is obtained by expanding all fields into modes of the internal manifold $X$. As an illustrative example, we discuss the Kaluza-Klein(KK) reduction \cite{Kaluza:1921tu,Klein:1926tv} of a ten dimensional scalar satisfying the ten dimensional equation of motion $\Delta_{10} \Phi=0$. Because of \eqref{SCds}, the Laplacian splits as $\Delta_{10}=\Delta_{3,1}+\Delta_{6}$. Since $X$ is compact, $\Delta_{6}$ has a discrete spectrum: $\Delta_6 f_n = m_n^2 f_n$. The ten dimensional scalar can be expanded as
\bee
 \Phi(x,y)=\sum_n \phi_n(x) f_n(y) \:.
\ee
Putting this into the equation of motion gives the four dimensional equations:
\bee
 \Delta_{3,1} \phi_n = m_n^2 \phi_n \:.
\ee
One ends up with an infinite tower of massive states, with masses quantized in terms of the eigenvalues of the Laplacian on $X$. The Laplacian $\Delta_6=g^{mn}\nabla_m \partial_n$ depends on the metric of $X$, so the low energy spectrum depends strongly on the geometry of the internal manifold. Roughly speaking the scale of the masses is given by 
$($Vol$X)^{-1/6}$. So the KK scale is strictly related to the compactification scale\footnote{However, if there are some dimensions that are much larger than others, there could be a hierarchy between the KK masses.A simple example is given by compactification on factorisable 6-torus with one radius, say $R_1$ much larger that the other two, say $R$; in this case there are modes that lead to four dimensional fields with a mass $n/R_1$ much smaller that the fields with mass $n/R$.}. Choosing the volume sufficiently small, the massive states become heavy and can be integrated out. So the effective four dimensional theory describes the dynamics of the fields $\phi_0^i$ related to the zero modes $f_0^i$ of the six dimensional Laplacian.

What we have described for a scalar field happens also to the other fields of the ten dimensional theory (for a review see \cite{Overduin:1998pn}). The surviving modes in the low energy effective theory are zero modes of some suitable six dimensional differential operator. Among these fields there is the metric, too. In particular, the massless fluctuations $\delta g_{mn}$ of the internal components $g_{mn}$ correspond to scalars in four dimensions. These massless scalars fields are called {\it geometric moduli} of the compactification.

\subsection*{CY Compactifications}\label{CYcomp}

We consider the case in which the ten dimensional spacetime is of the form $M^{3,1}\times X$. Due to this ansatz the Lorentz group of the ten dimensional space decomposes as $SO(9,1)\rightarrow SO(3,1)\times SO(6)$, where $SO(6)$ is the structure group of a six manifold. Demanding $X$ to preserve the minimal amount of supersymmetry gives the condition that the structure group of $X$ can be reduced to $SU(3)$. So $X$ admits a globally defined spinor $\eta$, since the $SO(6)$ spinor representation ${\bf 4}$ decomposes as ${\bf 1}\oplus {\bf 3}$. Further demanding $\eta$ to be covariantly constant tells that $X$ must have also holonomy group (with respect to the Levi-Civita connection) equal to $SU(3)$. These spaces are called Calabi-Yau manifolds and are complex K\"ahler manifolds, which are in addition Ricci flat (see for example \cite{Joyce:2001xt}. 

The existence of one covariantly constant spinor on a six dimensional manifold is equivalent to the existence of one covariantly constant 2-form $J$, the K\"ahler form, and one covariantly constant 3-form $\Omega$, the holomorphic 3-form. $\Omega$ defines a complex structure $I^i_j$ on the six manifold; $I^i_j$ and $J_{mn}$ defines a CY metric through $g_{mn}= - J_{mp} I^p_n$. In particular $SU(3)$ holonomy implies these forms to be harmonic.

\

The moduli parametrize continuous families of nearby vacua. Since a background consisting of a CY metric and zero field strengths for R-R and NS-NS fields is a solution of the equations of motion, the moduli parametrize the space of topologically equivalent CY manifolds. In other words, if $g_{mn}$ is a CY metric, one has to find deformations $\delta g_{mn}$ of this metric, such that the metric $g_{mn}+\delta g_{mn}$ is a CY metric too, with the same topology. By working out the linearized equations of motion, one finds that each modulus becomes a massless field.

A CY is a Ricci flat K\"ahler manifold. Therefore $g+\delta g$ must be Ricci flat too ($R_{mn}(g+\delta g)=0$). This implies that $\delta g$ satisfy the Lichnerowicz equation\cite{Candelas:1990pi}:
\bee
 \nabla^q \nabla_q \delta g_{mn} + 2 {R^q_m}^r_n \delta g_{qr} =0.
\ee
For K\"ahler manifolds the solutions to this equations are associated with either mixed ($\delta g_{m\bar{n}}$) or pure ($\delta g_{\bar{m}\bar{n}}$) deformations and are independent. These are in one-to-one correspondence with harmonic (1,1) and (2,1) forms respectively\footnote{
A ($p$,$q$)-form on a complex manifold is a ($p+q$)-form with $p$ holomorphic indices and $q$ antiholomprphic ones.}:
\bea
 \delta g_{m\bar{n}} &\leftrightarrow& \delta g_{m\bar{n}} dz^m\wedge dz^{\bar{n}} \, \in H^{1,1}(X)\\
 \delta g_{\bar{m}\bar{n}} &\leftrightarrow&  \Omega^{\bar{n}}_{k\ell}\delta g_{\bar{m}\bar{n}} dz^k\wedge dz^\ell \wedge dz^{\bar{m}} \, \in H^{2,1}(X)
\eea
and likewise for $\delta g_{mn}$ and (1,2) forms. As the structure of harmonic differential forms is isomorphic to that of tangent bundle cohomology classes, the number of geometric moduli in compactifications on $X$ is determined by the cohomology of $X$.

This is a general feature of string compactifications: the light particle spectrum is determined by topological considerations and the number of particles of given type is equivalent to the dimension of appropriate cohomologies.

Let us introduce a basis for different cohomology groups  by choosing the unique harmonic representative in each cohomology class. We denote the basis of harmonic 2-forms as $\{\omega_A\}$ and their dual harmonic 4-forms as $\{\tilde{\omega}^A\}$, which form a basis of $H^4(X)$. The harmonic 3-forms $(\alpha_{\hat{K}},\beta^{\hat{L}})$ give a real, symplectic basis of $H^3(Y)$. The non-trivial intersection numbers are given by:
\bea
 \int_X \omega_A\wedge\tilde{\omega}^B =\delta^B_A && \int_X \alpha_{\hat{K}} \wedge \beta^{\hat{L}} =\delta_{\hat{K}}^{\hat{L}}
\eea

The Hodge decomposition of the second and third cohomology group are given by 
\bea
 H^{2}&=&H^{2,0}\oplus H^{1,1}\oplus H^{0,2}\nn \\
 H^{3}&=&H^{3,0}\oplus H^{2,1}\oplus H^{1,2}\oplus H^{0,3}
\eea
For a CY, $h^{2,0}=h^{0,2}=0$ ($h^{p,q}\equiv \dim H^{p,q}$), so the basis $\{ \omega_A \}$ is also a basis of $H^{1,1}$. The same happens for $H^{4}=H^{2,2}$. As regard $H^3$, $h^{3,0}=h^{0,3}=1$ and $h^{2,1}=h^{1,2}$; $H^{2,1}$ has $h^{2,1}$ basis elements that we call $\chi_K$. The dimension of $H^{3}$ is $b_3=2(h^{2,1}+1)$, so the index $K$ runs from $1$ to $h^{2,1}$, while the hatted index $\hat{K}$ runs from $0$ to $h^{2,1}$.

The other non-trivial cohomology groups of a CY are $H^{0}=H^{0,0}$ with $h^{0,0}=1$ (constant function) and $H^{6}=H^{3,3}$ with $h^{3,3}=1$ (volume form).

\

The moduli associated with (1,1) harmonic forms are called {\it K\"ahler moduli}, while those associated with harmonic (2,1) forms are called {\it Complex Structure moduli}. This is because the former modify the K\"ahler form of the manifold whereas the latter alter the complex structure. This can be seen easily for the fist case, since under the transformation $g\mapsto g+\delta g$, the K\"ahler form transforms as:
\bee
 J=i g_{m\bar{n}} dz^m \wedge dz^{\bar{n}} \mapsto i (g_{m\bar{n}} +\delta g_{m\bar{n}})\, dz^m \wedge dz^{\bar{n}}  \:.
\ee
The deformations of the K\"ahler form can be expanded in the basis $\{\omega_A\}$
\bee
 J_{m\bar{n}} = v^a (\omega_A)_{m\bar{n}} 
\ee
In a KK compactification the $v^A$ are four dimensional scalars, whose expectation values give the K\"ahler form of the compact manifold. These real deformations are complexified by the $h^{1,1}$ real scalars $b^A$ arising in the expansion of the $B$-field present in all closed string theories. Its massless fluctuations are the harmonic 2-forms, so the KK expansion is given by:
\bee
 B=b^A \omega_A
\ee
The complex fields $t^A= b^A +i v^A$ parametrize the $h^{1,1}$-dimensional K\"ahler cone.
By the way, the moduli coming from antisymmetric form fields characteristic to string theory are called axions.

The second set of deformations are variations of the complex structure. To understand this, we first note that $g+\delta g$ is a K\"ahler metric. So its pure components can be put to zero by a change of coordinates. This cannot be a holomorphic change of coordinate, because this does not alter the pure components of the metric. Hence the complex structure under which the pure components are zero is different from the complex structure associated with the original metric $g$.
These deformations are parametrized by complex scalar fields $z^K$, where we expand the pure deformations on the forms $\chi_K$:
\bee
 \Omega^{\bar{n}}_{k\ell}\delta g_{\bar{m}\bar{n}} = z^K (\chi_K)_{k\ell \bar{m}}
\ee

Together, the complex scalars $z^K$ and $t^A$ span the geometric Moduli Space of the CY manifold. Its geometry has been nicely described in \cite{Candelas:1990pi}. Locally it is a product of two spaces $\MM=\MM_C \times \MM_K$; the first factor is associated with the complex structure deformations while the second with the complexified K\"ahler moduli. Both spaces are special K\"ahler manifolds of complex dimension $h^{2,1}$ and $h^{1,1}$ respectively. 

The metric on the space $\MM_C$ is given by:
\bee
 G_{K\bar{L}}= - \frac{\int_X \chi_K\wedge \bar{\chi}_{\bar{L}}}{\int_X \Omega \wedge \bar{\Omega}}
\ee
where $\chi_K$ is related to the variation of the 3-form $\Omega$ via Kodaira's formula:
\bee
 \chi_K(z,\bar{z}) = \partial_{z^K} \Omega(z) + \Omega (z) \partial_{z^K} K_C
\ee
From this expression, one can show that $G_{K\bar{L}}$ is a K\"ahler metric, since we can locally find complex coordinates $z^K$ and a function $K_C(z,\bar{z})$ such that:
\bea\label{CYKC}
 G_{K\bar{L}}= \partial_{z^K} \partial_{\bar{z}^L} K_C, && K_C= - \ln \left(i\int_X \Omega\wedge \bar{\Omega}\right) = 
  - \ln i\left( \bar{Z}^{\hat{K}} \mathcal{F}_{\hat{K}} - Z^{\hat{K}} \bar{\mathcal{F}}_{\hat{K}}\right)\nn
\eea
where the holomorphic periods are defined as:
\bea
 Z^{\hat{K}}(z) = \int_X\Omega(z)\wedge\beta^{\hat{K}}, && \FF_{\hat{K}}(z) = \int_X \Omega(z)\wedge\alpha_{\hat{K}} \:,
\eea
or equivalently:
\bee
 \Omega(z) = Z^{\hat{K}}(z) \alpha_{\hat{K}} - \FF_{\hat{K}}(z) \beta^{\hat{K}} \:.
\ee
The K\"ahler manifold $\MM_C$ is also special K\"ahler, since $\FF_{\hat{K}}$ is the first derivative with respect to $Z^{\hat{K}}$ of a prepotential $\FF = \frac{1}{2} Z^{\hat{K}} \FF_{\hat{K}}$. Hence the metric $G$ is fully determined in terms of the holomorphic function $\FF$.

$\Omega$ is only defined up to a rescaling by a holomorphic function $e^{-h(z)}$, which changes the K\"ahler potential by a K\"ahler transformation:
\bea
\Omega\mapsto e^{-h(z)}\Omega, && K_C\mapsto K_C +h +\bar{h}
\eea
This symmetry makes one of the period (conventionally $Z^0$) unphysical, as one can always choose to fix a K\"ahler gauge and set $Z^0=1$. The complex structure deformations can thus be identified with  the remaining $h^{2,1}$ periods, by defining the special coordinates $z^K=Z^K/Z^0$.

\

The metric on $\mathcal{M}_K$ is given by:
\bee
 G_{AB} = \frac{3}{2\KK} \int_X \omega_A \wedge \ast \omega_B = -\frac{3}{2} \left(\frac{\KK_{AB}}{\KK}-\frac{3}{2}\frac{\KK_A\KK_B}{\KK^2}\right) = \partial_{t^A}\partial_{\bar{t}^B} K_k
\ee
where $\ast$ is the six dimensional Hodge-$\ast$ on $X$ and $K_k$ is given by:
\bee\label{CYKk}
 K_k  = -\ln \left(\frac{i}{6}\KK_{ABC} (t-\bar{t})^A(t-\bar{t})^B (t-\bar{t})^C\right) = -\ln\frac{4}{3}\KK
\ee
where $\frac{1}{6}\KK$ is the volume of $X$, and the intersection numbers are:
\bea
 \KK_{ABC}=\int_X \omega_A\wedge \omega_B \wedge \omega_C, && \KK_{AB}=\int_X \omega_A\wedge \omega_B \wedge J = \KK_{ABC} v^C,\nn\\
 \KK_{A}=\int_X \omega_A\wedge J \wedge J = \KK_{ABC} v^B v^C, && \KK_{ABC}=\int_X J \wedge J \wedge J = \KK_{ABC} v^A v^B v^C\nn 
\eea
Also the manifold $\MM_K$ is special K\"ahler, since $K_k$ can be derived from a single holomorphic function $f(t) = -\frac{1}{6}\KK_{ABC} t^A t^B t^C$.

% \
% 
% As we have said before, the above are {\it geometric moduli}. In addition to geometric fields, string theory contains antisymmetric form fields, which also give rise to moduli. For example, heterotic compactifications contain a 2-form $B_2$, with field strength
% $H_3= dB_2 + \omega_3^G - \omega_3^L$ ($\omega_3$ are the Chern Simons 3-form). This is left invariant by a transformation
% \bee
%  B_2\mapsto B_2 + b^i \sigma_i,
% \ee
% where $\sigma_i$ are elements of $H^{2}(X,\mathbb{R})$. The fields $b^i(x)$ associated with internal 2-cycles do not affect the field strength and so the equation of motions and are in fact also moduli.

\

As we have said, all these moduli represent massless uncharged scalar particles. The existence of such massless scalars is inconsistent with experiments. Moduli couple gravitationally to ordinary matter and so can generate forces due to particle exchange. For a modulus of mass $m_\varphi$, the characteristic range of such force is $R\sim\mathcal{O}(1 /m_\varphi)$. As fifth force experiments have probed gravity to submillimetre distances, this requires that $m_\varphi > \mathcal{O}(10^{-3}) eV$ \cite{Hoyle:2004cw}.
Consequently the experiments require the existence of a potential giving mass to the moduli. 
In conclusion, given that massless moduli are a generic feature of string theory compactifications but are experimentally disallowed, we need techniques that will create a potential for these moduli, giving them mass. Fluxes are a powerfull example of this. In the next section we will describe their contribution.

\section{Flux Compactifications}\label{FluxCompactifications}

Each of the weakly coupled limits of string/M-theory has p-form gauge potentials  in its spectrum, that are sourced by the elementary branes. For example, all closed sting theories contain the NS 2-form potential $B$. Just as the 1-form Maxwell's potential can minimally couple to a point particle, the 2-form $B$ field minimally couples to the fundamental string world sheet. At least in a quadratic approximation, the spacetime action for $B$ is a direct generalization of the Maxwell's action:
\bee
 S= \int d^{10} x \sqrt{g} \left( R - H_{MNP} H^{MNP} \right)
\ee
where $H=dB$ is the field strength of $B$.
The resulting equations of motion are $\partial^M H_{MNP} = \delta_{NP}$, where $\delta$ is a source term localized on the worldsheets of the fundamental strings.

The analogy with Maxwell's theory goes further \cite{Douglas:2006es}. For example, some microscopic definition of Maxwell's theory contain magnetic monopoles, particles surrounded by a 2-sphere on which the total magnetic flux is non-vanishing. The monopole charge must satisfy the Dirac quantization condition ($e \, g = 2\pi \mathbb{Z}$). In the same way, closed string theories contain 5-branes (the so called NS5-branes), which are magnetically charged under $B$. A 5-brane, in a ten dimensional space, can be surrounded by a 3-sphere, on which the magnetic flux $\int H$ is non-vanishing. As in Maxwell's theory, this magnetic flux must be quantized in units of the inverse of the electric charge.

Beside the NSNS 2-form, the Type II theories contain $(p+1)$-form fields $C_{p+1}$ coming from the RR sector and sourced by the Dirichlet $p$-branes, with $p=0,2,4,6$ for Type IIA theory and $p=1,3,5$ for Type IIB theory. %Their field strength are $F_{p+2}$ and are not all independent, as they satisfy the ``self duality condition'': $\ast F_{p+2} = F_{10-p-2}+$non-linear terms.

The Type I theory has a RR $C_2$, but not a NSNS $B$-field, while M-theory has a 3-form $C_3$ coupled electrically to the M2-branes and magnetically to the M5-branes.

\

Now, suppose we compactify on a manifold $X$ with non-trivial homology group $H_{p+2}(X)$, and take a non-trivial $p+2$-cycle $\Sigma\in H_{p+2}(X)$. In this case, we can consider a configuration with a non-zero {\bf flux} of the field strength, defined by the condition:
\bee
 \int_\Sigma F_{p+2} = n \not =0
\ee 

To understand what is happening, we will follow \cite{Douglas:2006es} and review what happens taking six dimensional Maxell's theory and compactifying it on $X=\Ss^2$. In this case $H_2(X,\Zbb)\cong\Zbb$, and we can take as an element of it the sphere $\Ss^2$ itself. There is a field configuration that solves the equations of motion and that integrated over $\Ss^2$ gives a non-zero result: it is the ordinary magnetic monopole in $\Rbb_3$ restricted to $\Ss^2$:  
\bee
 F_{\theta\phi}=g \sin\theta d\theta d\phi
\ee
Note that we have defined a flux which threads a non-trivial cycle in the extradimensions, with {\it no charged source} on the $\Ss^2$. The monopole is just a pictorial device with which to construct it. 
The formal analogy with the monopole also allows to keep the Dirac's argument, to see that quantum mechanical consistency requires the flux $n$ to be integrally quantized.

The same construction applies to any $p$. Moreover, if we have a large cohomology group, we can turn on a flux for any basis element $\Sigma_i$:
\bee
 \int_{\Sigma_i} F_{p+2} = n_i 
\ee
where $i=1,...,b_{p+2}\equiv\dim H_{p+2}(X)$.

\

As in Maxwell's theory, turning on a field strength results in a potential energy proportional to the square of the flux. In compactifications we can turn on fluxes living in extradimension, without breaking four dimensional Lorentz invariance. 

The key point  is that since the fluxes are threading cycles on the compact geometry, the potential energy depends on the precise choice of the metric on $X$, generating a potential for the geometric moduli. If the potential is sufficiently generic, then minimizing it fixes all the moduli.

A generic $(p+2)$-field strength generates a potential of the form:
\bee
 V= \int_X F_{p+2} \wedge \ast F_{p+2}
\ee
the metric dependence is in the Hodge-$\ast$. If we write the CY metric in terms of $J$ and $\Omega$, substitute their expansions in terms of the moduli and do the integral, we obtain the explicit expression for $V(t,z)$ that we can minimize.

\

Let us take Freund-Rubin compactification \cite{Freund:1980xh} as an example of how fluxes generate a potential for the geometric moduli \cite{Douglas:2006es}.
We consider a six dimensional Einstein-Maxwell theory and compactify it on a 2-sphere $\Ss^2$. If one includes a magnetic field on $\Ss^2$, this flux can stabilize the radius of the sphere. 

The six dimensional action is:
\bee
 S=\int d^6x \sqrt{g_6} ( R_6 - |F_2|^2 )
\ee
This action is reduced to four dimension, by using a metric:
\bee
 ds^2 = \eta_{\mu\nu} dx^\mu dx^\nu + r^2 g_{mn}(y) dy^m dy^n
\ee
where $g_{mn}$ is the metric on a sphere of unit radius, and $r$ is the radius of $\Ss^2$.
On $\Ss^2$ there are $N$ units of $F_2$ flux:
\bee
 \int_{\Ss^2} F_2 = N
\ee
In the four dimensional description, $r(x)$ should be viewed as a field. After the reduction one has to go to the four dimensional Einstein frame (in which the four dimensional Einstein term is canonically normalized), by a Weyl rescaling. The resulting potential for the scalar $r(x)$ has two sources. One comes from the Einstein term: the positive curvature of $\Ss^2$ makes a negative contribution to the potential, which, after rescaling, is proportional to $1/r^4$. The other source is the magnetic flux through the $\Ss^2$, which gives a positive contribution proportional to $N^2/r^{10}$. Therefore, the potential takes the form:
\bee
 V(r) = \frac{N^2}{r^{10}} - \frac{1}{r^4}
\ee
By minimizing this function, one finds a minimum at $r\sim N^{1/3}$. So with a moderately large flux, one can get radii which are large in fundamental units, and curvatures which are small, making the found vacua reliable.

\subsection*{Calabi-Yau with Fluxes.}

The fact that fluxes allow the possibility of fixing (part of) the geometric moduli, made flux compactifications very attractive and much studied in the last years \cite{Grana:2005jc}. 
Fluxes cannot be turned on at will in compact spaces, as they give a positive contribution to the energy momentum tensor
\cite{Giddings:2001yu,Maldacena:2000mw}.
The first consequence is that one has to add negative tension sources (such as orientifold plains). The second one is that fluxes backreact on the geometry, and the CY manifold is no longer a solution of the equations of motion. However, in many cases it suffices to work in an approximation where the backreaction is ignored. One continues to treat the internal manifold as it were a CY, even after giving expectation values to the antisymmetric tensors along the internal directions. This situation is usually described as {\it Calabi-Yau with fluxes} even if it does not correspond to a true supergravity solution.

This approach is motivated partly by the fact that the physics community has grown particularly confidence of CY manifolds, on which one can use tools from algebraic geometry. This approximation is valid when the typical energy  scale of the fluxes is much lower than the KK scale: in this case we can assume that the spectrum is the same as that without fluxes, except that some of the massless modes acquire mass due to the fluxes. The energy scale of, for example, 3-form fluxes can be estimated using the quantization condition and is given by $\frac{N \alpha '}{R^3}$; the KK scale is $\frac{1}{R}$. $m_{flux} \ll m_{KK}$ when  the size of the compact manifold is much bigger than $\sqrt{N} \ell_s$ (where $\ell_s$ is the string length), which is in any case needed from the start in order to neglect $\alpha'$-corrections to the action.

\section{Four Dimensional Effective Theory}

After compactification, one gets a four dimensional effective theory \cite{Blumenhagen:2006ci}. It describes the physics that we can observe at low energy, below the compactification scale. If this scale is below the string scale, the only surviving string states are the massless ones.
While finding all light states of a given string vacuum can be rather straightforward, finding their interactions turns out to be really non-trivial. There are two ways to construct the effective interaction terms. The first is to start with the effective action of the underlying ten dimensional string theory and perform a dimensional reduction of all interaction terms. %This gives a four dimensional theory under certain approximations. First, the effective ten dimensional theory is only known up to a certain order in $\alpha'$. Moreover, this porcedure does not take into account truly stringy effects coming from string loops or from the worldsheet theory. Nevertheless this approache leads to important results. 
The second method uses the string S-matrix approach. This gives the relevant interaction term of the low energy theory at a given order in $\alpha'$ and $g_s$. It gives more quantitative results with respect to the previous method, but it requires the knowledge of the vertex operators and their interactions within the underlying conformal field theory.

\

In what follows, we will consider compactifications that give $\NN=1$ supergravity as the low energy four dimensional theory. Any $\NN=1$ supergravity action in four spacetime dimensions is encoded by three functions: the K\"ahler potential $K$, the superpotential $W$, and the gauge kinetic function $f$. The bosonic part is given by:
\bea
 \mathcal{L}_{eff}^{\NN=1} &=& \frac{1}{2\kappa_4^2} R - G_{\alpha\bar{\beta}}(\varphi,\bar{\varphi})D_\mu \varphi^\alpha D^\mu \varphi^{\bar{\beta}} - V(\varphi,\bar{\varphi})\nn\\
	&&-\frac{1}{8}\re f_{ab}(\varphi) F^a_{\mu\nu} F^{b\mu\nu} -\frac{1}{8}\im f_{ab}(\varphi)\epsilon^{\mu\nu\rho\sigma} F^a_{\mu\nu}F^b_{\rho\sigma}+...
\eea
The scalar fields $\varphi^\alpha$ are complex coordinates of the sigma-model target space with metric:
\bee
 G_{\alpha\bar{\beta}}=\frac{\partial^2K(\varphi,\bar{\varphi})}{\partial\varphi^\alpha\partial\bar{\varphi}^{\bar{\beta}}}
\ee
The gauge kinetic functions $f_{ab}$ has only off-diagonal elements for abelian factors in the gauge group, otherwise we can write $f_{ab}=f_a \delta_{ab}$. These functions are holomorphic in the $\varphi^\alpha$.

The general form of the scalar potential has two pieces which are called F-term and D-term:
\bee
 V(\varphi,\bar{\varphi}) = V_F + V_D
\ee
The two pieces are written in terms of $K$, $W$ and $f$. $W$ is a holomorphic function of the fields $\varphi^\alpha$.
The F-term potential is:
\bee\label{effpot}
 V = e^{\kappa_4^2 K} \left(G^{\alpha\bar{\beta}} D_\alpha W D_{\bar{\beta}}\overline{W} - 3\kappa_4^2 |W|^2\right)
\ee
The covariant derivative is given by
\bee
 D_\alpha W = \partial_\alpha W + \kappa_4^2  \partial_\alpha K  \, W \equiv F_\alpha
\ee
It indicates that $W$ is not a function but a section of a holomorphic line bundel over the sigma-model space. The $F_\alpha$ are the auxiliary complex scalars in the chiral multiplets, and a non-vanishing value indicates that the supersymmetry is spontaneously broken.

The D-term potential can be written in terms of the auxiliary fields $D^a$ in the vector multiplets as:
\bee
 V_D= \frac{1}{2}\left(\re f^{-1}\right)_{ab} D^a D^b = \frac{1}{8}(\re f_a)^{-1}(\partial_\alpha K T^a \varphi^\alpha + h.c.)^2
\ee
The last expression is valid when the scalars transform linearly and the gauge kinetic function takes a diagonal form. A non-vanishing value of $D^a$ means that supersymmetry is spontaneously broken.

The condition for unbroken supersymmetry are hence:
\bea
 F_\alpha =0 \:\,\, (\forall \alpha) & {\rm and} & D^a=0 \:\,\,(\forall a)
\eea
In a supersymmetric Minkowski vacuum, the vacuum energy has to vanish, implying also $W=0$. 

By the powerfull non-renormalization theorems, there are no perturbative corrections to the superpotential, and no perturbative correction beyond the one-loop to the gauge kinetic function. On the contrary, the K\"ahler potential can be corrected both by perturbative and non-perturbative contributions.

\

When these $\NN=1$ supergravities are effective theories of a higher dimensional string theory, the three functions $K$, $W$ and $f$ usually depend on the moduli field $\varphi^\alpha$ describing the background of the string model from which they are derived. It is useful to split the scalars $\varphi^\alpha$ into a set of neutral moduli fields $\MM$ and into a set of charged matter fields $\mathcal{C}$. While the set of fields in $\MM$ refers to the dilaton and the geometric moduli of the compactification manifold, the fields in $\mathcal{C}$ account for all kinds of charged chiral fields whose vev would change the gauge symmetry. These must vanish if the gauge symmetry is unbroken. We therefore may expand the superpotential and the K\"ahler potential with respect to small $\mathcal{C}$ fields. The coefficients of these expansions depend on the moduli in $\MM$ and give the physical couplings of the effective theory. If the moduli are stabilized at a given value, these couplings takes a specific value and do not vary continuously over the moduli space. 

\

In the next chapter, we will describe the four dimensional $\NN=1$ supergravities coming from compactification of the Type II theories (with some BPS objects included) and of the M-theory.

\chapter[Corners of the Landscape]{Corners of the Landscape}\label{E4DA}

In this chapter we will describe compactification of Type II theories and of M-theory with fluxes, working in the approximation in which the backreaction of the fluxes is neglected and the compact manifold is taken to be Ricci flat. In particular the compact manifold will be a CY for Type II compactifications and a $G_2$ holonomy manifold for M-theory compactification.

In each case we will find what are the effective potential for the geometric moduli that the fluxes generate and how this potential fixes part or all of the geometric moduli.

We will start presenting a part common to both Type II theories. Then we will concentrate on each one. Finally we will describe the very different case of M-theory.

\section{Type II: Common Facts}\label{E4DAComFacts}

At low energy and small string coupling the Type II theories are described by Type II supergravities. These theories have 32 supercharges. If we want to preserve the minimal amount of supersymmetries we must compactify them on a CY manifold. In this case we get an effective four dimensional theory with $\NN=2$ supersymmetry. 

In order to get a realistic spectrum, one requires at low energy $\NN \leq 1$. Hence we must introduce in Type II compactifications other sources of supersymmetry breaking. Fluxes can spontaneously breaks supersymmetry from $\NN=2$ to $\NN=1$. Another possibility is to introduce some BPS objects. String theory has objects of this kind, such as D-branes and Orientifold Planes. In what follows,
we will describe what these objects are and what are the constraints that they introduce. Then we will see what are the effective theories obtained compactifying Type II theories on CY with orientifold and fluxes.

\subsection{D-branes and Orientifold Planes}

In the middle of the 90's, the discovery of the D-brane opened a new perspective for String Theory\cite{Polchinski:1995mt,Johnson:2000ch}. On the one hand, D-branes were required to fill the conjectured web of string dualities \cite{Polchinski:1998rq}. Moreover, they led to the conjecture of various new connections between string theories and supersymmetric gauge theories, such as the famous AdS/CFT correspondence\cite{Maldacena:1997re,Witten:1998qj}. From a direct phenomenological point of view, they opened a whole new arena for model building \cite{Kiritsis:2003mc,Uranga:2003pz,Lust:2004ks,Ibanez:2004iv,Blumenhagen:2005mu,Marchesano:2007de}, since they are equipped with a  gauge theory.

More precisely, D-branes are extended objects defined as subspaces of the ten dimensional spacetime on which open strings can end \cite{Polchinski:1998rq,Johnson:2000ch}. Open strings with both ends on the same D-brane correspond to a $U(1)$ gauge field in the low energy effective actions. This gauge group gets enhanced to a $U(N)$ when putting a stack of $N$ D-branes on top of each other. At low energy this induces a Yang-Mills theory living on the D-brane worldvolume. This fact allows to construct phenomenologically attractive models from spacetime filling D-branes consistently included in a compactification of Type II String Theory. The basic idea is that the Standard Model, or rather its supersymmetric extensions, is realized on a stack of spacetime filling D-branes. The matter fields arise from dynamical excitations of the brane around its background configuration. 

The D-branes are also charged under the RR form potentials, so they contribute a source term in the Bianchi identities of these fields\cite{Polchinski:1998rq,Johnson:2000ch}. This is similarly true for non-trivial background fluxes. One can apply the Gauss law for the compact internal space such that consistency requires internal sources to cancel. In this respect, D-branes are the higher dimensional analog of charged particles. Putting such a particle in a compact space, the field lines have to end somewhere and we have to require for a source with opposite charge. In String Theory these negative sources are anti-D-branes and orientifold planes\cite{Johnson:2000ch}. To preserve supersymmetry, the second ones are usually chosen for model constructions.

Orientifold planes arise in String Theory constructed from Type II strings by modding out worldsheet parity plus a geometric symmetry $\sigma$ of $M^{3,1}\times X$ \cite{Dabholkar:1997zd,Acharya:2002ag}. In the effective supergravity description, the orientifolds break part or all of the supersymmetry of the low energy theory. By imposing suitable conditions on the orientifold projection and on the included D-branes, the setup can be adjusted to preserve exactly half of the original supersymmetry.

Summarizing, starting from Type II in ten dimensions, one compactifies on a CY to obtain $\NN=2$ theories in four dimensions. This $\NN=2$ can be further broken to $\NN=1$ if one adds to the background an orientifold plane (and possibly D-branes).

Also fluxes can break form $\NN=2$ to $\NN=1$. One can add to a CY background both orientifolds and fluxes, and if they break the same supercharges, the resulting background leads to an $\NN=1$ effective four dimensional theory

\

We now describe more precisely the D-branes and the orientifold planes, since they have been used in some works reviewed in this thesis.

\subsubsection{D-branes}

String Theory gives a low energy effective action for the gauge theory living on the D-brane worldvolume, as well as the couplings to the light closed string modes. More precisely, the gauge theory and the coupling to the NSNS sector is captured by the {\it Dirac-Born-Infeld(DBI) action} \cite{Polchinski:1998rq,Johnson:2000ch}. In the case of a single Dp-brane, it is given (in string frame\footnote{See appendix \ref{StrEinFrame}}) by:
\bee
  S_{DBI}= - T_p \int_\Sigma d^{p+1} \xi e^{-\phi} \sqrt{-\det (\varphi^\ast (g+B) + 2 \pi \alpha' F)},
\ee
$T_p$ is the brane tension. The integral is done over the $p+1$ dimensional worldvolume $\Sigma$ of the Dp-brane, which is embedded in the ten dimensional spacetime via the map $\varphi$. This DBI action contains a $U(1)$ field strength $F=dA$, which describes the $U(1)$ gauge theory to all order in $\alpha' F$. To leading order, the action reduces to the standard $U(1)$ gauge theory action. The dynamics of the Dp-brane is encoded in the embedding map $\varphi$. Fluctuations around a given $\varphi$ are parametrized by charged scalar fields, which provide the matter content of the low energy effective theory.

Dp-brane is charged under RR fields, so they couple as extended objects to the appropriate RR form \cite{Polchinski:1998rq,Johnson:2000ch}. More precisely, a Dp-brane couples naturally to the RR form $C_{p+1}$. Moreover, generically D-branes contain lower dimensional D-brane charges, and hence interact also with lower degree RR forms. All these couplings are described by the {\it Chern-Simons(CS) action}:
\bee
 S_{CS}=\mu_p \int_{\Sigma} \varphi^\ast \left(\sum_q C_q\wedge e^{-B}\right)\wedge e^{2\pi \alpha' F} 
\ee
$\mu_p$ is the Dp-brane charge. The lowest order terms in $S_{CS}$ in the RR fields are topological and represent the RR tadpole contributions to the low energy effective action. $S_{CS}$ encodes also the coupling of the gauge matter fields arising from perturbations of $\varphi$ to the RR fields.

In flat ten dimensional spacetime a static Dp-brane preserves half of the supersymmetry. In curved background the requirement of the Dp-brane to be a BPS object gives strong constraints on the possible embedding of the brane in the spacetime. A Dp-brane in the space $M^{3,1}\times X$ can fill Minkowski directions as well as the compact ones. The compact directions of the brane worldvolume must wrap a non-contractible cycle of the compact manifold $X$. 

The BPS condition demands that the brane tension $T_p$ and charge $\mu_p$ are equal. This ensures stability since the net force between BPS branes vanishes \cite{Johnson:2000ch}. Moreover, there are conditions on the cycles in the compact manifold $X$ wrapped by the branes. In a purely metric background with $X$ being a CY, the only allowed cycles are the so called calibrated cycles with respect to the invariant forms defining the CY ($J$, $\re\Omega$ and $\im \Omega$). These forms are actually calibrations. More precisely \cite{Harvey:1982xk}, one says that a closed p-form $\omega$ is a calibration if it is less or equal to the volume form on each oriented p-dimensional submanifold $\Sigma\in X$. If the equality holds for all points of one submanifold $\Sigma$, then $\Sigma$ is called a calibrated submanifold with respect to the calibration $\omega$. A calibrated submanifold has minimal volume in its homology class. The calibrated submanifolds are also called {\it supersymmetric cycles}, as the bound in volume becomes equivalent to the BPS bound.

\subsubsection{Orientifold Planes}

Similar to D-branes, orientifold planes are hyper-planes of the ten dimensional background. They arise when the string theory is divided out by a symmetry transformation that is a combination of $\Omega_p$, the worldsheet parity, and a transformation $\Ss$ that makes $\Ss \Omega_p$ a symmetry of String Theory \cite{Polchinski:1998rq,Johnson:2000ch}. The orientifold planes are the hypersurfaces left invariant by $\Ss$. They are charged under the RR potential and can have negative tension. This allows to construct consistent configurations with branes and orientifold planes. In particular, in $M^{3,1}\times X$ orientifold planes wrap cycles in $X$ arising as fix-point set of $\Ss$. If these are calibrated with respect to the same form as the cycles wrapped by D-branes, the brane-orientifold setup can preserve some supersymmetry. 

Let us be more precise on what is $\Ss$. In the simplest example, $\Ss$ only consists of a target-space symmetry $\sigma:M_{10}\rightarrow M_{10}$, such that $\Omega_p \sigma$ is a symmetry of the underlying string theory. This is the case for Type IIB orientifolds with $O5$ or $O9$ planes. However, Type IIB admits a second perturbative symmetry operation denoted by $(-1)^{F_L}$, where $F_L$ is the spacetime fermion number in the string left-moving sector. Under the action of $(-1)^{F_L}$ RNS and RR states are odd, while NSR and NSNS states are even. Orientifolds with $O3$ and $O7$ planes arise from projectors of the form $(-1)^{F_L}\Omega_p \sigma$. The transformation behavior of the massless bosonic states of Type II theories under $(-1)^{F_L}$ and $\Omega_p$ are:
\bee
 \begin{array}{c|cccccccc}\label{TIIorientproj}
 &\phi & g & B & C_0 & C_1 & C_2 & C_3 & C_4\\ \hline\\
 (-1)^{F_L} & + & + & + & - & - & - & - & - \\
 \Omega_p & + & + & - & - & + & + & - & -
 \end{array}
\ee
With these transformations, one can check that both $(-1)^{F_L}$ and $\Omega_p$ are symmetries of the ten dimensional Type IIB supergravity action. This is not the case for Type IIA. However, orientifolds with $O6$ planes arise if $\Ss$ includes $(-1)^{F_L}$ as well as some appropriately chosen target space symmetry that ensures that $\Ss\Omega_p$ leaves the effective action invariant.

\newpage

\section{Type IIB Vacua}\label{TypeIIBvacua}

We start the Type IIB section, by describing the ten dimensional picture of flux compactifications in the supergravity limits. We will follow the treatment of Giddings, Kachru and Polchiski (GKP) \cite{Giddings:2001yu} (see also \cite{Douglas:2006es} for a review).

The bosonic supergravity effective action of Type IIB string theory is given in Einstein frame (see appendix \ref{StrEinFrame}) by:
\bea
 S_{IIB}&=&\frac{1}{2\kappa_{10}^2} \int d^{10}x \sqrt{g} \left(R - \frac{\partial_M\tau \partial^M \bar{\tau}}{2 (\im\tau)^2}-\frac{G_3 \cdot \bar{G}_3}{12 \im\tau} - \frac{|\tilde{F}_5|^2}{4\cdot 5!}\right)\nn\\
 && + \frac{1}{8 i \kappa_{10}^2}\int \frac{C_4\wedge G_3 \wedge \bar{G}_3}{\im\tau} + S_{loc}
\eea
with $\kappa_{10}^2=(2\pi)^7\alpha'^4$. The fields involved are: the metric, an NSNS field strength $H$ (with potential $B$) and RR field strengths $F_1$, $F_3$ and $F_5$ (with potentials $C_0$, $C_2$ and  $C_4$). $G_3$ and $\tau$ (the axion-dilaton) are the complex combinations
\bea
 G_3= F_3 - \tau H_3 &&  \tau = C_0 + i e^{-\phi}
\eea
where $\phi$ is the dilaton. 

The 5-form $\tilde{F}_5$ is defined as 
\bee
 \tilde{F}_5 = F_5 -\frac{1}{2} C_2 \wedge H_3 + \frac{1}{2} B\wedge F_3
\ee
and one has to impose the selfduality condition $\tilde{F}_5 = \ast \tilde{F}_5$ by hand, when solving the equations of motion.

$S_{loc}$ includes the possibility that we add the action of any localized sources (such as D-branes and orientifold planes) in our background.

We start by looking for solutions with four dimensional Poincar\'e invariance, and so we choose the usual ansatz for the ten dimensional metric:
\bee
 ds^2= e^{2A(y)} \eta_{\mu\nu}dx^\mu dx^\nu + e^{-2A(y)} \tilde{g}_{mn}(y) dy^m dy^n
\ee
with $\mu,\nu=0,...,3$ and $m,n=4,...,9$. We have allowed the possibility of a warp factor. The four dimensional Poincar\'e invariance imposes constraints also on the other fields:
\bea
 \tau=\tau(y) && \tilde{F}_5 =(1+\ast) (d\alpha (y)\wedge dx^0 \wedge ... \wedge dx^3)
\eea
where $\alpha$ is a function on the compact manifolds. Moreover we can allow only compact components of the $G_3$ flux.

The equation of motion of $G_3$ forces $F_3$ and $H_3$ to be harmonic forms, which are thus determined in terms of their periods on a basis of 3-cycles:
\bea\label{TIIBH3F3}
 \int_{\Sigma_\alpha} F_3 = N_{RR}^\alpha, && \int_{\Sigma_\beta} H_3 = N_{NSNS}^\beta
\eea
Then one should impose the Dirac quantization condition on these 3-form fluxes, that makes the period to take quantized values (in suitable units), {\it i.e} $F_3,H_3\in H^3(X,\Zbb)$.

By taking the trace-reversed Einstein equations for the $M^{3,1}$ components of the metric, one gets the equation:
\bea\label{TIIBnogo}
 \tilde{\nabla}^2 e^{4A} \: = \:e^{2A}\frac{G_{mnp} \bar{G}^{mnp}}{12 \im \tau} &+& e^{-6A}\left(
				\partial_m\alpha \partial^m\alpha + \partial_m e^{4A} \partial^m e^{4A}\right)\nn\\
				& +& \kappa_{10}^2 e^{2A} (T_m^m - T^\mu_\mu)_{loc}
\eea
where the tilde objects are computed by using the $\tilde{g}$ metric. $T_{loc}$ is the stress-energy tensor of any localized source.

We note from this equation that the first two terms on the right hand side are positive definite. But on a compact manifold, the left hand side integrates to zero, being a total derivative. Therefore in compact models and in absence of localized sources, there is a {\it no-go theorem}: the only solutions have $G_3=0$ and $e^A=$constant. Therefore, Type IIB supergravity does not allow non-trivial warped compactifications \cite{Maldacena:2000mw}.
But String Theory allows localized sources such as D-branes and orientifold planes. In order to evade the global obstruction to solving \eqref{TIIBnogo}, given by the positive contribution of the first two terms, one needs:
\bee
 (T_m^m-T_\mu^\mu)_{loc} < 0
\ee

Another constraint comes from the Bianchi Identity for $F_5$:
\bee\label{TIIBBI}
 d \tilde{F}_5 = H_3\wedge F_3 + 2\kappa_{10}^2 T_3 \rho_3^{loc}
\ee
$T_3$ is the D3-brane tension, and $\rho_3^{loc}$ is the local D3-brane charge density on the compact space. Integrating this relation on the six dimensional compact manifold, one gets:
\bee\label{TIIBTadCanc}
 \frac{1}{2\kappa_{10}^2 T_3}\int_X H_3\wedge F_3 + Q_3^{loc} = 0
\ee
where $Q_3^{loc}$ is the total D3-brane charge arising from localized objects.

Writing \eqref{TIIBBI} in terms of $\alpha(y)$, $A(y)$ and $G_3$, and subtracting it from the equation \eqref{TIIBnogo},
one gets:
\bea\label{TIIBBIEOM}
 \tilde{\nabla}^2 (e^{2A} - \alpha) &=& \frac{e^{2A}}{24\im \tau}|iG_3 - \ast_6 G_3|^2 + e^{-6A}|\partial(e^{4A}-\alpha)|^2 \nn\\
	&& + 2\kappa_{10}^2 e^{2A} \left(\frac{1}{4}(T_m^m - T_\mu^\mu)_{loc} - T_3 \rho_E^{loc}\right)
\eea
We can restrict our attention to sources that satisfy the relation 
\bee\label{TIIBBPS}
(T_m^m - T_\mu^\mu)_{loc} \geq 4 T_3 \rho_3^{loc}
\ee
This inequality is saturated by D3-branes and O3-planes, as well as D7-branes wrapping supersymmetric cycles. It is satisfied by anti-D3-brane and it is violated by O5-planes and anti-O3-planes.

When we assume the relation \eqref{TIIBBPS}, from \eqref{TIIBBIEOM} it follows that $G_3$ must be imaginary selfdual ($\ast_6 G_3 = i G_3$) and that the warp factor is given by $e^{4A}=\alpha$. In this case, the relation \eqref{TIIBBPS} is saturated. Therefore solutions to the tree-level equations of motion should include only D3, O3 and D7 sources. 

Imposing the remaining equations of motion (namely the extradimensional Einstein and the dilaton-axion equations), one can find that this class of solutions describes the F-theory models \cite{Vafa:1996xn} in the supergravity approximation, including the possibility of background fluxes.

The simplest example of these solutions are perturbative Type IIB orientifolds. In this special case the metric $\tilde{g}$ is a CY metric and so the internal manifold is  conformally CY. In this particular class of solutions, neglecting the backreaction of fluxes on the geometry means neglecting the warp factor. This can be done if the warp factor is slowly varying through the compact manifold. This is the approximation assumed in the following derivation of the four dimensional effective theory of these vacua.

\subsubsection{Type IIB orientifolds: Four Dimensional Description}

In this section we will give a four dimensional description of the Type IIB orientifolds vacua. In particular we will give a formula for the effective potential depending on fluxes and geometric moduli.

\

If we compactify the Type IIB theory on a CY, this leads to $\NN=2$ supersymmetry in four dimensions. As we have seen at page \pageref{CYcomp} the geometric moduli of a CY are divided into $h^{1,1}$ K\"ahler moduli and $h^{2,1}$ Complex Structure moduli. The first are associated with the fluctuation of the 2-form $J$, while the second with the 3-form $\Omega$. These moduli fields represent scalar components of $\NN=2$ hyper and vector multiplets respectively. Together with the axion-dilaton hypermultiplet, they give $h^{2,1}+1$ hypermultiplets and $h^{1,1}$ vector multiplets.

The moduli are the coefficients of $J$ and $\Omega$ when expanded onto a basis of harmonic 2-forms $\{\omega_A\}$ and 3-forms $\{\alpha_{\hat{K}},\beta^{\hat{L}}\}$ respectively. 
%In particular, the expansion of $\Omega$ is given by:
% \bee
% \Omega(z) = Z^{\hat{K}}(z) \alpha_{\hat{K}} - \FF_{\hat{K}}(z) \beta^{\hat{K}}
% \ee
% One can collects the periods of $\Omega$ in the vector $\Pi(z)=(Z^{\hat{K}}(z),\FF_{\hat{L}}(z))$.

To arrive at $\NN=1$ supersymmetry in four dimensions, one introduces an orientifold projection $\OO$ \cite{Grimm:2004uq}. As described previously, the orientifold projection includes a reflection $\sigma$ in the internal Calabi-Yau $X$. Consistency requires $\sigma$ to act as an isometric and holomorphic involution on $X$. The transformation $\sigma$ leaves the K\"ahler form invariant, but may act non-trivially on the holomorphic 3-form $\Omega$. Due to its holomorphic action, $\sigma$ splits the cohomology groups $H^{p,q}(X)$ into a direct sum of an even eigenspace $H^{p,q}_+(X)$ and an odd eigenspace $H^{p,q}_-(X)$. Hence, this splits the $h^{1,1}$ harmonic $(1,1)$-forms of $X$ into a set of $h^{1,1}_+$ even forms and into a set of $h^{1,1}_-$ odd forms. Since the K\"ahler form is invariant under $\sigma$, it is expanded with respect to a basis of $H^{p,q}_+(X)$:
\bee
 J = \sum_{a=1}^{h^{1,1}_+} t^a \omega_a
\ee
The harmonic (2,1) forms are divided in an analogous way into even and odd. Here we consider Type IIB compactification with O3/O7 orientifolds planes (and D3/D7-branes). In this case $\OO=(-1)^{F_L}\Omega_p \sigma$. The holomorphic 3-form is odd under $\sigma$ and so it is expanded on a basis $\{\alpha_\lambda,\beta^\lambda\}$ of $H^3_-(X)$:
\bee
\Omega(z) = \sum_{\lambda=0}^{h^{2,1}_-} \left(Z^{\lambda}(z) \alpha_{\lambda} - \FF_{\lambda}(z) \beta^{\lambda}\right)
\ee
One can collects the periods of $\Omega$ in the vector $\Pi(z)=(Z^{\lambda}(z),\FF_{\lambda}(z))$.

\

As we have said before, the equations of motion forces the field strengths $H_3$ and $F_3$ to be harmonic. From the table \ref{TIIorientproj} one can see that the 2-forms $B$ and $C_2$, and consequently their field strengths, are odd under $(-1)^{F_L}\Omega_p$. Thus they are expanded on the basis $\{\alpha_{\lambda},\beta^{\lambda}\}$, that we rename for simplicity as $\{\hat{\Sigma}_\alpha\}$:
%, which can be taken as the dual of the basis of 3-cycles $\{\Sigma_\alpha\}$ used in \eqref{TIIBH3F3}, and that we will call $\{\hat{\Sigma}_\alpha\}$:
\bea
 H_3 = N_{NSNS}^\alpha \hat{\Sigma}_\alpha & F_3 = N_{RR}^\alpha \hat{\Sigma}_\alpha & \mbox{with } \alpha=1,...,2h_-^{2,1}+2
\eea
These fluxes generate a superpotential for the complex structure moduli as well as for the axion-dilaton \cite{Gukov:1999ya}:
\bee
 W = \int_X G_3 \wedge \Omega = N_{RR}\cdot \Pi(z) - \tau \, N_{NSNS}\cdot \Pi(z)
\ee

\

In order to write the K\"ahler potential for the scalars, one needs to identify the good K\"ahler coordinates, {\it i.e.} the complex coordinates such that the effective four dimensional action takes the canonical form and the potential is written as \eqref{TIIBeffpot}. For Type IIB, the surviving complex structure moduli and the axio-dilaton are good coordinates, while the surviving K\"ahler moduli are not. This implies that the form of $K_C$ remains the same as \eqref{CYKC}, with the only difference that the holomorphic 3-form is expanded on a smaller number of basis elements, {\it i.e} those that survive to the orientifold projection. 

The K\"ahler potential for the dilaton is
\bee\label{TIIBKdil}
 K_\tau = -\ln(-i(\tau-\bar{\tau}))
\ee

On the contrary the form of $K_k$ is sensitively modified and takes two different forms, corresponding to which orientifold projection is performed. For O3/O7 projections the good K\"ahler coordinates for $\MM_k$ are \cite{Grimm:2004uq}:
\bea
 G^\alpha &=& c^\alpha - \tau b^\alpha \nn\\
 T_a &=& \frac12 \int_X \omega_a \wedge J\wedge J+i \rho_a -\frac{i}{2(\tau-\bar{\tau})} 	
 	\int_X\omega_a\wedge\omega_\beta\wedge\omega_\gamma G^\beta(G-\bar{G})^c\nn
\eea
where $c^\alpha$ and $b^\alpha$ are the coefficients of $C_2$ and $B$ expanded on the basis $\{\omega_\alpha\}$ of $H^{1,1}_-$, and $\rho_a$ are the coefficients of $C_4$ expanded on the basis of $H^{2,2}_+$ dual to $\{\omega_a\}$.

In terms of these new coordinates the K\"ahler potential is:
\bee\label{TIIBKk}
 K_k = -2 \ln  \left(\frac16 \int_X J\wedge J\wedge J\right)
\ee
where $J$ is written in terms of the new coordinates.

\

Now we are able to write the effective potential for the geometric moduli and the axion-dilaton. Its expression in four dimensional $\NN=1$ supergravity takes the form \eqref{effpot}(we put $\kappa_4=1$):
\bee\label{TIIBeffpot}
 V = e^K \left(g^{i\bar{j}} D_iW \overline{D_jW} - 3|W|^2\right)
\ee
Here $K$ is the sum of \eqref{CYKC}, \eqref{TIIBKdil} and \eqref{TIIBKk}.
$D_iW$ is the K\"ahler covariant derivative $D_iW=\partial_iW+K_iW$, where $K_i=\partial_i K$. $g_{i\bar{j}}$ is the second derivative of $K$, {\it i.e.} $g_{i\bar{j}}=\partial_i\partial_{\bar{j}} K$.

The supersymmetric vacua of this potential are given by the solution $D_iW=0$ $\forall i$.

We note that $W$ does not depend on the K\"ahler moduli in Type IIB. Because of this, the piece $-3|W|^2$ in \eqref{TIIBeffpot} precisely cancels the term in $g^{i\bar{j}} D_iW \overline{D_jW}$ where $i,j$ run over the K\"ahler moduli. Therefore one can express the full tree-level flux potential as:
\bee
 V = e^K\left(g^{a\bar{b}} D_aW \overline{D_bW}\right)
\ee
where $a,b$ run over complex structure moduli and dilaton.
This potential is positive definite, with minima at $V=0$. Furthermore we see that generic vacua are not supersymmetric, as there are no constraints on $D_iW$, with $i$ running over K\"ahler deformations. This is precisely a realization of the cancellation that occurs in a general class of supergravities known as no-scale supergravities \cite{Cremmer:1983bf,Ellis:1983sf}. Unfortunately, the vanishing of the cosmological constant for non-supersymmetric vacua depends on the tree-level structure of the K\"ahler potential, which is not radiatively stable.

Let us consider the equations $D_\tau W=0$ and $D_L W=0$ ($L=1,...,h^{2,1}$). More explicitly they are given by:
\bea\label{TIIBfluxeq}
 (N_{RR}-\bar{\tau}N_{NSNS})\cdot \Pi(z) = 0 && (N_{RR}-\tau N_{NSNS})\cdot (\partial_L \Pi+\Pi \partial_L K_C)=0
\eea
These equations have a simple geometric interpretation. For a given choice of the internal fluxes for $G_3$, they require the metric to adjust itself (by motion in $\MM_C$) so that the (3,0) and (1,2) parts of $G_3$ vanish. It gives a solution where $G_3$ is imaginary self-dual. If one imposes also the remaining supersymmetry conditions $D_iW=0$, then the flux $G_3$ is forced to have only the (2,1) piece.

The system \eqref{TIIBfluxeq} is made up of $h^{2,1}+1$ equations in $h^{2,1}+1$ variables for each choice of integral fluxes. Thus it seems clear that generic fluxes will fix all of the complex structure moduli as well as the axio-dilaton. Furthermore one could suspect that the number of vacua diverges, as we have not given any constraint on the fluxes. But such a constraint exists. It comes from requiring tadpole cancellation for $\tilde{F}_5$. In fact we have seen that the 3-form fluxes induce a contribution to the total D3-brane charge:
\bee
 N_{flux}=\frac{1}{(2\pi)^4 \alpha'^2}\int_X F_3\wedge H_3
\ee
One can check that, for imaginary selfdual flux, $N_{flux}$ is positive definite \cite{Ashok:2003gk}. Moreover in a given orientifold of $X$, the tadpole cancellation condition \eqref{TIIBTadCanc} takes the form:
\bee
 N_{flux} + N_{D3} = L
\ee
where $(-L)$ is some total negative D3 charge which needs to be cancelled. It arises by induced D3-charge on D7 and O7 planes, and explicit O3 planes. For an orientifold limit of an F-theory compactification on elliptic CY fourfold $Y$ \cite{Vafa:1996xn}, one finds
\bee
 L = \frac{\chi(Y)}{24}
\ee
where $\chi(Y)$ is the Euler number of $Y$.

The allowed flux choices in an orientifold projection compactification on $X$, and hence the numbers of flux vacua, are stringently constrained by the requirement $N_{flux}\leq L$.\label{TIIBfluxconstr}

\

We finally note that in principle also open string fluxes can be turned on, when D7-branes are involved. This happens in general F-theory models, where one can turn on background field strength of the D7 gauge fields, generating additional contribution to the tadpole cancellation condition and the spacetime potential energy.
In this chapter we will concentrate on vacua where all these open string fluxes are null. 

\subsubsection{Warped Solutions and Stabilized Hierarchy}\label{WarpedSolutionsInTypeIIB}

We conclude this section with the famous example of the compact {\it conifold}. It has been presented in GKP work \cite{Giddings:2001yu}, following earlier work of \cite{Strominger:1986uh,Becker:1996gj,Verlinde:1999fy,Greene:2000gh}.
The starting point is the Klebanov-Strassler results \cite{Klebanov:2000hb}: locally in the vicinity of a conifold point, KS have found solutions with fluxes that generate smooth supergravity solutions with large relative warpings. GKP extended this work to the compact case.

CY spaces can develop singularities at special point of their moduli space. One famous example is the conifold. This can be described as the submanifold of $\Cbb^4$ defined by 
\bee\label{TIIBcon}
w_1^2+w_2^2+w_3^2+w_4^2=0
\ee
This manifold is singular at $w_i=0$ $\forall i$. This is a good singularity, {\it i.e.} String Theory makes sense in such a space. This singular space is a cone whose base has the topology $\Ss^3\times\Ss^2$. At the singular point both spheres shrink to zero size. The singularity can be resolved by deforming the equation \eqref{TIIBcon} into:
\bee
 w_1^2+w_2^2+w_3^2+w_4^2=\epsilon
\ee
This is equivalent to expand the $\Ss^3$ to finite size. $\epsilon$ is the parameter that controlls the size of $\Ss^3$. There are therefore two non-trivial 3-cycles: the A-cycle $\Ss^3$ just discussed and a dual B-cycle extending along the $\Ss^2$ times the radial direction of the cone.

This singularity arises locally in many compact CY spaces. In such manifolds, the B-cycle is also compact. The periods of $\Omega$ on these cycles are:
\bea
 \int_A \Omega = z && \int_B \Omega = \frac{z}{2\pi i} \ln z + \mbox{regular} = \FF(z)
\eea
Here $z\rightarrow 0$ is the singular point in the moduli space where the A-cycle $\Ss^3$ collapses.

We now add fluxes to this geometry:
\bea
 \frac{1}{(2\pi)^2\alpha'}\int_A F_3 = M && \frac{1}{(2\pi)^2\alpha'}\int_B H_3 = -K
\eea
These generate the superpotential:
\bee
 W = - K \tau z+ M \FF(z)
\ee
The K\"ahler potential is the one studied above. The equation $D_zW=0$ simplifies when $K/g_s$ is large:
\bee
 D_zW= \frac{M}{2\pi i}\ln z -i \frac{K}{g_s}+...=0
\ee
The solution is $z \sim e^{\frac{-2\pi K}{g_sM}}$. This means that there are flux vacua exponentially close to the conifold point in moduli space. In fact, due to the ambiguity arising from the logarithm when one exponentiates to solve for $z$, there are $M$ vacua, distributed in phase but with $|z|$ given by the expression above. The modulus $|z|$ is strictly connected to the minimal value that the warp factor takes \cite{Giddings:2001yu}:
\bee\label{TIIBwarpStab}
 e^{A_{min}}\sim |z|^{1/3} \sim e^{\frac{-2\pi K}{3 g_sM}}
\ee
In effect the fluxes produce a model similar to the Randall-Sundrum one \cite{Randall:1999ee}, in which the warp factor does not go to zero but to an exponentially small positive value. We will come back to this point in a later chapter.

\subsubsection{Complete Moduli Stabilization Through Quantum Corrections}

At classical level, the K\"ahler moduli of Type IIB CY orientifolds with fluxes remain exactly flat direction of the potential. However, quantum corrections can generically generate a potential for these moduli. There are two possible sources. The first one comes from corrections to the K\"ahler potential which depends on the K\"ahler moduli. 

Here we will describe the second one. It comes from non-perturbative corrections to the superpotential (it enjoys a non-renormalization theorem to all orders in perturbation theory). Such type of corrections can come from Euclidean D3 brane \cite{Witten:1996bn} wrapping some 4-cycle $\Sigma$ on the compact manifold. This can happen when the fourfold $Y$ used for F-theory compactification admits divisor of arithmetic genus one, which project to 4-cycles in the base $X$ \cite{Denef:2004dm}. The correction to the superpotential coming from such instantons is given by:
\bee
 W_{inst}=T(z_K) \, e^{i\rho}% - \mbox{Vol}(\Sigma)}
\ee
where the imaginary part of $\rho$ is $\mbox{Vol}(\Sigma)$ and where $T(z_K)$ is a complex structure dependent one-loop determinant. This superpotential depends on the K\"ahler moduli since  the volume of the 4-cycle depends on them. 

An analogous correction comes from gaugino condensation in the gauge theory living on a D7-brane wrapping a 4-cycle in the compact manifold and filling the four dimensional spacetime (see for example \cite{Tripathy:2002qw}). Since the square of the gauge coupling is proportional to the inverse of the volume of the 4-cycle, the contribution to the superpotential is given by
\bee
 W_{D7}=\Lambda^3_{N_c}=A(z_K)\, e^{i\rho}% - \mbox{Vol}(\Sigma)/N_c}
\ee
where $N_c$ is the number of color of the gauge theory living on the D7-brane.

As it was argued in the famous KKLT work \cite{Kachru:2003aw}, one can check that such corrections generically allow to find flux vacua with all the geometric moduli stabilized\footnote{One should also note that in order to stabilize the K\"ahler moduli at strictly positive radii, one needs a sufficient number of 4-cycles, which excludes the simplest case of internal manifold with $h^{1,1}=1$.}.

The complex structure moduli are fixed by the fluxes at a scale of order $\frac{\alpha'}{\sqrt{vol\, X}}$, while any K\"ahler modulus potential arising from the above non-perturbative corrections will be significantly smaller. Thus one can think to fix all the complex structure moduli neglecting the non-perturbative corrections in a controlled way, and then integrate them out. In this way one fixes the K\"ahler moduli using a potential where the complex structure dependent objects are substituted with them evaluated at the fixed point. This is the typical KKLT procedure by two steps and it leads to moduli stabilization in a supersymmetric minimum. When the non-perturbative corrections described here are added to the superpotential, one can find a supersymmetric minimum also for a non-zero (0,3) component of $G_3$.

This procedure has received several critiques. The first one is that the procedure of obtaining an effective potential for the light moduli via non-perturbative corrections after integrating out moduli that are assumed to be heavy at the classical level is not in general correct \cite{Choi:2004sx,deAlwis:2005tf}. In some cases, this two steps procedure can give rise to tachyonic directions. One should instead minimize the full potential, which has additional terms mixing the light and heavy modes.
Another critique is that the corrections to the K\"ahler potential, both perturbative and non-perturbative, have not been taken into account. In \cite{Balasubramanian:2005zx,Conlon:2005ki} it is shown that the $\alpha'$ corrections to the K\"ahler potential are subleading only when the flux superpotential is of the same order of the non-perturbative superpotential. Otherwise, the perturbative correction to the K\"ahler potential must be included when analyzing the details of the potential. Applying this, one finds a new minimum at exponentially large volume, that is not supersymmetric, contrary to the KKLT vacuum described above. Several works followed  \cite{Balasubramanian:2005zx,Conlon:2005ki}, trying to extract some predictions on the low energies quantities on this vacuum \cite{Conlon:2005jm,Conlon:2006us,Conlon:2006tj,Conlon:2006wz,Cremades:2007ig}.

\

After having fixed all moduli, KKLT outline the construction of de Sitter vacua \cite{Kachru:2003aw}. In order to get de Sitter solutions from Type IIB flux compactifications, one should uplift the AdS vacuum found after having fixed all moduli. KKLT do this by adding a small number of anti-D3-branes at the bottom of a region with an extremely non-trivial warp factor (there regions are called {\it throats}). This de Sitter vacuum cannot be the true vacuum in a theory of quantum gravity. On the other hand, the runaway behavior is a standard feature of all string theories. KKLT showed nevertheless that the lifetime of the dS vacuum is large in Planck times, and shorter than the recurrence time. Even the fastest decays have decay time much greater than the age of the universe \cite{Frey:2003dm}.

\newpage

\section{Type IIA Vacua}\label{TypeIIAvacua}

To derive a four dimensional description of the Type IIA orientifolds vacua with fluxes, one reduces to four dimension the ten dimensional action of massive Type IIA supergravity \cite{Romans}. In Einstein frame it is given by:
\bea \label{eq003b}
 S &=& \frac{1}{2\kappa_{10}^2} \int \left( R \ast 1 - \frac{1}{2} d\phi \wedge \ast d\phi - \frac{1}{2} e^{\phi/2} G \wedge \ast G - \frac{1}{2} e^{-\phi} H \wedge \ast H \right.\\
&&- \frac{1}{2} e^{3\phi/2} F \wedge \ast F - 2m^2 e^{5\phi/2} \ast 1 + \frac{1}{2}d C^2 \wedge B + \frac{1}{2} d C\wedge dA\wedge B^2 \nn\\
&&\left. + \frac{1}{6}dA^2\wedge B^3 + \frac{m}{3} d C\wedge B^3 + \frac{m}{4}dA\wedge B^4 + \frac{m^2}{10}B^5  \right)\:\:+ S_{loc}\nn
\eea
with $2\kappa_{10}^2=(2\pi)^{7}\alpha'^{4}$. $S_{loc}$ is the contribution of localized sources included in the compactification.
% \bea
%  S &=& \frac{1}{2\kappa_{10}^2}\int d^{10}x \sqrt{-g} \left(e^{-2\phi}(R+4(\partial_M\phi)^2-\frac{1}{2}|H|^2) - 
% 		(|\tilde{F}_2|^2 + |\tilde{F}_4|^2+m^2)\right) \nn\\\nn\\
% 	&& + S_{CS} \:\:+ S_{loc}
% \eea

The fields involved are: the metric, the dilaton, the NSNS field strength $H$ (with potential $B$) and RR field strengths $F_0=m$ that is not dynamical, and the 2-form $F$ and the 4-form $G$ (with potentials $A$ and  $C$).
The physical field strengths with their Bianchi Identities are:
\begin{equation} \label{eq023} \begin{split}
F &= dA + 2mB   \\
H &= dB               \\
G &= d C + B \wedge dA + m B^2
\end{split} \qquad \qquad
\begin{split}
dF &= 2m H \\
dH &= 0 \\
dG &= F \wedge H \:.
\end{split}
\end{equation} 
The gauge transformations which leave the action invariant are:
\begin{equation} \label{eq064}
\delta A = m \Lambda_1 \qquad
\delta B = -\frac{1}{2} d\Lambda_1 \qquad
\delta C =  \frac{1}{2} A\wedge d\Lambda_1 + \frac{1}{4} m \Lambda_1\wedge d\Lambda_1\:,
\end{equation}
as well as $\delta A = d \Lambda_0$ and $\delta C=d\Lambda_2$.

\

Here we consider the compactification with orientifold O6-planes \cite{Grimm:2004ua}. The introduction of orientifold planes cut the spectrum of Type IIA in a different way with respect to what happens in Type IIB. In Type IIB the orientifold projection cuts part of the complex structure moduli and part of the K\"ahler moduli; the former remain good K\"ahler coordinates while the latter are not the one appearing in the canonical form of the effective four dimensional action. For Type IIA the situation is different. The projected K\"ahler moduli are good K\"ahler coordinate, while the complex structure moduli $z^k$ are not.
 
In Type IIA, the complex structure moduli are promoted to quaternionic multiplets by combining them with the RR axions. The expansion of $C$ on the basis of harmonic 3-forms is given by:
\bee
 C = \xi^{\hat{K}}\alpha_{\hat{K}} - \tilde{\xi}_{\hat{K}} \beta^{\hat{K}}
\ee
We get $h^{2,1}+1$ complex axions. The axions coming from $\xi^0,\tilde{\xi}_0$ join the axion-dilaton, while the other  $h^{2,1}$ axions quaternionize the $z^K$. The orientifold projection cuts this quaternionic space and the K\"ahler potential is changed sensitively. Let us see the details. 

The orientifold projection is given by the operator $\OO=\Omega_p (-1)^{F_L} \sigma$, where $\sigma$ is an antiholomorphic involution of the CY. It acts on the forms $J$ and $\Omega$ as:
\bea
 \sigma^\ast J = - J, && \sigma^\ast\Omega=e^{2i\theta}\bar{\Omega}
\eea
with $\theta$ some arbitrary phase\footnote{
These are different from the Type IIB conditions where the transformation of $J$ and $\Omega$ under the holomorphic involution $\sigma$ are $\sigma^\ast J = J$ and $\sigma^\ast\Omega=-\Omega$.}. The fixed loci of $\sigma$ are special Lagrangian 3-cycles $\Sigma_n$ satisfying
\bea
 J|_{\Sigma_n} = 0, && \im (e^{-i\theta}\Omega)|_{\Sigma_n}=0 \:.
\eea
Orientifold O6-planes fill spacetime and wrap the $\Sigma_n$.

The orientifold involution splits $H^3=H^3_+ + H^3_-$. Each of these eigenspaces is of real dimesion $h^{2,1}+1$. We split the basis for $H^3$ into a set of even forms $\{\alpha_k,\beta^\lambda\}$ and a set of odd forms $\{\alpha_\lambda,\beta^k\}$; here $k=0,...,\tilde{h}$ while $\lambda=\tilde{h}+1,...,h^{2,1}$. Then the orientifold projections requires (taking $\theta=0$):
\bee\label{TIIAorientZF}
 \im Z^k = \re \FF_k = \re Z^\lambda = \im \FF_\lambda =0
\ee
Two of these conditions are constraints on the moduli, while the other two follow automatically for a space admitting the antiholomorphic involution $\sigma$. We see that for each complex $z^K$, only one real component survives the projection. The condition that $C$ must be even under $\sigma$ truncates the space of axions in half to $\xi^k,\tilde{\xi}_\lambda$. In addition, the orientifold projects in the dilaton and one of $\xi^0$ and $\tilde{\xi}_0$. So from each hypermultiplet, we get a single chiral multiplet, whose scalar components are the real or imaginary part of the complex structure modulus, and a RR axion.

We can summarize the surviving hypermultiplet moduli in terms of the object
\bee
 \Omega_c = C + 2 i \re (c\Omega)
\ee
Here, $c$ is a compensator which incorporates the dilaton dependence via
\bea
 c = e^{-D+K_C/2}, && e^D = \sqrt{8}e^{\phi+K_k/2}
\eea
One should think of $e^D$ as the four dimensional dilaton.
The good K\"ahler coordinate are then the coefficients of the expansion of $\Omega_c$ on a basis for $H^3_+$:
\bea
 N^k &=& \frac{1}{2}\int_X \Omega_c \wedge \beta^k = \frac{1}{2}\xi^k + i\re(c Z^k)\\
 T_\lambda &=& i \int_X \Omega_c \wedge \alpha_\lambda = i \tilde{\xi}_\lambda - 2 \re (c \FF_\lambda)\:.
\eea
The K\"ahler potential for these moduli is quite different from \eqref{CYKC}, and takes the form:
\bee
 K_Q = - 2 \ln \left(2 \int_X \re (c\Omega) \wedge \ast \re(c\Omega)\right) = \im(c Z^\lambda)\re(c\FF_\lambda) - \re(cZ^k)\im(c\FF_k)
\ee

\

The K\"ahler potential for the K\"ahler moduli remains of the same form as without orientifold \eqref{CYKk}. The only difference is that only odd fluctuations of $J$ survive. Actually it is complexified by $B$:
\bee
 J_c = B + i J
\ee
Since $J_c$ is odd under the orientifold projection, it is expanded on a basis $\omega_a$ of $h^{1,1}_-$ odd harmonic forms:
\bea
 J_c= t^a \omega_a, && t^a= b^a +iv^a
\eea

\subsubsection{Flux Superpotential and Moduli Stabilization}

We can now turn on the fluxes that are projected in by the antiholomorphic involution \cite{Grimm:2004ua}. It turns out that $H$ and $F$ must be odd, while $G$ should be even (see \ref{TIIorientproj}). So we can write:
\bea
 H^f=q^\lambda \alpha_\lambda-p_k \beta^k, & F^f = - m^a \omega_a, & G^f= e_a\tilde{\omega}^a
\eea
where $\tilde{\omega}^a$ are the 4-forms dual of the $\omega_a$. Since the volume form is odd, while the $\omega_a$'s are odd, the $\tilde{\omega}^a$ are even. There are in addition two parameters $m$ and $f$ parametrizing the $F_0$ and $F_6$ fluxes on $X$ \footnote{$F_6$ is the Hodge dual of the part of $G$ along the four dimensional spacetime}.

\

The background fluxes contribute to the total D6 charge, together with the orientifold O6-plane. Actually the Bianchi identity for $F$ is given by
\bee
 dF = 2 m H - 2 \mu_6 \delta_3
\ee
where $\delta_3$ is the Poincar\'e dual three-form of the 3-cycle wrapped by the O6-plane. Integrating this equation over any 3-cycle produces a cancellation condition between the combination $mH_3$ of the RR 0-form flux and the NSNS 3-form flux, and the background O6-plane charge. Adding D6-branes also would contribute. This is the analogue of the effective D3 charge of the 3-form fluxes in Type IIB. Differently from Type IIB, in Type IIA there are other RR fluxes that are not contrained.

\

The $\NN=1$ potential generated by these fluxes is determined (through \eqref{effpot}) by the K\"ahler potential
\bee
 K = K_k + K_Q
\ee
and by the superpotential:
\bea
 W &=& \int_X \left( \Omega_c \wedge H + F_6 + J_c\wedge G - \frac{1}{2} J_c \wedge J_c \wedge F - 
	\frac{m}{6} J_c \wedge J_c \wedge J_c \right)
\eea
This superpotential depends, in general, on all geometric moduli at tree-level. The system of equations governing supersymmetric vacua is:
\bee
  D_{t^a}W = D_{N^k}W = D_{T_\lambda}W = 0
\ee
In \cite{DeWolfe:2004ns} it was shown that under reasonable assumptions of genericity, one can stabilize all geometric moduli at tree-level in these constructions. The same considerations show that in the leading approximation, $h^{2,1}_+$ axions will remain unfixed. These solutions can moreover be brought into a regime where $g_s$ is arbitrary small and the volume is arbitrary large.

\newpage

\section{M-theory Vacua}\label{MthVac}

M-theory is locally supersymmetric and is well described at low energy by the eleven dimensional supergravity. Its action is given by:
\bee
 S = \frac{1}{2\kappa_{11}^2}\left( \int d^{11}x  \sqrt{-g}R -\frac{1}{2}\int  G\wedge \ast G - \frac{1}{6}\int  C\wedge G\wedge G\right)
\ee
The bosonic fields are the eleven dimensional metric and a 3-form $C$, whose field strength is $G=dC$.

To obtain a four dimensional theory, we have to compactify on a seven dimensional manifold $X$. Requiring $\NN=1$ supersymmetry in four dimension poses constraints on the holonomy group of $X$. One possibility is the Horava-Witten theory \cite{Horava:1996ma} compactified on a CY space $Y$ times an orbifold of a circle. A second possibility, that is the one studied in this work, is to compactify M-theory on a seven dimensional manifold with holonomy group given by $G_2$ \cite{Joyce}. A central point concerning such $G_2$ compactification is that, if $X$ is smooth, the four dimensional physics contains at most Abelian gauge group and no light charged fermions \cite{Acharya:2004qe}.

We will at first briefly describe the manifolds with holonomy group $G_2$ and then we will see the effective theory obtained compactifying on these manifolds. Then we will see what are the relevant physical singularities and at the end we will add fluxes.

\subsection{$G_2$-holonomy Manifolds}

$G_2$ is the automorphism group of the octonions $\Obb$. It is a simple Lie group, that is compact, connected, simply-connected and with dimension $D=14$ \cite{Joyce}.

It may be defined as the subgroup of $SO(7)$ that leaves invariant the following 3-form on $\Rbb^7$:
\bee
 \Phi_0 = \frac{1}{3!} \psi_{ijk} dx^i\wedge dx^j \wedge dx^k
\ee
where $x^i$ are coordinates of $\Rbb^7$ and $\psi_{ijk}$ are totally antisymmetric structure constants of the imaginary octonions.
%:
% \bea
%  \sigma_i \signa_j = -\delta_{ij} +\psi_{ijk}\sigma_k  && i,j,k=1,...,7
% \eea
In a particular choice of basis the non-zero structure constants are given by:
\bea
 \psi_{ijk}=+1 &&(ijk)=\{(123),(147),(165),(246),(257),(354),(367)\}
\eea

\

The spinorial representation ${\bf 8}$ splits into representations of $G_2$ when $SO(7)$ is reduced to this subgroup:
\bee
 {\bf 8} \rightarrow {\bf 1}\oplus{\bf 7}
\ee
We see that there is one singlet in this decomposition.\footnote{This shows, with a reasoning analogous to that of page \pageref{4DSusyReduced}, that there are four supercharges surviving after compactification and so that the four dimensional theory will have $\NN=1$ supersymmetry.}

One can so characterize a $G_2$ holonomy manifold as a {\it seven} dimensional manifold with one covariantly constant spinor, in the same way as a CY is a {\it six} dimensional manifold with one covariantly constant spinor. On the other hand special holonomy manifolds can be also characterized by the existence of certain invariants forms (for a CY, these are $J$ and $\Omega$) \cite{Joyce}.

Indeed, one can construct antisymmetric combinations of gamma matrices with the covariantly constant spinor $\eta$, to obtain tensor forms of various degree:
\bee
 (\omega_p)_{i_1...i_p}=\eta^\dagger \Gamma_{i_1...i_p}\eta
\ee
By construction, the p-form $\omega_p$ is invariant under the holonomy group. In order to find all possible invariant forms on a special holonomy manifold $X$, we need to decompose the space of differential forms on $X$ into irreducible representation of $G_2$ and identify the singlet components. For the case under consideration we have singlets in the decompositions of the 3-forms and of the 4-forms, since the ${\bf 35}$ representation of $SO(7)$ decomposes as:
\bee\label{G235dec}
 {\bf 35}\rightarrow {\bf 1} \oplus {\bf 7} \oplus {\bf27} \\
\ee
The other relevant representations split as:
\bea\label{G27e21dec}
 {\bf 7} &\rightarrow& {\bf 7} \\
 {\bf 21} &\rightarrow& {\bf 7} \oplus {\bf 14}\nn
\eea

From the decompositions \eqref{G235dec} and \eqref{G27e21dec} we see that on a $G_2$ manifold the invariant forms appear only in degree $p=3,4$. They are called respectively {\it associative} and {\it coassociative} forms and are denoted as $\Phi$ and $\ast\Phi$. In fact the coassociative 4-form is the Hodge dual of the associative 3-form.

The existence of a $G_2$ holonomy metric on $X$ is equivalent to the following conditions on the associative and the coassociative forms:
\bea\label{MthPhi}
 d \Phi =0 && d\ast\Phi=0
\eea
Actually, using the 3-form $\Phi$ one can reconstruct a metric, and if $\Phi$ is closed and coclosed, the Levi-Civita connection of this metric has holonomy group equals to $G_2$. In particular this metric is {\it Ricci flat}. With respect to it the conditions \eqref{MthPhi} say also that $\Phi$ must be harmonic.

Since the Laplacian of the metric $g$ on $X$ preserves the decomposition of the spaces of forms in $G_2$ representations, the harmonic forms can also be decomposed in this way. By knowing that for a $G_2$ holonomy manifold $H^1(X)=\{ 0 \}$ and that $\dim H^\ell_{\bf k}$ is independent of $\ell$, one can prove that the only non-trivial refined Betti numbers ($b_\ell^{\bf k}\equiv \dim H^\ell_{\bf k}$) are $b_2^{\bf14}$ and $b_3^{\bf 27}$, which satisfy $b_2=b_2^{\bf14}$ and $b_3=b_3^{\bf 27}+1$.

%\newpage

\subsection{Compactification of M-theory on $G_2$ Holonomy Manifolds}\label{MthCompG2holMan}

\subsubsection{Compactification on Smooth Manifolds}

In this section, we describe the KK reduction of the eleven dimensional supergravity on a smooth manifold with $G_2$ holonomy \cite{Acharya:2004qe}. 

As seen above, the supergravity theory has two bosonic fields, the metric $g$ and a 3-form $C$, with field strength $G=dC$. The equation of motion for $C$ is given by: $d\ast G = \frac{1}{2}G\wedge G$. We start from the usual ansatz for the higher dimensional metric that preserves four dimensional Poincar\'e invariance, {\it i.e.} a product of Minkowski spacetime times a compact seven dimensional manifold. This manifold is taken to have $G_2$ holonomy group, in order to have $\NN=1$ supersymmetry in four dimensions.

\

We now analyze the massless scalars coming from the KK reduction of the metric. We will follow the same procedure used for the CY manifolds, but we will be more precise in describing the KK reduction. The obtained scalars are associated with the moduli parametrizing the $G_2$ metric with a given topology. 

We begin with a $G_2$ holonomy metric $g(y)$ on $X$. $g$ obeys the vacuum Einstein equation $R_{mn}(g)=0$. To obtain the spectrum of modes originating from $g$ we look for fluctuations $\delta g(x,y)$ such that $g(y)+\delta g(x,y)$ is also Ricci flat. This implies that $\delta g(x,y)$ satisfies the Lichnerowicz equation:
\bee\label{MthG2Lich}
 \Delta_L \delta g_{mn} = -\nabla_{(11)}^2 \delta g_{mn} - 2 R_{minj}\delta g^{ij} + 2R^k_{(m}\delta g_{n)k} =0
\ee
Next we make the KK ansatz for the fluctuations as:
\bee
 \delta g_{mn}(x,y) = \sum_I h^I_{mn}(y)s^I(x)
\ee
where we $h^I_{mn}$ are the eigenvectors of the operator $\Delta_L$.
Since the full spacetime is taken to be a product of Minkowski times a seven dimensional space, the eleven dimensional Laplacian can be splitted as $\nabla_{(11)}^2 = \nabla_{(3,1)}^2 +\nabla_{(7)}^2$. So we can write the equation \eqref{MthG2Lich} as:
\bea
 \sum_I h^I_{mn}(y) \nabla_{(3,1)}^2 s_I(x) &=& - \sum_I (\Delta_L h^I_{mn}(y)) s_I(x) = - \sum_I \lambda_I h^I_{mn}(y) s_I(x)
\eea
Thus we see that the massless scalars $s_i$ are associated with the zero modes $h^{(i)}_{mn}$ of the Lichnerowicz operator $\Delta_L$. We will now show that they are associated with the harmonic 3-forms.

On a seven dimensional manifold of $SO(7)$ holonomy, the $h^{(i)}_{mn}$ transform in the ${\bf 27}$ dimensional representation. Under $G_2$ this representation remains irreducible. On the other hand, as we have seen before, the 3-forms decomposes as ${\bf 35} \rightarrow {\bf 1}\oplus {\bf 7} \oplus {\bf 27}$. Thus the $h^{(i)}_{mn}$ can also be regarded as 3-forms on $X$, using the 3-form $\Phi$:
\bee
 \Phi_{n[pq}(h^{(i)})^n_{r]} = \omega^{(i)}_{pqr}
\ee
The $\omega^{(i)}$'s are 3-forms in the same representation as $h^{(i)}_{mn}$, since $\Phi$ is in the trivial representation. The condition that $h^{(i)}$ is a zero mode of $\Delta_L$ is equivalent to $\omega^{(i)}$ being a zero mode of the Laplacian. But we know how many 3-forms there are that are zero modes of $\Delta$: they are the harmonic 3-forms, whose number is $b_3$. So the dimension of the moduli space of the $G_2$ holonomy manifolds is given by $b_3$, the dimension of the third cohomology group. This is analogous of what happens for a CY; in that case we have a correspondence between the moduli and harmonic 2-forms and 3-forms. As in that case, the moduli come from the expansion of the invariant 3-form on harmonic 3-forms and are called $s_i$:
\bee
 \Phi=\sum_i s_i(x) \phi^i(y)
\ee
with $\{\phi_i\}$ a basis for $H^3(X)$.

\

There are also scalars arising from the reduction of $C$. Let us make the KK ansatz:
\bee
 C = \sum_I \omega^I(y)t_I(x) + ...
\ee
The equation of motion for $C$ tells that $C$ must be harmonic. This happens because $C$ has components only along the internal directions and so the term $G\wedge G$ identically vanishes. This implies that massless scalars $t_i$ correspond to harmonic 3-forms. So again the number of massless scalars are given by $b_3$. Since $C$ is a $U(1)$ gauge 3-form potential, a gauge transformation add to $C$ a closed 3-form on $X$ of appropriately normalized period; so the fields $t_i$ takes values on a compact space. 

The scalars coming from the reduction of $C$ combine to the scalars coming from $g$ to give a massless complex scalars $z_i=t_i+is_i$:
\bee
 C+i\Phi = \sum_i z_i(x) \phi^i(y)
\ee
The $z_i$'s are the complex scalars appearing in the chiral multiplets. It is not surprising that the scalars $s_i$ and $t_i$ belong to the same four dimensional supersymmetry multiplet, as $g$ and $C$ are superpartners in eleven dimensions.

\

In addition to the massless chiral multiplets, we also get massless vector multiplets. The bosonic components of such multiplet is a massless Abelian gauge field, that arises from the field $C$ through the KK ansatz:
\bee
 C= \sum_\alpha \beta^\alpha(y) A_\alpha(x)+...
\ee
where $\beta^\alpha$'s are basis for the harmonic 2-forms and $A_\alpha$'s are 1-forms in Minkowski space. Again the equation of motion for $C$ implies that the $A_\alpha$'s are massless in four dimension. This gives $b^2(X)$ such gauge fields. As for the chiral multiplet introduced above, the fermionic superpartners arise from the gravitino field. 

Since on a manifold with $G_2$ holonomy there are no harmonic 1-forms ($b^1=0$), there are no four dimensional 2-forms arising from $C$.

Summarizing, the low energy four dimensional effective theory is an $\NN=1$ supergravity theory coupled to  $b^2(X)$ Abelian vector multiplets and $b_3(X)$ massless, neutral chiral multiplets. This theory is relatively uninteresting from a phenomenological point of view, since the gauge group is Abelian and there are no light charged particles. These features arise in the effective theory if we compactify on a singular $G_2$ holonomy manifold. 

In the next section we will briefly describe the theory arising from singularities.

\subsubsection{Compactification on Singular Manifolds}

Since compactification on smooth manifolds does not produce interesting physics, one has to study dynamics of M-theory on singular $G_2$ holonomy manifolds \cite{Acharya:2004qe}.

One simple and common kind of singularity is an {\it orbifold} singularity. Locally, it can be represented as a quotient of $\Rbb^n$ by some discrete group $\Gamma$: $\Rbb^n/\Gamma$. In perturbative string theory, the physics associated with such singularities can be systematically extracted from the orbifold conformal field theory. Typically, one finds new massless degrees of freedom localized at the orbifold singularity. However, CFT technique is not applicable for studying M-theory on singular $G_2$-manifolds. Moreover many interesting phenomena occur at singularities which are not of the orbifold type. One can study M-theory dynamics on singularities, by using the duality of this theory with Heterotic and Type IIA string theories \cite{Witten:1995ex,Townsend:1995kk}. In what follows we will use the duality with the Heterotic theory.

\vskip 7mm
\noindent
{\sf Non-Abelian Gauge Fields}

% \
% 
% \noindent
% {\it Non-Abelian Gauge Fields}

Non-Abelian gauge groups emerge from M-theory when the compact \begin{scriptsize}\begin{footnotesize}\end{footnotesize}\end{scriptsize}space has a so called ADE-singularity \cite{Acharya:2000gb,Acharya:1998pm}. One can learn this in the context of the duality between M-theory on $K3$ and the Heterotic string on a flat 3-torus $T^3$ \cite{Witten:1995ex}. 

Let us review this duality briefly \cite{Acharya:2004qe}. $K3$ is a four dimensional manifold with holonomy group $SU(2)$. Its metric moduli space has dimension 58. An $SU(2)$ holonomy metric admits two parallel spinors, which when tensored with the ${\bf 8}$ constant spinor of the seven dimensional Minkowski space give 16 global supercharges. This corresponds to minimal supersymmetry in seven dimensions. On a smooth point in the moduli space, we can use KK reduction to get 58 massless scalars. Additionally, since $b^2(K3)= 22$, reducing $C$ on a basis of harmonic 2-forms, we obtain a $U(1)^{22}$ gauge group in seven dimensions. 

This is the same spectrum of Heterotic compactification on $T^3$ at a general point of the moduli space. The Heterotic string theory in ten dimensions has a bosonic massless spectrum given by the metric $g$, the 2-form $B$, the dilaton $\phi$ and non-Abelian gauge fields with group $SO(32)$ or $E_8\times E_8$. There are 16 global supersymmetries that are preserved by compactification on $T^3$. One gets 6 scalars from the fluctuations of the metric, 3 scalars form $B$ and 1 from the dilaton. Then one gets scalars from the gauge fields with component along $T^3$. The condition for them to be supersymmetric on $T^3$ is that their field strengths vanish. They are parametrized by Wilson lines around the three independent cycles of $T^3$. These flat connections break the gauge group to the subgroup whose generators commutes with them. The most general unbroken group is $U(1)^{16}$. This gives rise to 48 more scalars. The total amount of scalars is actually 58 like in the M-theory dual. There are then 16 gauge fields coming from the unbroken gauge group, 3 coming from $g$ fluctuations and 3 from $B$, giving a total $U(1)^{22}$ gauge group in seven dimensions, as in M-theory.

At special points of the Heterotic moduli space, some of the eigenvalues of the flat connections will vanish and so the gauge group can be enhanced to a non-Abelian one. If M-theory on $K3$ is actually equivalent to the Heterotic string theory in seven dimensions, it too should exhibit non-Abelian symmetry enhancement at special point in the moduli space. These points are precisely the points where $K3$ develops orbifold singularities. These singularities can give rise to Lie groups of the type A, D or E, and so are called ADE-singularities. 

Locally these singularities can be described as $\Cbb^2/\Gamma_{ADE}$, where $\Gamma_{ADE}$ is a finite subgroup of $SU(2)\subset SO(4)$, in order to preserve supersymmetry. In absence of gravity, the low energy physics of M-theory on $\Cbb^2/\Gamma_{ADE}\times \Rbb^{6,1}$ is described by super Yang-Mills theory on ${\bf 0}\times \Rbb^{6,1}$ with ADE gauge group. ${\bf 0}$ is the singular point, so the gauge theory is localized on the singularity \cite{Acharya:2000gb}.

\

We have thus far restricted our attention to the ADE singularities in $K3\times \Rbb^{6,1}$. However, we can consider more complicated spacetimes $M^{10,1}$ with ADE singularities along more general spacetimes $Y^{6,1}$. In the context of $G_2$ compactification on $X\times \Rbb^{3,1}$, we want $Y$ to be of the form $Q\times \Rbb^{3,1}$, with $Q$ the locus of ADE singularities inside $X$. Near $Q\times \Rbb^{3,1}$, $X\times \Rbb^{3,1}$ looks like $\Cbb^2/\Gamma_{ADE}\times Q\times \Rbb^{3,1}$. In order to study the gauge theory dynamics without gravity, we can restrict our attention to the physics near the singularity. So we focus on seven dimensional super Yang-Mills theory on $Q\times \Rbb^{3,1}$.

In flat spacetime the SYM theory has a global symmetry group given by $SO(3)\times SO(6,1)$. The first factor is the R-symmetry,while the second one is the Lorentz group. The theory has a gauge field transforming in the $({\bf 1},{\bf 7})$, scalars in the $({\bf 3},{\bf 1})$ and fermions in the $({\bf 2},{\bf 8})$ representations. All these fields transform in the adjoint representation of the gauge group. The 16 supersymmetries transform in the $({\bf 2},{\bf 8})$ representation of the symmetry group.

On $Q\times \Rbb^{3,1}$ the symmetry group is broken to 
\bee\label{SingG2groups}
SO(3)\times SO(3)'\times SO(3,1). 
\ee
The supersymmetries transform as $({\bf 2,2,2})+({\bf 2,2,\bar{2}})$. For large enough $Q$ and at energy scales below the inverse size of $Q$, we can describe the physics in terms of a four dimensional gauge theory. In order to have a supersymmetric theory, we must require that the space $\Cbb^2/\Gamma_{ADE}\times Q$ admits a $G_2$ holonomy metric. When $Q$ is curved, this metric cannot be the product of the locally flat metric on $\Cbb^2/\Gamma_{ADE}$ and a metric on $Q$. Instead %the metric is warped and is more like the metric on a fiber bundle in which the metric on $\Cbb^2$ varies as we move around in $Q$. Thus in this case 
the seven dimensional $G_2$ manifold is a non-trivial $K3$ fibration over $Q$. The locus of singularity of $K3$ is a copy of $Q$.

The condition of supersymmetry requires that this fibration has holonomy group $G_2$. If it is the case, we must identify the $SO(3)$ group in \eqref{SingG2groups} with $SO(3)'$ \cite{Acharya:2004qe}. This breaks the symmetries to the diagonal subgroup of the two $SO(3)$'s and implies that the effective four dimensional field theory is classically supersymmetric. Identifying the two groups breaks the symmetry group down to $SO(3)''\times SO(3,1)$ under which the supercharges transform as $({\bf 1,2})+({\bf 3,2})+{\bf cc}$. We now have supersymmetry because the $({\bf 1,2})+{\bf cc}$ can be taken to be constant on $Q$.

We also stress that the locus of ADE-singularity is a calibrated cycle with respect to the associative form $\Phi$. 

\

Supposing we could find a $G_2$ manifold of this type, we can find the four dimensional supersymmetric gauge theory that it corresponds to. Assuming $Q$ to be smooth and large, we can do a standard KK reduction on it. Under $SO(3)''\times SO(3,1)$, the seven dimensional gauge fields transform as $({\bf 3,1})+({\bf 1,4})$, the three scalars give $({\bf 3,1})$ and the fermions give $({\bf 1,2})+({\bf 3,2})+{\bf cc}$. Thus the fields that are scalars under the four dimensional Lorentz group are two copies of the ${\bf 3}$ of $SO(3)''$, that can be interpreted as two one forms on $Q$. These will be massless if they are zero modes of the Laplacian on $Q$. Thus there will be precisely $b^1(Q)$ of these. Their superpartners are clearly the $({\bf 3,2})+{\bf cc}$ fermions, which will be massless by supersymmetry. These fields collect into $b^1(Q)$ chiral superfields.

The $({\bf 1,4})$ field is massless if it is constant on $Q$ and this gives one gauge field in four dimensions. The superpartners are the remaining fermions transforming as $({\bf 1,2})+{\bf cc}$.

All of these fields transform in the adjoint representation of the ADE gauge group. So the four dimensional spectrum theory is an $\NN=1$ SYM with $b^1(Q)$ massless adjoint chiral supermultiplets.

% \
% 
% \noindent
% {\it Chiral Matter Fields}

\vskip 7mm
\noindent
{\sf Chiral Matter Fields}

The theory described so far is not chiral. Therefore one has to introduce some new type of singularities, that are worse than orbifold ones \cite{Acharya:2001gy}.
Again we use the duality with Heterotic string theory to determine what kind of singularities are required. We describe it in a specific example, that will be usefull later on. We will consider the case $E_8\times E_8$ heterotic string with $SU(5)\subset E_8$ as the unbroken gauge group. Such model can have chiral ${\bf 5}$'s and ${\bf 10}$'s of $SU(5)$. Let us see how this happens in the region of moduli space in which the CY is $T^3$-fibered over $Q$, with small fibers, and then translate to M-theory on $X$.

Since the unbroken group is $SU(5)$, the structure group of the gauge bundle must be another copy of $SU(5)'$ (the commutant of $SU(5)$ in $E_8$). Massless fermions in the Heterotic theory transform in the adjoint of $E_8$. The part of the $E_8$ adjoint representation that transforms as ${\bf 5}$ under $SU(5)$, transforms as ${\bf 10}$ under $SU(5)'$. So to get chiral fermions in the ${\bf 5}$, we must look at zero modes of the Dirac equation on the CY with values in the ${\bf 10}$ of $SU(5)'$. 

The regime of validity of the duality between Heterotic theory on a $T^3$ fibration over $Q$ and M-theory on a $K3$ fibration over $Q$ is that the generic radius $R$ of $T^3$ is much smaller than the size of $Q$. For small $R$ we can split the Dirac operator as:
${\bf \not}D={\bf \not}D_{T^3}+{\bf \not}D_Q$. For a generic fiber  of CY$\rightarrow Q$ the eigenvalues of ${\bf \not}D_{T^3}$ are all non-zero and are of order $1/R$. This is much too large to be cancelled by the behavior of ${\bf \not}D_Q$. So the zero modes of ${\bf \not}D$ are localized near points in $Q$ above which ${\bf \not}D_{T^3}$ has zero modes. 

When restricted to a $T^3$ fiber, the $SU(5)'$ bundle can be described as a flat bundle with monodromies around the three cycles in $T^3$, {\it i.e.} we have three Wilson lines on each fiber. For generic Wilson lines, every vector in the ${\bf 10}$ of $SU(5)'$ undergoes non-trivial twists in going around some of the three cycles. When this is the case, the minimum eigenvalue of ${\bf \not}D_{T^3}$ is of order $1/R$. This is simply because for a generic flat gauge field on $T^3$ there will be no zero mode.

A zero mode of ${\bf \not}D_{T^3}$ above some point $P$ of $Q$ arises if for some vector in the ${\bf 10}$, the monodromies in the fiber are all trivial. This means that the monodromies lie in the subgroup $H$ of $SU(5)'$ that leaves fixed that vector. The commutant of $H$ in $E_8$ is a group $G$ larger than $SU(5)$. So over the point $P$, the monodromies commute not just with SU(5) but with $G$. The monodromies at $P$ give large masses to all $E_8$ modes except those in the adjoint of $G$.

The fact that, over $P$, the Heterotic string theory has unbroken $G$ means that, in the M-theory description, the fiber over $P$ has a $G$ singularity. Likewise, the fact that away from $P$, the Heterotic theory has only $SU(5)\times U(1)$ unbroken means that the generic fiber, in the M-theory description, must contain an $SU(5)$ singularity only, rather than a $G$ singularity. From this we can conclude that in M-theory the chiral fermions are localized at points in $Q$ over which the ADE-singularity gets worse \cite{Acharya:2001gy}. 

These singularities arise generally as conical singularities of the seven dimensional $G_2$ manifold $X$. One example is a cone over weighted projective space $\mathbb{W}\Cbb {\bf P}^3_{N,N,1,1}$. This six dimensional space has a family of A$_{N-1}$-singularities at points $(w_1,w_2,0,0)$. This set of points is a copy of $\Ss^2$. As we have said the $G_2$ manifold $X$ is a cone over $\mathbb{W}\Cbb {\bf P}^3_{N,N,1,1}$, so it has a family of  A$_{N-1}$-singularities which are a cone over this $\Ss^2$. This is a copy of $\Rbb^3$. Away from the origin in $\Rbb^3$, the only singularities are these orbifold singularities. At the origin however, the whole manifold develops a conical singularity. One can show that at this point the singularity becomes a A$_N$-singularity, giving the possibility of the appearance of chiral fermions \cite{Acharya:2001gy}.

\subsection{Flux Superpotential and Moduli Stabilization}\label{MthFluxSuperpotMdStab}

Turning on a background field strength $G$ for $C$ along the compact directions, induces the following superpotential into the four dimensional theory \cite{Gukov:1999gr,Acharya:2000ps,Beasley:2002db}:
\bee
 W = \int_X \left(\frac{1}{2}C+i\Phi\right)\wedge G
\ee
The relative factor of $1/2$ between the two terms is required by supersymmetry.

Since $G$ does not depend on the metric and is harmonic (from the equation of motion for $C$) and quantized, we can expand it in terms of a basis of harmonic 4-forms $\rho^j$ ($j=1,...,b_3(X)$) which are dual to the harmonic 3-forms $\phi^j$. This means that the superpotential can be written as 
\bee
 W = \sum_j z_j N_j
\ee
and is therefore linear in the moduli with coefficient which are the flux quanta. This is a standard form for the superpotential, as we have seen for Type II flux compactifications. Here and in what follows we have put $\kappa_4=1$.

The K\"ahler potential depends only on the $s_i$'s and not on the axions and it is given by\footnote{Our normalization conventions for $z$, $W$ and $K$ are slightly different from \cite{Beasley:2002db,Acharya:2000ps}. The value of the normalization coefficient appearing in (\ref{Kpot}) in front of $V_X$ is verified in appendix \ref{AppMth1}.} \cite{Beasley:2002db}:
% \bee
%  K = -3 \ln \left(\frac{1}{14\pi^2}\int_X\Phi\wedge\ast\Phi\right)
% \ee
\begin{equation} \label{Kpot}
 K(z,\bz) = - 3 \ln \left( 4 \pi^{1/3} V_X(s) \right),
\end{equation}
where $V_X=$Vol$(X)/\ell_M^7$; Vol$(X)$ is the volume of the seven dimensional manifold $X$ and $\ell_M$ is given by $1/\kappa_{11}^2 = 2\pi/\ell_M^9$.  
$V_X$ is given by
\bee
 V_X = \frac{1}{7}\int_X \Phi\wedge\ast\Phi
\ee
and it is a homogeneous function of the $s^i$ of degree $7/3$. This classical metric will receive quantum corrections, but at large enough volumes such corrections can be argued to be small. Since $\ast$ depends non-linearly on the metric, $e^K$ is a non-linear function of the moduli.

\

The $\NN=1$ supergravity potential is given by the formula \eqref{effpot}. Inserting the expressions above for $W$ and $K$ one gets a definite positive potential that runs off to zero at infinity. So one must find a new mechanism to fix all the moduli. This was found in \cite{Acharya:2002kv}. The idea is to turn on a backround value for the gauge fields and their bosonic superpartners living on the locus $Q$ of ADE-singularities of $X$. As we have seen in the previous section, the seven dimensional gauge field gives rise to a 1-form $A$ on the three dimensional submanifold $Q$ that is complexified by a 1-form $B$ coming from the scalar superpartners. We turn on background values for these 1-forms. Their contribution to the superpotential is given by the complex Chern-Simons functional 
\bee
 \omega = \int_Q\mbox{tr}(A+iB)\wedge d(A+iB)+\frac{2}{3}(A+iB)\wedge (A+iB)\wedge (A+iB)  
\ee
Its critical points are complex flat connections, that are the solutions to the supersymmetry equations \cite{Acharya:2002kv}. $\omega$ is a topologically invariant functional, independent of the geometric moduli\footnote{It is not obvious that $Q$ admits such topological invariant. The only known examples of flat connection with non-zero complex and non-real Chern-Simons invariant are in the cases in which $Q=\mathbb{H}^3/\Gamma$, {\it i.e} when $Q$ is diffeomorphic to a compact 3-manifold which admits a hyperbolic metric.}:%\footnote{We should note that we are assuming that the space of flat connections has isolated points, so that we do not have to consider further moduli.}
\bee
  \omega|_{A_{bkg}+iB_{bkg}} = c_1 + i c_2 
\ee 
In general this constant is complex. In particular the real part is only well defined modulo 1 in appropriate units and is essentially the more familiar real Chern-Simons invariant. Its imaginary part however can in general take any possibly large real number \cite{Acharya:2002kv}.

The resulting superpotential is then given by\cite{Gukov:1999gr,Acharya:2000ps,Beasley:2002db,Acharya:2002kv}:
\bee
 W = \sum_j z_j N_j + c_1+ic_2
\ee

This superpotential was demonstrated to give at least a supersymmetric vacuum in which all the geometric moduli are stabilized \cite{Acharya:2002kv}. We describe how this mechanism works in the simple example of $b_3(X)=1$. In this case the supersymmetry condition $D_z W=0$ gives the following equation for $s$:
\bee
 G = \frac{7}{2s}(N s + c_2)
\ee 
This has the unique solution $s=-\frac{7}{5}\frac{c_2}{N}$. The supergravity approximation is valid when the volume of $X$ is large and this corresponds to $s$ being large. Therefore as long as $c_2$ is large compared to the flux $N$, the minimum exists in a region of field space within the approximation.

The potential also has another critical point at finite $s$. This critical point is a de Sitter local maximum.

For $b_3$ generic, there can be several non-supersymmetric vacua, where all the geometric moduli are fixes. This is shown in the example studied in \cite{Acharya:2005ez} and reviewed in details in the section \ref{sec:statmodel}.

\vskip 35mm

\section*{Final Remarks}

In this chapter, we have seen three corners of String/M-theory that can give phenomenological viable compactifications. In the following part of this thesis, we will see how the various approaches to string phenomenology, that we described in the Introduction, can be applied to these sets of vacua.

First of all, we will describe the statistical approach to the enormous number of vacua arising in flux compactifications. Then we will deal with the ten dimensional description of such compactifications. Finally we will study some particular features of these vacua that can have important phenomenological consequences.

\part{Aspects of String Phenomenology}

\chapter{Statistics of Vacua}\label{Stat}

In this chapter we will describe the statistical approach to the Landscape, made natural by the enormous number of four dimensional string vacua. After a review of motivations for a statistical study and of techniques and results obtained with this approach, we will present the results obtained in \cite{Acharya:2005ez}, where we studied some ensembles of M-theory vacua.

\section{Statistics on the Landscape of String Vacua}

In the previous chapter we have described the effects, on the four dimensional effective theory, of turning on fluxes on the internal manifold. The most important qualitative feature of fluxes is that, since their contribution to the energy depends on the geometric moduli, minimizing this energy will stabilize moduli, eliminating undesired massless fields. Since coupling constants in the low energy theory depend on moduli, finding the values at which moduli can be stabilized is an essential step in determining low energy predictions.  

Taking into account the large number of possible discrete choices for the fluxes, one is led to a ``discretum'' of a large number of vacua, the so called ``Landscape''. A natural question is if one or more of them describe our universe. In fact there are many different scenarios for string phenomenology, each requiring different properties of the vacuum. Thus, rather than study individual vacua, M. Douglas and collaborators proposed to study the overall distribution of vacua in moduli space, and the distribution of quantities such as the cosmological constant and supersymmetry breaking scale. 
This approach may be useful to: 1) estimate the frequency with which SM-like models arise in Sting/M-theory; 2) get an idea in which regions of the Landscape to look for realistic models, giving a guide for model building; 3) find statistical evidence that SM-like properties are extremely rare in String/M-theory, getting evidence against it; 4) argue for a uniform distribution of certain physical quantities like for example the cosmological constant, giving a new approach to the so called fine tuning problems.

\

Guidance for model building provides the main motivation for statistical studies. Actually, before embarking on the search and construction of vacua describing our world, it would be useful to know what our chances of success are. For example it could save us a lot of time and effort if we found that large regions of the Landscape (or more precisely large classes of models) are excluded.
Traditionally the approach has been to focus on the most easily controllable constraints such as light charged particle spectra,
while ignoring issues such as moduli stabilization, supersymmetry breaking or the cosmological constant. However these constraints are
equally important, and it would be very useful to estimate how much they reduce the set of possibilities. 
%For instance,  one could propose a certain compactification with just the Standard Model living on a set of branes, and with very large compactification volume to explain the hierarchy between the electroweak and Planck scales. A model builder might then go on and try to turn on various fluxes to stabilize all moduli within the required region of moduli space. But this theorist will plausibly fail, finding not a single vacuum with volume even remotely close to what is required for this model to work. Indeed, flux vacua generally rapidly become more scarce at larger compactification volumes, and completely cease to exist even at very modest sizes. Clearly, our model builder would have benefited greatly from some simple estimates of distributions of actual vacua over the parameter space of the model.

In \cite{Bousso:2000xa}, it was further pointed out that a  sufficiently fine ``discretum'' of string theory vacua could naturally accommodate an extremely small but nonzero cosmological constant. More precisely, in an ensemble of $\NN_{\rm vac}$ vacua with roughly uniform distributed cosmological constant $\Lambda$, one expects that there exist vacua realizing a cosmological constant as small as $M_{pl}^4/\NN_{\rm vac}$. To obtain the observed $\Lambda\sim 10^{-122}M_{pl}^4$ one should have $\NN_{\rm vac}\gtrsim 10^{122}$.
To find out whether this idea can be realized in a given ensemble of vacua, and more generally to analyze how strongly constraints on various parameters reduce the number of possibilities within a given class of models, one would like to have an estimate of how many vacua with certain properties lie in a given region of the Landscape. In other words, one needs to study the statistics of vacua in this region (where ``statistics'' does {\it not} refer to any probability measure, but simply to number distributions on parameter space).

Finally, one could take things one step further and altogether discard the idea that some dynamical mechanism has
uniquely selected our vacuum and in particular picked the extremely nongeneric scale hierarchies which just happen to be
also necessary to make structure formation and atoms other than hydrogen or helium possible \cite{Weinberg:1987dv,Vilenkin:1994ua,Agrawal:1997gf,ArkaniHamed:2004fb,ArkaniHamed:2005yv}.
Instead, one could start with the hypothesis that a ``multiverse'' exists in which a huge number of vacua is actually realized, and in
particular that we observe ourselves to be in a vacuum with such large scale hierarchies simply because this is needed for structure
and atoms, and therefore observers, to exist. To make direct predictions from just string theory in such a framework, one would
need to know the probability measure on at least the part of the Landscape compatible with a number of basic requirements. There is
no established way of computing these probabilities at this time, but as an additional working hypothesis one might assign for example
in a given ensemble of flux vacua approximately the same probability to every choice of flux. One could refine this by restricting this equal probability postulate to subsets of vacua with fixed values of parameters relevant for cosmology, such as the vacuum energy, as one does for microstates with equal energy in the microcanonical ensemble of statistical mechanics. Different choices of these relevant parameters might then be weighted by cosmological considerations (up to the extent that this is needed, as some will
be effectively fixed by environmental requirements). Under such hypotheses, suitable number distributions can be interpreted as
probability distributions, and one can test the hypotheses that went in by Bayesian inference.

It should be emphasized that this framework is significantly more predictive than the traditional model building approach of simply
considering any model compatible with current observations. Under these simple hypotheses, the Landscape picture together with a few
rough environmental principles gives a new notion of naturalness for effective field theories, which translates into a set of rules for model building. This turns out to lead to \emph{very distinct} models which do not need contrived engineering to fit known data,
and which moreover give very specific predictions, including many unambiguous signatures at LHC \cite{ArkaniHamed:2004fb,ArkaniHamed:2004yi,ArkaniHamed:2005yv}.

% More theoretical data on number distributions obtained from string theory would obviously be very useful to make further progress in
% this area, and although as we discussed there are several other motivations for studying the statistics of string and M-theory
% vacua, we considered this to be a very important one.

In conclusion, %the main justification of 
the vacua statistical program will improve our knowledge of the vacua of String/M-theory. It is important both from a theoretical point of view, being a step through a complete understanding of the theory, and from a phenomenological point of view, giving hints of how face the problem of finding the Standard Model inside String/M-theory.

\subsection*{Statistical Methods}

\begin{figure}
\begin{center}
  \epsfig{file=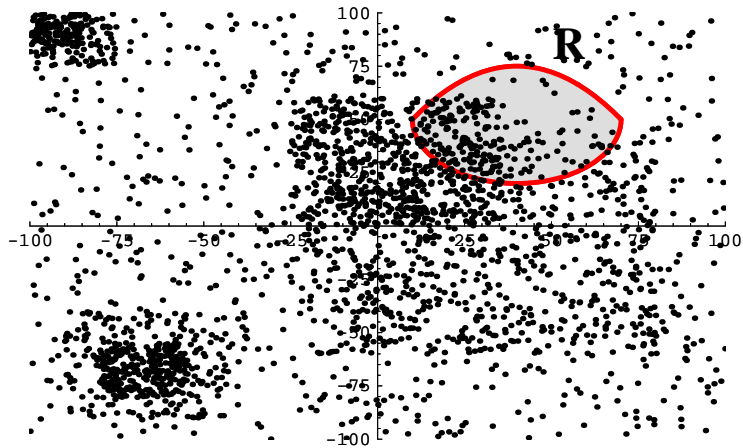,height=6cm,angle=0,trim=0 0 0 0}
  \caption{\small Example of discretum of vacua over the moduli space. The circle represents a region selected by some  requirements on the physical quantities.}
  \label{lansc}
\end{center}
\end{figure}

From String/M-theory, one can extract different $\NN=1$ effective four dimensional supergravity theories with a given configuration (moduli) space. Each of these theories includes discrete data $\vec{N}$, such as fluxes, brane charges, etc. This defines ensembles of four dimensional effective supergravity theories $\{T_{\vec{N}}\}$, with Lagrangians $\{\mathcal{L}_{\vec{N}}\}$, whose potential on the moduli space is given by:
\bea
	V_{\vec{N}}(t)&=&e^K(|DW_{\vec{N}}|^2-3|W_{\vec{N}}|^2)
\eea
Minimizing this potential, one finds the vacua of the theory. Varying the discrete data, the potential changes and one finds different vacua. Collecting all these solutions, one gets a discretum of vacua $\{t^a_{\vec{N}}\}$, that can be represented as in 
fig.\ref{lansc}. The physical quantities depend on the moduli and so one can study the distribution of them over the discretum of vacua. In particular phenomenological or theoretical constraints on physical observables select a region in the moduli space and one can compute how many vacua there are that realize these constraints, just counting the points inside the region (see fig.\ref{lansc}).

% Let us consider the data which goes into explicitly specifying a particle physics EFT. This will include both discrete and continuous choices. Discrete choices include the gauge group $G$ and matter representtion ${\bf r}$ of fermions and bosons. Choices involving parameters include the effective potential, Yukawa couplings, kinetic terms and so on. We will denote the vector containing such parameters as $\vec{g}$. We can regard all these choices as defining a point $P_i=(G,{\bf r},\vec{g})$ in a theory space $T$.

\

Conceptually, the simplest distribution one could consider is the density of supersymmetric vacua \cite{Denef:2004ze}. As we have seen, a supersymmetric vacuum is a solution of $D_{z^i}W=0$, thus the density of such vacua is given by:
\bee
 d\mu_s(z) = \sum_{\vec{N}} \delta_z\left(DW_{\vec{N}}(z)\right)
\ee
where $\delta_z(f)$ is a delta function at $f=0$, with normalization factor such that each solution of $f=0$ contributes unit weight in an integral over $d^{2n}z$. Thus, integrating this density over a region of the moduli space, gives the number of vacua which stabilize the moduli in that region.

\newpage

One can also define joint distributions such as the distribution of supersymmetric vacua with a given cosmological constant, that for a supersymmetric minimum is given by $-3e^K|W|^2$:
\bee
 d\mu_s(z,\Lambda) = \sum_{\vec{N}} \delta_z\left(DW_{\vec{N}}(z)\right) \delta\left(\Lambda-(-3e^K_{\vec{N}}|W_{\vec{N}}|^2) \right)
\ee

\

If we have a finite list of supergravity theories $\{T_{\vec{N}}\}$, and if in each the number of vacua is finite, such a density will be a sum of delta functions. This is hard to study and for many purposes it is enough to use a continuous approximation to this density, a function $\rho(z)$ whose integral over a region $R$ approximates the actual number of vacua in this region.

Now, suppose we have an explicit string construction of the Standard Model, and we are trying to reproduce the actual value for some physical quantities, such as Yukawa couplings. As we have said, these quantities depend on the moduli of the compactification. When we have their explicit formula in terms of the moduli, we can use it to identify a region of the moduli space where their values is compatible with the experiments (see fig.\ref{lansc}). The string construction "works" if this region contains vacua. 
If we can find an approximate density $\rho(z)$ and integrating it on this region gives a non-zero value, we cannot immediately conclude that there exists a vacuum in that region.  The result could be that the number of vacua is $10^{-20}$; in this case one has to be careful in interpreting the result, and should find other methods to prove if there exists or not a vacuum in that region. However if there are other string constructions that lead to different regions of the Landscape, with a much bigger number of vacua, then it is more natural to search the SM there. This criterium of naturalness is called {\it statistical selection} and it is described in \cite{Douglas:2004zg}. To understand how this mechanism works, we report the example given there. Let us first assume that the distribution of the cosmological constant (cc) is {\it uniform} on the space of vacua. Then suppose there are $10^{160}$ vacua with the property $P$, and which realize all known physics, except for the observed cosmological constant (cc). We would expect that out of this set, $10^{40}$ vacua reproduce the observed cc. Suppose furthermore that there exist diffrent $10^{100}$ vacua with the property $\bar{P}$, that realize the known physics except for the value of cc. Then we expect that the observed cc is realized on $10^{-20}$ vacua. As we have said above we will not conclude that there are no vacua realizing the property $\bar{P}$. On the other hand we have another set of vacua that work much better. So we have reasonable grounds for expecting the property $P$ instead of $\bar{P}$. This is not an exact prediction, but if $\bar{P}$ was realized in nature, we could say to have an evidence against string theory.  In a systematic approach, one would take all aspects of the physics resulting from each choice of vacuum, not just the cc, but couplings and matter content as well, and make analogous arguments. Clearly, if the total number of vacua is too large, the statistical selection just described cannot work.

One can also profit by the hypothesis that some distributions are statistically independent. In this case, one can argue that the fraction of vacua which realize both properties is the product of the fractions which realize each, without explicitly finding the vacua that realize both. Naturally, before multiplying the distribution one should first verify that this hypothesis is true. 
%For example, it seems very likely that the value of the cc is independent of number of Standard Model generation $N_g$, {\it i.e.} if we restrict attention to the vacua having a definite value for $N_g$, the cc distribution is still uniform. So one could study the population of vacua with arbitrary $N_g$ and check if this hypothesis is verified. 
On the other hand, if one found that the hypothesis is false, this would be surprising and even more interesting, as it could happen that it is peculiar of string theory.

Multiplying the fractions of vacua that realize the experimental constraints leads to an estimate of the fraction of vacua which agree with the Standard Model. This ratio is so small that the task of finding the vacuum which actually realizes all of its properties simultaneously is almost impossible. A statistical study of the distributions seems to be a more approachable problem and could give hints to find the vacua describing the SM with all its properties realized.

\subsubsection{Finiteness of Vacua}

In counting vacua, one is implicitly assuming that the number of quasi-realistic vacua of String/M-theory is finite. Since it is easy to write down effective potential with an infinite number of local minima, this is a non-trivial hypothesis, which must be checked. Actually there are many well established infinite series of compactifications, such as Freund-Rubin ones. 

A basic reason to want a finite number of quasi-realistic vacua, is that if this is not true, one runs a real risk that the theory can match any set of observables, and so it would not be falsifiable. However one must say that this is not automatic: one can have infinite series which would lead to a definite prediction. This happens for example if the series has an accumulation point: almost all the vacua give the same prediction, and one could further say that this accumulation point is the preferred prediction. One could also have that the measure factor suppresses infinite series. In any case the question of finiteness of the number of vacua is a very important issue to solve, and at present there is no completely general argument for it. To see a discussion look at \cite{Acharya:2006zw}, where the authors presented evidence that the number of string/M-theory vacua consistent with experiments is a finite number. They did that both by explicit analysis of infinite sequences of vacua and by applying various mathematical finiteness theorems.

\section{Ensembles of Vacua}

Up until \cite{Acharya:2005ez} most statistical Landscape studies had focused on Type IIB flux vacua. Because of string duality, one might optimistically hope that such IIB flux vacua could be representative, in the sense that they constitute a significant fraction of all string vacua, perhaps constrained to have some additional properties such as supersymmetry in the UV. However, it should be noted that the studies in \cite{Ashok:2003gk} - \cite{DeWolfe:2004ns} are limited to vacua described as Calabi-Yau orientifolds at moderately large volume and moderately weak string coupling, and strictly speaking these represent only a corner of the string theory Landscape. In principle, distributions of observables could change dramatically as one explores different regions of the Landscape. It was suggested in \cite{Acharya:2004nq} that the set of four dimensional string and M-theory vacua with $\NN=1$ or no supersymmetry is a disconnected space whose different components represent qualitatively different low energy physics, and this could translate into very different statistical properties. This gave us additional motivation to study the statistics of $G_2$ vacua and compare to the IIB case.

Another branch of the Landscape whose statistical analysis we initiated in \cite{Acharya:2005ez} is the set of Freund-Rubin vacua \cite{Freund:1980xh}, {\it i.e.} M-theory compactifications on Einstein manifolds with positive scalar curvature. Their properties are quite different from more familiar compactifications on special holonomy manifolds, and this is reflected in their vacuum distributions which are very different as well.

\

In this work, we will present in details the statistical study of ensembles of M-theory compactifications on $G_2$ holonomy manifolds with fluxes. For comparison, we will briefly review the result obtained in the statistical study of Type IIB flux vacua \cite{Ashok:2003gk,Denef:2004ze,Denef:2004cf} and of M-theory Freund-Rubin vacua\cite{Acharya:2005ez}. \footnote{Analogous statistica studies have also been done on the Heterotic String Landscape \cite{Dienes:2006ut,Lebedev:2006tr,Dienes:2007ms}.}

\subsection*{Type IIB Statistics}

Here we present what is known about the statistics of IIB flux vacua of Calabi-Yau orientifolds. There is a natural splitting of K\"ahler and complex structure moduli in this context. We have seen that turning on fluxes induces a superpotential which only depends on the complex structure moduli. We have also described how, in suitable circumstances, the K\"ahler moduli can be stabilized by nonperturbative effects. It is reasonable to ignore the K\"ahler moduli altogether as far as vacuum statistics is concerned, because: 1) the main contribution to vacuum multiplicities comes from the huge number of different fluxes, 2) at sufficiently large volume the K\"ahler moduli do not influence the positions in complex structure moduli space significantly, and 3) practically, the
K\"ahler sector is less under control and more difficult to treat systematically.

In what follows we describe the results for the distributions of moduli, cosmological constants, volumes and supersymmetry breaking scales for IIB flux vacua.
\begin{itemize}
	\item As seen at page \pageref{TIIBfluxconstr}, the fluxes are constrained by the condition $N_{flux}\leq L$, where $L$ denotes the contribution of the negative D3-charge objects and the 3-form fluxes D3-charge is given by $N_{flux}=\frac{1}{(2\pi)^4\alpha'^2}\int_X F_3\wedge H_3 $. The number of supersymmetric vacua in a region $\RR$ of dilaton and complex structure moduli space and satisfying $0\leq N_{flux}\leq L$ is given by:
	\bea
	 \NN_{susy} &=& \sum_{{\rm susy vac}\in\RR} \theta(L-N_{flux})  
	 	= \sum_{{\rm susy vac}\in\RR} \frac{1}{2\pi i}\int_{\mathcal{C}}\frac{d\alpha}{\alpha}\,e^{\alpha(L-N_{flux})}\nn\\
			&=& \frac{1}{2\pi i}\int_{\mathcal{C}}\frac{d\alpha}{\alpha}\,e^{\alpha L}\left(
				\sum_{{\rm susy vac}\in\RR} e^{-\alpha N_{flux}} \right)
	\eea
	As argued in the introduction of this chapter, we can approximate the discrete sum over the flux quanta by an integral. So the sum over all vacua can be written as:
	\bea
	  n(\alpha) &\equiv&  \sum_{{\rm susy vac}\in\RR} e^{-\alpha N_{flux}} \\
	  	&=& \int_\RR d^{2m}z\int d^{4m}N \, e^{-\alpha N_{flux}(N)} \delta^{2m}(DW)|\det D^2W|\nn	  
	\eea
	where $N$ denotes the 3-form fluxes, and $m=b_3/2$.\\
	Using this formula, $\NN_{susy}$ was estimated by \cite{Ashok:2003gk}, and the result is:
	\begin{equation}
 	  \NN_{susy}= \frac{(2 \sqrt{\pi} L)^{2m}}{(2m)!} \int_{\cal
 		R}  \det(R+\omega {\bf 1})
	\end{equation}
	where $R$ is the curvature form on the moduli space and $\omega$ the K\"ahler form.\footnote{Actually this expression gives an index rather than an absolute number: it counts vacua with signs, so it is strictly speaking a lower bound.} Essentially this expression implies that vacua are uniformly distributed over moduli space, except when the curvature part becomes important, which is the case near conifold degenerations. %The above result was verified by Monte Carlo experiments in \cite{Giryavets:2004zr, Conlon:2004ds}.
	\item The cosmological constant for supersymmetric vacua is $\Lambda = - 3 e^K
|W|^2$. Its distribution for values much smaller than the string scale was found in \cite{Denef:2004ze} to be essentially uniform, {\it i.e.}\
\begin{equation} \label{IIBccdistr}
 d\NN \sim  \NN_{\rm tot} \, d\Lambda.
\end{equation}
Here $\NN_{\rm tot} \sim L^{2m}/(2m)!$ is the total number of flux vacua.
	\item The compactification volume $V$ is stabilized by nonperturbative D3-instanton effects and/or gaugino condensates, 
both of which give contributions $\sim e^{-c V^{2/3}}$ to the superpotential, where $c<1$ decreases when $b_3$ increases. These have to balance against the contribution $W_0$ from the fluxes. That is, at sufficiently large $V$ (or equivalently sufficiently small $W_0$), $e^{-c V^{2/3}} \sim W_0$. Since $|W_0|^2 \sim \Lambda$ is uniformly distributed according to (\ref{IIBccdistr}), this gives the volume distribution 
\begin{equation}
 d\NN \sim \NN_{\rm tot} \, e^{-2 c V^{2/3}} d (V^{2/3}).
\end{equation}
Large volumes are therefore exponentially suppressed, and for reasonable values of $L$ and $b_3(=2m)$, the maximal volume will be of
the order $V_{\rm max} \sim \left(\frac{\log \NN_{\rm tot}}{2 c}\right)^{3/2} \sim (b_3/c)^{3/2}$.\footnote{One could also interpret the IIB complex structures fixed by the fluxes to be mirror to IIA K\"ahler moduli. Then the distribution of IIA compactification
volumes can be shown to be 
\begin{equation}
 d \NN \sim \frac{(k L)^{b_3}}{b_3!} \, d(V_{IIA}^{-b_3/6})
\end{equation}
for $V_{IIA}\gg 1$. Here $k$ is a constant weakly decreasing with increasing $b_3$. Again, large volumes are suppressed, now bounded
by $V_{IIA}<(e k L/b_3)^6$.} 

	\item The flux potential has nonsupersymmetric minima as well. The supersymmetry breaking scale 
$F=e^{K/2}|DW|=M_{\rm susy}^2$ for $F \ll 1$ is distributed as \cite{Denef:2004cf}
\begin{equation}
 d \NN \sim \NN_{\rm tot} \, dF
\end{equation}
if no further constraints are imposed, and 
\begin{equation}
 d\NN \sim \NN_{\rm tot} \, F^5 dF \, d\Lambda
\end{equation}
if one requires the cosmological constant $\Lambda$ to be much smaller than $F^2$. Scenarios in which supersymmetry breaking is
driven by D-terms were also considered in \cite{Denef:2004cf}, and it was pointed out that in the special case of supersymmetry
breaking by an anti-D3 brane at the bottom of a conifold throat, low scales are more natural. Since we work in the large radius regime, there is no counterpart of this scenario in the M-theory compactifications we will study, so we will not get into details
here. We should also point out that large classes of string compactifications have been proposed in \cite{Saltman:2004jh} where
supersymmetry is broken at the KK scale.
\end{itemize}

\subsection*{Freund-Rubin Statistics}

Freund-Rubin vacua of M-theory have geometry ${\rm AdS}_4 \times X$, with $X$ a positively curved seven dimensional Einstein manifold, and can be understood as arising from the near-horizon geometry of $N$ coincident M5-branes, which become $N$ units of $G$-flux in the AdS-space. The compactification volume is fixed and depends on $N$ and the choice of $X$. Typically, these geometries cannot really be considered as compactifications on $X$ in the usual sense, because the Kaluza-Klein scale tends to be of the same order as the AdS scale.

Nevertheless one can study the distributions of AdS cosmological constants $\Lambda$ and compactification volumes $V$. We did this in \cite{Acharya:2005ez} for a model ensemble with $X = X/\Zbb_k$, where we vary $k$ and $N$. This ensemble is extremely simple, and is therefore additionally useful as a simple toy model for testing counting methods.

At fixed $k$, {\it i.e.} for a fixed topology of the extra dimensions, we found the following distributions for $\Lambda$ and
$V$:
\begin{equation}
 d \NN(\Lambda) \sim d \Lambda^{-2/3}, \qquad d \NN(V) \sim d
 V^{6/7}.
\end{equation}
This already shows a dramatic difference compared to IIB flux vacua. First, there are an infinite number of vacua, since $N$ is arbitrary. Secondly, the distribution of $\Lambda$ is not uniform near zero, but diverges. Finally, large volumes are not suppressed, as the larger $N$, the larger the volume becomes.

Allowing both $k$ and $N$ to vary, {\it i.e.} by sampling the topology of the extra dimensions as well, these results significantly
change. Now
\begin{equation}
 d \NN(\Lambda) \sim d \Lambda^{-2}, \qquad d \NN(V) \sim d
 V^6.
\end{equation}
\begin{figure}
\begin{center}
  \epsfig{file=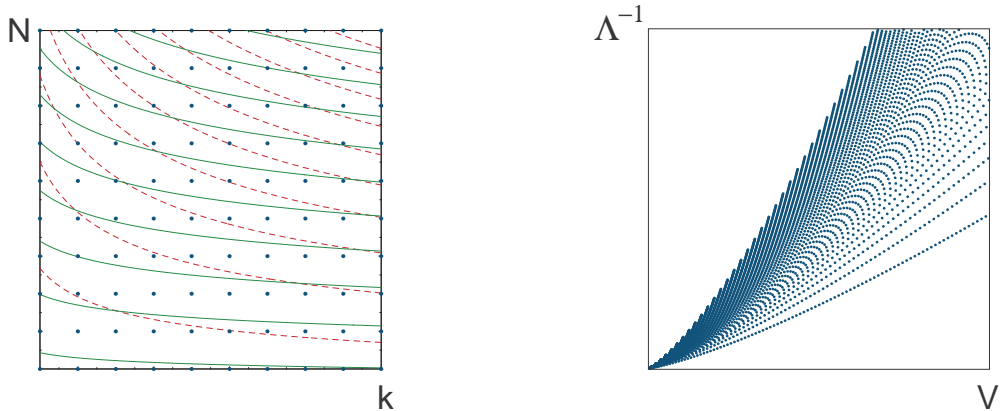,height=6.5cm,angle=0,trim=0 0 0 0}
  \caption{\footnotesize \emph{Left}: Lattice of vacua in $(k,N)$-space.
  The green solid lines have constant $V$ and the red dashed lines constant
  $\Lambda^{-1}$. Both are increasing with $N$ and $k$.
  \emph{Right}: Vacua mapped to $(V,\Lambda^{-1})$-space with $V \leq 200,\Lambda^{-1} \leq 2000$.
  The lower and upper boundaries correspond to $k=1$ resp.\ $N=1$.}
  \label{vacpoints}
\end{center}
\end{figure}
The qualitative features are the same though: smaller cosmological constants and larger volumes are favored.\footnote{Perhaps we should stress that by ``favored'' we do not mean ``more probable''. As emphasized earlier, we are computing \emph{number} distributions at this level, not probability distributions.} We also obtain the joint
distribution for $V$ and $\Lambda$:
\begin{equation}
 d\NN(V,\Lambda) \sim \Theta(V,\Lambda) \, \frac{dV}{V^4} \, \frac{d\Lambda}{\Lambda^4}
\end{equation}
where $\Theta(V,\Lambda)=1$ when $V^{-3} \leq \Lambda \leq V^{-9/7}$ and zero otherwise (see also fig.\ \ref{vacpoints}). One interesting feature that can be read off from this distribution is that at fixed $\Lambda$, $V$ actually accumulates at \emph{smaller} values, opposite to what we found for the unconstrained case. This is possible because the step function $\Theta$ allows $\Lambda$ to vary over a larger domain when $V$ increases. This illustrates the importance of constraints for statements about which parameters are favored. Such issues become especially important if one wishes to interpret number distributions as probability distributions, since through such correlations, the dependence of these probabilities on one parameter may strongly influence the likelihood of values of the other.

\section{$G_2$ holonomy Statistics}

As we have seen in the previous chapter, $G_2$ holonomy vacua are compactifications of M-theory to four dimensions which, in the absence of flux classically give four dimensional ${\cal N}=1$ vacua with zero cosmological constant. These classical vacua have $b_3(X)$ complex moduli, of which the real parts are axions $t_i$ and the other half $s_i$ are the massless fluctuations of the metric on $X$.

The addition of fluxes when $X$ is smooth does {\it not} stabilize these moduli, as the induced potential is positive definite and runs down to zero at infinite volume. However, if $X$ has an orbifold singularity along a three dimensional manifold $Q$, additional non-Abelian degrees of freedom arise from massless membranes \cite{Acharya:1998pm}. Non-Abelian flux for these degrees of freedom then gives an additional contribution to the potential which {\it can} stabilize all the moduli if $Q$ admits a complex, non-real Chern-Simons invariant \cite{Acharya:2002kv}.

The vacua studied in \cite{Acharya:2002kv} were supersymmetric with negative cosmological constant. In fact, it was shown that in the
large radius approximation, for a given flux within a certain range, there is a single supersymmetric vacuum (in addition to an unstable de Sitter vacuum). In principle however there could be other, non-supersymmetric, vacua and one of our aims here is to study this possibility. One might wonder if any of these vacua could be metastable de Sitter. We answered this question to a certain extent. 

One of the difficulties in studying $G_2$ holonomy compactifications is that $G_2$ holonomy manifolds are technically very difficult to produce. For instance, we still do not know whether or not there exists a $G_2$ holonomy manifold with a non-real Chern-Simons
invariant. So how can we hope to study the statistics of such vacua? As we have seen, the superpotential of these $G_2$ compactifications with flux is very simple and does not contain much information about $X$. Instead, this information comes through the K\"ahler potential on moduli space, which could be a quite complicated function in general, of which very little is concretely known. One approach then is to try to obtain general results for an arbitrary K\"ahler potential. We will show this in the next section by extending some of the general techniques which were developed in the context mainly of IIB flux vacua in \cite{Ashok:2003gk,Denef:2004ze,Denef:2004cf}. Secondly we could study particular ensembles of model K\"ahler potentials and hope
that the results are representative in general. We follow this approach in section \ref{sec:statmodel}, where we study a particular class of model K\"ahler potentials which allow explicit construction of all supersymmetric and nonsupersymmetric vacua.

\subsection{General Results}
\label{sec:general}

\subsubsection{$\mathbf{G_2}$ Compactifications with Fluxes and Chern-Simons Invariants}

We remind that the complexified moduli space $\MM$ of a $G_2$ holonomy compactification of M-theory has dimension $n=b^3(X)$ and has holomorphic coordinates $z^i$, defined by 
\begin{equation}
 z^i = t^i + i s^i = \int C + i \Phi,
\end{equation}
where $\Phi$ is the $G_2$-invariant 3-form on $X$. The metric on $\MM$ is K\"ahler, derived from the K\"ahler potential
%\footnote{In terms of $\ell_M$, we have $T_2 = {2 \pi / \ell_M^3}$, $T_5 = {2 \pi / \ell_M^6}$, and $\int G / \ell_M^3 \in \Zbb$. Thus, the instanton action is $e^{2 \pi i z}$ and $t$ has periodicity 1.} The metric on $\MM$ is K\"ahler, derived from the K\"ahler potential
\begin{equation} \label{Kpot2}
 K(z,\bz) = - 3 \ln \left( 4 \pi^{1/3} V_X(s) \right),
\end{equation}
where $V_X \equiv \mbox{Vol}(X)$ is a homogeneous function of the $s^i$ of degree $7/3$. 

We turn on 4-form flux
\begin{equation}
 G  = N_i \rho^i,
\end{equation}
where $N_i \in \Zbb$ and $\rho^i$ is a basis of $H^4(X,\Zbb)$, and also assume the presence of a complex Chern-Simons contribution as
described in section \ref{MthFluxSuperpotMdStab} . This induces the superpotential 
\begin{equation}
 W(z) =  N_i z^i + c_1 + i c_2
\end{equation}
on $\MM$, where $c_1$ and $c_2$ are the real and imaginary parts of the complex Chern-Simons invariant.

The corresponding potential is obtained from the standard four dimensional supergravity expression \eqref{effpot}:
\begin{equation} \label{potential}
 V =  e^K \left( g^{i\bj} F_i \bF_{\bj} - 3|W|^2 \right),
\end{equation}
where
\begin{equation}
 F_i \equiv D_{z^i} W \equiv (\partial_{z^i} + \partial_{z^i} K) W = N_i
 + \frac{1}{2i} \partial_{s^i} K \, W.
\end{equation}
Here and in what follows, we have put $\kappa_4=1$.

Since we are working in the large radius regime, the axions $t^i$ essentially decouple from the moduli $s^i$, and all nontrivial
structure resides in the latter sector. This is seen as follows. Writing
\begin{equation}
 W \equiv W_1(t) + i \, W_2(s), \qquad K_i \equiv \partial_{s^i} K(s),
 \qquad K_{ij} \equiv \partial_{s^i} \partial_{s^j} K = 4 g_{i \bj}
\end{equation}
and so on, the potential (\ref{potential}) becomes
\begin{eqnarray}
 V &=& e^K \left( 4 K^{ij} (\re F_i) (\re F_j) +  K^{ij} K_i K_j W_1^2 - 3 W_1^2 -
 3 W_2^2 \right) \nonumber \\
 &=& e^K \left(4 K^{ij} (\re F_i) (\re F_j) - 3 W_2(s)^2  + 4 W_1(t)^2 \right).  \label{Vsplit}
\end{eqnarray}
In the last line we used the fact that the volume $V_X$ is homogeneous of degree 7/3, which implies the following identities:
\begin{equation} \label{Kid}
 K_i s^i = -7, \qquad K_{ij} s^j = - K_i.
\end{equation}
The second is obtained from the first by differentiation.

Since $W_1 = N_i t^i + c_1$ and everything else in (\ref{Vsplit}) depends only on $s$, it is clear that any critical point of $V$ will
fix
\begin{equation}
 N^i t_i + c_1 = 0
\end{equation}
and therefore $W_1 = \im F =0$. Apart from this, the $t^i$ are left undetermined, and they decouple from the $s^i$. From now on we will work on this slice of moduli space, so we can write
\begin{equation} \label{Vreal}
 V = e^K \left( 4 K^{ij} F_i F_j  - 3 W_2^2 \right)
\end{equation}
with 
\begin{equation} \label{Fdef}
 F_i = D_{s^i} W_2 \equiv (\partial_{s^i} + \frac{1}{2} K_i) W_2 = N_i + \frac{1}{2} K_i (N_j s^j + c_2).
\end{equation}
The geometry of the real moduli space $\MM$ parametrized by the $s^i$ is the real analog of K\"ahler, often called Hessian, with
metric $K_{ij}$.

\subsubsection{Distribution of Supersymmetric Vacua over Moduli Space}
\label{sec:susydistr}

The aim of this section is to find the distribution of vacua over $\MM$, along the lines of \cite{Ashok:2003gk,Denef:2004ze,Denef:2004cf}. As seen before, the condition for a supersymmetric vacuum in the above notations is 
\begin{equation} \label{susycrit}
 D_{s^i} W_2(s) = 0
\end{equation}
In what follows we will drop the index $2$ to avoid cluttering. The number of solutions in a region $\RR$ of $\MM$, for all possible
fluxes $G$ but at fixed $c_2$, is given by 
\begin{equation}
 N_{\rm susy}(c_2,\RR)= \sum_{N \in \Zbb^n} \int_{\RR} d^n s \, \delta^n(D_i W) |\det
 \partial_i D_j W|.
\end{equation}
The determinant factor ensures every zero of the delta-function argument is counted with weight 1. As in Type IIB  we approximate expression for $N_{\rm susy}$ by replacing the discrete sum over $N$ by a continuous integral, which is a good approximation if the number of contributing lattice points is large (which will be the case for sufficiently large $c_2$). Thus
\begin{equation} \label{approxint}
 \NN_{\rm susy} = \int_{\RR} d^n s \int d^n N \, \delta^n(D_i W) |\det
 \partial_i D_j W|.
\end{equation}
Differentiating (\ref{Fdef}) and using (\ref{susycrit}) and (\ref{Kid}), one gets 
\begin{equation}
 \partial_i D_j W = \biggl( \frac{1}{2} K_{ij} - \frac{1}{4} K_i K_j \biggr)
 W = \frac{1}{2} K_{ik} \biggl( \delta^k_j + \frac{1}{2} s^k K_j
 \biggr) W
\end{equation}
One eigenvector of the matrix in brackets is $s^j$, with eigenvalue $-5/2$ (this follows again from (\ref{Kid})). On the orthogonal
complement $\{ v | K_j v^j = 0 \}$, the matrix is just the identity, so all other eigenvalues are $1$. Thus,
\begin{equation}
 |\det \partial_i D_j W| = \frac{5 |W|^n}{2^{n+1}} \det K_{ij}
\end{equation}
Furthermore, by contracting (\ref{Fdef}) with $s^i$, we get at $F_i=0$:
\begin{equation} \label{susyWval}
 W = -\frac{2}{5} c_2.
\end{equation}
To compute (\ref{approxint}), we change variables from $N_i$ to $F_i$. The Jacobian is
\begin{equation} \label{jac}
 {\rm Jac} =|\det \partial_{N_i} F_j|^{-1} = |\det (\delta^i_j + \frac{1}{2} K_j
 s^i)|^{-1} = 2/5.
\end{equation}
Putting everything together, we get 
\begin{eqnarray}
 \NN_{\rm susy} &=& \int_{\RR} d^n s \int d^n F \, \delta^n(F) \,
 \biggl(\frac{c_2}{5}\biggr)^n \det K_{ij} \\
 &=& \biggl(\frac{c_2}{5}\biggr)^n \int_{\RR} d^n s \det K \\
 &=& \biggl(\frac{4 c_2}{5}\biggr)^n \mbox{Vol}(\widehat{\RR})
 \label{susycount}
\end{eqnarray}
where $\widehat{\RR}$ is the part of the complexified moduli space projecting onto $\RR$ ({\it i.e.}\ the direct product of $\RR$ with the
$n$-torus $[0,1]^n$ swept out by the axions $t^i$), and the volume is measured using the K\"ahler metric $g_{i\bj} = K_{ij}/4$ on this
space.

Thus, supersymmetric vacua are distributed \emph{uniformly} over moduli space. This result is similar to the Type IIB orientifold case studied in \cite{Denef:2004ze}, but simpler: the Type IIB vacuum number density involves curvature terms as well, and in order
to get a closed form expression for $n>1$, it was necessary there to count vacua with signs rather than their absolute number.

Note that in any finite region of moduli space, the number of vacua is finite, because $c_2$ is finite despite the absence of a tadpole cutoff on the fluxes. In particular, the total number of vacua in the large radius region of moduli space (where our computation is valid) is finite. For IIB vacua on the other hand, finiteness is only obtained after imposing the tadpole cutoff $\int F \wedge H \leq L_*=\chi/24$ on the fluxes. In a way, the Chern-Simons invariant $c_2$ plays the role of $L_*$ here.

\

\noindent The simplest example is the case $n=1$. Then homogeneity fixes $V_X \sim s^{7/3}$, so $K_{ss}=\frac{7}{s^2}$, and
\begin{equation} \label{nisone}
 \NN_{\rm susy}(c_2,s \geq s_*) = \frac{c_2}{5} \int_{s_*}^\infty ds
 \, \frac{7}{s^2} = \frac{7}{5} \frac{c_2}{s_*}.
\end{equation}
Thus, vacua become denser towards smaller volume of $X$, and from this equation one would estimate there are no vacua with $s>7
c_2/5$, which is where $\NN_{\rm susy}$ drops below 1. Indeed, it is easily verified that the exact critical point solution is $s=-7c_2/5 N$, so the largest possible value of $s$, obtained at $N=-1$, is precisely $7 c_2/5$. Better even, from the explicit solution it easily follows that the approximate distribution (\ref{nisone}) becomes in fact exact by rounding off the right hand side to the nearest smaller integer.

\subsubsection{Large Volume Suppression} \label{sec:largevolsuppr}

Although the precise form of the metric for $n>1$ is unfortunately not known, one general feature is easy to deduce: flux vacua with
large compactification volume are suppressed, and strongly so if the number of moduli is large. This follows from simple scaling. If $s \to \lambda s$, $K$ shifts with a constant, so $K_{ij} \to \lambda^{-2} K_{ij}$ and the measure $d^n s \det K \to \lambda^{-n} \, d^n s \det K$. Thus we have for example
\begin{equation}
 \NN_{\rm susy}(c_2,s^i \geq s_*) = (c_2/s_*)^n \, \NN_{\rm susy}(1,s^i \geq 1),
\end{equation}
and
\begin{equation} \label{largevolsuppr}
 \NN_{\rm susy}(c_2,V_X \geq V_*) = k_n \left(c_2/V_*^{3/7}\right)^n,
\end{equation}
where $k_n$ is independent of $c_2$ and $V_*$.\footnote{Even though the obvious metric divergence at $V_X = 0$ is avoided by bounding
$V_X$ from below, it might still be possible that this bound alone does not determine a finite volume region in $s$-space. Then the
left hand side will be infinite and the scaling becomes meaningless. In this case, additional cutoffs should be imposed, which will
complicate the dependence on $V_*$, but large volume suppression is still to be expected.} Hence large compactification volume is
strongly suppressed when $n$ is large.

To get an estimate for the absolute number of vacua with $V_X \geq V_*$, one would need to estimate $k_n$ in (\ref{largevolsuppr}).
This would be quite hard in general even if the metric was explicitly known. 
It is expected that the volume has an upper bound as we confirm in section \ref{sec:statmodel}. In general, large extra dimension
scenarios with, say, micrometer scale compactification radii are therefore excluded in these ensembles unless $c_2$ is exceedingly
large.

\subsubsection{Distribution of Supersymmetric Cosmological Constants}

The vacuum energy in a supersymmetric vacuum is, in four dimensional Planck units, using (\ref{susyWval}):
\begin{equation} \label{susycc}
 \Lambda = -3 e^K |W|^2 = - \alpha \frac{c_2^2}{V_X^3}
\end{equation}
where $\alpha = 3 / 400 \, \pi \sim 0.002$. Therefore, the only way to get a small cosmological constant is to have large
compactification volume. This contrasts with the ensemble of type IIB flux vacua, where very small cosmological constants can be
obtained at arbitrary complex structure. The underlying reason is the fact that in IIB, there are
four times as many fluxes as equations $D_i W=0$, so at a given point, there is still a whole space of (real) fluxes solving the
equations. This freedom can be used to tune the cosmological constant to a small value. Here on the other hand, there is only one
flux per equation, so at a given $s$, the fluxes are completely fixed, and no freedom remains to tune the cosmological constant.
This would likely change however if more discrete data were turned on, such as the M-theory duals of IIA RR 2-form flux or IIB NS
3-form flux. Unfortunately these are difficult to describe systematically in M-theory in a way suitable for statistical analysis.

%In any case this illustrates the obvious fact that one must be
%careful to draw general conclusions from a single ensemble, even if
%it looks fairly general.

Let us compute the distribution of cosmological constants more precisely. Equation (\ref{susycc}) implies that this follows
directly from the distribution of volumes. Using (\ref{largevolsuppr}), we get 
\begin{equation}
 \NN_{\rm susy}(c_2,|\Lambda| \leq \lambda_*) = k_n \biggl(
 \frac{c_2^5 \lambda_*}{\alpha}
 \biggr)^{n/7}.
\end{equation}
The corresponding distribution density is therefore 
\begin{equation}
 d\NN/d\lambda \sim \lambda^{(n-7)/7}.
\end{equation}
In particular, for $n<7$, the distribution diverges at $\lambda=0$, while for $n>7$, the density goes to zero. We will see that one gets a lower bound for $|\Lambda|$. For large $n$, this turns out to be much larger than the naive $1/\NN_{\rm susy}$ which was found to be a good estimate in the Type IIB case, where the cosmological constants of supersymmetric vacua are always distributed uniformly near zero \cite{Denef:2004ze}.

Because $G_2$ manifolds with many moduli are much more numerous than those with only a few, we can thus conclude that small cosmological constants are (without further constraints) strongly suppressed in the ensemble of all supersymmetric $G_2$ flux vacua.

\subsubsection{Nonsupersymmetric Vacua}

A vacuum satisfies $V'=0$, and metastability requires $V''>0$. Thus, the number of all metastable vacua in a region $\RR$ is given by
\begin{equation} \label{Nvacnonsusy}
 N_{vac} = \sum_N \int_{\RR} d^n s \, \delta(V') \, |\det V''| \,
 \Theta(V'').
\end{equation}
In principle one could again approximate the sum over $N$ by an integral and try to solve the integral by changing to appropriate
variables, as we did for the supersymmetric case in section \ref{sec:general}, and as was done in \cite{Denef:2004cf} for
supersymmetry breaking scales well below the fundamental scale. In practice, we encounter some difficulties doing this for $G_2$
vacua.

By differentiating equation (\ref{Vreal}), one gets
\begin{equation}
 \partial_i V = e^K(8 (D_i D_j W) D^j W - 6 \, W D_i W)
\end{equation}
where $D_i$ denotes the Hessian Levi-Civita plus K\"ahler covariant derivative. The matrix $M_{ij} \equiv D_i D_j W$ is related to the
fermionic mass matrix, and with this notation the critical point condition becomes
\begin{equation} \label{nonsusycrit}
 M_{ij} F^j = \frac{3}{4} W F_i.
\end{equation}
In \cite{Denef:2004cf}, a similar equation was interpreted at a given point in moduli space as a linear eigenvalue equation for the
supersymmetry breaking parameters $F$, assuming the matrix $M$ (denoted $Z$ there) to be independent of $F$. This is indeed the
case for the Type IIB flux ensemble, but it is not true for the $G_2$ ensemble. The reason is again the fact that there are only as
many fluxes as moduli here, so a complete set of variables parametrizing the fluxes $N_i$ is already given by the $F_i$ (affinely related to the $N_i$ as expressed in (\ref{Fdef})). Therefore, all other quantities such as $W$ and $M_{ij}$ and so on
must be determined by the $F_i$. Indeed, a short computation gives:
\begin{eqnarray}
 W &=& -\frac{2}{5} (s^i F_i + c_2) \label{WF} \\
 M_{ij} &=& \frac{1}{2} \bigl(
 (K_{ij} - \frac{1}{2} K_i K_j) W + K_i F_j + K_j F_i - K^k_{ij} F_k
 \bigr) \label{MF}
\end{eqnarray}
This means that at a given point in moduli space, (\ref{nonsusycrit}) is actually a complicated system of quadratic equations in $F$, with in general only one obvious solution, the supersymmetric one, $F=0$. In total one can expect up to $2^n$ solutions for $F$. The nonsupersymmetric solutions will generically be of order $c_2$.\footnote{The physical supersymmetry breaking scale has an additional factor $e^{K/2}$.} In particular it is not possible to tune fluxes to make $F$ parametrically small, so it is not possible to use the analysis of \cite{Denef:2004cf} here, and there is no obvious perturbation scheme to compute (\ref{Nvacnonsusy}).

\

As a simple example we again take the case $n=1$. As noted in section \ref{sec:susydistr}, homogeneity then forces $K=-7 \ln s$, hence
$M_{ss} = (7 c_2 - 5 F s)/2s^2$, and (\ref{nonsusycrit}) (considered as an equation for $F$ at fixed $s$) has solutions $F = 0$ and
$F=\frac{14 \, c_2}{s}$. Neglecting the metastability condition, the continuum approximated number density of nonsupersymmetric vacua is given by
\begin{eqnarray}
 d\NN_{\rm nonsusy} &=& \int dN \, \delta(F-\frac{14\,c_2}{s}) \,
 |\partial_s (F - \frac{14\,c_2}{s})| \, ds \nonumber \\
 &=& \int \frac{2}{5} \, dF \, \delta(F-\frac{14\,c_2}{s}) \, \frac{35 \,
 c_2}{2 \, s^2} \, ds \nonumber \\
 &=& \frac{7 \, c_2}{s^2} \, ds.
\end{eqnarray}
We changed variables from $N$ to $F$ in the integral using the Jacobian (\ref{jac}). This result should be compared to the supersymmetric density implied by (\ref{nisone}):
\begin{equation} \label{nonsusyvsssuy}
 d\NN_{\rm nonsusy} = 5 \, d \NN_{\rm susy}.
\end{equation}
This does \emph{not} mean that for a given flux, there are on average five nonsupersymmetric critical points for every supersymmetric one. In fact, as we will see below, and as was already pointed out in \cite{Acharya:2002kv}, for $n=1$, there is exactly one nonsupersymmetric critical point for every supersymmetric one: the supersymmetric minimum is separated from $s=\infty$ by a barrier, whose maximum is the (de Sitter, unstable) nonsupersymmetric critical point. Equation (\ref{nonsusyvsssuy}) only expresses that \emph{in a given region} of moduli space (at large $s^i$), there are on average 5 times more nonsupersymmetric vacua then supersymmetric ones. This is simply because in the region of moduli space under consideration, the nonsupersymmetric critical
points are located at five times the value of $s$ of the supersymmetric critical points.

\subsubsection{Supersymmetry Breaking Scales}
\label{sec:gensusybreakingscales}

For more moduli, things become more complicated. A few useful general observations can be made though. If we require $V = 0$
(which remains a good approximation for what follows as long as $|V| \ll e^K c_2^2$) we can make a fairly strong statement about the
value of the supersymmetry breaking scale. Contracting (\ref{nonsusycrit}) with $s^i$ and using (\ref{WF})-(\ref{MF}) and
(\ref{Kid}), we get
\begin{equation}
 25 \, F^2 - 3 (s\cdot F)^2 - 8 \, c_2 (s \cdot F) = 0.
\end{equation}
Together with $V \sim 4 F^2 - 3W^2 =  0$, this gives a system of two equations in two variables, $s \cdot F$ and $F^2$, with solutions:
\begin{equation}
 F^2 = \frac{3 c_2^2}{4}, \qquad s \cdot F = \frac{3 c_2}{2}.
\end{equation}
The physical supersymmetry breaking scale for such vacua (assuming they exist) is
\begin{equation} \label{susyscalepred}
 M_{\rm susy}^2 \equiv m_p^2 \sqrt{4 e^K K^{ij} F_i F_j} = \frac{\sqrt{3} \, c_2 \, m_p^2}{8 \sqrt{\pi} \, V_X^{3/2}}
 \sim \frac{M_p^3 c_2}{m_p},
\end{equation}
where $m_p \equiv 1/\kappa_4$ and $M_p$ are the four and eleven dimensional Planck scales.

Let us plug in some numbers to get an idea of the implications. If we identify $M_p$ with the unification scale $M_{\rm unif} \sim
10^{16} \, {\rm Gev}$, this means that $M_{\rm susy} \sim \sqrt{c_2} \, 10^{14.5} \, {\rm GeV}$, and the gravitino mass $M_g \sim M_{\rm susy}^2/m_p \sim c_2 \, 10^{10} \, {\rm GeV}$. Since $c_2$ is at least of order 1, this estimate implies (under the given
assumptions) that in this ensemble supersymmetry is always broken at a scale much higher than what would be required to get the
electroweak scale $M_{\rm ew} \sim 100 \, {\rm GeV}$ without fine tuning.

In fact a slight extension of this calculation shows that even if one allows the addition of an arbitrary constant to the potential to
reach $V=0$ (e.g.\ to model D-terms or contributions from loop corrections), the supersymmetry breaking scale in this ensemble is
still bounded from below by the scale $c_2 M_p^3/m_p$.

One could of course question the identification $M_p \sim M_{\rm unif}$. Lowering the 11d Planck scale down to $M_p \sim 10^{13} \,
{\rm Gev}$ for example (which requires $V_X \sim 10^{12}$) gives a supersymmetry breaking scale $M_{\rm susy} \sim \sqrt{c_2} \,
10^{10} \, {\rm Gev}$ and gravitino mass $m_g \sim c_2 \, 10 \, {\rm Gev}$, which for $c_2$ not too large would give low energy
supersymmetry.

Note however that in analogy with the supersymmetric case, and based on general considerations, we expect suppression of vacua at large volume and therefore also suppression of low energy supersymmetry breaking scales in this ensemble. This is further confirmed by the exactly solvable models we will present in section \ref{sec:statmodel}. In particular we get a lower bound on the
supersymmetry breaking scale from the upper bound on the volume $V_X$, which in general we expect to be of the form $V_X < (c_2/r_n)^{7/3}$, with $r_n$ weakly growing with $n$. Using the relation (\ref{susyscalepred}) between
$M_{\rm susy}$ and $V_X$, this implies the following lower bound for the supersymmetry breaking scale 
\begin{equation} \label{susybreakingbound}
 M_{\rm susy}/m_p > \frac{r_n^{7/4}}{c_2^{5/4}}.
\end{equation}
Getting $M_{\rm susy}$ below $10^{12} \, {\rm Gev}$ in the case of many moduli would thus require $c_2$ to be at least of order $10^6$.

Using the volume distribution (\ref{largevolsuppr}), we furthermore get an estimate for the distribution of supersymmetry breaking
scales (in four dimensional Planck units): 
\begin{equation} \label{Fdistr}
 d\NN_{\rm nonsusy} \sim \tilde{k}_n \, c_2^{5n/7} d(M_{\rm susy}^{4n/7})/m_p^{4n/7},
\end{equation}
where $\tilde{k}_n$ is a constant independent of $c_2$ and $M_*$.

Here we have not yet taken into account the tuning required to get a tiny cosmological constant: $|\Lambda| \sim |4F^2-3W^2|/V_X^3<
\Lambda_*$. For a given volume $V_X$ (or equivalently a given supersymmetry breaking scale), this requires tuning the fluxes such
that $|4F^2-3W^2|<\Lambda_* V_X^3$, which can be expected to at least add another suppression factor $\Lambda_* V_X^3 \sim
\Lambda_*/M_{\rm susy}^4$. The suppression may in fact be stronger, if the distribution of cosmological constants is not uniform but
more like a Gaussian sharply peaked away from $\Lambda=0$. Such distributions are quite plausible in these ensembles, as will be
illustrated by the model ensemble we will study in section \ref{sec:statmodel}. Another potentially important factor which we
are neglecting in this analysis is the metastability constraint (this was found in \cite{Denef:2004cf} to add another factor $M_{\rm
susy}^4$ to the distribution in the ensembles studied there).

%On the other hand since there are up to $2^n$ nonsupersymmetric
%solutions $F$ to (\ref{nonsusycrit}), one could expect an additional
%$2^n$ enhancement. But naively at least one could expect this factor
%to be largely canceled by requiring metastability (at positive
%$V$).\footnote{In \cite{Denef:2004cf} it was noted that the naive
%$2^{-n}$ suppression from the metastability constraint was actually
%not present for the ensembles studied there. However this relied on
%being in a regime with small $F$, which is not the case here.} At
%any rate, this factor would only effectively multiply $c_2$ in
%(\ref{Fdistr}) by $2^{7/5} \sim 2.6$, which is not significant given
%the uncertainty in $c_2$ values.

Finally, when one also takes into account the observed value of the electroweak scale $M_{\rm ew}$, there is an additional expected
tuning factor presumably of order $M_{\rm ew}^2 m_p^2 / M_{\rm susy}^4$ (in the region of parameter space where this is less than
1) \cite{Susskind:2004uv,Douglas:2004qg,Dine:2004is}. Putting everything together, this gives (for $m_p^2 > M_{\rm susy}^2 >
M_{\rm ew} m_p$):
\begin{equation}
 d \, \NN \sim \tilde{k}_n \, \Lambda_* \, M_{\rm ew}^2 \, c_2^{5n/7}
|d \, M_{\rm susy}^{4(n/7-2)}|/m_p^{4n/7-2}
\end{equation}
So we see that for $n<14$, the Higgs mass and cosmological constant tunings tilt the balance to lower scales, while for $n>14$ higher
scales are favored, and strongly so if $n$ is large. For $n<14$ we should keep in mind however that there is an absolute lower bound on the supersymmetry breaking scale, given by (\ref{susybreakingbound}), which will further be increased by the additional tunings of cosmological constant and Higgs mass. We should also not forget that this is only a naive analysis; in principle a full computation of the measure should be done along the lines of \cite{Denef:2004cf}, but as we discussed, this does not seem possible in the present context, because of the absence of a small parameter.

Nevertheless, the above consideration indicate clearly that low energy supersymmetry is typically disfavored in $G_2$ flux
ensembles, and even excluded if $c_2$ is less than $10^6$.

\subsection{Model K\"ahler Potentials and Exact Solutions} \label{sec:statmodel}

In the previous section we gave a number of general results about distributions of $G_2$ flux vacua, independent of the actual
form of the K\"ahler potential. For nonsupersymmetric vacua the results were less detailed, mainly because the constraint $V'=0$ is
quadratic in $F$, and, unlike the situation in \cite{Denef:2004cf}, no regime exists in which the equations can be linearized. To make
further progress, we study a class of model K\"ahler potentials for which all vacua can be computed explicitly.

In general, at large volume, the K\"ahler potential is given by (\ref{Kpot}): $K = -3 \log (4 \pi^{1/3} V_X)$, where $V_X$ is the
volume of $X$ regarded as a function of the moduli $s_i$. Unlike the case of a Calabi-Yau, where the volume function is always a third
order homogeneous polynomial in the K\"ahler moduli, no strong constraints on $V_X$ are known for $G_2$ holonomy manifolds, just
that the volume function is homogeneous of degree $7/3$ and that minus its logarithm is convex, {\it i.e.} the second derivative of $K$, which gives the kinetic energies of the moduli, is positive definite. In general it is difficult to find simple candidate volume
functions which satisfy this positivity constraint. The most general homogeneous degree $7/3$ function is of the form
\begin{equation}\label{RFvolgen}
   V_X =\prod_{k=1}^n s_k^{a_k} f(s_i)
\end{equation}
with $\{a_k\}$ such that 
\begin{equation}
 \sum_{k=1}^n a_k=\frac{7}{3}.
\end{equation}
and $f(s_i)$ invariant under scaling. If we now suppose that we are in a region of moduli space where $f(s_i)$ is approximately constant then we can take 
\begin{equation}\label{RFvol}
   V_X =\prod_{k=1}^n s_k^{a_k},
\end{equation}
and this in fact gives a positive moduli space metric. This justifies this particular choice of K\"ahler potentials.

The above choice of $V_X$ gives a simple geometry to the moduli space which is quite natural. The K\"ahler metric is
\begin{equation}
   ds^2 = \sum_{i=1}^n \frac{3a_i}{4s_i^2} dz_i d\bar{z}_i
        = \sum_{i=1}^n \frac{3a_i}{4s_i^2} (dt_i^2+ds_i^2)
\end{equation}
This is locally the metric of the product of $n$ hyperbolic planes $\mathbb{H}^2$, which is:
\begin{equation}
   ds^2 = \frac{1}{\ell^2 x^2} (dx^2+dy^2)
\end{equation}
where $\ell$ is connected to the curvature tensor by
\begin{equation}
   \hat{R}_{12}=-\ell^2 e_1\wedge e_2
\end{equation}
So locally the moduli space is $\mathbb{H}^{2n}$. Globally it is given by $\mathbb{H}^{2n}/\mathbb{Z}^n$, because the axions $t_i$
are periodic variables. In this class of $n$-parameter K\"ahler potentials labeled by $a_i$, all the information about $X$ is
contained in the values of the $a_i$. Since the moduli space metric (equivalently moduli kinetic terms) is singular if $a_i = 0$ and the moduli have the wrong sign kinetic terms if $a_i < 0$, we take $a_i > 0$. We will not restrict to any other particular values for the $a_i$ if it is not necessary to do so\footnote{To find examples which realize these K\"ahler potentials, consider the case $n=7$ and $a_i = 1/3$, Then this K\"ahler potential correctly describes the seven radial moduli of $X = T^7$ and certain orbifolds thereof
\cite{Lukas:2003dn}.}.

The potential on the moduli space is given by (\ref{Vsplit}): 
\begin{equation}\label{RpotentialF}
   V = \frac{c_2^2}{48 \pi \, V_X^3} \biggl(
                3+\sum_{j=1}^n a_j \nu_j s_j(\nu_j s_j -3) \biggr)
        +\frac{1}{48 \pi\, V_X^3}(\vec{N}\cdot\vec{t}+c_1)^2.
\end{equation}
where $\nu_j \equiv -\frac{N_j}{c_2 a_j}$.

\subsubsection{Description of the Vacua} \label{sec:descrvac}

We now describe the vacua {\it i.e.} the critical points of $V$. The equations for the axions give
\begin{equation}\label{RSolEq2}
   \vec{N} \cdot \vec{t}+c_1=0
\end{equation}
which fixes this particular linear combination of axions. We do not concentrate on fixing the remaining axions, since they are compact fields and are fixed by any non-perturbative corrections. Our interest is in the moduli $s_i$. The equations of motion for the $s_i$ reduces to a system of $n$ quadratic equations. For the case at hand these are equivalent to:
\begin{equation}\label{RSolEq3}
   \sum_{j=1}^n 3a_j h_j(h_j-3)-2h_i^2+3h_i+9=0,
\end{equation}
where we defined $h_j\equiv \nu_j s_j$  (no sum). Note that this system separates in $n$ quadratic equations in one variable $h_i$. The solutions are therefore of the form 
\begin{equation} \label{solhi}
 h_i = \frac{3}{4} + m_i H
\end{equation}
where $m_i=+1$ or $-1$, and $H$ is determined by substituting this in (\ref{RSolEq3}). This results in a single quadratic equation:
\begin{equation}\label{RSolEq8}
    5 H^2 -\frac{9}{2} A H -\frac{27}{16}=0.
\end{equation}
where 
\begin{equation}
 A \equiv \vec{a}\cdot\vec{m}.
\end{equation}
A priori therefore, $H$ can take two possible values: 
\begin{equation}\label{RSolEq12}
   H^{\pm}_{(\vec{m})}=\frac{3}{20} \left( 3A \pm
                   \sqrt{9 A^2+15} \right)
\end{equation}
However, because
\begin{equation}
   H^+_{(\vec{m})}= - H^-_{(-\vec{m})}
\end{equation}
we only ever need to consider, say, the negative branch of the square root to get all solutions in (\ref{solhi}). In total, therefore, the number of vacua for a fixed choice of fluxes is $2^{n}$.
%Equivalently we could consider both branches of the square
%root but all $h_i = {3 \over 4} + H$. F: I see what you mean,
% but this may be confusing since we introduced H as independent of i.
We choose the following parametrization: take all $2^n$ choices for $\vec{m}$. Then
\begin{equation} \label{hsol}
 h^{(\vec m)}_i = \frac{3}{4} + m_i H_{(\vec m)}
\end{equation}
with $H \equiv H^-$.

We can thus think of the vacua as the states of a system with $n$ ``spins'' $m_i$. When all spins are aligned with the ``external
field'' $\vec{a}$, that is if all $m_i=+1$, we have
\begin{equation}
 A = 7/3, \quad h_i = \frac{3}{5}
\end{equation}
When all spins are anti-aligned ($m_i=-1$), this becomes
\begin{equation}
 A = -7/3, \quad h_i = 3.
\end{equation}
The first of these can be shown to be the supersymmetric AdS vacuum discussed in \cite{Acharya:2002kv} whilst the second is the unstable de Sitter vacuum also discussed there. The remaining $2^n-2$ are all non-supersymmetric and could be either de Sitter or anti de Sitter. The metastability of these vacua will be analyzed in a following section.

\begin{figure}
\begin{center}
  \epsfig{file=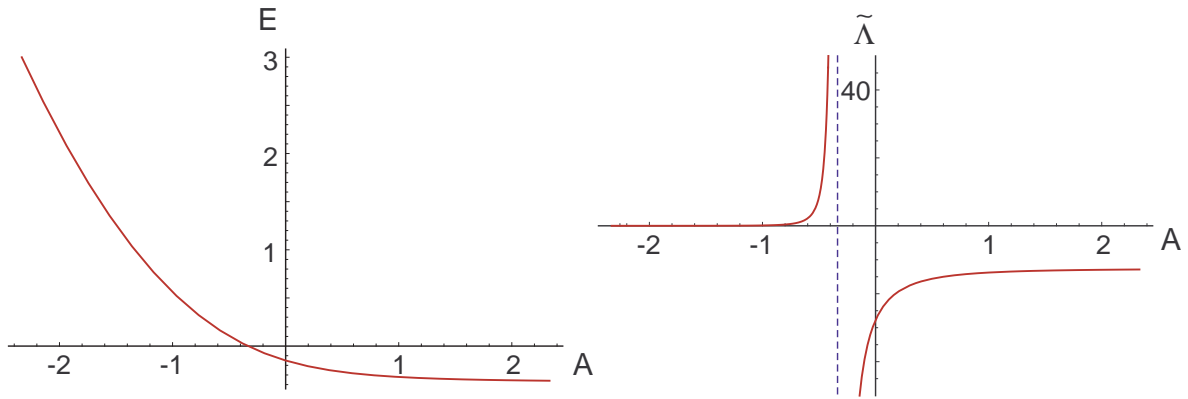,height=6cm,angle=0,trim=0 0 0 0}
  \caption{\small \emph{Left}: $E$ as a function of $A \equiv
  \vec{a}\cdot\vec{m}$. Here $E(-7/3)=3$ and $E(7/3)=-9/25$.
  \emph{Right}: dependence of $\tilde{\Lambda}$ on $A$. At $A=-7/3$, $\tilde{\Lambda} \approx 10^{-3}$,
  and at $A=7/3$, $\tilde{\Lambda} \approx -13$. The divergence at $A=-1/3$ is due to the vanishing of
  the volume there.}
  \label{vacE}
\end{center}
\end{figure}
Substituting (\ref{hsol}) in (\ref{RpotentialF}), we get that the energy of these vacua is given by $V = \frac{c_2^2}{48 \pi V_X^3} E$
with\footnote{The normalization is chosen such that $E$ equals the term inside the big brackets in (\ref{RpotentialF}). Also, $c_2^2 E
= \frac{3}{4} \left(|F|^2 - 3|W|^2\right)$.} $E = \frac{2}{3} H^2 - \frac{3}{8}$. At fixed volume, the vacuum energy varies only through $E$. The dependence of $E$ on $A$ is shown on the left in fig.\
\ref{vacE}. However, the volume depends on $\vec{m}$ as well:
\begin{equation} \label{volexpr}
 V_X = \prod_{i=1}^n s_i^{a_i} = \prod_{i=1}^n \left( \frac{c_2
 a_i}{|N_i|} \right)^{a_i} \prod_{i=1}^n |h_i|^{a_i},
\end{equation}
so the total vacuum energy is, up to $\vec{m}$-independent factors:
\begin{equation} \label{tildeLambda}
 \Lambda \sim \tilde{\Lambda} \equiv E \, \prod_{i=1}^n |h_i|^{-3a_i} =
 E \, \bigl|\frac{3}{4} + H\bigr|^{-\frac{7}{2}-\frac{3A}{2}} \,
 \bigl|\frac{3}{4} - H\bigr|^{-\frac{7}{2}+\frac{3A}{2}}.
\end{equation}
The dependence of this on $A$ is shown on the right in fig. \ref{vacE}. The divergence at $A=-1/3$ is due to the vanishing of
the volume there, as all $h_i$ with $m_i=+1$ vanish at this point. Obviously the supergravity approximation breaks down in this regime. One notable fact is further that the smallest positive cosmological constant is obtained when all spins are down, while the smallest negative cosmological constant (in absolute value) is obtained when all spins are up, {\it i.e.} at the supersymmetric critical point.

\begin{figure}
\begin{center}
  \epsfig{file=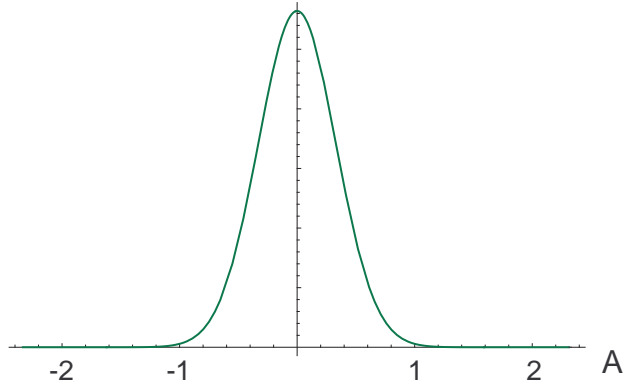,height=6cm,angle=0,trim=0 0 0 0}
  \caption{\small Distribution of $A$ values for $a_i=7n/3$, $n=50$.}
  \label{Adistrplot}
\end{center}
\end{figure}

At large $n$, the vast majority of vacua will be ``halfway'' the extrema. More precisely, if say all $a_i = a = 7/3n$, the variable
$A$ will be binomially distributed around $A=0$, as illustrated in \ref{Adistrplot}. At large $n$ this distribution asymptotes to the
continuous normal density
\begin{equation} \label{Adistr}
 d \NN[A] \approx \frac{2^n}{\sqrt{2\pi} \sigma} \exp\bigl(-\frac{A^2}{2 \sigma^2}\bigr) \,
 dA; \qquad \sigma=\frac{7}{3 \sqrt{n}}.
\end{equation}
For large $n$ this is sharply peaked around $A=0$, for which $E=-3/20$ and $|F|^2 = \frac{119}{80} c_2^2$, so the majority of
vacua are AdS and break supersymmetry at a scale $M_{\rm susy}^2 = |F|/V_X^{3/2} \sim c_2 M_p^3/m_p$.

Thus far we have considered solutions in terms of $h_i$. The actual values of the moduli are given by 
\begin{equation} \label{sfromh}
 s_i = -\frac{c_2 a_i h_i}{N_i}.
\end{equation}
Since the moduli fields $s_i$ are {\it positive} in the supergravity approximation, the signs of the $h_i$ and flux quanta $N_i$ must be correlated. Without loss of generality we take $c_2$ to be positive. Then $N_i$ and $h_i$ must have opposite signs. Now as long as
$A>-1/3$, every $h_i$ is automatically positive, so any such vacuum must have negative $N_i$. When $A<-1/3$, $h_i$ is positive if
$m_i=-1$ and negative if $m_i=+1$, so these vacua must have $\mbox{sign } N_i = \mbox{sign } m_i$. Recall that the condition $A<-1/3$ is also the condition to have $E>0$. 

Thus, for any given $\vec{m}$, there is a unique choice of sign for each $N_i$ which renders all $s_j$ positive for all choices of
$|N_i|$.

On the other hand, not all given, fixed fluxes $N$ gives rise to the same number of vacua. The following cases can be distinguished:
\begin{itemize}
 \item All $N_i<0$: set of vacua = $\{ \vec{m} | A \equiv \vec{a}\cdot\vec{m}>-1/3 \} \cup \{ (-1,-1,\cdots,-1)
 \}$. All vacua in the first set are AdS. The additional one is dS
 (but is unstable, as we will discuss in the next section).
 There are of order $2^n$ such vacua (between $2^{n-1}$ and
 $2^n$ for example when all $a_i$ are equal as for the distribution
 of fig.\ \ref{vacE}).
 \item Some $N_i>0$, and $\vec{a} \cdot \mbox{sign}(\vec{N}) < -1/3$: just one
 vacuum, given by $m_i = \mbox{sign}(N_i)$. This vacuum is dS (but
 again  will turn out to be unstable).
 \item Some $N_i>0$, and $\vec{a} \cdot \mbox{sign}(\vec{N}) > -1/3$: no vacua.
\end{itemize}

As noted before, the choices of $\vec{m}$ for which $A$ is at or near $-1/3$ do not correspond to vacua within the region of validity
of our computations for reasonable values of $c_2$, because some of the moduli, and hence the volume, will be at or near zero then.

%Then if the flux quantum $N_j < 0$ we see find that positivity  of
%$h_j$ requires $A > -{1 \over 3}$ if $m_j = 1$ whereas
%$h_j$ is automatically positive if $m_j = -1$. If {\it all} the
%fluxes are negative then these solutions include the supersymmetric
%case $(h_i = {3 \over 5})$ and the de Sitter maximum $(h_i = 3)$.
%Apart from the de Sitter maximum, all of these `same sign' flux
%vacua have a negative cosmological constant.

%In the generic situation of mixed sign flux the supersymmetric
%vacuum is absent and positivity of the moduli vevs requires
%$A < -{1 \over 3}$ for those negative $h_j$ with $m_j =
%1$. Since $h_j$ can never be negative if $m_j = -1$, in these cases
%the number of solutions (for a fixed choice of mixed sign fluxes)
%with positive moduli is $1$, where $\vec{m}$ is simply given by the
%vector whose entries are the signs of the fluxes. All these mixed
%sign solutions have a positive cosmological constant but are
%unstable since the radial modulus whose vev sets the overall volume
%of $X$ is at a maximum. Therefore we will restrict attention to the
%same sign flux vacua with $N_j < 0$.

%This does not affect our analysis, because we concentrate only on
%points corresponding to $A>-\frac{1}{3}$.

\subsubsection{Stability Analysis} \label{sec:metastab}
 
The sign of the potential for a particular solution depends on the
sign of $E$, with $E$ as defined in section \ref{sec:descrvac}. It
is negative when $A \equiv \vec{a}\cdot\vec{m}>-1/3$ and it is
positive when $A<-1/3$. We discuss these cases separately.

\begin{description}

\item {\bf [AdS vacua]}
An AdS critical point $s^0$ does not have to be a local minimum to
be perturbatively stable. It suffices that the eigenvalues of the
Hessian of $V$ are not too negative compared to the cosmological
constant, and more precisely that Breitenlohner-Freedman bound is
satisfied \cite{Breitenlohner:1982bm}:
\begin{equation} \label{RStabCond1app}
   \hat{\partial}_{i}\hat{\partial}_{j} V(s^0)-\frac{3}{2} V(s^0) \delta_{ij} \geq
   0,
\end{equation}
{\it i.e.} this matrix should be positive definite.

The derivatives are done with respect to the canonically normalized
scalars. The relevant kinetic term expanded around the critical
point $s^0$ is (in four dimensional Planck units):
\begin{eqnarray}
   g_{i\bar{j}}\partial_\mu z^i\partial^\mu \bar{z}^{\bar{j}} &=&
   \sum_{i}\frac{3a_i}{4(s_i^0)^2}(\partial_\mu t_i\partial^\mu t_i
   +\partial_\mu s_i\partial^\mu s_i) + \cdots
   \\
   &&\equiv \sum_{i} \frac{1}{2}(\partial_\mu \hat{t}_i\partial^\mu \hat{t}_i
   +\partial_\mu \hat{s}_i\partial^\mu \hat{s}_i) + \cdots
\end{eqnarray}
Hence the canonically normalized scalars are
\begin{equation}
 \hat{s}_i= \sqrt{\frac{3a_i}{2}}\,\frac{s_i}{s_i^0}.
\end{equation}
With this redefinition, the condition (\ref{RStabCond1app}) becomes
\begin{equation}\label{RStabCond2}
      \frac{2}{3} \frac{s_i^0 s_j^0}{\sqrt{a_i a_j}}
      \partial_i\partial_j V(s^0)-\frac{3}{2}V(s^0) \delta_{ij} \geq 0
\end{equation}
where
\begin{eqnarray}
   \partial_i\partial_j V&=&\frac{c_2^2 }{48\pi V_X^3}\left(\frac{3a_ia_j}{s_i s_j}
                           \left(3 E -2(\nu_i^2 s_i^2+\nu_j^2 s_j^2)
               +3(\nu_i s_i+\nu_js_j)\right)\right.\nonumber\\
               &&+\left.\frac{a_i}{s_i^2}\left(3 E
                 + 2\nu_i^2 s_i^2\right)\delta_{ij}\right)\\
   V           &=&\frac{c_2^2 }{48\pi V_X^3} E\nonumber\\
   E       &=& 3+\sum_{j=1}^n a_j \nu_j s_j(\nu_j s_j -3) \nonumber
\end{eqnarray}
By substituting these expressions in (\ref{RStabCond2}), using the
variables $h_i=\nu_i s_i$ and factoring out the common positive term
$\frac{c_2^2}{48\pi V_X^3}$, the stability condition becomes:
\begin{equation}\label{RStabCond3}
   M_{ij}\equiv \sqrt{a_i a_j}(6 E -4(h_i^2 +h_j^2)
      +6(h_i +h_j)) + \delta_{ij} (\frac{E}{2}
    +\frac{4}{3}h_i^2)\geq 0
\end{equation}
where
\begin{equation}
       E=\frac{2}{3}H^2-\frac{3}{8} = -\frac{3}{100}\left(5-9 A^2
                     + 3 A \sqrt{9A^2 +15}\right).
\end{equation}
Let us evaluate (\ref{RStabCond3}) on our solutions.
\begin{itemize}
  \item supersymmetric solution:
        \begin{equation}
       \mathbf{M}=\frac{54}{25} \, \mathbf{q}+\frac{3}{10} \, \mathbf{1}
    \end{equation}
    where $q_{ij}=\sqrt{a_ia_j}$ is a rank$=1$ positive definite matrix.
    We see that $\mathbf{M}\geq0$ and so that the supersymmetric solution is
    always stable.
  \item Other AdS ($A>-1/3$) solutions:
        \begin{equation}
       \mathbf{M}=Q \, \mathbf{q}+S \, \mathbf{1}+T \, \mathbf{t}
    \end{equation}
    where
    \begin{equation}
        Q = -6 E, \qquad
        S = \frac{E}{2}+\frac{4}{3} \left(H+\frac{3}{4}\right)^2, \qquad
        T = -4H
    \end{equation}
    \begin{equation}
        t_{ij}=\frac{(1-m_i)}{2} \delta_{ij}.
    \end{equation}
    The following relations are valid: $Q>0$, $T>0$ and $S+T>0$;
    so $Q\,\mathbf{q}$ and $T\,\mathbf{t}$ are always positive definite.
    $S$ has no definite sign. We have two cases:
    \begin{enumerate}
      \item when $A\geq \frac{1}{3}$, $S\geq 0$; then
        $S \, \mathbf{1}$ and consequently $\mathbf{M}$ is positive,
        and the corresponding solutions are perturbatively stable;
      \item when $-\frac{1}{3}< A  <\frac{1}{3}$, $S<0$; in this case the
            $\mathbf{M}$ matrix is not positive definite when
        $\vec{m}$ has more than one entry equal to $+1$ and the
        corresponding solutions are not stable.
        Actually, if $m_h=+1$ and $m_k=+1$, $\mathbf{M}$ has the $2\times 2$ minor
        \begin{equation}
           Q\left(\begin{array}{cc}
            a_h &\sqrt{a_ha_k}\\ \sqrt{a_ha_k}& a_k
                  \end{array}\right) +
           \left(\begin{array}{cc}
            S & 0\\ 0 & S
                  \end{array}\right)
        \end{equation}
        which has $S<0$ as one of the two eigenvalues.\\
        On the other hand, when $\vec{m}$ has exactly one entry equal to $+1$, the solution
        is a local minimum and so it is stable.
    \end{enumerate}
\end{itemize}

Let us prove that when $A> -\frac{1}{3}$ and exactly one $m_i$ is
equal to $+1$, the corresponding solution is a local minimum.

    Without loss of generality we choose $i=1$.
    $\partial^2 V$ is proportional to
    $\mathbf{\hat{M}} \equiv \mathbf{q}+\frac{S'}{Q}\mathbf{1}+\frac{T}{Q}\mathbf{t}$, where
    $S' \equiv S+\frac{3}{2}E$ is negative for all considered values of $A$.
    Let us call $B=\frac{S'+T}{Q}>0$, $C=-\frac{S'}{Q}>0$, $v_i=\sqrt{a_i}$.
    So $|v|^2=7/3$, $v_1>1$ and
    \begin{equation}
       \mathbf{\hat{M}}= v\,v^t + \left(\begin{array}{cccc}
                    -C & & & \\ &B& &\\ & &\ldots&\\ & & & B
                      \end{array}\right)
    \end{equation}
    By direct study of its eigenvalues, it can be shown that this matrix is positive definite for
    $n=2$. Now we prove by induction that this matrix is positive definite for each
    $n$.

    Let us assume that $\forall y\in \Rbb^n$
    \begin{equation}
       y^t \mathbf{\hat{M}}_n y = (v_1 y_1 + \tilde{v}\cdot\tilde{y})^2+ B\tilde{y}^2-C y_1^2>0
    \end{equation}
    where $V=(V_1,\tilde{V})$.
    Let $u$ be the extension of $v \in \Rbb^n$ to $\Rbb^{n+1}$, {\it i.e.}\ $u_i=\sqrt{a_i}$. Then
    $\forall x\in\Rbb^{n+1}$ there exists $y\in\Rbb^n$, such that
    \begin{eqnarray}
       y_1 =x_1 & \tilde{y}\cdot\tilde{v} = \tilde{x}\cdot\tilde{u}&\tilde{y}^2= \tilde{x}^2
    \end{eqnarray}
    It follows that
    \begin{equation}
       x^t \mathbf{\hat{M}}_{n+1} x = (u_1 x_1 + \tilde{u}\cdot\tilde{x})^2+ B\tilde{x}^2-C x_1^2>0\:.
    \end{equation}
\item 

\item {\bf [dS vacua]}
The solutions with positive potential ({\it i.e.} $A<-1/3$) must be local
minima to be metastable, and so the matrix $\mathbf{M'} \equiv Q \,
\mathbf{q}+S' \, \mathbf{1}+T \, \mathbf{t}$ (in the notations
introduced previously) must be positive
definite. We now show that this is never the case.

We have $\mathbf{q}=v v^t$, where $v_i=\sqrt{a_i}$ and so
$|v|^2=7/3$. Let us consider the $n\times n$ matrix defined by
$\mathbf{M''} \equiv Q \, \mathbf{q}+S'\, \mathbf{1} + T \,
\mathbf{1}$. If $\mathbf{M'}$ is positive definite then
$\mathbf{M''}$ is positive too. Clearly, its eigenvalues are $S'+T$,
which is positive for $A<-1/3$, and
\begin{equation}
   \lambda=Q |v|^2+S'+T=\frac{7}{3}Q+S'+T.
\end{equation}
By studying $\lambda$ as a function of $A$, one finds that it is
positive for $A >-17/21$. So $\mathbf{M''}$ is not positive definite
for $A<-17/21 \approx -0.8095$ and so $\mathbf{M'}$ is not positive
for the same values.

Moreover one can study the $k\times k$ submatrix of $\mathbf{M'}$
obtained by restriction to the subspace on which $m_j=+1$ ($k$ is
the number of such $m_j$'s). This is equal to $Q \, v v^t+S'\,
\mathbf{1}_{k}$, where we are considering only $v_j$ corresponding
to $m_j=+1$. Its eigenvalues are $S'$, which is positive for
$A<-1/3$, and
\begin{equation}
   \mu=Q \sum_{i=1}^n \frac{(1+m_i)}{2}a_i + S'=
   Q(\frac{7}{6}+\frac{A}{2})+S'.
\end{equation}
By studying $\mu$ as a function of $A$, one finds that it is negative
for $\alpha<A<-\frac{1}{3}$, where $\alpha \approx -1.44075$. For
these values of $A$ we have a nonpositive minor and so $\mathbf{M'}$
is not positive definite.

By combining these two results, we see that for no value of
$A<-\frac{1}{3}$ the corresponding solution is a local minimum.
Thus, all dS critical points are unstable in our model ensembles.
\end{description}

We have found that all dS vacua ({\it i.e.} the vacua with $A<-1/3$) for our model ensembles have a tachyon and hence are perturbatively unstable. We therefore focus on AdS vacua in what follows. We have found that exponentially large numbers of the $2^n$ vacua are in fact metastable. Specifically, vacua for which $A > \frac{1}{3}$ are always metastable. Vacua with $-\frac{1}{3} < A < \frac{1}{3}$ can in principle also be metastable (and actually turn out to be local minima), but this is rather exceptional: they correspond to having only one of the $m_i$ equal to $+1$. In particular, since $\sum_i a_i = 7/3$, this means that the corresponding $a_i$ must be greater than $1$, and thus there cannot be more than two such solutions.

Some lower bounds on the numbers of metastable vacua with $A \geq 1/3$ can be derived as follows. For simplicity, but without loss of
generality, we put
\begin{equation}\label{Rorderai}
   a_n \geq a_{n-1} \geq \ldots \geq a_1.
\end{equation}
When $a_n\geq \frac{4}{3}$, all solutions with $m_n=+1$ correspond
to $A\geq \frac{1}{3}$, and we have
\begin{equation}
    N_{\rm stab} \geq 2^{n-1}
\end{equation}
one of which is the supersymmetric solution. When $a_n+a_{n-1}\geq \frac{4}{3}$ the number of stable vacua is at least
\begin{equation}
    N_{\rm stab} \geq 2^{n-2}
\end{equation}
and so on. So in a model with $a_n+a_{n-1}+\ldots+a_{n-j+1}\geq \frac{4}{3}$, but with $a_n+a_{n-1}+\ldots+a_{n-j} < \frac{4}{3}$,
the number of stable vacua is at least
\begin{equation}
    N_{\rm stab}\geq 2^{n-j}
\end{equation}
Because of (\ref{Rorderai}) and the fact that $\sum a_i = \frac{7}{3}$, we cannot have $j>4n/7$. So for a model
with $n$ moduli, the number of stable vacua is surely bigger than
\begin{equation}
   N_{\rm stab} \geq 2^{n-4n/7} = 2^{3n/7},
\end{equation}
which is exponentially smaller than $2^n$ but still exponential in $n$.

Actually this number is very hard to reach and for generic models the number of stable vacua is much bigger than this. Take for
example the case $a_i=3/7n$ of the figure, for which $\tilde{A}=\frac{3}{7} A$ is distributed according to (\ref{Adistr})
in the large $n$ limit. The number of vacua with $A>1/3$ is then given by integrating the distribution (\ref{Adistr}) from $A=1/3$ to
$A=\infty$. At large $n$ this gives asymptotically 
\begin{equation}
 N_{\rm stab}/2^n \approx \frac{7}{\sqrt{2 \pi n}} \exp(-n/98).
\end{equation}
Again for large $n$ this is an exponentially small fraction, but still exponentially large in absolute number. In fact for $n=100$
the stable fraction is still about $10\%$. For $n=1000$, this goes down to about $10^{-6}$, but this is not a small number compared to
the total number of vacua, which is $2^{1000} \sim 10^{300}$.

\subsubsection{Distributions over Moduli Space}

Let us fix $c_2$ and $\vec{m}$. We want to study the distributions
of physical quantities over the space of vacua parametrized by
$\vec{N}$. As discussed in section \ref{sec:descrvac}, the sign of
each $N_i$ is completely determined by $\vec{m}$. We can therefore
restrict to counting positive $\tilde{N}_i \equiv |N_i|$.

Let us start by finding the number of such vacua in a region
$\mathcal{R}$ given by $s_i\geq s_i^*$. By the substitution $s_i=
c_2 a_i |h_i^{(\vec m)}|/\tilde{N}_i$, this condition becomes
\begin{equation}
   \tilde{N}_i \leq \frac{c_2 a_i |h_i|}{s_i^*}.
\end{equation}
So, in the large N approximation, the number of vacua at fixed
$\vec{m}$ in this region is
\begin{equation}\label{RNsmag}
   \mathcal{N}_{(\vec m)}(s_i\geq s_i^*) = c_2^n
    \prod_{i=1}^n \frac{a_i |h_i|}{s_i^*} = \left(\frac{4c_2}{3}\right)^n
    \prod_{i=1}^n |h_i| \, \mbox{vol}(\hat{\mathcal{R}}).
\end{equation}
As in (\ref{susycount}), $\hat{\mathcal{R}}$ is the region of the
complexified moduli space projecting to $\RR$. In particular, the
number of vacua in any finite region of moduli space is finite. Note
that in the supersymmetric case $h_i=3/5$, this reproduces the
general formula (\ref{susycount}). More generally, we see that also
nonsupersymmetric vacua are distributed uniformly with respect to
the volume form in the supergravity approximation, but that their
density relative to the supersymmetric vacua, given by $\prod_i (5
h_i/3)$, is higher. Moreover, the density grows with increasing
numbers of anti-aligned spins. The highest density is that of the dS
vacua with all $m_i=-1$, which is $5^n$ higher than the density of
supersymmetric vacua. Obviously, this does \emph{not} mean that in
total there are $5^n$ times more dS maxima as supersymmetric
critical points, since we know there is a one-to-one correspondence
between them (in the supergravity approximation). The density in a
given region is higher simply because the nonsupersymmetric vacua
sit at larger radii. Integrated over the entire moduli space in the
supergravity approximation, we do not run into a paradox, because
both numbers are then infinite. In the fully quantum corrected
problem, the numbers presumably will be finite (by analogy of what
happens for type II flux vacua due to worldsheet instanton
corrections), but then of course also the relative densities will
change.

\vspace{5mm}

In order to have a meaningful four dimensional effective theory,
decoupled from the KK modes, we need the Kaluza-Klein radius to be
smaller than the AdS radius. Taking all $s_i \sim s$, we have
\begin{eqnarray}
 R^2_{\rm AdS} &\sim& \frac{m_p^2}{\Lambda} \sim \frac{V_X^3}{c_2^2 m_p^2} \sim \frac{s^7}{c_2^2 m_p^2} \\
 R^2_{\rm KK} &\sim& \frac{s^{2/3}}{M_p^2} \sim \frac{s^3}{m_p^2},
\end{eqnarray}
so $R_{\rm KK} < R_{\rm AdS}$ iff $s > c_2^{1/2}$.\footnote{The
assumption that all $s_i \sim s$ can be relaxed. Then one can prove
that $R_{\rm KK} < R_{\rm AdS}$ is guaranteed if $s_i>c_2^{4/7}$.
However for most vacua, $s_i>c_2^{1/2}$ will be sufficient to have
the required scale hierarchy, so we stick to this estimate.}

This also ensures the validity of the supergravity approximation.
The number of vacua at fixed $\vec{m}$ satisfying this condition is
given by (\ref{RNsmag}):
\begin{equation}\label{RNsmag2}
   \mathcal{N}_{(\vec m)}(s_i\geq c_2^{1/2}) = c_2^{n/2}\prod_{i=1}^n a_i
   |h_i|
\end{equation}

\vspace{5mm}

Finally, to get the total density for all possible $\vec{m}$ as
well, we must sum (\ref{RNsmag}) over all $\vec{m}$. Because of the
absolute values and the nonlinear dependence of $h_i$ on $\vec{m}$,
this is not easy to compute analytically even in special cases. But
one can get numerical results without much effort. For example in
the case with all $a_i=7/3n$, we get numerically that
\begin{equation}
 \sum_{\vec m} \prod_i |h_i| \approx (3.328)^n.
\end{equation}
This approximation becomes better for large $n$, but is quite good
for smaller values as well. For $n=1$, the exact result is
$3+3/5=3.6$, which is already not too far from this expression.

\subsubsection{Distributions of Volumes and Cosmological Constants: Fixed $\vec{m}$}

Now, let us consider the distributions of the volumes $V_X$. %and of the cosmological constant.
For a given $\vec{m}$ the volume is given by (\ref{volexpr}):
\begin{eqnarray}
   V_X &=& {c_2}^{7/3}
    \prod_{j=i}^n\biggl(\frac{a_j |h_j| }{\tilde{N}_j}\biggr)^{a_j}%\\
%   \Lambda_{(\vec{m})}\equiv V(\vec{s}(\vec{h}^{(\vec{m})})) &=& \frac{\pi^2}{6c_2^5}B_{(\vec{m})}
%                          \prod_{j=1}^n\left(\frac{2\pi \tilde{N}_j}{a_j
%             h^{(\vec{m})}_j}\right)^{3a_j}\nonumber
\end{eqnarray}
We see that as the $\tilde{N}_j$ go to infinity, $V_X$ tends to
zero, so the density of vacua (strongly) increases with decreasing
volume, {\it i.e.} large volumes are suppressed. This agrees with what we
found earlier for the general supersymmetric case in section
\ref{sec:largevolsuppr}.

\begin{figure}
\begin{center}
  \epsfig{file=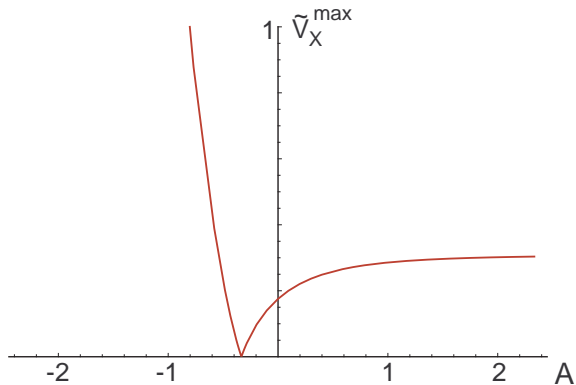,height=6cm,angle=0,trim=0 0 0 0}
  \caption{\small Dependence of $\tilde{V}_X^{\rm max}$ on $A$. At $A=7/3$,
  $\tilde{V}_X^{\rm max} = (3/5)^{7/3} \approx 0.3$, at $A=-7/3$, $\tilde{V}_X^{\rm max} = 3^{7/3} \approx 13$, and
  the zero is at $A=-1/3$.}
  \label{vols}
\end{center}
\end{figure}

The maximal value that $V_X$ can assume is obtained when all
$\tilde{N}_j=1$:
\begin{equation} \label{modvolbound}
  V_X^{\rm max} = c_2^{7/3}\prod_j a_j^{a_j} |h_j|^{a_j}
  \equiv c_2^{7/3} \prod_j a_j^{a_j} \, \tilde{V}_X^{\rm max}.
\end{equation}
Here we isolated the $\vec{m}$-dependent product $\prod_j
|h_j|^{a_j}$:
\begin{equation} \label{tVmaxdef}
 \tilde{V}_X^{\rm max} = \bigl|\frac{3}{4} + H\bigr|^{\frac{7}{6}+\frac{A}{2}} \,
 \bigl|\frac{3}{4} - H\bigr|^{\frac{7}{6}-\frac{A}{2}}.
\end{equation}
The variation of $V_X^{\rm max}$ over different choices of $\vec{m}$
is entirely given by the dependence of this function on
$A=\vec{a}\cdot\vec{m}$. This is shown in fig.\ \ref{vols}.

When we take all $a_i = 7/3n$, and we consider say the susy case
$m_i=+1$ so $h_i = 3/5$, (\ref{modvolbound}) becomes
\begin{equation}
 V_X^{\rm max}|_{\rm susy} = \biggl( \frac{7 \, c_2}{5 \, n} \biggr)^{7/3}\:.
\end{equation}
%which is similar to the toy model result (\ref{volbound}).

%So the more sensible question is how many vacua there are with
%$V_X>V_*$.
%|\Lambda_{(\vec{m})}|<\lambda^*
%In our solutions this condition becomes
%\begin{equation}\label{RNumbVol}
%   \prod_{j=i}^n\tilde{N}_j^{a_j}< 1/v_*
%\end{equation}
%with
%\begin{equation}\label{Rv*}
%   v_*\equiv \frac{V_*}{c_2^{7/3}
%    \prod_{j=i}^n\left(a_j |h_j|\right)^{a_j}}
%\end{equation}
%%\begin{equation}
%%   \rho_*^{(\vec{m})}=l_*^{(\vec{m})}\equiv c_2^{5/3}{\lambda^*}^{1/3}
%%   \left(\frac{6}{\pi^2 |B_{(\vec{m})}|}\right)^{1/3}
%%   \prod_{j=i}^n\left(\frac{a_j}{2\pi}h^{(\vec{m})}_j\right)^{a_j}
%%\end{equation}
%We see that large volumes are suppressed. Actually the maximal value
%that $V_X$ can assume ($\tilde{N}_j=1$ $\forall j$) is
%\begin{equation} \label{modvolbound}
%  V_X^{\rm max} = c_2^{7/3}\prod_j a_j^{a_j} |h_j|^{a_j}.
%\end{equation}
%This bound is of the general form obtained for supersymmetric vacua
%in section \ref{sec:largevolsuppr}. When we take all $a_i = 7/3n$,
%and consider the susy case $m_i=+1$ so $h_i = 5/3$,
%(\ref{modvolbound}) becomes
%\begin{equation}
% V_X^{\rm max} = \biggl( \frac{35 \, c_2}{9 \, n} \biggr)^{7/3},
%\end{equation}
%which is similar to the toy model result (\ref{volbound}).

%If $V_*>V_X^{\rm max}$, no vacuum satisfies the condition above.
%Moreover, as one can clearly see from (\ref{RNumbVol}) and
%(\ref{Rv*}), as $V_*$ tends to $V_X^{\rm max}$, $v_*$ tends to $1$,
%so the nuber of $\vec{N}$ satisfying the condition decreases.

Using the usual continuum approximation for the fluxes, it is
possible to get explicit expressions for the volume distribution of
vacua at fixed $\vec{m}$. Taking as example again the case
$a_i=7/3n$, we show in appendix \ref{AppMth2} that for $v \equiv
V_X/V_X^{\rm max} \leq 1$, at fixed $\vec{m}$ (or fixed $A$), the
vacuum number density is
\begin{equation} \label{vdistr}
 d \NN_A[v] = \frac{\left(\frac{3n}{7}\right)^n}{(n-1)!} (-\ln
 v)^{n-1} \, v^{-\frac{3n}{7}-1} \, dv.
\end{equation}
Therefore the total number of such vacua with $V_X \geq V_*$ goes as
$V_*^{-3n/7}$, in agreement with the general estimates of section
\ref{sec:largevolsuppr}. Note that in addition here, the density of
vacua vanishes to order $(n-1)$ near the cutoff $v=1$.

%\begin{equation}
% \CN(V_X \geq V_*) \approx  |\frac{\Gamma(n,\frac{3n}{7} \ln
% v_*)}{\Gamma(n)}-1|
% = |v_*^{-3n/7} \sum_{k=0}^{n-1} \frac{(\frac{3n}{7} \ln
% v_*)^k}{k!} -1 |
%\end{equation}
%where $\Gamma(n,x)$ is the incomplete Gamma function. For $v_* \ll
%1$, this behaves as $v_*^{-3n/7}$, in agreement with the general
%estimates of section \ref{sec:largevolsuppr}. For $v_* = 1 -
%\epsilon$ with $\epsilon \ll 1$, this goes as $\epsilon^n$, so the
%number density $d\CN(V_X)$ of vacua near the cutoff vanishes to
%order $(n-1)$.

Thus, in accordance with general expectations, we see that large
volumes are strongly suppressed, and more so when there are more
moduli.

Analogous considerations can be made about the distribution of
cosmological constants at fixed $\vec{m}$:
\begin{equation}
   \Lambda = \frac{c_2^2 \, E}{48 \pi V_X^3},
\end{equation}
with $E(A)$ as defined in section \ref{sec:descrvac}. Clearly, the
distribution of $\Lambda$ is completely determined by the
distribution of $V_X$. Thus, because large volumes are suppressed,
we see that small cosmological constants are suppressed. In
particular there is a lower bound on $|\Lambda|$:
\begin{equation}
 |\Lambda|_{\rm min} = \frac{1}{48 \pi} \, c_2^{-5} \prod_j a_j^{-3a_j} \, \frac{E}{(\tilde{V}_X^{\rm
 max})^3} = \frac{1}{48 \pi} \, c_2^{-5} \prod_j a_j^{-3a_j} \,
 \tilde{\Lambda}
\end{equation}
with $\tilde{\Lambda}$ as given by (\ref{tildeLambda}) and plotted
in figure \ref{vacE}.

\subsubsection{Distributions of Volumes and Cosmological Constants: All $\vec{m}$}

So far in this section, we have studied distributions at fixed
$\vec{m}$, or equivalently at fixed $A$. To get complete statistics
of all vacua, we need to combine these results with the distribution
of solutions over values of $A$.

One general observation one can make is that since there are no
metastable vacua at $A \leq -1/3$, the largest possible volume of a
metastable vacuum is obtained at $A=7/3$ (and $N_i=1$), the
supersymmetric solution. This can be seen from fig.\ \ref{vols}.
Therefore, the maximal volume for a metastable vacuum is $c_2^{7/3}
\, \prod_j a_j^{a_j} \, (3/5)^{7/3}$. It is not hard to show further
that $(7/3n)^{7/3} \leq \prod_j a_j^{a_j} \leq (7/3)^{7/3}$. The
former corresponds to all $a_i$ equal, the latter to the limiting
case $a_i \to 0$ for all but one $a_i$. Similar considerations hold
for the cosmological constant. Thus we arrive at the result that for
our ensembles:
\begin{equation}
 V_X \leq 2.2 \, c_2^{7/3} \, \hat{n}^{-7/3}, \qquad
 |\Lambda| \geq 2.3 \times 10^{-4} \, c_2^{-5} \, \hat{n}^7
\end{equation}
where $1 \leq \hat{n} \leq n$.

To get more refined results on the actual distributions, we need the
precise distribution of solutions over $A$. This depends on the
values of $a_i$. When all $a_i=7/3n$ for example, the distribution
is binomial, peaked around $A=0$, which for sufficiently large $n$
is well approximated by the normal distribution given in
(\ref{Adistr}). When all $a_i$ are approximately equal, the
distribution will still be approximately binomial, and
(\ref{Adistr}) is still a good approximation for large $n$. When say
$a_1=1$ and all other $a_i=4/3n$ with $n$ large, the distribution
will have two peaks, one at $A=1$, corresponding to $m_1=+1$, and
the other one at $A=-1$, corresponding to $m_1=-1$. In the following
we will work with the distribution given in (\ref{Adistr}).

\begin{figure}
\begin{center}
  \epsfig{file=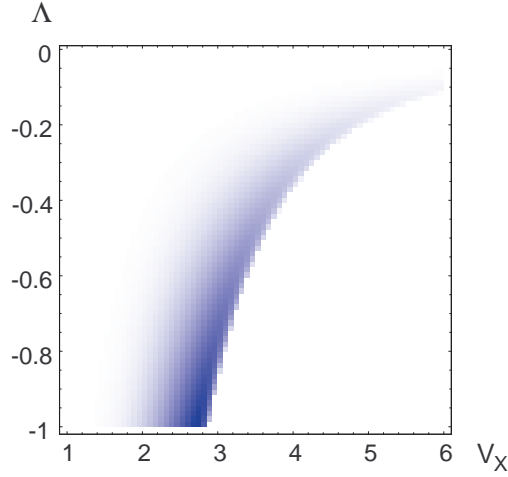,height=7cm,angle=0,trim=0 0 0 0}
  \caption{\small Density plot of the joint distribution of AdS vacua over $V_X$ and $\Lambda$, for
  $c_2=100$, $n=20$.}
  \label{ccvol}
\end{center}
\end{figure}

The joint distribution for $A$ and $V_X$ is then given by
multiplying the distributions (\ref{Adistr}) and (\ref{vdistr}).
Note that (\ref{vdistr}) depends on $A$ through $V_X^{\rm max}$. One
can also change variables from $A$ to $E$ and thus write down a
joint distribution for $E$ and $V_X$, or equivalently (and
physically more relevantly) for $\Lambda$ and $V_X$, as we did for
the Freund-Rubin ensemble. The explicit expressions are not very
illuminating, so we will not get into details here. An example is
plotted in fig.\ \ref{ccvol}.

Clearly, $V_X$ and $\Lambda$ are correlated, as in the Freund-Rubin
case. This follows directly of course from the relation $\Lambda
\sim E/V_X^3$. In particular given one variable, we get a roughly
Gaussian distribution of the other variable. Qualitatively this is
somewhat similar to the Freund-Rubin ensemble
(compare\footnote{Notice that fig.\ \ref{vacpoints} shows
$|\Lambda|^{-1}$ instead of $\Lambda$ on the vertical axis.} with
fig.\ \ref{vacpoints}), although the details are different. For
example in the Freund-Rubin case, at fixed $\Lambda$, the vacua
accumulate near the lower bound on $V_X$, whereas in the $G_2$ case
they accumulate near the upper bound. Without constraint on
$\Lambda$ this gets reversed for both.

\begin{figure}
\begin{center}
  \epsfig{file=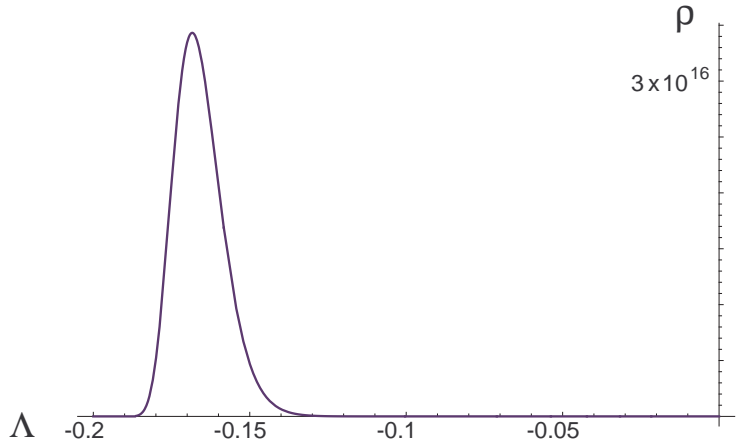,height=7cm,angle=0,trim=0 0 0 0}
  \caption{\small Distribution of cosmological constants $d\NN = \rho(\Lambda) \, d\Lambda$.
  Here we took $a_i=7/3n$, $n=50$, $c_2=100$, and we restricted to stable vacua
  with $V_X \geq 5$.}
  \label{ccdistr}
\end{center}
\end{figure}

Finally, one can get the distribution of cosmological constants for
vacua with volume above some cutoff value by integrating the joint
density over $V_X$. Again for the case with all $a_i=7/3n$, we
obtain the distribution for $\Lambda$ shown in fig.\ \ref{ccdistr}
for $n=50$, $c_2=100$ and $V_X \geq 5$. The cutoff at smaller values
of $|\Lambda|$ appears because of the lower cutoff we impose on the
volume; the lower we take the volume cutoff, the lower the value of
$\Lambda$ at which the density peaks. This is because $1/V_X^3$ sets
the scale for $\Lambda$. The cutoff for small $|\Lambda|$ appears
because the ensemble does not contain vacua with arbitrarily large
volume, as discussed at length before.

\subsubsection{Supersymmetry Breaking Scales}

Let us consider the distribution of the supersymmetry breaking scale, which for a fixed $\vec{m}$
is given by
\begin{eqnarray}
    M_{{\rm susy}(\vec{m})}^2 &=& \left(\sqrt{e^K g^{i\bar{j}}DW_i\overline{DW_j}}\right)_{\vec{m}}\nonumber\\
             &=& \frac{c_2}{(48\pi)^{1/2}}\frac{G_{\vec{m}}^{1/2}}{V_X^{3/2}}
\end{eqnarray}
where
\begin{equation}
    G_{\vec{m}}\equiv \left(\frac{7}{3}+\frac{9}{4}A^2
    \right)H^2+\frac{15}{8} A H+\frac{21}{64}
\end{equation}
$G_{\vec{m}}$ is a decreasing function of $A$, which
is positive for $-\frac{7}{3}\leq A < \frac{7}{3}$ and zero in the supersymmetric solution
($A=\frac{7}{3}$).

\begin{figure}
\begin{center}
  \epsfig{file=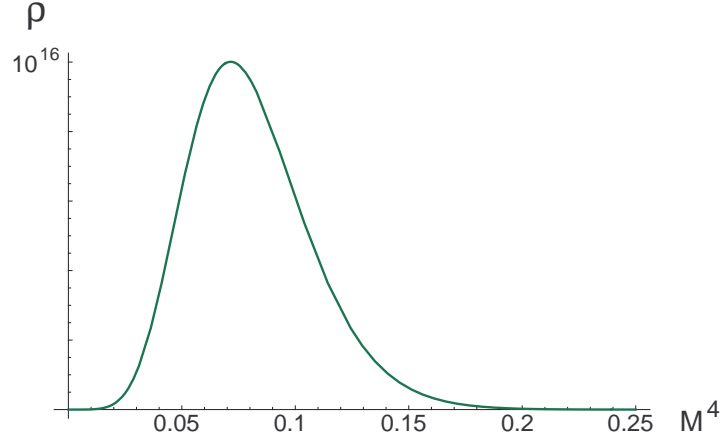,height=7cm,angle=0,trim=0 0 0 0}
  \caption{\small Distribution of supersymmetry breaking scales $d\NN = \rho(M^4) \, dM^4$, with
  $M \equiv M_{susy}/m_p$. Here we took $a_i=7/3n$, $n=50$, $c_2=100$, and we restricted to stable vacua
  with $V_X \geq 5$.}
  \label{susybrdistr}
\end{center}
\end{figure}

We see that the expression for $M_{\rm susy}^4$ is the same as that
of the cosmological constant, but with $G_{\vec{m}}$ instead of
$E_{\vec{m}}$.  So the distribution of the supersymmetry breaking
scale at fixed $\vec{m}$ is very similar to that of the cosmological
constant:
\begin{itemize}
    \item it is completely determined by the volume distribution;
    \item low supersymmetry breaking scales are suppressed (in line with the general expectations of section
    \ref{sec:gensusybreakingscales} \footnote{The situation here is a bit different
    than the case on which we focused there, namely $\Lambda \sim 0$, which cannot be obtained in the
    present ensemble.});
    \item $M_{\rm susy}^2$ has a lower bound given by
        \begin{equation}
           (M_{\rm susy}^2)_{\rm min}=\frac{1}{(48\pi)^{1/2}} \, c_2^{-5/2} \prod_j a_j^{-3a_j/2} \,
           \frac{G^{1/2}}{(\tilde{V}_X^{\rm max})^{3/2}}
        \end{equation}
  with $\tilde{V}_X^{\rm max}$ defined in (\ref{tVmaxdef}).
\end{itemize}

As for the $V_X$ and $\Lambda$ distributions, in order to get the
complete statistic of all vacua, we need to consider the
distribution of solutions over $A$, which depends on the $a_i$'s.
Similar to what we did for $\Lambda$ in the previous subsection, we
can compute the distribution of $\Lambda$ for vacua with volume
bounded by some lower cutoff. This is illustrated in fig.\
\ref{susybrdistr}. The cutoff at large values of $M_{\rm susy}$
appears because of the lower cutoff we impose on the volume, again
because $1/V_X^3$ sets the scale for $M_{\rm susy}^4$. The lower
cutoff is there because large volumes are absent.

\section{Conclusions and Discussion}

The potential importance of the emerging ideas surrounding the
Landscape, {\it e.g.} for the notion of naturalness, is clear. These
ideas should therefore be tested both experimentally by
verifying specific predictions and theoretically within the
framework of string theory.
To make progress in the latter, {\it one
should further scrutinize} the proposed ensembles of string theory
vacua, to establish whether or not these vacua truly persist after
taking into account the full set of subtle consistency requirements,
quantum corrections, and cosmological constraints
\cite{Banks:2004xh}. Parallel to that 
one should develop techniques
to analyze large classes of (potential)%\footnote{One should further scrutinize the proposed ensembles of string theory vacua, to establish whether or not these vacua truly persist after taking into account the full set of subtle consistency requirements, quantum corrections, and cosmological constraints \cite{Banks:2004xh}.}
vacua without actually having to go through their detailed constructions. 
In \cite{Acharya:2005ez} we have contributed to this program. We analyzed the statistics of Freund-Rubin and $G_2$ flux
compactifications of M-theory, and compared this to known IIB results. Here we have presented in details only the second ensemble. 

\

From what we have briefly said about the Freund-Rubin vacua, one can note that their statistics is very different from that of
flux compactifications on special holonomy manifolds. Most notably,
 they can have arbitrarily high compactification volume, and
 accumulate near zero cosmological constant. On the other hand, these
 vacua {\it typically} do not have a large gap between the KK scale
 and the four dimensional AdS curvature scale, so they are not really
 compactifications in the usual sense. This is no longer true
 \cite{Acharya:2003ii} if (in units in which the scalar curvature is
 one) the Einstein metric has all length scales much smaller than 1,
 but it remains a challenge to construct such manifolds with more
 than a modest scale hierarchy. Alternatively, one can imagine adding
 positive energy sources to lift the cosmological constant, perhaps
 even to small positive values, but no viable controlled scenario
 that would accomplish this is presently known. The problem is clear:
 in order to lift the cosmological constant to a positive value, one
 would have to add an energy source at or near the Kaluza-Klein
 scale, which makes four dimensional effective field theory unreliable and could
 easily destabilize the compactification. Adding such effects could
 also drastically change the distributions of vacua over parameter
 space. Therefore, the results for Freund-Rubin vacua
 should not be interpreted as showing there are parts of the
 Landscape \emph{compatible with rough observational requirements}
 that for example strongly favor large volumes. But our results do
 show that this is possible in principle, and constructing
 Freund-Rubin-like vacua which overcome the above mentioned problems
 would therefore be all the more interesting.

\
 
In this chapter we have presented in more details our statistical analysis of flux compactification of M-theory. We have seen that the statistics of $G_2$ flux vacua has a number of universal features. One is that the distribution of vacua over moduli space is uniform with respect to the K\"ahler metric on moduli space. In essence,  because the large volume region of moduli space is small when its dimension $b_3$ is large, this implies that large compactification volumes are suppressed, and strongly so if $b_3$ is large. In particular there is an upper bound to the volume in a given ensemble, set by the Chern-Simons invariant $c_2$. This is true as well for IIB flux ensembles and their IIA mirror counterparts, where the maximal size is set by the D3 tadpole cutoff $L$. Since both the IIB and the $G_2$ ensembles contain flux degrees of freedom that are not dual to flux degrees of freedom in the other ensemble, but are possibly dual to discrete geometrical deformations away from special holonomy, this suggests this behavior will persist when extending the ensembles to also sample non-flux discrete compactification data away from special holonomy. Our results on Freund-Rubin vacua on the other hand show that departure from special holonomy can produce the opposite behavior. So before general conclusions can be drawn, this question needs to be investigated in more general ensembles. 
% Since the full ten dimensional geometries of such extended ensembles tend to be hard to describe, an effective four dimensional field theory approach, as used for example in \cite{Kachru:2004jr} to find $\NN=2$ IIA flux vacua, would probably be the best way to make progress in this direction.

Essentially because of the limited discrete tunability of $G_2$ flux vacua, the scale of the cosmological constant is set by the volume, $\Lambda \sim m_p^4/V^3$. Since large volumes are strongly suppressed at large $b_3$, it follows that small cosmological
constants are strongly suppressed as well. This contrasts with type IIB ensembles, where the distribution of cosmological constants is
uniform near zero. Similarly, the scale of supersymmetry breaking is set by $m_p^2/V^{3/2}$, so small supersymmetry breaking scales are suppressed. For $b_3$ not too small, this remains true even when taking into account tuning of Higgs mass and cosmological constants. F-breaking IIB flux vacua similarly favor high susy breaking scales, although $M^2_{\rm susy}$ can be tuned to be small there, and the suppression was found to be independent of the number of moduli in the regime where $M_{\rm susy}^2$ is much smaller than the fundamental scale \cite{Denef:2004cf}. It is plausible that extending the $G_2$ ensembles as discussed above would allow the
supersymmetry breaking scale to be tuned small as well, reproducing the statistics of the generic ensembles of \cite{Denef:2004cf} in
this regime, but this would still favor higher scales.

\

In section \ref{sec:statmodel} we have presented and studied a class of models defined by K\"ahler potentials which give a direct product metric on moduli space. Though very rich, the semi-classical, supergravity vacuum structure of these models can be solved exactly, allowing us to explicitly verify our more general results just summarized. Minimizing the potential we obtained one supersymmetric vacuum and $2^{b_3-1}$ nonsupersymmetric ones. As the fluxes are varied, we found that
these are uniformly distributed over the moduli space and that there are roughly an equal number of de Sitter {\it vs} anti de Sitter vacua. Not all of these vacua exist within the supergravity approximation and we analyze the conditions under which they do, finding that an exponentially large number survive. Finally, after analyzing the stability of these vacua we found that {\it all de Sitter vacua are classically unstable} whilst an exponentially large number of non-supersymmetric anti de Sitter vacua are metastable. As expected on general grounds, the supersymmetry breaking scale is typically high.

We finally note that these model ensembles are reminiscent of the effective field theory Landscapes recently considered in \cite{ArkaniHamed:2005yv}. In particular, at large $b_3$, the distributions we found are sharply peaked, see {\it e.g.} figure \ref{ccdistr} and figure \ref{susybrdistr} for examples of distributions of cosmological constants and supersymmetry breaking scales. The cosmological constant in these ensembles does not scan near zero; there is a cutoff at $\Lambda \sim -c_2^{-5}$. It is conceivable though that this would change for more complicated K\"ahler potentials which do not lead to direct product metrics (see for example \cite{Milanesi:2007ss}).

\chapter[Ten Dimensional Description of Type IIA]{Ten Dimensional Description of Type IIA Flux Vacua}\label{TDD}

In this chapter we will study the ten dimensional description of Type IIA vacua. In the chapter 3 we have given the ten dimensional description of the Type IIB CY orientifold vacua, but not of Type IIA vacua. 

The standard ten dimensional approach consists in finding supersymmetric solutions of the ten dimensional theory. From this requirement one usually gets some constraints that relate the geometry of the internal manifold with the flux background.
As we have seen in the second chapter, if the fluxes are zero, the requirement of four dimensional supersymmetry translates into requiring reduced holonomy on the internal manifold; in particular we get CY manifolds if the holonomy is reduced to $SU(3)$. When the fluxes are turned on, the structure group of the internal manifold is still reduced but it does not coincide with the Levi-Civita holonomy group any more. The special class of ten dimensional Type IIB solutions studied in section \ref{TypeIIBvacua}, is of this type: the internal manifold is no more a CY, but a conformal CY (that has not Levi-Civita holonomy group $SU(3)$). However the difference with respect to the CY vacua are only in the conformal factor and the vacua can be studied using what is known about CY geometry. 

On the other hand, Type IIA has not such kind of tractable solutions. The solutions of the supersymmetry equations are all far from a CY geometry. In this chapter we will review  the results of \cite{Acharya:2006ne}, where we found an approximation that allows to obtain a CY internal manifold in presence of fluxes in the Type IIA context. Before doing this, we will review the connection between the structure group of the internal manifold and the background required by four dimensional supersymmetry.

%\newpage

\section[Supersymmetric Backgrounds with Fluxes]{Supersymmetric Backgrounds with Fluxes: Beyond the Ricci-flatness Approximation}

In this section we will review what the supersymmetry condition requires for Type II compactifications without and with fluxes \cite{Grana:2005jc}.

We always use the most general ansatz for the ten dimensional metric with four dimensional maximal symmetry:
\bea
 ds^2= e^{2A(y)} \hat{g}_{\mu\nu} dx^\mu dx^\nu  + g_{mn} dy^m dy^n,  & \mu,\nu=0,...,3 & m,n=1,...,6 
\eea
where $A(y)$ is a function of the internal coordinates and is called {\it warp factor}, $\hat{g}_{\mu\nu}$ is a Minkowski, $dS_4$ or $AdS_4$ metric, and $g_{mn}$ is a six dimensional metric.

A supersymmetric vacuum where only bosonic fields have non-vanishing vacuum expectation values should obey $< \delta_\epsilon \chi>=0$, where $\epsilon$ is the supersymmetry parameter and $\chi$ any fermion field. In Type II theories, the fermionic fields are two gravitinos $\psi^A_M$ ($A=1,2$) and two dilatinos $\lambda^A$. In the supergravity approximation, the bosonic parts of their supersymmetry transformations in string frame are 
\bea
 \delta \psi_M &=& \nabla_M \epsilon + \frac{1}{4} {\bf \not}  H_M \PP \epsilon +\frac{1}{16}e^\phi\sum_n {\bf \not} F_n \Gamma_M \PP_n \epsilon \nn \\
 \delta \lambda &=& (\not\!\partial \phi +\frac{1}{2}{\bf \not}  H \PP+ \frac{1}{8}e^\phi\sum_n (-1)^n (5-n) {\bf \not} F_n  \PP_n \epsilon
\eea
where $M=0,...,9$, $\psi_M \equiv \left(\begin{array}{c} \psi_M^1 \\ \psi_M^2  \end{array}\right)$ ( $\psi^{1,2}_M$  are the two Majorana-Weyl spinors with the same chirality in Type IIB and with opposite chirality in Type IIA); similar definitions hold for $\lambda$ and $\epsilon$. For Type IIA $\PP=-\sigma^3$ and $\PP_n=\sigma^1$ when $\frac{n+1}{n}$ is even while $\PP=-\sigma^3$ and $\PP_n=i\sigma^2$ when $\frac{n+1}{n}$ is odd; for Type IIA $\PP=\Gamma_{11}$ and $\PP_n=\Gamma_{11}^{n/2}\sigma^1$. 
${\bf \not} V_m \equiv V_{M_1...M_m}\Gamma^{M_1...M_m}$. $F_n$ are the physical RR field strength and $n=0,...,6$ as we want fluxes only along the internal directions.

\subsubsection{Supersymmetric Fluxless Backgrounds}

As we have seen for Heterotic compactifications, when no fluxes are present, demanding zero vev for the gravitino variation requires the existence of a covariantly constant spinor on the ten dimensional manifold, {\it i.e.} $\nabla_M \epsilon =0$.
The spacetime component of this equation is 
\bee
 \hat{\nabla}_\mu \epsilon + \frac{1}{2} (\hat{\gamma}_\mu \gamma_5 \otimes \not\! \nabla A)\epsilon =0
\ee
where we have used the standard decomposition of the ten dimensional gamma matrices and the hat-objects means that they are computed using $\hat{g}_{\mu\nu}$.

The integrability condition for this equations implies that
\bee
 \kappa + \nabla_m A \nabla^m A = 0
\ee
where the constant $\kappa$ is negative for $AdS_4$, zero for Minkowski and positive for $dS_4$. The only constant value for $(\nabla A)^2$ on a compact manifold is zero, which implies that the warp factor is constant and the four dimensional manifold can only be Minkowski spacetime.

\

As explained with more details in the second chapter, the  internal component of $\nabla_M \epsilon =0$ says that there should exist at least one covariantly constant spinor on the six manifold. This is a very strong requirement from the topological and differential point of view. It forces the manifold to have reduced holonomy. When the compact manifold has dimension six, the requirement to have one covariantly constant spinor implies the compact manifold to be CY. If there are more than one covariantly constant spinor, the holonomy group has to be a proper subgroup of $SU(3)$ and this results in a larger number of preserved supersymmetries.

In Type II theories, when there is one covariantly constant internal spinor, the internal gravitino equation tells us that there are two four dimensional supersymmetry parameters. This compactifications therefore preserve eight supercharges, that means an effective $\NN=2$ four dimensional theory.

Fluxes can break the $\NN=2$ supersymmetries spontaneously to $\NN=1$ or even completely in a stable way. We will see this in the next section by including also their backreaction on the geometry.

\subsubsection{Supersymmetric Background with Non Zero Fluxes}

In this section we consider compactifications preserving the minimal amount of supersymmetry, {\it i.e.} $\NN=1$ in four dimension. In order to have some supercharges preserved, or even in the case when all of them are completely broken spontaneously by fluxes, we need to have globally well defined supercurrents. This requires to have globally well defined spinors on the compact manifold, which is only possible if its structure group is reduced. 

In the absence of fluxes, supersymmetry requires a covariantly constant spinor on the internal manifold. This condition actually splits into two: the existence of a nowhere-vanishing globally well defined spinor, and the condition that it is left invariant by the holonomy group (of the Levi-Civita connection). The first condition is topological and implies an effective $\NN=2$ four dimensional action, while the second is a differential requirement on the metric connection and implies that there exists a $\NN=2$ Minkowski vacuum.

\

A globally well defined nowhere-vanishing spinor exists only on manifolds that have {\it reduced structure group}. The structure group of a manifold is the group of transformations required to patch the orthonormal frame bundle. A $d$-dimensional Riemann manifold
has automatically structure group $SO(d)$. All vector, tensor and spinor representations can be decomposed in representations of $SO(d)$. If the manifold has structure group $G \subset SO(d)$, all the representations can be further decomposed in $G$-representations.

Let us consider a six dimensional manifold. If it has $SU(3)$ structure group, then the spinorial representation of $SO(6)$ decomposes as ${\bf 4} \rightarrow {\bf 1} \oplus {\bf 3}$. There is therefore an $SU(3)$ singlet in the decomposition, which means that there is a spinor that depends trivially on the tangent bundle of the manifold and is so well defined globally and nowhere-vanishing.

We can decomposes also other $SO(6)$ representations; for example the 1-, 2- and 3-forms decomposes as:
\bea
 {\bf 6} \rightarrow {\bf 3} \oplus {\bf \bar{3}},
 &{\bf 15} \rightarrow {\bf 8} \oplus {\bf 3} \oplus {\bf \bar{3}} \oplus {\bf 1},
 &{\bf 20} \rightarrow {\bf 6} \oplus {\bf \bar{6}} \oplus {\bf 3} \oplus {\bf \bar{3}} \oplus {\bf 1}\oplus {\bf 1}\nn
\eea  
From here, we can see that there are singlets also in the decomposition of 2-forms and 3-forms. Therefore there is also a nowhere-vanishing globally defined real 2-form and complex 3-form, that are called respectively $J$ and $\Omega$. We also see that there are no singlets in the vector decomposition; this in particular implies $J\wedge \Omega =0$. On the contrary a six form is a singlet and there is only one of them, so $J\wedge J \wedge J \propto \Omega \wedge \bar{\Omega}$. We note that these conditions are valid also for the K\"ahler form and the holomorphic 3-form on a CY. In fact a CY is a special case of $SU(3)$ structure manifold; in addition the holonomy group of its Levi-Civita connection coincides with the structure group.

The invariant forms $J$ and $\Omega$ determine a metric: $\Omega$ says what are the holomorphic and antiholomorphic coordinates and in these coordinates the metric takes the form $g_{m\bar{n}}=-iJ_{m\bar{n}}$.

Raising one of the indices of $J$ we get an almost complex structure, {\it i.e.} a map that satisfies $J_m^p J_p^n = - \delta_m^n$. The existence of an almost complex structure allows to introduce local holomorphic and antiholomorphic vectors. If their dual forms are integrable and if the transition functions between different patches are holomorphic, then the structure is integrable and the manifold is a complex manifold.

The $SU(3)$ structure is determined equivalently either by the invariant spinor $\eta$ , or by the forms $J$ and $\Omega$. Actually these are related by:
\begin{gather}\label{SBJOeta}
J_{mn} \equiv i \eta_{-}^\dagger \gamma_{mn}\eta_{-} = -i \eta_{+}^\dagger \gamma_{mn}\eta_{+} \\
\label{SBJOetaa}
\Omega_{mnp} \equiv \eta_{-}^\dagger \gamma_{mnp}\eta_{+} \qquad
\Omega^*_{mnp} = -\eta_{+}^\dagger \gamma_{mnp}\eta_{-} \:,
\end{gather}
$J$ is a (1,1)-form with respect to the almost complex structure $J^m_n$, while $\Omega$ is a (3,0)-form.

\

Let us now pass to the differential condition coming from the supersymmetry requirement. It is given by the gravitino variation, which gives a differential condition on the invariant spinor, that is schematically given by:
\bee
 \nabla \eta + \Upsilon [H,F_n] \eta =0
\ee
where $\Upsilon [H,F_n]$ is the piece proportional to the fluxes. If the fluxes are zero, we recover the condition that $\eta$ is covariantly constant with respect to the Levi-Civita connection. If, on the other hand, the fluxes are different from zero, we get that $\nabla \eta \not = 0$. This means that the spinor is not invariant under the Levi-Civita Holonomy group and so that Hol$(\nabla)\not \subset SU(3)$ ({\it i.e} the Levi-Civita connection is not compatible with the structure group). In the case of manifolds with $SU(3)$ structure, one can show that there is always a metric compatible connection ($\nabla' g =0$), possibly with torsion, such that 
\bee
 \nabla'\eta = 0
\ee
The torsion is defined by the relation
\bee
 [\nabla'_m,\nabla'_n] V_p = - {R_{mnp}}^q V_q + 2 {T_{mn}}^q \nabla'_q V_p
\ee
and belongs to $\Lambda^1\otimes (su(3)\oplus su(3)^\perp )$, where $\Lambda^1$ is the space of 1-forms, while $so(6)=su(3)\oplus su(3)^\perp$ is the Lie algebra of $SO(6)$. Acting on $SU(3)$ invariant forms, the $su(3)$ piece drops. The corresponding torsion is called the {\it intrinsic torsion} ${T^0_{mn}}^p \in \Lambda^1\otimes su(3)^\perp$; it can be decomposed in $SU(3)$ representation as follows:
\bea
 {T^0_{mn}}^p &\in& ({\bf 3} \oplus {\bf \bar{3}}) \otimes ({\bf 1} \oplus {\bf 3} \oplus {\bf \bar{3}})\nn\\
 	&&\begin{array}{cccccccccc} = &({\bf 1} \oplus {\bf 1})& \oplus& ({\bf 8} \oplus {\bf 8})& \oplus &({\bf 6} \oplus {\bf \bar{6}})& \oplus& ({\bf 3} \oplus {\bf \bar{3}})& \oplus &({\bf 3} \oplus {\bf \bar{3}})\\
	& W_1 && W_2 && W_3 && W_4 && W_5
 	  \end{array}\nn
\eea
where the $su(3)^\perp$ is obtained by subtracting the adjoint ${\bf 8}$ representation of $SU(3)$ from the adjoint ${\bf 15}$ representation of $SO(6)$ (and using ${\bf 15}\rightarrow {\bf 1} \oplus {\bf 3} \oplus {\bf \bar{3}} \oplus {\bf 8}$).

The $W_i$'s are the five {\it torsion classes} that appear in the covariant derivatives of the spinor and of the forms $J$ and $\Omega$. $W_1$ is a complex scalar, $W_2$ is a primitive (1,1) form ({\it i.e.} $(W_2)_{mn} J^{mn} =0$), $W_3$ is a real primitive (2,1)+(1,2) form and $W_4$ and $W_5$ are real vectors. Because of the supersymmetry condition, the torsion classes are determined by the fluxes. This can be easily seen by the matching the conditions $\nabla \eta + \Upsilon [H,F_n] \eta =0$ and $\nabla'\eta =0$, and considering that $\nabla'$ can be written as a sum of the Levi-Civita connection and a piece depending on the torsion. The last piece is completely determined by the torsion classes when $\nabla'$ is applied to the invariant spinor $\eta$.

The supersymmetry condition allows also to compute the Levi Civita covariant derivative of the $\eta$ in terms of the torsion classes. Because of $\nabla' \eta=0$, one can express the Levi-Civita connection of $\eta$ in terms of the torsion classes. Then, by using the expressions \eqref{SBJOeta}, one can find what are the covariant derivatives of $J$ and $\Omega$. Antisymmetrizing them, one gets the expressions for their differential in terms of the torsion classes:
\begin{equation}\label{eq008}
\begin{split}
dJ &= -\frac{3}{2} \mbox{Im}(\mathcal{W}_1\Omega^*) +\mathcal{W}_4\wedge J +\mathcal{W}_3 \\
d\Omega &= \mathcal{W}_1 J\wedge J + \mathcal{W}_2\wedge J +  \mathcal{W}_5^*\wedge\Omega
\end{split} \end{equation}

A manifold of $SU(3)$ structure is complex if $W_1=0=W_2$. This condition comes from the fact that in a complex manifold the differential of a ($p,q$) form is a ($p+1,q$)$+$($p,q+1$) form. But if one of $W_1$ and $W_2$ is different from zero, $d\Omega$ contains a (2,2) form. One can show that this condition is also sufficient for the manifold to be complex.
In a symplectic manifold the 2-form $J$ is closed. This corresponds to vanishing $W_1$, $W_3$ and $W_4$. A K\"ahler manifold is complex and symplectic. So the only possible non-vanishing torsion class is $W_5$. In this case the Levi-Civita connection has holonomy $U(3)$ (we note that actually the Levi-Civita holonomy group is not compatible with the structure group). Finally, for a CY all the torsion classes are zero.

\section[Ten Dimensional Description of Type IIA]{Ten Dimensional Description of Type IIA with Fluxes}

The ten dimensional description of the Type IIA vacua we described in section \ref{TypeIIAvacua} is less well understood than the Type IIB case. This is because in Type IIB one special class of solutions is conformally CY: the fluxes drive the internal manifold away from $SU(3)$ holonomy, but their effects results in a conformal factor in front of a CY metric. For Type IIA there are not solutions of this type. We will see in what follows that the supersymmetric compactifications are {\it half-flat} manifolds with $SU(3)$ structure.

It is natural to wonder what relation these solutions have with the CY flux vacua discussed in the four dimensional language at page \pageref{TypeIIAvacua}, where fluxes are viewed as a perturbation of a Type IIA CY compactification. In that case the massive Type IIA was compactified on a CY threefold. Switching on the RR fluxes gives rise to a potential which depends on the K\"ahler moduli. In order to stabilize the complex  structure moduli one could introduce NSNS 3-form flux, H. However this leads to a tadpole for the D6-brane charge, which is cancelled by introducing orientifold O6-planes. The full system of fluxes and
O6-planes then stabilizes all the moduli, essentially at leading order in $\alpha'$ and $g_s$.
This was done by using the effective four dimensional potential for the moduli in the large volume limit, when the backreaction of the fluxes on Einstein's equations can be ignored (since their contribution to the stress tensor is volume suppressed). This class of vacua is an excellent arena to study aspects of moduli stabilization in detail, since the vacua are essentially classical solutions of ten dimensional IIA supergravity. 
In the work \cite{Acharya:2006ne} we studied these classical solutions from a ten dimensional perspective. 

We proved that the exact ten dimensional solution is
{\it not} Calabi-Yau. The precise modification of the Calabi-Yau geometry can be described by a particular
type of {\it half-flat} SU(3) structure \cite{Chiossi:2002tw}. Though we were unable to find the full solution (for which we will have to await further developments in the mathematical literature), in the approximation that the O6-plane source is smoothed out, we found an exact solution. This solution is CY and by studying the moduli stabilization from the ten dimensional point of view, we found the same results as \cite{DeWolfe:2005uu}.

\

In what follows we shortly review a class of solutions of Type IIA supergravity found in \cite{Behrndt:2004km,Behrndt:2004mj} and \cite{Lust:2004ig}.  These form the basis of the solutions with O6-planes. They describe compactifications on an internal $SU(3)$ structure manifold down to four dimensional $AdS_4$. Then we discuss the introduction of orientifold O6-planes in supergravity, the issue of supersymmetry preserving configurations and how the original solutions are modified by their presence. In particular, we present an exact ``smeared'' solution in which the orientifold charge is smoothed out. Finally moduli stabilization is studied. We show that all the geometrical moduli are lifted at tree level in supersymmetric vacua. 

\subsection{Massive Type IIA Supergravity on $AdS_4$} \label{LustSol}

%Recently, a large class of supersymmetric four dimensional smooth compactifications of massive Type IIA supergravity have been classified \cite{Lust:2004ig}.

% The massive IIA theory has bosonic fields consisting of a metric $g$, an RR 1 form potential $A$ (with field strength $F$)
% and 3-form potential $C$ (with field strength $G$), a NSNS 2-form potential
% $B$ (with field strength $H$) and a dilaton $\phi$. 
% 
% 
% \medskip

We are interested in the ten dimensional description of the supersymmetric vacua with non-zero cosmological constant discussed by de Wolfe et al from an effective field theory point of view in \cite{DeWolfe:2005uu}. Therefore, without loss of generality, we can take the ten dimensional spacetime to be a warped product $AdS_4 \times_{\Delta} X$, where $X$ is a compact manifold and the ten dimensional metric is given by
\begin{equation} \label{eq021}
ds^2 = \Delta^2(y) \hat{g}_{\mu\nu}(x) dx^\mu dx^\nu +  g_{mn}(y) dy^m dy^n \:,
\end{equation}
where $x$ and $y$ are coordinates for $AdS_4$ and $X$ respectively and the warp factor is $\Delta$($=e^{2A}$ in previous conventions).
All the fluxes have non-zero $y$-dependent components only along the compact directions, except for $G$ which has a non-zero four-dimensional component\footnote{It can be seen as an $F_6$ background on the compact manifold $X$.} 
\begin{equation}
G_{\mu\nu\rho\sigma} = \sqrt{g_4} f(y) \epsilon_{\mu\nu\rho\sigma} \:,
\end{equation}
and $f$ is a function on $X$. These assumptions are dictated by local Poincar\'e invariance on $AdS_4$.

As we have seen in the first section of this chapter, $\mathcal{N}=1$ supersymmetry in four dimensions implies that the compact manifold $X$ has a globally defined spinor, $\eta$. The structure group of $X$ reduces (at least) to $SU(3)$ and 
the spinor $\eta$ is related to the globally defined 2-form $J$ and 3-form $\Omega$ by \eqref{SBJOeta}.
% \begin{gather}\label{SBJOeta}
% J_{mn} \equiv i \eta_{-}^\dagger \gamma_{mn}\eta_{-} = -i \eta_{+}^\dagger \gamma_{mn}\eta_{+} \\
% \label{SBJOetaa}
% \Omega_{mnp} \equiv \eta_{-}^\dagger \gamma_{mnp}\eta_{+} \qquad
% \Omega^*_{mnp} = -\eta_{+}^\dagger \gamma_{mnp}\eta_{-} \:,
% \end{gather}
These forms completely specify an $SU(3)$ structure on $X$ and as described previously, from the $SU(3)$ decomposition of their differentials $dJ$ and $d\Omega$, one can read off the torsion classes which characterize the $SU(3)$ structure, as shown in \eqref{eq008}.
% \begin{equation}\label{eq008}
% \begin{split}
% dJ &= -\frac{3}{2} \mbox{Im}(\mathcal{W}_1\Omega^*) +\mathcal{W}_4\wedge J +\mathcal{W}_3 \\
% d\Omega &= \mathcal{W}_1 J\wedge J + \mathcal{W}_2\wedge J +  \mathcal{W}_5^*\wedge\Omega
% \end{split} \end{equation}

\

By requiring the fluxes to preserve precisely $\mathcal{N}=1$ SUSY in four dimensions,
the ten dimensional supersymmetry parameter has to be of the form \cite{Lust:2004ig}:
\begin{align} \label{eq004}
\begin{split}
\epsilon &= \epsilon_+ + \epsilon_- \\
     &= (\alpha \theta_+ \otimes \eta_+ - \alpha^* \theta_- \otimes \eta_-)+
        (\beta \theta_+ \otimes \eta_- - \beta^* \theta_- \otimes \eta_+) \:.
\end{split}
\end{align}
Here $\theta_+$ and $\theta_-$ (with
$\bar \theta_+= \theta_-^T C$) are the two Weyl spinors on $AdS_4$, satisfying the Killing spinor equations
\begin{equation}
\hat{\nabla}_\mu \theta_+ = W \hat{\gamma}_\mu \theta_- \qquad
\hat{\nabla}_\mu \theta_- = W^* \hat{\gamma}_\mu \theta_+  \:,
\end{equation}
where $W$ is related to the scalar curvature $\hat{R}$ of $AdS_4$ through $\hat{R}=-24|W|^2$. On the other hand, $\eta_+$ and $\eta_-$ are chiral spinors on $X$ related by charge conjugation, so that $\epsilon$ is a Majorana spinor.

\

Now, we can solve the supersymmetry equations $\delta \Psi_M =0$, $\delta \lambda=0$, where:
\begin{align} \label{eq009}
\begin{split}
\delta \Psi_M &= \bigg[ \nabla_M - \frac{m \, e^{5\phi/4}}{16} \Gamma_M -
		\frac{e^{3\phi/4}}{64} F_{NP} ({\Gamma_M}^{NP} - 14{\delta_M}^N \Gamma^P) \Gamma_{11} \\
	&\quad +\frac{e^{-\phi/2}}{96} H_{NPQ} ({\Gamma_M}^{NPQ} - 9{\delta_M}^N \Gamma^{PQ}) \Gamma_{11} \\ 
	&\quad + \frac{e^{\phi/4}}{256} G_{NPQR} ({\Gamma_M}^{NPQR}
		- \frac{20}{3}{\delta_M}^N \Gamma^{PQR}) \bigg] \epsilon
\end{split} \\
\begin{split}
\delta\lambda &= \bigg[ -\frac{1}{2} \Gamma^M \nabla_M \phi - \frac{5m \, e^{5\phi/4}}{4} + 
                \frac{3 \, e^{3\phi/4}}{16} F_{MN} \Gamma^{MN} \Gamma_{11} \\ 
	&\quad +\frac{e^{-\phi/2}}{24} H_{MNP} \Gamma^{MNP} \Gamma_{11}  
                -\frac{e^{\phi/4}}{192} G_{MNPQ} \Gamma^{MNPQ} \bigg] \epsilon
\end{split}
\end{align}
In order to solve this, one substitutes the ansatz for $\epsilon$ \eqref{eq004}, for the metric and for the forms and contracts the resulting six dimensional equations with $\eta^\dagger_\pm \gamma^{(n)}$. In this way, one obtains separate equations for every $SU(3)$ representation in the decomposition of forms \cite{Lust:2004ig}: one can decompose the tensors $F$, $H$ and $G$ in terms of irreducible $SU(3)$ representations.
For example, for $F$ one gets:
\begin{equation}
F_{mn}=\frac{1}{16}{\Omega^*_{mn}}^s F_s^{(1,0)}+\frac{1}{16}{\Omega_{mn}}^s F_s^{(0,1)}+
	(\tilde{F}_{mn}+\frac{1}{6}J_{mn}F^{(0)}) \:,
\end{equation}
where the different pieces can be extracted through
\begin{equation}
F^{(0)}=F_{mn}J^{mn} \sim {\bf 1} \qquad \qquad F_m^{(1,0)}={\Omega_{m}}^{np}F_{np} \sim {\bf 3}
\end{equation}
and $\tilde F\sim \mathbf{8}$ is such that
\begin{equation}
\tilde F_{mn} J^{mn} = \tilde F_{mn} {\Omega^{mn}}_p = \tilde F_{mn} {(\Omega^*)^{mn}}_p = 0 \:.
\end{equation} 
By different contractions one has a set of equations, and then recasting together the various pieces one obtains two cases, depending whether $|\alpha|\not = |\beta|$ or $|\alpha| = |\beta|$ \cite{Lust:2004ig}.

If  $|\alpha|\not = |\beta|$, one gets the usual Calabi-Yau supersymmetric compactification, {\it i.e.} $X$ is a Calabi-Yau manifold, all the fluxes vanish and $W=0$, so the four dimensional space is Minkowski.

If $|\alpha| = |\beta|$, one can, without loss of generality, choose $\alpha = \beta$ and get the following expressions that relate the fluxes to the geometry:
\bea \label{eq006}
F &=& \frac{f}{9} e^{-\phi/2} J + \tilde{F}\nn \\
H   &=& \frac{4m}{5} e^{7\phi/4} \re \Omega \nn\\
G   &=& f d\mbox{Vol}_4 + \frac{3m}{5} e^{\phi} J\wedge J  \\
W   &=& \Delta \left(\frac{\alpha}{|\alpha|}\right)^{-2} (-\frac{1}{5}m \, e^{5\phi/4} + \frac{i}{6}f \, e^{\phi/4}) \nn\\
&&\phi,\Delta,f, \mbox{Arg}(\alpha) = \mbox{constant} \nn\:.
\eea
Here $\tilde F$ is the $\mathbf{8}$ component in the $SU(3)$ decomposition of $F$, as explained above, and it
is not determined by supersymmetry. 

From contraction of the supersymmetry equations, one can also get the covariant derivative of the forms $J$ and $\Omega$. By antisymmetrizing the resulting expressions in all indices, one obtains the differential of $J$ and $\Omega$:
\bea\label{dJdOmega}
 dJ &=& -J\wedge d\ln|\alpha|^2 +\frac 23 f e^{\phi/4} \re \Omega \\
 d\Omega &=& -\Omega \wedge d\ln|\alpha|^2-ie^{3\phi/4} J\wedge \tilde{F} -\frac{4i}{9}g e^{\phi/4} J\wedge J\nn
\eea

\

A solution of the supersymmetry equations is also a solution of the Einstein equations, if the form fields satisfy the Bianchi identities and the equations of motion \cite{Lust:2004ig}. 

Therefore we impose the Bianchi identities on the supersymmetry solution \eqref{eq006}. The BI for $H$ gives $d\re \Omega =0$ that implies $|\alpha|=$constant.
On the other hand, by imposing the Bianchi identity for $F$, one finds a constraint on the differential:
\begin{equation}
d\tilde{F} = - \frac{2}{27}e^{-\phi/4} \left( f^2 - \frac{108}{5} m^2 \, e^{2\phi} \right) \re\Omega \:.
\end{equation}
From the last equation, using the fact $d(\Omega\wedge \tilde{F}) = 0$ and the expression \eqref{dJdOmega} for $d\Omega$,
one can in particular compute:
\bea\label{eq007}
|\tilde{F}|^2 &=& \frac{8}{27} e^{-\phi} \left( f^2 - \frac{108}{5} m^2 \, e^{2\phi} \right) \\
 && f^2 \geq \frac{108}{5} m^2 \, e^{2\phi} \:. \label{eqfm}
\eea
Note that the Bianchi identities are crucial to obtain a solution of all the equations of motion.

\

From these results we can obtain a characterization of the $SU(3)$ structure of these backgrounds:

\begin{equation} \label{eq045}
\begin{split}
dJ &= \frac{2}{3} f e^{\phi/4} \re\Omega \\
d\Omega &= -\frac{4i}{9}f e^{\phi/4} J\wedge J -i \, e^{3\phi/4} J\wedge \tilde{F} \:,
\end{split}
\end{equation}
Thus, the nonvanishing torsion classes of $X_6$ are:
\begin{equation} \begin{split} \label{eq018}
\mathcal{W}_1^- &= -\frac{4i}{9}fe^{\phi/4} \\
\mathcal{W}_2^- &= -i e^{3\phi/4}\tilde{F} \\
\end{split} \end{equation}
A manifold with such an $SU(3)$ structure is a special case  of a so-called \emph{half-flat} manifold. (Compactifications on \emph{half-flat} manifolds are considered in \cite{Gurrieri:2002wz,Gurrieri:2002iw,House:2005yc}).

\

From these results we can see that
the only Calabi-Yau solution (which has zero torsion) is the standard one with zero fluxes and zero cosmological constant.
The only other special class of solutions which can be considered have $\mathcal{W}_2^- = 0$ (because of \ref{eqfm}).
This requires $f^2 = \frac{108}{5}m^2 e^{2\phi}$. These manifolds are called \emph{nearly-K\"ahler}, and solutions of this kind were obtained in \cite{Behrndt:2004km,Behrndt:2004mj}.

\subsection{IIA Supergravity with Orientifolds} \label{orientifold}

Our main result in \cite{Acharya:2006ne} has been the ten dimensional description of the vacua discovered in \cite{DeWolfe:2005uu} (an example of such vacua is also given in \cite{Ihl:2006pp}). Since these vacua must also have O6-planes we need to understand how the solutions of \cite{Lust:2004ig} change in the presence of the O6. The O6-plane is not a genuine supergravity object, but rather something defined by the superstring compactification. Nevertheless, the supergravity action can be enriched with terms that describe the interactions of such an object with the low energy fields.

\

As seen in section \ref{E4DAComFacts}, in IIA String Theory an O6-plane is obtained by modding out the theory by the discrete symmetry operator $\mathcal{O}$:
\begin{equation} \label{eq060}
\mathcal{O} \equiv \Omega_p (-1)^{F_L} \sigma
\end{equation}
where $\Omega_p$ is the world-sheet parity, $(-1)^{F_L}$ is the left-moving spacetime fermion number, while $\sigma$ is an isometric involution of the original manifold.
The fixed point locus of $\sigma$ is the orientifold O6-plane. It is a BPS object, which preserves half of the supersymmetries: those such that $\epsilon_\pm = \mathcal{O} \, \epsilon_\mp$, where $\epsilon_\pm$ are the two Majorana-Weyl supersymmetry parameters \eqref{eq004}.

We add an O6-plane filling the $AdS_4$ factor and wrapping a 3-cycle in the internal manifold. Since the background preserves only four supercharges, in general an O6-plane will break all of them. On the other hand, in order to get an $\NN=1$ four dimensional theory, we must take the O6 such that it preserves the same supercharges as the background. As in the case of a D6-brane, this is achieved by wrapping the plane on a supersymmetric (calibrated) 3-cycle.

The operator $\mathcal{O}$ does not act on the four dimensional spinors $\theta_{\pm}$ while it exchanges $\eta_+$ and $\eta_-$.%
\footnote{Note that $\Omega_p (-1)^{F_L}$ acts trivially on the supersymmetry parameters, since they have the same parity properties of the metric.}
Thus
\begin{align}
\label{eq057}
J_{mn} = -i\eta_+^\dagger \gamma_{mn} \eta_+ &\stackrel{\sigma^*}{\longrightarrow} -i\eta_-^\dagger \gamma_{mn} \eta_- = -J_{mn} \\
\label{eq026}
\Omega_{mnp} = \eta_-^\dagger \gamma_{mnp} \eta_+ &\stackrel{\sigma^*}{\longrightarrow} \eta_+^\dagger \gamma_{mnp} \eta_-
        = -\Omega^*_{mnp}
\end{align}
Supersymmetry forces $\sigma$ to be antiholomorphic with respect to the almost complex structure $J_i^j$.

The fixed locus of the isometry $\sigma$ (if any) on the internal manifold is the supersymmetric 3-cycle $\Sigma$ the O6 wraps. In particular, we get for the pull-back to the plane:
\begin{equation}
 J|_\Sigma = 0 \qquad \qquad \re\Omega |_\Sigma = 0 \:,
\end{equation}
which implies
\begin{equation}
J \wedge \delta_3 = 0 \qquad \qquad \re\Omega \wedge \delta_3=0 \:.
\end{equation}
$\delta_3$ is a singular $\delta$-like 3-form that will be usefull later and that is defined by:
\bee
 \int_\Sigma \omega_3 = \int_X \omega_3\wedge \delta_3
\ee

$\Omega$ is a calibration and $\Sigma$ is calibrated with respect to $-\im\Omega$. In fact one can compute
\begin{equation} \label{eq052}
\int_\Sigma \im\Omega = \int_X \im\Omega \wedge \delta_3 = - \int_X \frac{\delta^{(3)}(\Sigma)}{\sqrt{g_3^t}}\, d\mbox{vol} = - \mbox{Vol}_\Sigma \:.
\end{equation}
% These indeed show that $\Sigma$ is a supersymmetric \hyph{3}cycle (in fact special Lagrangian) \cite{CascUranga}.

One obtains the spatial parity of the other form fields by considering their worldsheet origin and imposing them to be invariant under the orientifold operator \eqref{eq060}: so, under $\sigma^*$, $F$ and $H$ are odd as well as $\delta_3$, while $G$ is even.

Now consider the modifications to the equations of motion (EOM) and the Bianchi identities (BI) given by the O6-plane to Type IIA massive supergravity. The bosonic action is, at leading order in $\alpha'$:
\begin{equation} \label{eq002}
S_{O6} = 2 \mu_6 \int_{O6} d^7\xi e^{3\phi/4} \sqrt{-g_7} - 4 \mu_6 \int_{O6} C_7 \:,
\end{equation}
where the first piece comes from the DBI action, the second one from the CS action.%
\footnote{This action is directly derived from the one of a D6-brane noticing that the orientifold projection forces $B$ to vanish on the plane, and O-planes do not support gauge fields.}
Moreover $g_7$ is the pulled-back metric determinant on the plane, $\mu_6 = 2\kappa_{10}^2 \bar\mu_6= 2 \pi \sqrt{\alpha'}$, while $\bar\mu_p=(2\pi)^{-p}\alpha'^{-(p+1)/2}$ is the Dp-brane charge and tension, and we have taken into account that the charge of an Op-plane is $-2^{p-5}$ times that of a Dp-brane.

These terms are only the first ones in an infinite expansion in $\alpha'$. Keeping just them and working with the leading supergravity action \eqref{eq003b} is consistent. In $\NN=2$ ten dimensional supergravity theories, the first corrections coming from string theory are of order ${\alpha'}^3 R^4$, where $R^4$ stands for various contractions of four Riemann tensors, to be compared to the leading term $R$.%
\footnote{For $\NN=1$ ten dimensional theories the first corrections are of order $\alpha' R^2$.}
The orientifold leading action is instead of order $\sqrt{\alpha'}$. Classical solutions will be reliable only in regions where $\alpha' R \ll 1$.

The DBI term gives a contribution to the Einstein and dilaton equations, while the CS term represents an electric coupling to $C_7$.
The DBI term brings a localized contribution to the energy momentum tensor
\begin{equation}
 T_{MN}^{loc} \equiv -\frac{2}{\sqrt{-g}} \, \frac{\delta S_{O6}}{\delta g^{MN}}
 = 2\mu_6 \, e^{3\phi/4} \, \Pi_{MN} \, \frac{\delta^{(3)}(O6)}{\sqrt{g_3^t}} \:,
\end{equation}
where $\Pi_{MN}$ is the projected metric on the plane and $g_3^t=g_{10}/g_7$ is the determinant of the transverse metric. In case of a warped product metric as in \eqref{eq021} and for a submanifold wrapping the four-dimensional factor, $\Pi_{\mu\nu} = g_{\mu\nu}$.

The equations of motion are%
\footnote{Remember: ${F_p}^2= p! |F_p|^2$. Moreover the equation of motion for $A$ is given by the differential of \eqref{eq051}.}
\begin{align}
\begin{split}
0 &= R_{MN} - \frac{1}{2}\partial_M\phi\partial_N\phi -\frac{1}{12}e^{\phi/2}G_M\cdot G_N
    +\frac{1}{128}e^{\phi/2}g_{MN} G^2  \\
 &\quad -\frac{1}{4}e^{-\phi}H_M\cdot H_N +\frac{1}{48}e^{-\phi}g_{MN} H^2
    -\frac{1}{2}e^{3\phi/2}F_M\cdot F_N+\frac{1}{32}e^{3\phi/2}g_{MN} F^2
       \\
 &\quad -\frac{1}{4}m^2e^{5\phi/2}g_{MN} -\mu_6 e^{3\phi/4} \Pi_{MN} \frac{\delta^{(3)}(O6)}{\sqrt{g_3^t}}
    +\frac{7}{8}\mu_6 e^{3\phi/4} g_{MN} \frac{\delta^{(3)}(O6)}{\sqrt{g_3^t}}
\end{split} \\
\label{eq056}
\begin{split}
0 &= \nabla^2\phi -\frac{1}{96}e^{\phi/2} G^2
        +\frac{1}{12}e^{-\phi}H^2
 -\frac{3}{8}e^{3\phi/2} F^2  -5m^2e^{5\phi/2} \\
 &\quad  +\frac{3}{2}\mu_6 e^{3\phi/4} \frac{\delta^{(3)}(O6)}{\sqrt{g_3^t}}
\end{split} \\
\label{eq051}
0 &= d(e^\phi \ast H) - \frac{1}{2} G\wedge G +  e^{\phi/2} F\wedge \ast G + 2m e^{3\phi/2} \ast F \\
\label{eq030}
0&= d(e^{\phi/2} \ast G) - H\wedge G \:.
\end{align}
Here $X_M\cdot X_N$ means contraction on all but the first index.
Notice that the only equations that get modified with respect to \cite{Lust:2004ig}, due to the presence of an orientifold plane, are the Einstein and dilaton equations.

The CS term in \eqref{eq002} describes the coupling of the plane to $C_7$, which is the gauge potential dual to $A$, and so the O6 is a magnetic source for $A$. This term does not modify the equations of motion, but only the Bianchi identity. The way this modification can be evaluated is taking the dual description in terms of $F_8$, so that the BI is obtained by varying with respect to $C_7$. We obtain
\begin{equation} \label{eq010}
dF = 2m H - 2 \mu_6\, \delta_3 \qquad \qquad dH = 0 \:.
\end{equation}
The other BI is $dG =F\wedge H$ and it is satisfied.%
\footnote{Looking at the complete CS term for a D6-brane, one could have suspected a localized modification to the BI for $G$ like $\delta_3\wedge F$. But the orientifold projection forces the pull-back of $F$ on the plane to vanish. This would not necessarily be true for D6-branes.}

In the derivation it has been convenient to express integrals on the plane as integrals on the whole space, through the 3-form $\delta_3$, transverse to the plane and localized on it:
\begin{equation} \label{eq050}
\int_{O6} C_7 = \int C_7 \wedge \delta_3 \:.
\end{equation}
In local coordinates $y^M$, where the O6-plane is located ad $y^7=...=y^9=0$, we have
$\delta_3 = \delta^{(3)} (y^7,y^8,y^9) \: dy^7 \wedge dy^8 \wedge dy^9$
expressed through a usual delta function.
Notice the closure
\begin{equation}
d\delta_3 = 0 \:,
\end{equation}
which means nothing more than charge conservation. A precise treatment of distributional forms would be to consider the embedding of a seven dimensional manifold $M_7$ into the target space $f:M_7 \to Z$, so that $\int_{M_7} f^* C_7$ is a nondegenerate linear map from 7-forms to real numbers. The Poincar\'e dual to $f(M_7)$ is now, by definition, an object $\delta_3$ which realizes \eqref{eq050} as a linear map on 7-forms. It turns out that the differential $d\delta_3$ is defined by $\int C_6 \wedge d\delta_3 = - \int_{\partial M_7} f^* C_6$ on 6-forms. In our case the O6-plane has no boundary, hence closure.

\

Summarizing, the introduction of the O6-plane does not modify the SUSY variations in \eqref{eq009}; it changes the Bianchi identity for the 2-form field-strength and induces some additional terms in the Einstein and dilaton equations of motion.

In order to find the new solution, we follow the same procedure as in \cite{Lust:2004ig}, i.e. we solve the SUSY equations
$\delta\psi_M=0$ and $\delta\lambda=0$, and then we impose BI's and EOM's for form fields.
In fact, one can show that the Einstein and dilaton equations are automatically satisfied (with the minor requirement on the Einstein equation $E_{0M}=0$ for $M\neq 0$, which is granted with the ansatz \eqref{eq021}). We will partly verify it in the appendix \ref{AppTIIA1}.

The system of relations \eqref{eq006}  solve also the form field equations \eqref{eq051}, \eqref{eq030} and the BI for $G$. So we are left with only the modified BI for $F$ \eqref{eq010}. Substituting the solution (\ref{eq006}) into the modified BI and using the expression
\eqref{eq045} for $dJ$, one gets
\begin{equation} \label{eq027}
 d\tilde{F}= - \frac{2}{27} e^{-\phi/4}\left( f^2 - \frac{108}{5} m^2 e^{2\phi} \right)
            \re\Omega - 2 \mu_6 \, \delta_3 \:.
\end{equation}
From this, through the same procedure used to obtain \eqref{eq007}, we can compute $|\tilde F|^2$. Start from $0 = d(\Omega\wedge \tilde F)$, use again \eqref{eq007} and \eqref{eq052} to get
\begin{equation} \label{eq028}
 |\tilde{F}|^2 = \frac{8}{27} e^{-\phi}\left( f^2 - \frac{108}{5} m^2 e^{2\phi} \right)
        + 2 \mu_6 e^{-3\phi/4} \frac{\delta^3(\Sigma)}{\sqrt{g_3^t}} \:.
\end{equation}
The first term is constant on $X$, while the second one has support on the cycle
$\Sigma$. $|\tilde{F}|^2$ is positive definite, so we find two conditions:
\begin{equation} \label{eq019}
f^2 \geq \frac{108}{5}m^2 e^{2\phi} \qquad \mbox{and} \qquad \mu_6 \geq 0 \:.
\end{equation}
Note that the latter is perfectly expected: changing the sign of the charge of the O6-plane gives an anti-O6-plane, which however preserves orthogonal supersymmetries incompatible with the background.
The discussion of the possibility of getting a Calabi-Yau geometry is parallel to section \ref{LustSol}. One would have to put $f$ and $\tilde F$ to zero, but this would also imply $m$ vanishing. The massless limit has to be taken with care, and one finds Calabi-Yau without fluxes. Moreover, as long as the localized contribution is present, there will always be a singular behavior on it, captured by \eqref{eq045}.

\subsubsection{A Smeared Solution} \label{smeared}

To find exact solutions in presence of localized objects is not easy, mainly because, as we saw, {\it in no case with non-vanishing mass parameter does the geometry reduce to Calabi-Yau}. Nevertheless, as a first step, we can consider a long-wavelength approximation in which this situation is realized. In a Calabi-Yau metric the torsion classes vanish. This happens if:
\begin{equation}
    f=0 \qquad \tilde F=0 \qquad F=0 \qquad m^2 > 0 \:.
\end{equation}
In the long-wavelength approximation the charge of the orientifold plane, localized on $\Sigma$, is substituted with a smeared distribution, obviously keeping the total charge the same.
Thus the 3-form describing the new charge distribution must be in the same cohomology class as $\delta_3$. Integrating the Bianchi identity \eqref{eq010} on 3-cycles gives the tadpole cancellation conditions. Actually, requiring $F=0$ and imposing the supersymmetry equation for $H$ \eqref{eq006} implies the smeared charge distribution to be:
\begin{equation} \label{eq053}
    \mu_6 \, \delta_3^{\text{smeared}} =  \frac{4 m^2}{5} \, e^{7\phi/4} \re\Omega \:.
\end{equation}
Direct inspection of \eqref{eq027} shows that in fact we can consistently put $f$ and $\tilde F$ to zero.

Requiring the further condition that the total charge of the $O6$ is actually $\mu_6$,
% that the form $\delta_3^{\text{smeared}}$ is dual to the homology class of the 3-cycle $\Sigma$ that the $O6$ wraps, 
one gets a relation for the value of the dilaton:
\begin{equation} \label{eq054}
\frac{4 m^2}{5} \, e^{7\phi/4} = \frac{\mu_6}{\sqrt{4 \mbox{vol}}} \:.
\end{equation}
Using the last equation in \eqref{eq006}, the relation \eqref{eq054} fixes also the value of the four-dimensional cosmological constant (as it depends on $W$). Summarizing, the solution is completely described by the internal Calabi-Yau manifold defined by $SU(3)$ invariant forms $J$ and $\Omega$, with an anti-holomorphic isometrical involution $\sigma$: the background fields $G$ and $H$ are determined by \eqref{eq006} with $f=0$, $F=0$; the dilaton is given by \eqref{eq054} where in turn the volume is set by $J$. Further constraints come from the integral quantization of fluxes, and this mechanism provides the stabilization of geometrical moduli in the geometry. Thus $J$ and $\Omega$ are (completely) determined by the integer fluxes. This will be analyzed in the next section.

% It would be of interest to establish in even more detail how the smeared and localised exact solutions
% are related.

\subsubsection{Tadpole Cancellation and Topology Change}

In the exact localized solution, the fact that $\re \Omega$ is exact implies that
$H$ must be exact.%
\footnote{Actually the exact forms are $e^{\phi/4}\re \Omega$ and $e^{-3\phi/2}H$ (as one reads from the equations \eqref{eq006} and \eqref{eq045}). But $\phi$ is constant.}
The most important consequence is that the modified BI implies that $mH - \sum_i \mu_6 \delta^{(i)}_3$ must vanish in cohomology; here $i$ runs over all the localized sources. Therefore from the tadpole cancellation conditions one gets that the possible configurations of localized charges are constrained: charge cancellation must work among localized charges only. Specifically, it must be that:
\begin{equation} \label{eq049}
\int \sum_i \delta^{(i)}_3 = 0
\end{equation}
on all closed 3-cycles. This is different from the smeared CY solution (in which $f=0$), where a non-trivial closed $H$ was allowed by the supersymmetry equations and could be used to cancel the $O6$ charge.

In the case of a single source we see that $\delta_3$ is exact. Since $\delta_3$ is the Poincare dual
of the homology class of the O6-plane, we learn that the 3-cycle that the O6-plane wraps is contractible.
This is in stark contrast to the smeared Calabi-Yau case in which the O6-plane is necessarily non-trivial
in homology. Therefore, we learn that the transition from the Calabi-Yau approximation to the exact solution necessarily involves a topology change.

\subsection{Moduli Stabilization} \label{stabilization}

In this section we will describe from the point of view of ten dimensional supergravity, how the introduction of the fluxes
stabilizes the moduli which are present in the zero flux, Calabi-Yau limit. After a brief general discussion, we will first discuss the
moduli vevs in the examples studied in \cite{DeWolfe:2005uu} and then go on to discuss the general case.

\

We begin with the axions. A background value for the field strength of a gauge form potential can be separated into two pieces:
\begin{equation} \label{eq031}
H = H^f + dB \:.
\end{equation}
The former, cohomologically nontrivial, when integrated on cycles gives the integer amounts of flux, whilst the second term is globally exact. 
$H^f$ must be closed (so that the flux depends only on cohomology), and we can {\it choose} an harmonic representative of the integral cohomology class. Note however that this separation is arbitrary. From the exact solution the total field strength $H$ is harmonic so that $dB=0$.
We can then use the gauge freedom $B \to B + d\lambda$ to choose $B$ harmonic. The internal harmonic components of $B$ are four dimensional
axions. This shows that all the other Kaluza-Klein modes have a zero vacuum expectation value and are hence massive.

In the same way, we split the other field-strengths:%
\footnote{Notice that the field strengths $F$ and $G$ are not automatically closed. They are indeed closed in the smeared solutions we are considering, as it turns out from the BI's \eqref{eq010}.}
\begin{align}
\label{eq036}
F &= F^f +  dA + 2 m\,B \\
\label{eq032}
G &= G^f + f d\mbox{Vol}_4 + dC + B\wedge dA + mB^2 \:.
%G &= G^f + f d\mbox{Vol}_4 + dC + B\wedge (dA + {F_2}^f) + mB^2 \:.
\end{align}
Arguing as before, ${F}^f$ is the integrally quantized flux of the gauge potential $A$ while $G^f$ is the flux of $C$; all of them can be taken harmonic exploiting the gauge redundancy. Note that being $A$ harmonic, it is actually vanishing on our Calabi-Yau solution because of the vanishing of $H^1(CY, {\mathbb R})$.

So one simply expands the fluxes (quantized), the gauge potentials and the $SU(3)$ structure forms defining the metric. The right basis is dictated by the exact solution, and by the constraints imposed by the orientifold projection. In the special example at hand, everything is harmonic. On the other hand, we can only study the vacuum and can not go off-shell, so can not see any superpotential.

In order to discuss the stabilization of axions coming from $C$, we need to consider the BI for $\tilde F_6 \equiv e^{\phi/2} \ast_E G$, or equivalently the EOM \eqref{eq030}. Splitting the field strength according to \eqref{eq031} and \eqref{eq032} and recalling that $A=0$ one can recast it in the form of an exact differential:
\begin{equation}
d\big( e^{\phi/2} \ast G + H\wedge C - B \wedge G^f - \frac{1}{3} m\, B^3 \big) = 0 \:.
%d\big( e^{\phi/2} \ast G + H\wedge C - B \wedge G^f -\frac{1}{2} B^2 \wedge {F_2^f} - \frac{1}{3} m\, B^3 \big) = 0 \:.
\end{equation}
When $f \neq 0$, $C$ must contain also a four-dimensional piece $C_M$ such that $dC_M = f d\mbox{Vol}_4$. Being a BI, the term in parenthesis is recognized as the closed component of $\tilde F_6$, which can be further split into flux and an exact piece:
\begin{equation} \label{eq037}
F_6^f + dC_5 = e^{\phi/2}\ast G + H \wedge C - B \wedge G^f - \frac{1}{3}m\, B^3 \:.
%F_6^f + dC_5 = e^{\phi/2}\ast G + H \wedge C - B \wedge G^f -\frac{1}{2} B^2 \wedge {F_2^f} - \frac{1}{3}m\, B^3 \:.
\end{equation}

\subsubsection{Example: the $T^6 / (\mathbb{Z}_3)^2$ Orientifold}

The smeared solution in the long-wavelength approximation can be exploited to compare results with another widely used approximation: what is called Calabi-Yau with fluxes. In the latter, one keeps the contribution of fluxes small compared to the curvature of the compactification manifold. Note that fluxes can not be taken arbitrarily small; Dirac quantization condition puts a lower bound $F_p \sim (\alpha')^{\frac{p-1}{2}}$ to the amount for a p-form field-strength. So one requires the contribution of fluxes to the action to be small compared to the Einstein term $R$, which is of order $L^{-2}$ with respect to the characteristic length of the manifold. This gives $(\alpha'/L^2)^{p-1} \ll 1$. In other words, we must be in the limit of large compactification manifold with respect to the string length, which anyway is the regime of applicability of supergravity. Under these conditions, one can neglect the backreaction of fluxes on geometry, and work with the Calabi-Yau metric. Of course one has to be careful to remember that in the action there are factors of the dilaton, and both the dilaton and the volume are (possibly) determined by fluxes themselves, so it is not always possible to keep the fluxes to their minimal amount while increasing the volume. 

A simple example studied in detail by \cite{DeWolfe:2005uu} is the $T^6 / {\mathbb{Z}_3}^2$ orientifold and will be
useful as a concrete model.
The model is constructed by compactifying Type IIA supergravity on a six dimensional manifold which is (the singular limit of) a Calabi-Yau: a torus $T^6$ firstly orbifolded by ${\mathbb{Z}_3}^2$ and then orientifolded. It has Hodge numbers $h^{2,1}=0$ and $h^{1,1}=12$, where 9 of the 12 K\"ahler moduli arise from the blow-up modes of 9 $\mathbb{Z}_3$ singularities. There are no complex structure moduli.
The O6-plane wraps a special Lagrangian 3-cycle and is compatible with the closed $SU(3)$ structure of the CY. The resulting theory has 4 preserved supercharges. The number of moduli from the form fields are: 3 from the NS-NS 2-form potential $B$ (odd under $\sigma$), no one from the R-R 1-form potential $A$ and 1 from the R-R 3-form potential $C$ (even).
Fluxes are switched on as described above.

In \cite{DeWolfe:2005uu} the stabilization of the moduli, due to the fluxes, is analyzed by a computation of the four dimensional effective moduli potential. We are going to apply to this model the machinery previously developed, in the long-wavelength approximation.

\

Let us introduce an integer basis of harmonic forms for the even cohomology groups. The 2-forms (odd under $\sigma$) $w_i$:
\begin{equation}
w_i \propto \frac{i}{2}dz_i\wedge d\bar z_i \qquad \qquad  \int w_1 \wedge w_2 \wedge w_3 = 1 \:.
\end{equation} 
The 4-forms (even under $\sigma$)
\begin{equation}
\tilde w^i = w_j \wedge w_k \qquad \Rightarrow \qquad \int w_a \wedge \tilde w^b = \delta_a^b
\end{equation} 
where $j$ and $k$ are the two values of $1,2,3$ besides $i$.

Start with the decomposition of $F$ \eqref{eq036}. Expand the fields on harmonic forms (of correct parity)
\begin{equation} \label{eq034}
{F}^f = f^i \, w_i \qquad \qquad B = b^i \, w_i \:,
\end{equation} 
where $f^i$ are quantized in units of $\mu_6$. Imposing the smeared solution $F=0$, we get
\begin{equation}
b^i = - \frac{f^i}{2m} \:.
\end{equation} 
The ``moduli''%
\footnote{We call them moduli because they are so in the Calabi-Yau compactification without fluxes, but here the exact solution fixes completely $B$, and so there are no moduli at all.}
$b^i$ corresponding to four dimensional axions are fixed by the fluxes $f^i$. We can take for simplicity ${F}^f=0$, as in \cite{DeWolfe:2005uu}, then $B=0$ and the axions are fixed to $b^i=0$. The general case is dealt with in the next section.

Then expand the 4-form flux $G$ and the $SU(3)$ structure fundamental form
\begin{align}
\label{eq035}
G^f &= \sum_i e_i \, \tilde w^i \\
\label{eq033}
J &= e^{-\phi/2} \sum_i v^i \, w_i \qquad v^i > 0 \:,
\end{align} 
where $e_i$ are quantized in units of $\mu_4$, and we put a power of the dilaton for later convenience. Note in particular
\begin{equation} \label{eq016}
v^1 v^2 v^3 = e^{3\phi/2} \, \mbox{vol} = \mbox{vol}^\textit{String frame} \:.
\end{equation}
Substituting into the decomposition of $G$ \eqref{eq032} and in the solution \eqref{eq006} with $f=0$ and $b^i=0$, we get
\begin{equation}
\frac{6m}{5} \, v^j v^k = e_i \:,
\end{equation}
where, as before, $j$ and $k$ are the two values of 1,2,3 besides $i$.
 
We find a series of relations on the possible fluxes that characterize a supersymmetric vacuum: $\mbox{Sgn} (m\,e_1e_2e_3) = \mbox{Sgn} (m\, e_i) = +$ and the sign of $e_i$ is independent on $i$. These are in agreement with \cite{DeWolfe:2005uu}. Moreover we can invert to
\begin{equation} \label{eq015}
v^i = \frac{1}{|e_i|} \sqrt{\frac{5}{6} \, \frac{e_1e_2e_3}{m}} \:.
\end{equation} 
So the K\"ahler moduli are fixed. In the more general case $b^i \neq 0$ they are still fixed, apart from changing the range of fluxes for which the supergravity approximation is reliable.

The stabilization of the dilaton comes from the decomposition of $H$ \eqref{eq031}. Expand $H$ in a basis of harmonic forms for the third cohomology group, odd under the spatial orientifold operation $\sigma^*$. In the present example there is only $\re\Omega$. Note that this is consistent with the solution \eqref{eq006}. So let us put
\begin{equation}
H = H^f = p \frac{1}{\sqrt{4 \mbox{vol}}} \re\Omega \:.
\end{equation} 
The normalization comes from $\int_\Gamma \delta^{\text{smeared}}_3 = 1$ %(see also the discussion after \eqref{eq053})
, so $p$ is integrally quantized in units of $\mu_5$.
Integrating the BI for $F$ on the cycle $\Gamma$ we get the only nontrivial tadpole cancellation condition
\begin{equation}
\int_\Gamma m\,H = m\,p= \mu_6
\end{equation} 
whose only two solutions are%
\footnote{Note, in quantizing $m$, that it is not canonically normalized in the action \eqref{eq003b}; then it is quantized in units of $\mu_8/2$.}
$(m,p) = \pm (\mu_8/2, 2 \mu_5)$ and $\pm(\mu_8, \mu_5)$. Comparing with the solution, the dilaton gets stabilized to
\begin{equation}
e^\phi = \frac{3}{4} \mu_6 \left( \frac{5}{6} \, \frac{1}{m^5\,e_1e_2e_3} \right)^{1/4} \:.
\end{equation} 

The last issue is the stabilization of possible axions coming from the 3-form potential $C$. Being it odd under $\sigma^*$ and harmonic, there is only one axion:
\begin{equation}
C = - \xi \frac{\im \Omega}{\sqrt{4 \mbox{vol}}} \:.
\end{equation} 
This must be substituted into the decomposition of the field-strength $\tilde F_6$ dual to $G$ \eqref{eq037}, with quantized flux $\int F_6^f = e_0$. We get:
\begin{equation}
-p \, \xi = e_0 
\end{equation} 

The result is that, in this simple model, all the K\"ahler moduli, the dilaton and the only axion are geometrically stabilized, whilst there are no complex structure moduli. {\it All the results found in this section are in precise agreement with those found in \cite{DeWolfe:2005uu}}. Really one should discuss the moduli associated to the 9 resolved singularities as well, which are one K\"ahler modulus each. One would find that the singularities are blown up to a finite volume. In the next section will discuss how this example generalizes to any Calabi-Yau, of which the orbifold is just a singular limit.

\

We can determine also the four-dimensional cosmological constant, that is the vacuum energy in $AdS_4$. The exact solution \eqref{eq006} gives the scalar curvature $\hat{R}=-24 |W|^2$ of the $AdS_4$ factor in ten dimensional Einstein metric (note that the constant $\Delta$ cancells out). Then we must express it in four dimensional Einstein frame, through
\begin{equation}
R^{4DE} = M_P^2  \kappa_{10}^{2} \frac{1}{\mbox{vol}} \hat{R} = -\frac{24}{25} M_P^2 \kappa_{10}^{2} m^2 \frac{e^{5\phi/2}}{\mbox{vol}} \:.
\end{equation} 
Eventually, choosing conventions for the Einstein equation $R_{\mu\nu} -\frac{1}{2} g_{\mu\nu} R = -\frac{1}{2} g_{\mu\nu} \Lambda$:
\begin{equation}
\Lambda = - (2\pi)^{11} \left( \frac{3}{4} \right)^4  \left( \frac{6}{5} \frac{\alpha'^4}{m\,e_1e_2e_3}  \right)^{3/2} \, M_P^2 \:.
\end{equation} 

\subsubsection{General Calabi-Yau with Fluxes}

The generalization of this example to any Calabi-Yau model with an orientifold projection is straightforward. 
We will continue to adopt the long-wavelength approximation as done in the previous section.
First of all the antiholomorphic involutive isometry $\sigma$ divides the cohomology groups of the internal manifold into even and odd components. In particular, $H^{1,1}= H^{1,1}_+ \oplus H^{1,1}_-$ with dimensions $h^{1,1} = h^{1,1}_+ + h^{1,1}_-$. Let $\{ w_i \}$ be an integer basis for $H^{1,1}_-$, with intersection numbers
\begin{equation}
\kappa_{abc} = \int w_a \wedge w_b \wedge w_c \:,
\end{equation} 
and $\{ \tilde w^i \}$ the dual basis for $H^{2,2}_+$ (since $J^3$ is odd):
\begin{equation}
\int w_i \wedge \tilde w^j = \delta_i^j \:.
\end{equation}
The third cohomology group $H^3 = H^3_+ \oplus H^3_-$ is halved in two spaces of real dimension $h^{2,1}+1$. We choose an integer symplectic real basis for $H^3$: $\{\alpha_K, \beta^L\}$ with $K,L:0,\ldots,h^{2,1}$, such that $\alpha_K$ are even under the projection $\sigma^*$ while $\beta^L$ are odd. It satisfies $\int \alpha_K \wedge \beta^L = \delta_K^L$. Let the Poincar\'e dual basis of integer cycles be $\{ \Sigma_A, \Gamma^B \}$ so that $\Sigma_A \cap \Gamma^B = \delta^B_A$. It satisfies $\int_{\Sigma_A} \alpha_K = \delta^A_K$, $\int_{\Gamma^B} \beta^L = \delta^L_B$ while the other vanishing. The orientifold homology class $\Sigma$ will be a combination of $\Sigma_A$'s.

Then we expand the various fields and forms on these basis, according to their behavior under the orientifold operation $\mathcal{O}$. The K\"ahler form $J$, the field $B$ and the flux $F^f$ are odd and follow \eqref{eq033}, \eqref{eq034}.%
\footnote{A possible axion coming from $B$ lying on the four dimensional space is forbidden by the orientifold projection.}
In particular
\begin{equation} \label{eq038}
\mbox{vol} = \frac{1}{6} e^{-3\phi/2} \, v^a v^b v^c \, \kappa_{abc} \:.
\end{equation} 
The flux $G^f$ is even and follows \eqref{eq035}. The treatment of the holomorphic 3-form needs a little bit more of care. On a Calabi-Yau it can be expanded on the full $H^3$:
\begin{equation}
\Omega = Z^K \alpha_K + \FF_L \beta^L \:.
\end{equation} 
We can take $Z^K$ as projective coordinates on the complex structure moduli space of the Calabi-Yau, while $\FF_L$ as functions of $Z^K$ on this space. Nonetheless, we choose the particular normalization $\Omega\wedge \bar\Omega = -8i d\mbox{vol}$, and this fixes the overall factor. The orientifold projection requires $\re\Omega$ and $\im\Omega$ to be respectively odd and even under $\sigma$; this translates to\footnote{One could note a difference with respect to \eqref{TIIAorientZF}. It is because here we have done a different choice for $\theta$.}
\begin{equation}
\re Z^K = \im \FF_L = 0 \:.
\end{equation}
Notice that while the first set of relations really cuts out half of the moduli space, the second set is automatically guaranteed on a CY manifold which admits the antiholomorphic isometry $\sigma$.
The flux $H^f$ is odd and the gauge potential $C$ is even, so
\begin{equation}
H=H^f= p_L \beta^L \qquad \qquad C= \xi^K \alpha_K \:.
\end{equation}

The stabilization proceeds on the same track as before. We substitute the expansions given above in the equations determining the solution. From \eqref{eq036} and \eqref{eq032} we get
\begin{gather} \label{eq065}
b^i = - \frac{f^i}{2m} \\
\frac{3m}{5} v^i v^j \, \kappa_{ija} = e_a + m \, b^i b^j \, \kappa_{ija} \:.
\end{gather} 
The axions $b^i$ are all fixed, as well as the K\"ahler moduli $v^i$. For these last ones we have as many quadratic equations as unknowns (provided that there is no $a$ such that $\kappa_{aij}$ is always zero), and, as pointed out in \cite{DeWolfe:2005uu}, one has only to check that the solution lies in the supergravity regime (among the others, one asks for large positive volumes $v^i$). Integrating the BI for $F$ on the cycles $\Gamma_L$ yields
\begin{equation}
m\,p_L = \mu_6 \frac{\re \FF_L}{\sqrt{4\mbox{vol}}} \:.
\end{equation}
This fixes all the remaining complex structure moduli%
\footnote{The equations are not invariant under scaling (what one would have expected for the projective coordinates), but this relies on the fact that a normalization for $\Omega$ is chosen, for example in \eqref{eq025}.}%
. Then subsituting in the solution \eqref{eq006} we find the dilaton
\begin{equation} \label{eq043}
e^\phi = \frac{5}{8} \frac{\mu_6}{m^2} \sqrt{ \frac{6}{v_a v_b v_c \, \kappa_{abc}} } \:.
\end{equation} 
Eventually, by direct application of \eqref{eq037} follows
\begin{equation}
-p_L \, \xi^L = e_0 + b_i e_i + \frac{1}{3} m \, b_a b_b b_c \, \kappa_{abc} \:.
\end{equation} 
Note that only this particular combination of the axions can be fixed, while for the other ones non-perturbative effects and $\alpha'$ corrections must be invoked. Anyway, the stabilization of axions is a minor problem, because their configuration space is periodic and compact, so any contribution which generate a nonconstant potential fixes them at a finite value.

\

As noted in \cite{DeWolfe:2005uu}, there is a gauge redundancy in the solutions described above, i.e. solutions which are transformed into each other by the gauge transformations \eqref{eq064} and following, are equivalent. In the four-dimensional low energy theory those translate in Peccei-Quinn symmetries that shift the axions:
\begin{equation}
b^i \to b^i + 1 \qquad \mbox{or} \qquad \xi^K \to \xi^K + 1 \:.
\end{equation}
These are accompanied by translations of the fluxes, and the correct transformation rules are obtained by \eqref{eq036}, \eqref{eq032} ,\eqref{eq037} by noticing that $F$, $G$ and $F_6$ are gauge-invariant.
The point is that one can always reduce to the case of $b^i$ and $\xi^K$ of order unity, and the large volume limit (the one reliable in supergravity) is controlled just by the fluxes $e_i$. This simplifies considerably the equations in the limit.

\

As in the particular case studied in the previous section, we have found the same results as \cite{DeWolfe:2005uu}: all the geometric moduli and the axions coming from $B$ are fixed, whilst only one combination of the $C$ axions is fixed. 

\section{Summary and Comments}

The flux compactifications have been largely studied during the last years. As we have already said a lot of times during this thesis, the main reason is that their contribution to the total energy depends on the moduli of compactification manifolds. Minimizing this energy fixes the value of the geometrical moduli (see section \ref{FluxCompactifications}). But this energy also contributes to the energy-momentum tensor, giving contribution to the Einstein equations. This contribution backreacts on the geometry, giving a solution that is no more Ricci flat. 

In the first part of this chapter we have described this departure from CY geometry given by the fluxes. We have seen it for supersymmetric solutions of the Einstein equations. This is because it is more simple to solve the supersymmetry equations ($\delta \psi = 0$) than the Einstein equation itself. One has to solve the supersymmetry equations and then impose the Bianchi identities and the equations of motion for the form fields. The result is that if one wants the minimal supersymmetry\footnote{{\it i.e.} $1/4$ of the original supercharges are preserved by the geometry.} in four dimension, the structure group of the six dimensional manifold must be reduced form $SO(6)$ to $SU(3)$. In any
case that the fluxes are turned on, the holonomy group of the Levi-Civita connection is not included in the structure group. This tells us that the compact manifold is not a CY. The departure from CY structure is encoded in the torsion classes, that vanish for a CY (actually a CY has a torsionless $SU(3)$ structure). 

In the work that we have presented here \cite{Acharya:2006ne}, we concentrated on the supersymmetric solutions that give rise to four dimensional theories on AdS$_4$ spacetime. These supergravity solutions have been classified in \cite{Lust:2004ig}. We added to these setup an orientifold O6-plane. This leaves the supersymmetry equations invariant, but changes the Bianchi identities. So the solutions are modified and include localized terms. To find an explicit solution we took the so called "smeared approximation", {\it i.e.} the orientifold charged is smoothed out through the compact manifold, by substituting a smooth 3-form to the singular 3-form $\delta_3$ that gives the location of the fixed point locus.

In the smeared case we can put the parameter $f$ and the flux $F$ to zero and get a torsionless solution, {\it i.e.} a CY, but with some fluxes turned on. These fluxes allow to stabilize all the CY moduli. To see this, we have substituted the KK ansatz in the supersymmetry equations and found the values that the moduli take.

Before \cite{Acharya:2006ne}, this ensemble of vacua had been studied only from a four dimensional point of view by \cite{DeWolfe:2005uu}, as described in section \ref{TypeIIAvacua}. In that work they studied the four dimensional effective potential, in the small flux approximation and they found complete moduli stabilization by minimizing the four dimensional effective potential. In \cite{Acharya:2006ne} we have given a ten dimensional description of the Type IIA CY with fluxes that was missing before.

\chapter{Warped Models in String Theory}

%\section{Introduction}

In the previous chapters we have seen various aspects of flux compactifications. In discussing Type IIB we have seen that the fluxes can generate a non-trivial warp factor (this is a main difference with respect to Type IIA case that we have studied in the last chapter). The {\it warp factor} is a factor in front of the four dimensional spacetime metric, that depends on the compact coordinates. It can so take very different values on different points of the compact space, generating a hierarchy of scales in the effective four dimensional theory.

Theories with strongly warped extradimensions have revealed novel features compared with the standard factorised compactifications. Such theories have been recently applied to phenomenological model building beyond the SM to address a variety of questions, such as the hierarchy problem and the fermion masses. The prototypical example of such applications is the Randall-Sundrum model \cite{Randall:1999ee}. Since this seminal paper, the state of the art five dimensional models have evolved somewhat \cite{Davoudiasl:1999tf,Chang:1999nh,Gherghetta:2000qt,Grossman:1999ra,Huber:2000ie,Agashe:2003zs,Agashe:2004rs,Huber:2000fh,Csaki:2002gy,Hewett:2002fe,Pomarol:2000hp,Randall:2001gb,Randall:2001gc,Goldberger:2002cz,Goldberger:2002hb,Agashe:2002bx,Contino:2002kc,Goldberger:2002pc,Agashe:2004bm}
(see, for instance, \cite{Gherghetta:2006ha} for a review). Moreover, there are potentially very interesting signals for the LHC, since these models are dual descriptions of `compositeness' \cite{ArkaniHamed:2000ds,Rattazzi:2000hs}.
Their most basic features are:
\begin{itemize}
\item[a)] for every standard model field, there is a five dimensional bulk field;
\item[b)] to solve the hierarchy problem, the Higgs is localized in a region of large warping; 
\item[c)] turning on bulk and boundary masses localizes the fermion zero modes and hence one obtains hierarchical Yukawa couplings since the fermions can have varying degrees of overlap with the Higgs.
\end{itemize}
Since these models have arbitrary parameters {\it e.g.} the bulk and boundary masses, in \cite{Acharya:2006mx} we decided to investigate the realization of these models in String Theory. This perspective offers a framework for explaining the parameters of the five dimensional models and some new insights:
\begin{itemize}
\item To realize a warped geometry we considered warped string compactifications which arise naturally in the IIB string theory with fluxes \cite{Giddings:2001yu,Verlinde:1999fy}, as we have seen in section \ref{WarpedSolutionsInTypeIIB}.
\item Matter and gauge fields in the bulk arise as strings which end on D7-branes in the bulk\footnote{
Previous studies of warped models in string theory had tended to have the standard model on D3-branes
\cite{Cascales:2003wn} See however \cite{Gherghetta:2006yq}.}.
\item To have several standard model generations, we turned on a topologically non-trivial (``instanton'') background field on the D7 worldvolume. 
\item Fermion zero modes then naturally localize near the instantons and/or by warping.
%This led to new features: a) the zero modes can be {\it localised anywhere} in the extra dimension and b) the scale of the topologically non-trivial background (instanton size) can also be used to suppress Yukawa couplings, in addition to the usual mechanism of separating the fermion zero modes in the extra dimension.
\end{itemize}

Our main results were explicit formulae for the profile of the fermion zero modes in the fifth dimension and their Yukawa couplings. 
These formulae show in particular how the physical size of the topologically non-trivial ``instanton'' background field can give rise to hierarchies of Yukawa couplings. They also show that the large Yukawa coupling is associated with a ``small instanton'' in the extra dimension.

\

In this chapter we will firstly give a brief review of the five dimensional warped models. In particular we will focus on the aspects that we realized in our string theory construction. Then we will explain the arising of warping in string theory and at the end we will describe the realization of the five dimensional features that we constructed in \cite{Acharya:2006mx}.

\section{Five Dimensional Models}

Five dimensional warped models are very interesting from a phenomenological point of view. They are based on the Randall-Sundrum idea of warping \cite{Randall:1999ee}: a non-factorisable geometry gives a chance to address the hierarchy problem between the electroweak and the Planck scale. It is really different with respect to the usual extradimensional models, where the metric is of the form $ds^2(x,y)= ds_{3,1}^2(x)+ds_C^2(y)$. In that case the hierarchy problem is addressed by taking the size of the extradimension very large: the fundamental scale is the (five dimensional) Planck scale, that is suppressed with respect to the four dimensional one. 

The RS models live in a five dimensional spacetime. The fifth dimension $y \in [0,2\pi R]$ is compactified on an orbifold of a circle $\Ss^1/\Zbb_2$. The orbifold action is $y\mapsto -y$. We have two fixed points at $y=0$ and $y=\pi R$. At each boundary there is a 3-brane. The one at $y=0$ is called the UV-brane, while the one at $y=\pi R$ is called IR-brane. 
The metric between the two branes is non-factorisable and takes the form:
\bee\label{5DmodelsMetric}
 ds^2= e^{-2\kappa y}\eta_{\mu\nu}dx^\mu dx^\nu + dy^2\:.
\ee
This is the $AdS_5$ metric and so the spacetime between the two 3-branes is simply a slice of $AdS_5$ geometry. The four dimensional metric is multiplied by a function depending on the extradimensional coordinate $y$, that is called {\it warp factor}. It is an exponential of $y$ where $y$ is a good coordinate to measure distances in the extradimension. It is equal to 1 on the UV-brane, while it is exponentially small ($e^{-2\kappa \pi R}$) on the IR-brane. $\kappa$ is the AdS curvature.

The four dimensional Planck mass is given by:
\begin{equation}
M_4^2=\frac{M_5^3}{\kappa}(1-e^{-2\pi\kappa R})
\end{equation}
It depends only weakly on the size of the extradimension $R$. Moreover the exponential warp factor has very little effect in determining the Planck scale. One can naturally take $M_5\sim \kappa$, obtaining a four dimensional Planck scale of the order of the fundamental scale.

On the other hand, the warp factor plays an important role in determining the four dimensional masses on the IR brane. In fact a generic mass scale $M$ in the five dimensional theory is scaled down to $e^{-\pi \kappa R}$ on the IR-brane. So with non-large extradimensions one can get large hierarchy of scales. In particular, if the Higgs field is localized on the IR-brane (at $y=\pi R$), the weak scale is exponentially suppressed with respect to the Planck scale. 

The first proposal of Randall-Sundrum \cite{Randall:1999ee} was to put all the Standard Model fields on the IR-brane. But to address the hierarchy problem, it is not necessary to localize the matter fields on the IR-brane. Moreover, this would introduce problems with operators associated to proton decay, neutrino masses and FCNC, that would be suppressed by a small mass, giving predictions inconsistent with experiments.

So in \cite{Gherghetta:2000qt} it was proposed to consider models in which the SM fermions and the gauge fields live in the five dimensional bulk. We will concentrate on the fermion fields. We include in the five dimensional Lagrangian both the kinetic term and a mass term. The five dimensional Dirac equation is given by:
\begin{eqnarray}\label{WMST5DDirEq}
 (g^{MN}\gamma_M D_N + m_\Psi) \Psi &=& 0
\end{eqnarray}
the mass $m_\Psi = c\,\kappa \,\epsilon(y)$ is an odd function of $y$. This is because $\bar{\Psi}(y)\Psi(y)$ is odd under the orbifold action and we want an even mass term $m_\Psi \bar{\Psi}(y)\Psi(y)$.
The covariant derivative $D_N$ contains the warp factor.  Making it explicit, we can write the equation \eqref{WMST5DDirEq} as:
\begin{eqnarray}
 e^{\kappa y}\eta^{\mu\nu}\gamma_\mu \partial_\nu \Psi_{(-)} +\partial_5\Psi_{(+)} + (m_\psi-2k)\Psi_{(+)}&=&0\nn\\
 e^{\kappa y}\eta^{\mu\nu}\gamma_\mu \partial_\nu \Psi_{(+)} -\partial_5\Psi_{(-)} + (m_\psi+2k)\Psi_{(-)}&=&0\nn
\end{eqnarray}
where the five dimensional Dirac spinor can be splitted into even and odd eigenvectors with respect to $\gamma_5$: $\Psi=\Psi_{(+)} + \Psi_{(-)}$ with $\gamma_5 \Psi_{(\pm)}=\pm \Psi_{(\pm)}$.

One then does the usual KK ansatz
\begin{eqnarray}
\Psi(x,y)&=&\sum_n \chi^{(n)}(x)\psi^{(n)}(y) %& {\rm {\tiny with}} & {\tiny\eta^{\mu\nu}\gamma_\mu \partial_\nu\chi^{(n)}=m_n\chi^{(n)}}\nn
\end{eqnarray}
where $\chi^{(n)}(x)$ are the KK modes satisfying $\eta^{\mu\nu}\gamma_\mu \partial_\nu\chi^{(n)}=m_n\chi^{(n)}$, and $\psi^{(n)}(y)$ is the profile of the KK mode in the bulk. Substituting this ansatz in the five dimensional Dirac equation \eqref{WMST5DDirEq}, one can find the expression for the zero modes:
\begin{equation}\label{5DmodelsFermProf}
d_{\psi 5d} \, \psi_{5d} \sim \sqrt{\frac{k (1-2c)}{e^{(1-2c)kR} -1}} e^{(2-c)ky} \;,
\end{equation}
where $d_\psi$ is a normalization constant. We see that the profile is not constant in the extradimensional coordinate $y$.

The Standard Model Yukawa coupling interactions are promoted to five dimensional interactions in the warped bulk:
\begin{equation}
  \int d^4x\int dy \sqrt{-g}\,\lambda^{(5)}_{ij} \bar{\Psi}_i(x,y)\,\Psi_j(x,y)\, H(x)\,\delta(y-\pi R)
\end{equation}
The Higgs is a four dimensional field localized on the IR-brane, and its profile is a delta function in the coordinate $y$.
If we insert the expression for the zero mode \eqref{5DmodelsFermProf} and for the metric \eqref{5DmodelsMetric} in the five dimensional Yukawa coupling term, we get the four dimensional Yukawa coupling:
\begin{eqnarray}
  \lambda_{ij} &\sim&  \left\lbrace \begin{array}{lr}
  \lambda^{(5)}_{ij} \kappa &c_{i,j}<1/2\\ \\
  \lambda^{(5)}_{ij} \kappa \,\,e^{(1-c_i-c_j)\pi \kappa R}&c_{i,j}>1/2\\
  \end{array}\right. 
\end{eqnarray}
$\lambda^{(5)}_{ij}$ is the dimensionfull five dimensional Yukawa coupling and it is taken to be of order of the fundamental scale, {\it i.e.} $\lambda^{(5)}_{ij}\kappa\sim 1$. 
The parameter $c_i$ determines how the fermion profile is localized in the compact direction. Depending on it we can have a large overlap of the fermion profile with the Higgs one ($c_{i,j}<1/2$), giving a top-like Yukawa coupling, or a small overlap ($c_{i,j}>1/2$), that gives an exponentially small Yukawa coupling. 

Summarizing, introducing a bulk mass term for the fermions gives localized profiles. The Higgs is localized on the IR-brane in order to have a suitable weak scale. Taking fermion profiles with different overlaps with the Higgs realizes the Yukawa hierarchy.

\section{String Realization}

In this section, we will see how to realize the interesting features that the five dimensional models in a string theory setup.

We will start by reviewing how to get warp compactifications in String Theory, then we will present our result in \cite{Acharya:2006mx}, {\it i.e.} how to realize the Yukawa hierarchy in this context.

\subsection{Warped String Compactifications}

In section \ref{WarpedSolutionsInTypeIIB} we have seen that in Type IIB String Theory, there are solutions with non-factorisable metric of the form:
\begin{equation}
 ds^2=e^{-4A(z)}\eta_{\mu\nu}dx^\mu dx^\nu + e^{4A(z)}\tilde{g}_{mn}dz^m dz^n\
\end{equation}
%Both D-branes and fluxes are sources of non-trivial warp factor. 
The regions of the compact manifold where $e^{-4A(z)}\ll 1$ are called {\it throats}. This is because these regions are small with respect to the metric $\tilde{g}$, but  become large with respect to the warped metric. So one can draw the picture in which the compact manifold is the manifold described by $\tilde{g}$ with some throats attached to it, where the warp factor is sensitively different from 1. In the throats, the four dimensional energy of the phenomena is redshifted by a factor of $e^{-2A}$ with respect to regions of negligible warping.

Both D-branes and fluxes are sources of non-trivial warping. One typical example is an orientifold $T^6$ compactification of Type IIB with a stack of $N$ D3-branes \cite{Verlinde:1999fy}  on one point (that we will choose to be $z=0$). These D3-branes and the O3-planes backreact on the geometry, giving the metric:
\begin{equation}\label{D3brMetric}
 ds^2 = \frac{1}{f(z)^{1/2}}ds_{3,1}^2 + f(z)^{1/2} dz^2
\end{equation}
Let us define  $r \equiv |z|$ and $L^4\equiv 4\pi N \,g_s \alpha'^2$. When $r \gtrsim L$ then  $f(z) \sim 1$ and locally the space is the product of the Minkowski spacetime and the six-torus. On the other hand, when $r \lesssim L$ then $f(z) \sim \frac{L^4}{r^4}$ and the geometry reduces to $AdS_5\times \mathcal{S}^5$:
\begin{equation}\label{WarpAdS5S5metric}
 ds^2 = \frac{r^2}{L^2}ds_{3,1}^2 + \frac{L^2}{r^2}dr^2 + L^2 d\Omega_5^2
\end{equation}

Inserting the background \eqref{D3brMetric} in the gravitational action, one gets the relation between the ten dimensional and the four dimensional Plack scales. The four dimensional reduced Planck mass $M_4$ is given by:
\begin{eqnarray}
 M_4^2=M_{10}^8 V_6^{w} &\mbox{ with }&V_6^{w}\equiv \int_{T^6} d^6z \, f(z) 
\end{eqnarray}
As in the five dimensional models, the warp factor modifies only weakly the relation between the Planck scales, that can so be taken of the same order. On the other hand, the warp factor generates a hierarchy of four dimensional scales. To see this, take the action of a scalar in the background \eqref{WarpAdS5S5metric}:
\begin{equation}
 S_H^{(p)}=-\frac{1}{2}\int d^4 x \int d^{p-4}z \sqrt{-g}f^{1/2}[(\partial H)^2 + \frac{1}{f^{1/2}}M ^2 H^2] \:. \nn
\end{equation}
If a scalar field is localized in a region with warp factor $f_0^{-1/2}$, then its mass is suppressed (with respect to the ten dimensional mass $M$) to $f_0^{-1/4}M$. This is the effect that we found also in five dimensional models. Hence we can address the hierarchy problem if we take the Higgs localized in a region of the compact space, where the warp factor is large.

One can generalize this setup and take as a compact manifold a CY, with some 3-form fluxes turned on and some D-branes and orientifold planes. All of these generate the warp factor. In some constructions (see page \pageref{TIIBwarpStab}) there is a point where the warp factor takes its minimal value, different from zero. In this case the throat is not infinite. This does not happens in the case studied above, where the warp factor $f^{-1/2}$ goes to zero when $r\rightarrow 0$.

\

The situation on the throat resembles what happens in a slice of $AdS_5$:
\begin{itemize}
\item There is a warp factor depending on an extra dimensional coordinate and that  generates hierarchy of four dimensional scales.
\item The role of the UV-brane is played by the bulk compact manifold (where $e^{-4A}\sim 1$).
\item There are string mechanisms to end the throat at $r_0>0$, avoiding vanishing warp factor. The IR-brane is associated with $r=r_0$. 
\end{itemize}
The question we tried to answer in \cite{Acharya:2006mx} is if other features of the five dimensional models can be realized in a string setup. In particular we find a setup where the matter lives in the bulk and the fermion profiles are localized in extradimensions, giving Yukawa hierarchy.

\subsection{The Setup: A Simple Example}

In this section we will describe a simple example which illustrates the setup we considered in \cite{Acharya:2006mx}.

As said above, our interest is understanding how various features of the five dimensional phenomenological models are realized in string theory vacua, with the motivation that this might lead to additional insights about the phenomenology. The three basic features which we aimed to understand better are:
\begin{itemize}
\item[a)] The five dimensional warped models tend to have the standard model gauge fields propagating in the bulk of $AdS_5$.
\item[b)] For each standard model fermion there is a five dimensional bulk fermion field with both bulk and boundary mass parameters which determine whether or not the fermion is localized in the UV or IR end of $AdS_5$.
\item[c)] The hierarchy amongst standard model Yukawa couplings is realized by the varying degrees of overlap
between these localized wavefunctions and the Higgs.
\end{itemize}
We study the string theory realization of these features within the context of Type IIB 
string theory vacua with fluxes, since this class of vacua realizes warped extra dimensions in a natural
way. In such vacua, non-Abelian gauge fields can reside on D3 and D7-branes, so in order to realize
property a) the only possibility is to put the standard model gauge fields on the D7-branes. Recall that
the ten dimensional spacetime is a warped product of four dimensional Minkowski spacetime $M^{3,1}$ and
a compact Calabi-Yau manifold $X$ \cite{Giddings:2001yu}. The metric takes the form of a D3-brane metric, where the D3-branes span the Minkowski spacetime. The D7-branes have a world-volume which is a warped product of $M^{3,1}$ and a four dimensional cycle $\Sigma \subset X$.

Now we turn to property b). The physics behind the introduction of bulk and boundary masses is that, before
symmetry breaking, the standard model fermions are all
zero modes of the Dirac operator on $M^{3,1}$. We thus need to study the Dirac equation on
the D7-brane in the
warped background. For the ten dimensional geometries described in \cite{Giddings:2001yu} the metric induced on
the D7-branes is of the form:
\begin{equation}
ds_8^2 = f(z)^{-1/2} \eta_{\mu\nu} dx^\mu dx^\nu + f(z)^{1/2} \,g_{\alpha\beta}\,dz^\alpha dz^\beta \qquad (\alpha,\beta=1,...,4) \:,
\end{equation}
where the warp factor $f$ is a function of the coordinates $z^\alpha$ on the 4-cycle $\Sigma$, which 
the D7-brane wraps
and $x^{\mu}$ are coordinates on $M^{3,1}$. For simplicity, we study the
warped geometry induced by D3-branes in flat spacetime. %Most of the interesting features we observe are not very sensitive to the geometry of $\Sigma$ as we will see.
In this case
\begin{equation}
ds_8^2 = f(r)^{-1/2} \eta_{\mu\nu} dx^\mu dx^\nu + f(r)^{1/2} \,\delta_{\alpha\beta}\,dz^\alpha dz^\beta \qquad (\alpha,\beta=1,...,4) \:,
\end{equation}
where $f(r) = 1+L^4/(r^2+d_0^2)^2$, $r^2= |\vec z|^2$ and $d_0$ is the separation between the D7 and the D3-branes. For simplicity, in this example, we set $d_0=0$.

%and consider the near horizon limit. The conclusion will not change with $d_0\neq 0$. 
We also use an almost ``flat'' radial coordinate $y$ defined by
\begin{equation}
r = L \, e^{-ky} \qquad \qquad k=\frac{1}{L}
\end{equation} 
For illustration, the near horizon geometry in these coordinates is
\begin{gather}
ds_8^2 = e^{-2ky} \eta_{\mu\nu} dx^\mu dx^\nu + dy^2  + L^2 d\Omega_3^2 \\
%f(y) = e^{4ky} \:,
\end{gather}
which is an $AdS_5 \times S^3$ contained in $AdS_5 \times S^5$.
In these coordinates, $y\to\infty$ is the tip of the throat while $y=0$ is its origin.
 
The low energy spectrum of the D7-brane modes includes massless fermions in the adjoint representation of the gauge group:
$\D_8 \Psi = 0$. Under the splitting induced by the D3-brane background, the fermions factorise as products
of fermions on $M^{3,1}$ and $\Sigma = \mathbb{R}^4$:
\begin{equation} \label{fermions splitting}
\Psi = \sum\nolimits_k \chi_k(x) \otimes d_{\psi_k} \, \psi_k(z)  \;,
\end{equation}
where $d_{\psi_k}$ is a normalization constant.

The Dirac equation can be written as\footnote{In our conventions, $\Gamma^\mu$ are the gamma matrices relative to the background metric, while $\gamma^\mu$ are relative to the flat metric.} 
\begin{equation}
\D_8 \Psi = \biggl( f^{1/4}\tilde{\D}_{3,1} + \frac{1}{f^{1/4}}\tilde{\D}_4 - \frac{1}{8f^{1/4}} \frac{f'}{f} \, 
		\gamma_r \biggr) \, \Psi =0\;,
\end{equation}
where $\tilde{\D}_{3,1}$ and $\tilde{\D}_4$ are respectively the Dirac operator on $M^{3,1}$ and on flat $\mathbb{R}^4$. Massless fermions in $M^{3,1}$ are the zero modes of $\bigl( \tilde{\D}_4 - \frac{f'}{8f} \, \gamma_r \bigr)$. As shown in the appendix \ref{AppWMST2} these are given by: 
\begin{equation}\label{eqRfermzeromd}
\psi = f^{1/8} \tilde{\psi} \:, \nn
\end{equation}
where $\tilde{\psi}$ are the zero modes of the operator $\tilde{\D}_4$. This means that in the warped background, the {\it four-dimensional zero modes are conformally equal to the zero modes in an unwarped geometry}.

\

The simplest possibility in this example is to take $\tilde{\psi}$ to be the constant zero modes
of the flat Euclidean Dirac operator $\tilde{\D}_4$ on the extra dimensions. Whilst this indeed would give us
a four dimensional fermion zero mode, it raises two problems:
\begin{enumerate}
\item since the fermion field $\Psi$ on the
D7-brane is in the adjoint representation, the four dimensional zero mode $\psi$ is also in the adjoint
representation;
\item there would be  {\it four} such fermion zero modes (since there are four constant spinors), whilst the standard model requires three generations of zero modes in representations which are certainly not adjoint.
\end{enumerate}
In principle, there is an elegant
solution to both of these problems, which also elucidates the string theory description of property b):
the gauge covariant Dirac operator $\tilde{\D}_4$ can have multiple non-trivial zero modes in the presence of topologically non-trivial gauge
field backgrounds. This is a standard mechanism to generate light fermion generations in string theory, however the novelty here is the presence of the warp factor in $\psi$ and that we will be quite explicit about
the profile of the wavefunction.

The background field strength should be a solution of the equations of motion. These come from the YM theory living on the D7-brane:
\begin{equation}
 S_{D7}= -\frac{1}{2g^2} \, \int d^8X \, \sqrt{-G}\, \mbox{Tr} \,( F\wedge\ast_8 F - F\wedge F\wedge C_4)\nn
\end{equation}
$G_{MN}$ and $C_4$ are the D3 background induced on the D7 worldvolume.
Among the solutions of the eight dimensional equations of motion there are gauge fields living in the Euclidean (4)-space and satisfying:
\begin{equation}
\ast_4 F = - F  \nn
\end{equation}			
This is the {\it instanton} anti-selfduality condition. So we will turn on a background instanton gauge field, living only in (4)-space. This also breaks the gauge group living on the D7 worldvolume.

As is well known from gauge theory instanton physics, gauge field-strengths satisfying the condition $F=-\ast_4 F$ in four Euclidean dimensions can be topologically non-trivial and support multiple
fermion zero modes which are not in the adjoint representation. Depending on the topological charge
(or instanton number) one can have different numbers of fermion zero modes.
One can check that such gauge field configurations also solve the equations of motion on the D7-brane, so are acceptable backgrounds.

The zero mode wave functions $\tilde{\psi}$ have been computed explicitly long ago 
for many different $F=-\ast_4 F$
backgrounds \cite{Osborn:1978rn}. 
If we take the simplest known solution to these equations \cite{tHooft:1976fv}, then we obtain
a zero mode which depends on the {\it size} of the instanton $\rho$,  as well as its position $\vec Z_\psi$
in the Euclidean space (see also appendix \ref{AppWMST1}): 
\begin{equation}
\psi(\vec z)= f^{1/8} \frac{\rho}{\big[ \rho^2+( \vec z - \vec Z_\psi)^2 \big]^{3/2}} \; \eta \;,
\end{equation}
here $\eta$ is a constant spinor normalized as $\eta^\dag \eta=1$.

These fermion zero modes have to be normalized properly. Consider the kinetic term:
\begin{multline}
-\int d^8x \sqrt{-G}\: G^{\mu\nu} \: \bar{\Psi}\Gamma_\mu \partial_\nu \Psi + ...\\
= -\int d^4x \, \eta^{\mu\nu}\,\bar{\chi}(x) \gamma_\mu \partial_\nu \chi(x)
\int d^4z \: d_\psi^2 \,f^{1/4} (z) \: \psi(z)^\dagger \psi(z) + ...
\end{multline}
where the normalization constant $d_{\psi}$ was introduced in the Kaluza-Klein ansatz \eqref{fermions splitting}
and we used $\Gamma_\mu = f^{-1/4}\gamma_\mu$.
In order to have a canonical kinetic term, we require:
\begin{eqnarray}
 d_\psi^2 \int d^4z\, f^{1/4}\,\psi^\dagger \psi&=& 1
\end{eqnarray}
In regions of negligible warping, this condition is realized for $d_\psi\sim 1$, whilst when the warp factor is large (for instance in the near horizon region) the normalization is given by:
\begin{align}
d_\psi^{-2} &= \int r^3dr \, 4\pi\sin\theta \, d\theta \, f(r)^{1/2}\, \frac{\rho^2}{(\rho^2+r^2 +Z_\psi^2-2r Z_\psi\cos\theta)^{3}} \\
&= \frac{\pi^2}{2}\left(\frac{\rho^2}{L^2}+e^{-2kY_\psi} \right)^{-1} \:,
\end{align}
where $|\vec Z_\psi|/L \equiv e^{-kY_\psi}$ is the radial position of the instanton in almost flat radial coordinates.
When $\rho/L < e^{-kY_\psi}$, we get $d_\psi \simeq (\sqrt{2}/\pi) \, e^{-k Y_\psi}$.

We see that in string theory
the instanton scale size is important in determining the profile of the fermion zero modes. Putting
all the factors together, the {\it normalized} zero mode wave function is:
\begin{equation}
d_\psi \, \psi \sim e^{-k Y_\psi} e^{\frac{k}{2}y} \frac{\rho}{\Bigl[ \rho^2 + (\vec z - \vec Z_\psi)^2 \Bigr]^{3/2}} \: \eta
\end{equation}
We can compare this wave function with the five dimensional profile \eqref{5DmodelsFermProf}. From this we learn that the zero mode wavefunction in string theory is quite different from the five dimensional models. Note that there is a dependence on the instanton scale size, $\rho$. In particular, in string theory {\it the zero mode can be localized anywhere} in the fifth dimension.

\subsubsection{Instantons as D3-branes}

As is well known, gauge field backgrounds on D7-branes with $F\wedge F \neq 0$ carry D3-brane
charge \cite{Douglas:1995bn}. 
In fact, smooth instanton backgrounds such as those we are considering here, are ``fat D3-branes''
with size $\rho$. Therefore, we can also say that the fermion zero modes are localized on fat D3-branes. 
The fermion zero modes are therefore 3-7 strings.
Note however that, in order to trust the metric we have been using, we should consider the number of such fat D3-branes to be small compared to the large number of ordinary D3-branes
and fluxes which generate the bulk geometry.

The parameters $\rho$ and $\vec Z_{\psi}$ are therefore moduli field vevs which arise in the open string
sector. It would be interesting to investigate mechanisms which stabilize these moduli. Presumably
closed and open string fluxes generate a potential for these fields. 

\

\subsubsection{Yukawa Couplings}

The zero mode profiles are crucial for computing the four dimensional Yukawa couplings, and clearly the answer will depend on $\rho$. In order to determine the Yukawa couplings, we need to identify the Higgs field in string theory. Essentially, with
only D3 and D7-branes the Higgs can be a 7-7 or a 3-7 string, since it must be charged under the standard model gauge group. The simplest case to consider is that the Higgs is a 3-7 string state. The 7-7 case will be described later.
In this case its wavefunction will be localized near a point $\vec Z_H$ in $\Sigma$ and we  will simply model this by a delta-function.
This choice is very similar to the standard five dimensional proposal \cite{Gherghetta:2006ha}.

We must first determine the correctly normalized four dimensional Higgs field from its kinetic term by imposing
\begin{multline}
-\int d^8x \sqrt{-\hat G_{3,1}}\: G^{\mu\nu} \: d_H^2 \: \partial_\mu \bar{H}(x)\partial_\nu H(x) \, \delta(\vec z - \vec Z_H) = \\
= - \int d^4x \, \partial_\mu \bar{H}(x)\partial^\mu H(x)
\end{multline}
which gives $d_H= f(|\vec Z_H|)^{1/4}$.

The four dimensional Yukawa coupling is obtained by direct dimensional reduction of the eight dimensional one (remembering localization of the Higgs):
\begin{multline}
\int d^8x \, \sqrt{-\hat G_{3,1}} \, \lambda^{(8)} \, d_H \, \bar \Psi \Psi H \, \delta(\vec z - \vec Z_H) = \\
= \lambda^{(8)} d_H d_\psi^2 f(|\vec Z_H|)^{-1} \psi(\vec Z_H)^2 \int d^4x \, \bar \chi(x) \chi(x) H(x) \:,
\end{multline}
so that
\begin{equation}
\lambda = \lambda^{(8)} d_\psi^2 \, \left. \frac{\psi^2(z)} {f(z)^{3/4}} \right|_{\vec Z_H} \:.
\end{equation} 
Remember that the eight dimensional  Yukawa has dimension of (length)$^4$. We see therefore that the Yukawa
coupling in the standard model is determined by several factors: the fermion zero mode evaluated at
the Higgs position, the warp factor at the Higgs position and the normalization constant $d_{\psi}$
(which itself depends on $\rho$ and $Y_{\psi}$).

\

Let us analyze the four dimensional Yukawa coupling further. For simplicity we study the case when
the fermion zero mode is localized in a region of large warping and  $\rho/L < e^{-kY_\psi}$. Then the four dimensional Yukawa coupling is given by:
\begin{equation}\label{Yukawa}
\lambda = \frac{2}{\pi^2} \lambda^{(8)} e^{-2k(Y_H + Y_\psi)} \frac{\rho^2}{\big[ \rho^2 + (\vec Z_H - \vec Z_\psi)^2 \big]^3} \:, \nonumber
\end{equation}
where again we used almost flat radial coordinates $|\vec Z_H|/L \equiv e^{-kY_H}$. In the standard model
the Yukawa couplings of the charged fermions range from order one for the top quark to $10^{-6}$ for the
electron, and clearly (\ref{Yukawa}) is rich enough to span this range. In more detail,
the top quark Yukawa coupling ($\lambda\sim 1$) can arise when the top wave function peaks at the
location of the Higgs {\it i.e.}  $Y_H = Y_\psi$:
\begin{equation}\label{eqR019}
\lambda = \frac{2}{\pi^2} \frac{\lambda^{(8)}}{\rho^4} e^{-4kY_H}
\end{equation}
Notice that, due to the warping in the spacetime,
$\rho$ is not the physical size $\rho_{phys}$ of the instanton, which depends upon its location in $AdS_5$:
\begin{equation}\label{physsize}
\rho_\text{phys} = \int_{|\vec Z_\psi|-\frac{\rho}{2}}^{|\vec Z_\psi|+\frac{\rho}{2}} ds = \int_{|\vec Z_\psi|-\frac{\rho}{2}}^{|\vec Z_\psi|+\frac{\rho}{2}} f^{1/4}(r) dr \simeq e^{kY_\psi} \rho \:,
\end{equation}
where the last result is valid when $\rho < Le^{-kY_\psi}$. The same can be seen by evaluating the instanton displacement in the almost flat radial coordinate: $\Delta y = e^{kY_\psi} \rho$. 
Note that in terms of the physical size, this is simply $\rho_\text{phys} < L$: 
the instanton is physically smaller than the $AdS_5$ radius, which is a natural requirement.
Substituting in \eqref{eqR019}, one gets:
\begin{equation}
\lambda = \frac{2}{\pi^2} \frac{\lambda^{(8)}}{\rho_{phys}^4}\:.
\end{equation}
In general, we expect $\lambda^{(8)}$ to be of order $\ell^4$, with $\ell$ the string scale, we obtain $\lambda \sim 1$ when $\rho_\text{phys}$ is of order of the string scale. In other words, {\it the instanton which localizes the top quark is a small instanton.} We therefore might expect strong quantum corrections to the top sector.
On the other hand, when $\rho_\text{phys}$ is larger than the fundamental scale, $\lambda$ is smaller than 1 and we can also realize smaller Yukawa couplings by localizing the corresponding fermions on large
instantons.
%For example, we could get the electron one by taking $\rho_{phys}\sim 10^{3/2} \ell$.

\

The smaller Yukawa couplings are actually better obtained in the case when $Y_\psi < Y_H$, which means that the fermion zero mode is localized far from the Higgs, and again when $\rho < L e^{-kY_\psi}$. The Yukawa coupling is then given by:
\begin{equation}
\lambda = \frac{2}{\pi^2} \lambda^{(8)} \, e^{-2k(Y_H -2Y_\psi)} \, \frac{\rho^2}{L^6} \:.
\end{equation}
This can be written as 
\begin{equation}
\lambda = \frac{2}{\pi^2} \, \frac{\lambda^{(8)}}{\rho_\text{phys}^4} \, \frac{\rho_\text{phys}^6}{L^6} \, e^{-2k(Y_H-Y_\psi)} \:.
\end{equation}
We see that even when the $AdS_5$ radius $L$ is just a little bigger than the instanton size, 
that {\it both the instanton scale size and the warp factor suppress the generic Yukawa coupling.}

%In particular, in order to suppress the Higgs mass from the 4d Planck scale $10^{19}$ GeV to the weak scale %1 TeV, we need $e^{-kY_H}\sim 10^{-16}$ which means $kY_H \simeq 37$. To get the electron Yukawa coupling %$\lambda_e\sim 10^{-6}$, it is enugh to put the instanton at $kY_\psi\simeq 30$, which is actually a %\textit{natural} configuration.

\subsection{The Higgs as a Vector Zero Mode}

In this section we will study the case that the Higgs is a 7-7 string which is a zero mode
of the 8-dimensional gauge field on the D7-brane. 
We will see that such zero modes are not affected by the presence of the warping and can also be computed in the instanton background. 

In the eight dimensional kinetic term, all the fields are in the adjoint representation of the gauge group $G$. The background instanton gauge field breaks this group, leaving a (3+1)-dimensional gauge theory, whose gauge group is a subgroup $G'$ of $G$. The adjoint representation of $G$ splits into irreducible representations of $G'\times SU(2)$, where $SU(2)$ is chosen as the gauge group of the instanton. Thus, an eight dimensional field in the Adj rep of $G$ can be written as a sum of products of fields in $M^{3,1}$ and $\Sigma$ in various representations
of $G' \times SU(2)$. In order to reproduce a GUT theory at low energy, we could take $G'$ to contain some GUT group as a subgroup.

Let us see some details. The eight dimensional kinetic term is:
\begin{eqnarray}
  &&\int d^8X \sqrt{-G} \bar{\Psi}\D \Psi 
\end{eqnarray}
and contains the term
\begin{eqnarray}
  g\int d^8X \sqrt{-G} \bar{\Psi}\not\!\! \delta \! A \Psi
  	&\supset& g\int d^4x \,\bar{\chi}_i(x) \chi_j(x) H_k(x)\,\int d^4y\, \psi_i^\dagger(y)\not\!\!\delta \! a_k(y) \psi_j(y)\nn
\end{eqnarray}
where $g$ is the eight dimensional gauge coupling (of order $\ell^2$, with $\ell$ the string length) and  where we have used the splitting \eqref{fermions splitting} of the fermion fields and that of the vector:
\begin{equation}
 A(x,y)_m dy^m =A_{bkg}(y)+\sum_k H^k(x) \delta a_k(y)\:.
\end{equation}
We see that the effective Yukawa coupling in $(3+1)$-dimensions is given by:
\begin{equation}\label{eqR008}
 g \, d_{\psi_i} d_{\psi_j} d_H \int d^4y \, \tilde{\psi}_i^\dagger(y) \not\!\!\delta \! a_k(y) \tilde{\psi}_j(y)\:.
\end{equation}
where we have substituted the expression \eqref{eqRfermzeromd} for the fermion zero modes $\psi$.
Note that the warp factor has disappeared; it only enters in the fermion normalization constants\footnote{For this particular choice for the Higgs, its normalization is not affected by the warping and will be put $d_H=1$}. The zero modes $\delta a_k(y)$ are warp factor independent
because the Yang-Mills action on $\Sigma$ is conformally invariant.

The fields $\psi_i$, $\psi_j$ and $\delta a_k$ are in the $SU(2)$ representations dictated by the splitting of Adj$G$ and by the $G'$ representations that one wants $H$, $\chi_i$ and $\chi_j$ to belong to.

We compute the integral \eqref{eqR008} in the simple case in which the two fermions are in the fundamental representation of $SU(2)$, while the vector zero mode is in the adjoint. We will see that the coupling can be highly suppressed in the usual approximation of well separated instantons, and that this suppression is due to the localization of the zero modes near individual single instantons. This justifies this simple choice of representations, since the localization is characteristic of the zero modes in any representation. This is important, because the suppression works whatever $SU(2)$-representations are associated (by the splitting of Adj$G$) with the particular GUT-representations that one wants to find in the GUT Yukawa interaction terms. It would be interesting to compute the integral exactly, since new phenomena might arise.

\

We consider the 't Hooft solution with instanton number $k=2$. This solutions has $5k=10$ explicit parameters: $\rho_1$, $\rho_H$, $\vec Z_1$ and $\vec Z_H$. The zero mode profiles when $k>1$ are given in the appendix \ref{AppWMST1}. We also choose both the fermion zero modes in \eqref{eqR008} to be localized around $\vec Z_1$, while the vector one (the Higgs) is to be localized around $\vec Z_H$. We put $\vec Z_H$ in a region of large warping, in order to address the hierarchy problem.
% We consider the zero modes in the integral \eqref{eqR008} as follows: the $\psi_i$ are the fermion zero mode $(v^\dagger b f)_1$(given by \eqref{eqR014} with $k=2$), while $\delta a$ is the vector zero mode relative to traslations of $X_H$\eqref{eqR015}. 
We will see that, in order to have a sufficiently large top Yukawa coupling, one must have $\delta a$ sharply localized around $\vec Z_H$. 

We substitute the expressions \eqref{eqR014} and \eqref{eqR015} in \eqref{eqR008} and estimate it in several asymptotic regions of the parameter space of the $k=2$ solution. With more than one instanton, we find a new suppression mechanism:  due to the localization of wavefunctions at well separated points, suppression can also occur due to a hierarchy in the two instanton sizes $\rho_1$ and $\rho_H$. The maximal value of the integral is actually obtained when $|\vec Z_H-\vec Z_1|\ll \rho_1,\rho_H$ and $\rho_1\sim \rho_H$. 

Actually when  $|\vec Z_H-\vec Z_1|\ll \rho_1,\rho_H$, the parameter $X\equiv |\vec Z_1-\vec Z_H|$ disappears from the result, that is:
\begin{equation}\label{eqRYukVec}
 g \, d_\psi^2 \int  \psi_i^\dagger \sigma^\mu \delta \! A_\mu^\Phi \psi_j \simeq g \, d_\psi^2 \alpha^\Phi \int r^3 dr\,
	 \frac{\rho_1^2\rho_H^2}{(r^2+\rho_1^2+\rho_H^2)^4} = \, d_\psi^2 \frac{g \alpha^\Phi}{24} 
	\frac{\rho_1^2 \rho_H^2}{(\rho_1^2+\rho_H^2)^3}
\end{equation}
where $\delta \! A_\mu^\Phi$ is defined in \eqref{eqR015}, and where $\alpha^\Phi$ is a constant of order one.
The expression \eqref{eqRYukVec} takes its maximal value when $\rho_1\sim\rho_H$:
\begin{equation}\label{eqR016}
 g \, d_\psi^2 \int  \psi_i^\dagger \sigma^\mu \delta \! a_\mu^\Phi \psi_j \sim \, \frac{2}{\pi^2}\frac{g}{\rho_H^2} e^{-2\kappa Y_H} 
\end{equation}
The same result as \eqref{eqR016} is obtained taking $k=1$. %\footnote{Or equivalently in the case of general $k$ and in the approximation of well separated instantons, the same resul is got taking the top as the fermion zero mode $(v^\dagger b f)_2$).}.
Then one has to substitute the physical size in this formula (see \eqref{physsize}). The final result is:
\begin{eqnarray}
 \lambda &=& \frac{2}{\pi^2}\frac{g}{\rho_{H\text{phys}}^2}
\end{eqnarray}

From here, we see that if one wants the top coupling to be of order one, the top zero mode must be localized close to the Higgs and the $\rho$-parameter of the corresponding instanton has to be of the order of the Higgs one. %Moreover, the physical instanton size corresponding both to the Higgs and to the top quark has to be of order of string length. In particular, that means that the profile of the Higgs zero mode (and consequently of the top) is very sharp (as Gherghetta suggested taking it to be a delta function). 

\

The Yukawa hierarchy can then be obtained by varying the instanton parameters in such a way as
to have different overlaps of the zero modes. One can approximate the integral giving the Yukawa
couplings in different asymptotic regions of the instanton moduli space.
We summarize the results in Table \ref{tab}. In order to get the actual Yukawa coupling, this integral has to be multiplied by $d_\psi^2$ and the instanton `sizes' have to be substituted with their physical sizes. Let us consider some relevant cases, which turn out to be similar to the result found in the simple example of the previous section.
\begin{itemize}
\item When the fermions are localized around the same position of the Higgs:
	\begin{equation}
	\lambda = \frac{g}{\rho_{\psi\text{phys}}^2}\left( \frac{\rho_H}{\rho_\psi}\right)^2\nn
	\end{equation}
\item When the fermions are far from the Higgs:
	\begin{eqnarray}
	\frac{X}{\rho_\psi}\frac{\rho_H^2}{\rho_\psi^2}\gg 1 &\rightarrow& \lambda = \frac{g}{\rho_{\psi\text{phys}}^2}\left( 	
		\frac{\rho_{\psi}}{X}\right)^4\nn\\
	\frac{X}{\rho_\psi}\frac{\rho_H^2}{\rho_\psi^2}\ll 1 &\rightarrow& \lambda = \frac{g}{\rho_{\psi\text{phys}}^2}\left(	
		\frac{\rho_{\psi}}{X}\right)^3 e^{-2\kappa (Y_\psi-Y_H)}\nn
	\end{eqnarray}
\end{itemize}
 
\

\begin{table}[ht]
\begin{center}
\begin{tabular}{|c|cl||}
\hline&&\\
limits & & $g \int d^4z \, \tilde{\psi}_i^\dagger(z)  \Phi_H(z) \tilde{\psi}_j(z)$ \\ && \\ \hline&&\\
$\rho_H\sim\rho_\psi\ll X$ &   $\frac{g}{\rho_{H}^2}$&$\left( \frac{\rho_H}{X}\right)^3$  \\ && \\
$\rho_H\ll\rho_\psi\sim X$ &  $\frac{g}{\rho_{H}^2}$ & $\left( \frac{\rho_H}{X}\right)^2$   \\ &&\\
$\rho_H\ll\rho_\psi\ll X$ &  $\frac{g}{\rho_{H}^2}$ & $\left( \frac{\rho_H}{X}\right)^2 \left( \frac{\rho_\psi}{X}\right)^2 
				\left[ 1+\frac{X}{\rho_\psi}\left(\frac{\rho_H}{\rho_\psi}\right)^2\right]$  \\ &&\\
$\rho_H\ll X \ll \rho_\psi$ &  $\frac{g}{\rho_{H}^2}$ & $\left( \frac{\rho_H}{\rho_\psi}\right)^4  
				\left[ 1+\left(\frac{X}{\rho_H}\right)^2\left(\frac{X}{\rho_\psi}\right)^2\right]$  \\ &&\\
$X \lesssim \rho_H \ll \rho_\psi$ &  $\frac{g}{\rho_{H}^2}$ & $ \left( \frac{\rho_H}{\rho_\psi}\right)^4$  \\ &&\\\hline
\hline
\end{tabular}\caption{Various limits of the integral giving the Yukawa coupling. \label{tab}}
\end{center}
\end{table}
%\newpage

\section{Summary and Comments}

We have seen that there is a rather intricate string theory picture underlying many of the important
features of the five dimensional warped phenomenology models. 
The hierarchy problem is addressed in the same way: a non-factorisable geometry is taken, in which the four dimensional metric is multiplied by a function of the extradimensional coordinates, the warp factor. It generates a naturally exponential hierarchy between four dimensional scales. In string theory both fluxes and D-brane configurations generate a warp factor. Regions of the compact manifold with large warping are called throats. One solves the hierarchy problem, by localizing the Higgs in such regions. 

In five dimensional models, the matter fields live in higher dimensions and the fermion zero modes are localized through the introduction of five dimensional mass terms. We realized this situation by considering fields living on the eight dimensional  worldvolume of a D7-brane. We mimic the mass terms by turning on a non-trivial background gauge field (an instanton in the four euclidean extradimensions). It indeed gives localized profile for the fermion zero modes. The new feature of the string construction is that the zero modes can be {\it localized anywhere} in the extra dimensions.

Finally exponentially Yukawa hierarchy is generated by localizing the fermions far from the Higgs position, while the top Yukawa coupling is obtained when the top zero mode is localized near the Higgs. This mechanism works both in five dimensional models and in our string construction. The new feature of the last one is that,  the scale of the topologically non-trivial background (instanton size) can also be used to suppress Yukawa couplings, in addition to separating the fermion zero modes in the extra dimensions. In particular this implies that to realize the largest Yukawa coupling the top must be associated to a small size instanton.

\

A natural question arises: can we distinguish the string theory models from the five dimensional phenomenology?

Obviously, yes in principle: the spectrum of the five dimensional models consists of the zero modes
which become the standard model particles after symmetry breaking; then in addition, 
for each standard model particle there is an infinite Kaluza-Klein tower of resonances 
with the {\it same spin} as its associated
standard model cousin. These particles are also present in the string spectrum, but the string
theory has more: for each standard model particle, there is also an {\it infinite tower of string states
of increasing spins}. So, measuring even part of the spectrum could be enough to distinguish them.%
\footnote{Usually, in the holographic limit \cite{Maldacena:1997re} we decouple these massive open string states, but
here we cannot since the string length and string coupling is finite.}

In the five dimensional models, the masses of the Kaluza-Klein modes are typically quantized in units of a TeV.
Therefore, the LHC will only be sensitive to the first or second resonance. What about the string states?
The $AdS_5$ scale is of order $m_p$ so, for weak string coupling the string scale is below this.
However, the D7-branes fill the entire $AdS_5$ and hence, the 7-7 strings which are in the
infrared end of $AdS_5$ will have a TeV scale or lower mass: 
hence only the first or second of these will be directly
accessible at the LHC. Since these states have the same gauge quantum numbers as the Kaluza-Klein
modes, they could only be distinguished by their decay patterns or their spins. For example, there
might be a spin 3/2 colored particle which is a string excitation of the gluon. If produced,
this particle must eventually decay into jets, the angular distributions of which will be sensitive
to its spin. It would be interesting to study to what extent these events can be selected
and the discovery reach for the LHC.

\

We conclude with a discussion of some additional issues which deserve further investigation.
Firstly there is the issue of supersymmetry breaking. In five dimensional models, one does not {\it a priori} need
supersymmetry at all, since the electroweak scale is generated through the warped extra dimension.
But in Type IIB string theory, there is certainly local supersymmetry in the UV, and one needs to
break it. One possibility is to choose the background fluxes and geometry to explicitly break supersymmetry,
such as was recently considered in \cite{Burgess:2006mn}. 
However, backgrounds which explicitly break supersymmetry in string theory can often be unstable; thus,
it would be good to investigate this further.

Secondly, there is the issue of fermion chirality. With one collection of parallel D7-branes,
even though the backgrounds we have considered generate multiple copies of the same standard model
representations, the representations include both fermion chiralities. This can be avoided by
the introduction of another set of D7-branes intersecting the first set along a surface in $\Sigma$,
but we have not investigated this in detail. Also, in five dimensional models, the chirality problem is resolved by
considering a $\mathbb{Z}_2$ orbifold and perhaps such a mechanism can also be realized in string theory.
Finally, it could be interesting to extend this construction to non-flat background, such as the Klebanov-Strassler throat.

%\part{Phenomenology}

\chapter{Proton Decay in Theories with Localized Fermions}

In this chapter we %is a bit disconnected from the rest of the thesis. Here we 
will study the decay of the proton in theories coming from String/M-theory. In particular we will focus on theories where the fermions are localized in the extradimensions while the gauge bosons can propagate in them. We have seen an example of this in compactifications of M-theory on singular $G_2$ manifolds (see section \ref{MthCompG2holMan}).

We will first review the four dimensional GUT theory and how the decay of the proton arises in it. Then we will go to higher dimensional theories and see how GUT arises in M-theory compactifications on $G_2$ manifolds, concentrating on the proton decay. Finally we will describe the results of our paper \cite{Acharya:2005ks}, {\it i.e.} a mechanism that suppresses some proton decay channels with respect to the four dimensional prediction.

\section{Four Dimensional GUT and Proton Decay}

The basic idea in a Grand Unified Theory (GUT) is that the Standard Model gauge group $G_{SM}=SU(3)_c\times SU(2)_L\times U(1)_Y$ is embedded in a larger underlying group $G$. In this case the additional symmetries may restrict some of the features that are arbitrary in the Standard Model. The group $G$ is broken spontaneously, giving at low energy the SM gauge group.

A typical consequence of this embedding is that the new symmetry generators and their associated gauge bosons involve both flavor and color. So the new interactions generally violate the conservation of the baryon number and in most models lead to {\it proton decay}. 

The proton is observed to be stable in nature. In fact the experimental limit on its lifetime is extremely restrictive: $\tau_p\gtrsim 10^{32}$years. This requires that the baryon number violating interactions must be weak, and gives bounds on the scale of the interaction and on the masses of the massive gauge bosons: $M_X\gtrsim10^{15}GeV$. 

If $G$ is a simple group, there is only one coupling constant at scales larger than $M_X$. At these energies the spontaneous symmetry breaking (SSB) effects are negligible and the strong, weak and electromagnetic interactions are unified; the quarks and leptons behave very similarly and are put together in representations of $G$. At energies smaller than $M_X$, the SSB becomes important  and the running of the three coupling constant relative to $SU(3)_c$, $SU(2)_L$ and $U(1)_Y$ become different; the $G$ representations decomposes in $G_{SM}$ representations and the different terms behave differently. Taking the running back, one can predict the scale $M_{GUT}$, that is the scale where the running coupling constants meet each others. It is of the same order as the gauge boson mass $M_X$. Thus, one can check if it is consistent with the bound given by the proton lifetime.

\

Let us be more specific, and choose $G=SU(5)$. This will be the GUT group that we will study in the higher dimensional theories.
The field content is given by:
\begin{description}
\item {\bf Gauge Bosons}. They are in the adjoint representation of $SU(5)$ (${\bf 24}$). When $SU(5)$ is broken to $G_{SM}$, this representation is splitted into the sum of SM representations $({\bf r}_{SU(3)},{\bf r}_{SU(2)})^{Q_{U(1)}}$:
\bee \begin{array}{ccccccccc}
  {\bf 24} & \rightarrow & ({\bf 8,1})^0 & \oplus & ({\bf 1,3})^0 & \oplus & ({\bf 1,1})^0 & \oplus& ({\bf 3,2})^{-5/3}\oplus({\bf \overline{3},2})^{5/3}\\
   & & & & & & & &\\
  A_\mu^A && G_\mu^\alpha&& W_\mu^{\pm},W_\mu^0 && B_\mu&&  X_\mu^a,Y_\mu^a\, ;\, \bar{X}_\mu^a,\bar{Y}_\mu^a  \\
  & & & & & & &&\\
%   ({ 5\times 5})&&{\left(\baa{cc} 3\times 3&0\\0&0\\ \eaa\right)} &&{\left(\baa{cc} 0&0\\0&2\times 2\\ \eaa\right)} &&{
%   \left(\baa{cc} -\frac{\lambda}{3}{\bf 1}_3&0\\0&\frac{\lambda}{2}{\bf 1}_2\\ \eaa\right)}&&
%   {\left(\baa{cc} 0& 3\times 2  \\2\times 3&0\\ \eaa\right)} \\
  \end{array}\nn
  \ee
  The first three terms are the gauge bosons of the Standard Model, while the bosons $X_\mu,Y_\mu$ are the massive gauge bosons mediating the proton decay, and are called {\it lepto-quark bosons}.
\item {\bf Fermions}. Each family of 15 fields is placed in a ${\bf \bar{5}} \oplus {\bf 10}$ representation. Their decompositions are:
  \bee \baa{ccccccc}
  {\bf \overline{5}} &\rightarrow & ({\bf \overline{3},1})^{1/3} &\oplus& ({\bf 1,2})^{-1/2}&&\\
  && ( d^c & ; & \nu , e^- )_L&&\\
  &&&&&&\\
  {\bf 10} &\rightarrow & ({\bf \overline{3},1})^{-2/3} &\oplus& ({\bf 3,2})^{1/6} &\oplus& ({\bf 1,1})^{1}\\
  && ( u^c & ; & u , d &;& e^+)_L\\
  \eaa\nn\ee
\item {\bf Higgs}: The Higgs doublet field responsible for the SSB $G_{SM} \rightarrow SU(3)_c\times U(1)_{el}$ is embedded in the ${\bf 5\oplus \overline{5}}$ representation of $SU(5)$:
  \bee \baa{ccccc}
  {\bf 5} &\rightarrow & ({\bf 3,1})^{-1/3} &\oplus& ({\bf 1,2})^{1/2}\\
  && ( H_t & ; & \Phi )\\
  \eaa\nn\ee
  $H_t$ is a color triplet, while $\Phi$ is the SM Higgs. $H_t$ can also mediate proton decay so they are constrained to be very massive  by the bound on $\tau_p$. On the other hand $\Phi$ has a weak scale mass. It is the so called {\it Doublet-Triplet splitting problem}: one has to find a mechanism that suppresses the mass operators for $\Phi$ but not those for $H_t$. We will see that there is a natural such mechanism in M-theory context \cite{Witten:2001bf}. 
\end{description}

The $SU(5)$ GUT models have both attractive and less attractive features. Among the first ones, we have:
\begin{itemize}
\item $SU(5)$ incorporates the SM gauge group as  a maximal subgroup.
\item The electric charge is quantized. This comes from the fact that the electric charge operator $Q_e$ is a generator of $SU(5)$ and so traceless. For example the condition Tr$Q_e=0$ in the ${\bf \bar{5}}$ representation implies $3q_d + q_e =0$, and hence $q_d=-\frac{1}{3}q_e$.
\item The $B-L$ charge is conserved.
\item The $B$-violating operators can explain the asymmetry between baryons and antibaryons.
\item The ${\bf 5}$ Higgs gives $m_b/m_\tau\gtrsim 3$ for three families (the result is different for a different number of families).
\item There are no FCNC effects associated with the light gauge bosons.
\end{itemize}
Among the less attractive features there are:
\begin{itemize}
\item Each family is in a reducible representation.
\item There are difficulties with the predictions for $m_s$ and $m_d/m_s$.
%\item ...  ADD ...
\end{itemize}

\subsection{Proton Decay}

In this section we will concentrate on the GUT interactions that drive the decay of the proton. There are different operator contributing to the nucleons decay. In supersymmetric theories the D=4 and D=5 operators give the most important contributions. The  D=6 operators coming from lepto-quark bosons exchange are the most important contributions in non-supersymmetric theories, but also in supersymmetric one if the D=4 and D=5 operators are suppressed (we will see an example of this in the extradimensional models we will study). The D=6 operators coming from Higgs exchange are less important.

Our main interest is in the D=6 operators coming from massive gauge bosons exchange. Their contribution comes from the matrix elements of an operator product:
\bee\label{ProtDecAmplOp}
 g_{\rm GUT}^2 \int d^4x J^\mu(x) J_\mu(0) D(x,0)
\ee
where $D(x,0)$ is the propagator of the heavy lepto-quark gauge bosons. Because the proton is so large compared to the range of $x$ that contributes appreciably in the integral, we can replace $J^\mu(x)$ by $J^\mu(0)$ and use
\bea
 (\Delta + M_X^2)D(x,0) = \delta^4(x) &\Rightarrow& \int d^4x\,D(x,0) = \frac{1}{M_X^2}
\eea
Replacing in \eqref{ProtDecAmplOp}, one gets an effective 4-fermions interaction:
\bee\label{ProtDecAmplOpApp}
 \frac{g_{\rm GUT}^2}{M_X^2} J^\mu J_\mu(0)
\ee
The current is given by $J \equiv {\bf J}^{\bf \bar{5}\oplus 10}$. From this interaction term one obtains that the lifetime of the proton is given by
\bee\label{protonlifetime4d}
 \tau_p \sim \frac{1}{\alpha_{\rm GUT}^2}\frac{M_X^4}{m_P^5}
\ee
where $m_P$ is the proton mass.

From \eqref{protonlifetime4d}, we see that having a bound on $\tau_p$ (from experiments) gives a bound on $M_X$. As we have said before, $M_X$ can be predicted independently from the running of the SM coupling constants, that meet each other at $M_{\rm GUT}\sim M_X$ and one can check if these two results are consistent with each other.

\

Let us be more precise on the operators governing the proton decay in the $SU(5)$ GUT theory. The possible D=6 operators coming from \eqref{ProtDecAmplOpApp} and giving contribution to the proton decay are:
\bea
 \OO_I &=& \frac{g_{\rm GUT}^2}{2M^2_X}\overline{u^c_{\alpha L}}\gamma^\mu Q_{\alpha L}^T (i\sigma^2) \overline{e^c_{\beta L}} \gamma_\mu Q_{\beta L} \leftrightarrow J^{\bf 10}  J^{\bf 10}\\
 \OO_{II} &=& \frac{g_{\rm GUT}^2}{2M_X^2}\overline{u^c_{\alpha L}}\gamma^\mu Q_{\alpha L}^T (i\sigma^2)\overline{d^c_{\beta L}} \gamma_\mu L_{\beta L} \leftrightarrow J^{\bf 10}  \tilde{J}^{\bf \bar{5}}
\eea 
In the above expressions $M_X$ and $g_{\rm GUT}$ are the mass of the lepto-quark bosons and the coupling constant at the GUT scale.
$Q_L=(u_L,d_L)$ and $L_L = ( \nu_L, e_L)$ are the $SU(2)$ doublets; $\sigma^2$ is the Pauli matrix and acts on the doublets $Q_L$ and $L_L$. $\alpha,\beta$ are family indices, while the color indices are suppressed.  $J^{\bf 10}$ and $\tilde{J}^{\bf \bar{5}}$ are the fermion currents associated to the two matter representations; the current associated with one family is ${\bf J}^{\bf \bar{5}\oplus 10} =\tilde{J}^{\bf \bar{5}} + J^{\bf 10}$. The operators above are written in the interaction basis.

As we have said before, all these operators preserve the $B-L$ charge; this means that the proton always decays into an antilepton. A second rule is satisfied by these operators: $\Delta S/\Delta B=-1,0$.

The two different operators contribute to different decay channels:
\begin{description}
\item $\OO_I \leftrightarrow J^{\bf 10}  J^{\bf 10}$ gives in the final state a left-handed antilepton ({\it i.e.} an $SU(2)$ singlet); a typical example of such decays is  $p^+\rightarrow \pi^0{e^+}_L$. Feynman diagrams contributing to this process are given in figure \ref{PrDec1}.
\begin{figure}[ttt]
\begin{center}
  \epsfig{file=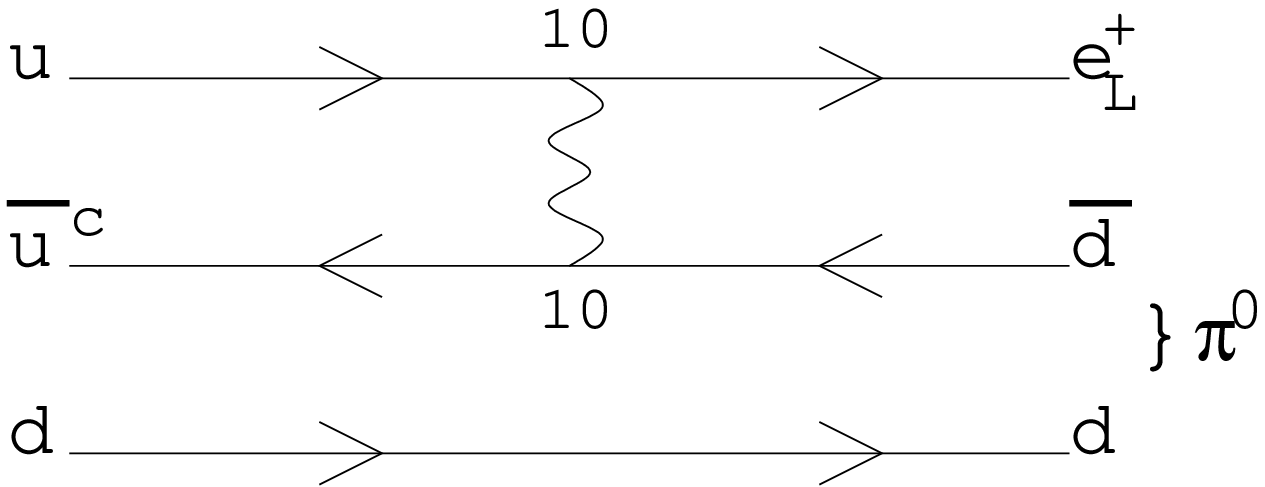,height=25mm,angle=0,trim=0 0 0 0} \hspace*{7mm}
  \epsfig{file=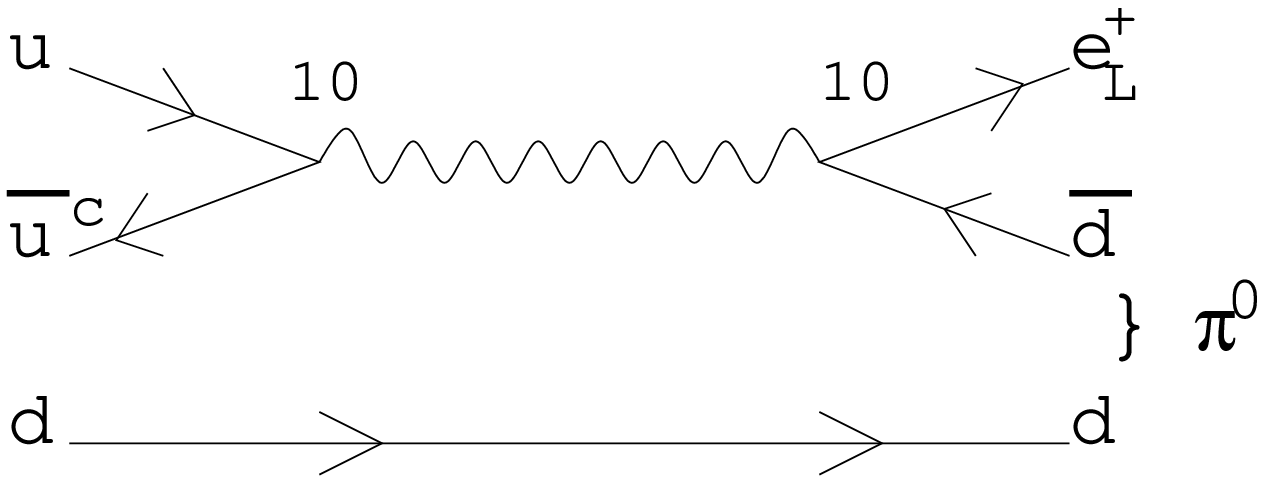,height=25mm,angle=0,trim=0 0 0 0}\\ \vspace*{5mm}
  \epsfig{file=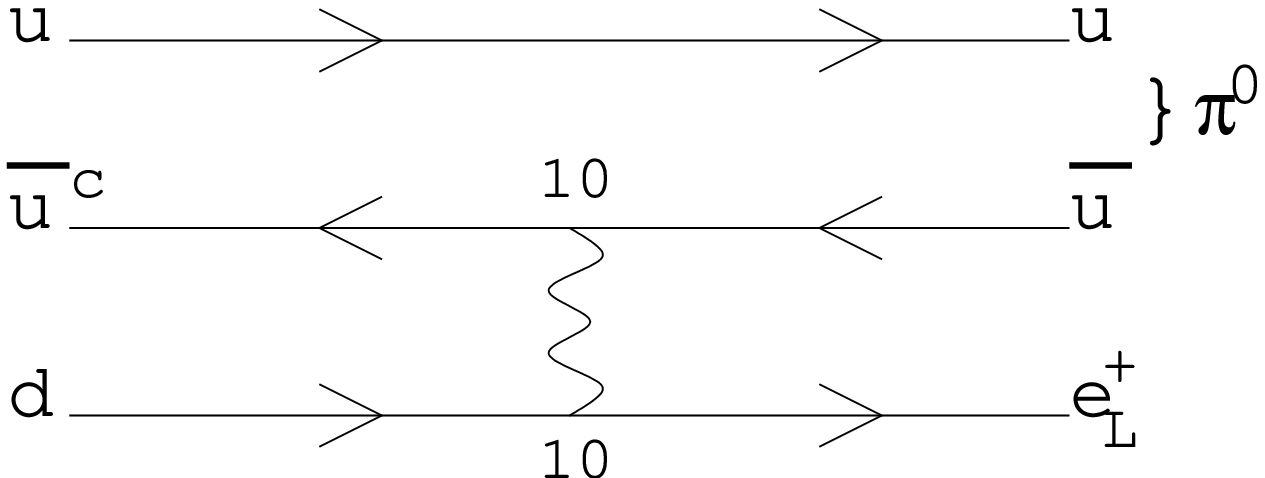,height=25mm,angle=0,trim=0 0 0 0}
  \caption{\small Feynman diagrams giving the decay channel $p^+\rightarrow \pi^0{e^+}_L$.}
  \label{PrDec1}
\end{center}
\end{figure}
\item $\OO_{II} \leftrightarrow J^{\bf 10}  \tilde{J}^{\bf \bar{5}}$ gives a right-handed antileptons ({\it i.e.} an $SU(2)$ doublet); as examples we have $p^+\rightarrow \pi^0{e^+}_R$ and $p^+\rightarrow \pi^+{\bar{\nu}}_R$. Feynman diagrams contributing to this processes are given in figures \ref{PrDec2} and \ref{PrDec3}.
\begin{figure}[tt]
\begin{center}
  \epsfig{file=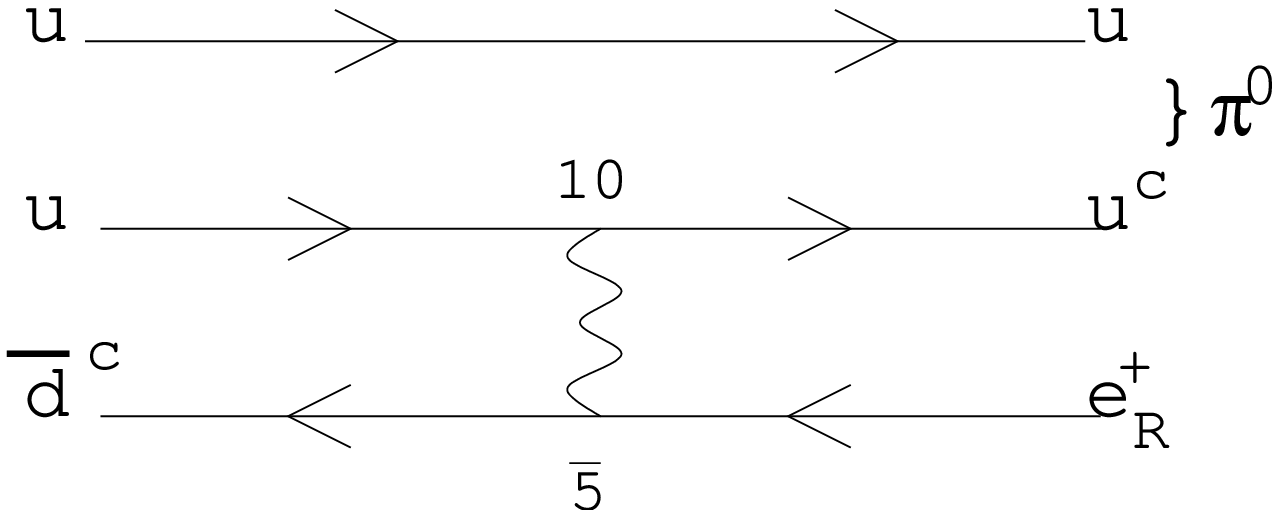,height=25mm,angle=0,trim=0 0 0 0} \hspace*{7mm}
  \epsfig{file=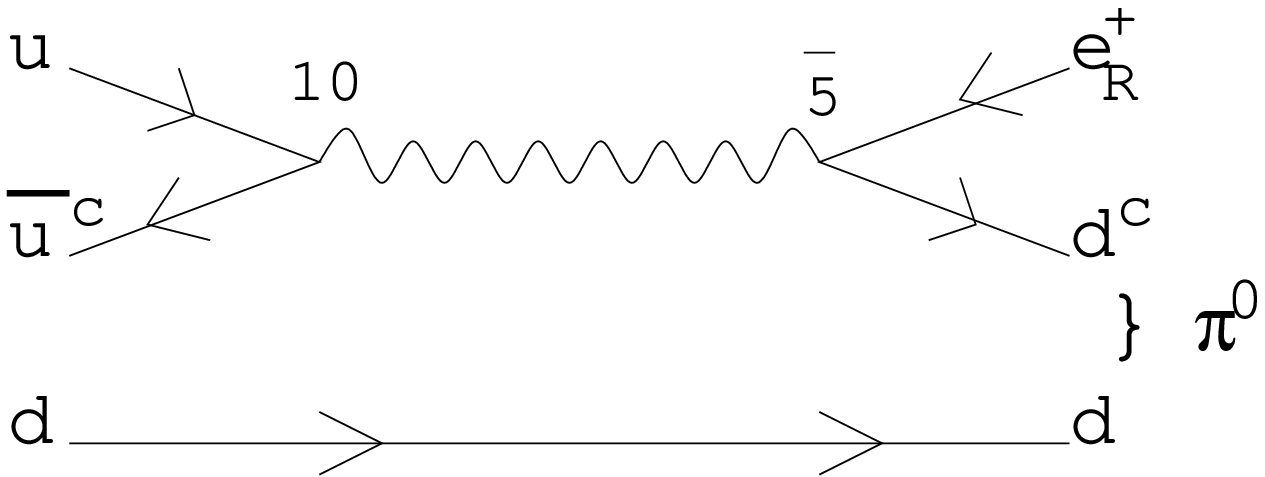,height=25mm,angle=0,trim=0 0 0 0}
  \caption{\small Feynman diagrams giving the decay channel $p^+\rightarrow \pi^0{e^+}_R$.}
  \label{PrDec2}
\end{center}
\end{figure}
\begin{figure}[tt]
\begin{center}
  \epsfig{file=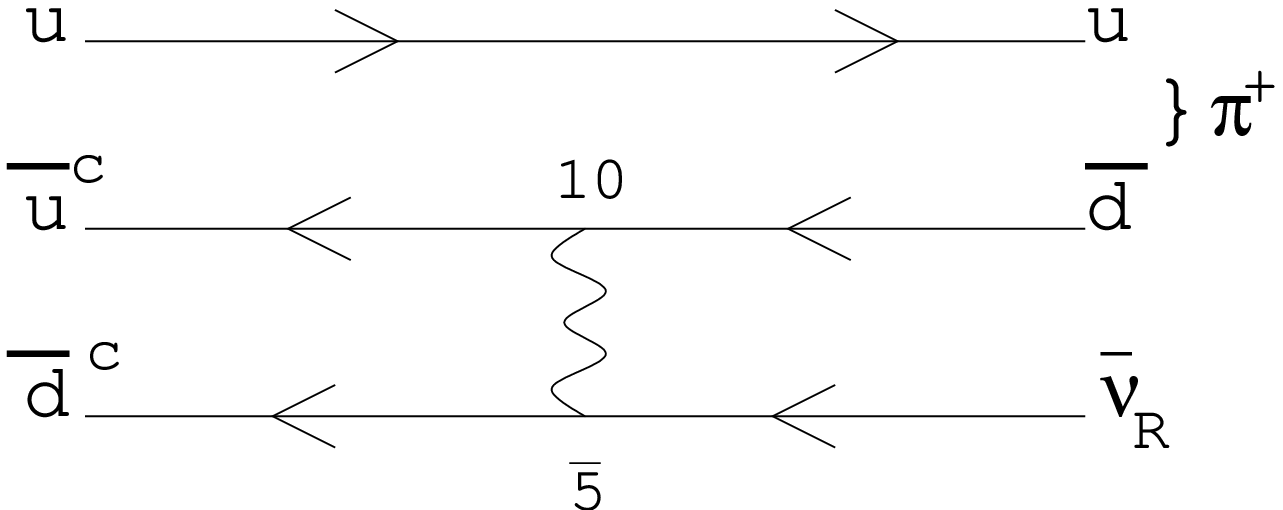,height=25mm,angle=0,trim=0 0 0 0} \hspace*{7mm}
  \epsfig{file=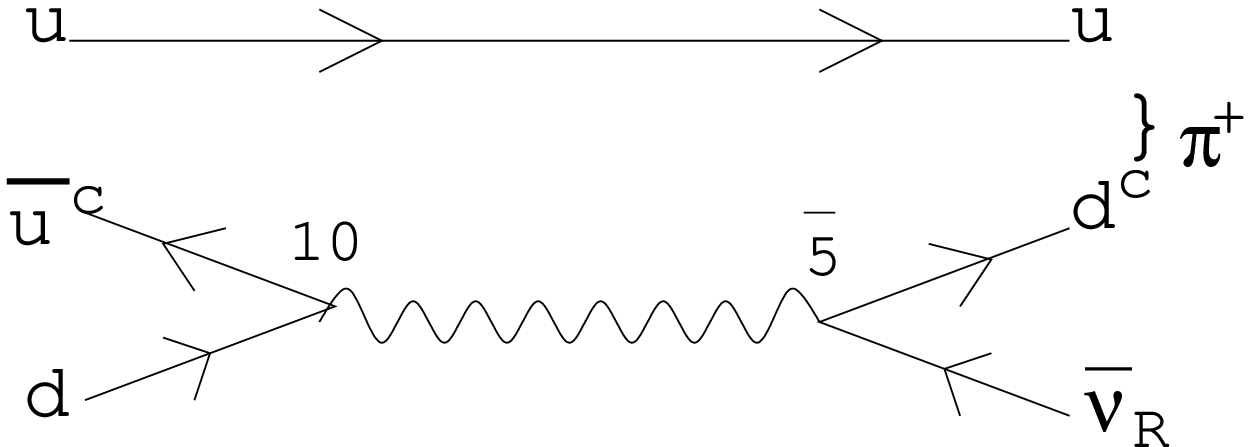,height=25mm,angle=0,trim=0 0 0 0}
  \caption{\small Feynman diagrams giving the decay channel $p^+\rightarrow p^+\rightarrow \pi^+{\bar{\nu}}_R$.}
  \label{PrDec3}
\end{center}
\end{figure}
\end{description}
We note that the operator $\tilde{J}^{\bf \bar{5}}\tilde{J}^{\bf \bar{5}}$ does not contribute to the proton decay, because at least a u-quark must be involved in the interaction. 

Above we have presented decays that involve fermions only in the first family. Actually there are other decay channels that give also a muon or a Kaon in the final state. What is important for our future treatment is the chirality of fermions that are in the final state that says if the decay has been driven by $\OO_I$ or $\OO_{II}$. It is not important the family they belong to. 

\

The D=6 operators coming from Higgs exchange are quite model dependent and are generally less important than the gauge contribution. In fact they are suppressed with respect to them by a factor $\lambda_i\lambda_j$ where $\lambda_i$ are Yukawa couplings of quarks involved in the proton decay (and so exponentially small).

\

All we have said so far is valid both for non-supersymmetric and for supersymmetric GUT theories. We are more interested in the supersymmetric ones, because in this case the SM running coupling constants meet all together at the same point giving the GUT scale. Moreover the models we will study in the following are all derived in a supersymmetric context. The main problem with the supersymmetric GUT theories is that the proton stability is more difficult relative to the non-supersymmetric case. This is because proton decay arises from D=4 and D=5 operators in addition to D=6 gauge bosons contributions. These new operators generally give too large decay rate and must be suppressed in order to give acceptable models. The D=4 operators can be eliminated by imposing R-parity conservation. R-parity is defined by $R=(-1)^{2S}M$, where $S$ is the spin and $M=(-1)^{3(B-L)}$ is the matter parity, which is $-1$ for all the matter superfields and $+1$ for Higgs and gauge superfields. The D=5 operators can be eliminated imposing some discrete symmetries. We will see an example in the M-theory context.

\subsubsection*{Experimental Tests of Proton Decay.}

In the 80's there were large scale experiments for the detection of proton decay\cite{Nath:2006ut}. They were mainly of two kinds: they use tracking calorimeter ({\it e.g.} SOUDAN\cite{Thron:1989cd}) or Cherenkov effect ({\it e.g.} Kamiokande\cite{Hirata:1989kn}). These experiments yielded null results but set lower bounds on various proton decay modes. 

In the 90's Super-Kamiokande\cite{Viren:1999pk} came on line. It is the currently most sensitive proton decay experiment. It is a ring imaging water Cherenkov detector containing 50 kton of ultra pure water held in a cylindrical tank 1km underground in Japan. The surface of the cylinder are covered by photomultilier tubes. When a relativistic particle pass through the water, they emit a cone of Cherenkov light in the particle direction of travel. By measuring the charge produced in each photomultiplier tube and time at which it is collected, it is possible to reconstruct the position and energy  of the event as well as the number, identity and momenta of the individual charged particles in the event.
The signature of {\it e.g.} $p^+\rightarrow e^+\pi^0$ is given by three cones: one generated by the positron and the other two generated by the two photons coming from the pion decay. Naturally, this process must be distinguished by other processes giving the same particles, such as the scattering of atmospheric or solar neutrinos with nucleons. So far no clear signal of proton decay has been observed. This experiment has however improved the bounds on proton decay rates\cite{Yao:2006px}. Here we report the rates of the decay channels that are interesting for our work:
\begin{equation}
\begin{array}{|l|c|}
\hline
{\bf\mbox{Channel}} & \tau_p (10^{30}years)\\
%\hline 
&\\
p\rightarrow e^+\pi^0 & 1600 \\
p\rightarrow \mu^+\pi^0 & 473 \\
p\rightarrow \bar{\nu} \pi^+ & 25 \\
p\rightarrow e^+K^0 & 150 \\
p\rightarrow \mu^+K^0 & 120 \\ 
p\rightarrow \bar{\nu} K^+ & 670 \\ \hline  
\end{array}\nn
\end{equation}

\

Other more sophisticated experiments are going to start in the next years. They will use Cherenkov detector with water ({\it e.g.} Hyper-Kamiokande\cite{Nakamura:2003hk}) or with noble gases ({\it e.g.} ICARUS which will use liquid Argon\cite{Rubbia:2004yq}). These experiments will either find proton decay or at least improve significantly the lower bounds and eliminate many models. For example, Hyper-Kamiokande is to explore the proton lifetime at least up to $\tau_p/B(p^+\rightarrow e^+\pi^0)>10^{35}years$ in a period of about 10 years\cite{Nakamura:2003hk}.

%$\clubsuit$ Yuk Unification pag16Wit.$\clubsuit$

\section{GUT Theories in Extradimensions}

In this section we study GUT theories living in more than four spacetime dimensions. Most of these models do not precisely lead to four dimensional GUT's, since unification takes place in higher dimensions. This leads among the other things to the possibility for GUT symmetry breaking by discrete Wilson lines and to higher dimensional mechanisms for doublet-triplet splitting, as we will see in the models we will study.

%We describe the proton decay in these models and we concentrate on the result presented in [...]. In this work we found a mechanism that suppresses the rate of the decay channels induced by the operator $\OO_{II}$. It is based on the fact that the gauge bosons propagate in extradimension, while fermions are four dimensional object, localized in the extradimensions. M-theory compactified on singular $G_2$ manifolds gives naturally such kind of models, as we have seen in section \ref{MthVac}.
In particular, we will consider theories in which the fermions and Higgs particles of the Standard Model are localized in the extra dimensions, but in which the gauge fields propagate in (part of) the bulk. The full spacetime is thus of the form $M^{3,1}\times X$ with $M^{3,1}$ our four dimensional spacetime and $X$ the compact extra dimensions. The Standard Model matter particles are localized at points on $X$ and the gauge fields propagate along a submanifold $Q$ of $X$ times the four dimensional spacetime. In the GUT context, the GUT gauge group could be broken to $SU(3) \times SU(2) \times U(1)$ by a Wilson loop of the gauge field on $Q$. We will also restrict our attention to theories in which the leading contribution to the violation of baryon number comes from dimension six operators (the analog of the gauge boson contribution in the original non-supersymmetric four-dimensional GUT's).

Although our results are more generally applicable, we will for concreteness focus on the case of M-theory compactifications on manifolds of $G_2$-holonomy described in section \ref{MthVac}. As we have seen, they provide an explicit realization of theories of this kind. Here $X$ is a seven dimensional manifold with $G_2$ holonomy, $Q$ is a three dimensional submanifold along which $X$ has a particular orbifold singularity, and the chiral fermions are localized at particular kinds of conical singularity. 
In such models the D=4 and D=5 baryon number violating operators are naturally suppressed \cite{Witten:2001bf}.

Also, for definiteness we will restrict attention to the case where the GUT gauge group is $SU(5)$. It is realized by taking $Q$ to be the three dimensional locus of $\Zbb_5$ orbifold singularities inside $X$. 

Before %turning on the Wilson loop which breaks 
breaking the gauge group $SU(5)$ to the Standard Model gauge group, each generation of (supersymmetric) Standard Model matter resides in the ${\bf \bar{5} \oplus 10}$ with Higgs particles in the ${\bf {\bar 5}  \oplus 5}$. So with the minimal field content there are eight points $P_i \subset Q$ where matter is localized: two for the Higgs multiplets, three for the ${\bf 10}$ matter and three for the anti-fundamental generations.

If $Q$ has incontractible loops, so that its fundamental group $\pi_1(Q)$ is non-empty, it is possible to break $SU(5)$ to the 
Standard Model gauge group by a Wilson line in the vacuum. This modifies the Kaluza-Klein spectrum with respect to zero background gauge field; for example the lightest modes of the gauge fields corresponding to the unbroken generators remain massless,  while the others generically get a non-zero mass.

For an example, we take $Q=\Ss^3/\Zbb_p$ \cite{Friedmann:2002ty}. This space has non-contractible circles which correspond to open curves in $\mathcal{S}^3$ that connect two points identified by the elements of $\Zbb_p$. The background gauge field can be taken to be a Wilson line around such cycles $\gamma_\Gamma$ (relative to the generator $\Gamma\in\Zbb_p$). For instance, the following Wilson line breaks $SU(5)$ to the Standard Model gauge group (as long as $5q$ is not a multiple of $p$):
\bee\label{UG} 
   U_\Gamma=P \,e^{i\oint_{\gamma_\Gamma} A_{\rm bkg}}=\left(
	\begin{array}{ccccc}
		e^{4\pi i q/p}&&&&\\& e^{4\pi i q/p}&&&\\&&e^{4\pi i q/p}&&\\&&&e^{-6\pi i q/p}&\\&&&&e^{-6\pi i q/p}\\
	\end{array}\right).
\ee
The introduction of discrete Wilson lines, together with discrete symmetries of $Q$, gives the possibility to solve the doublet-triplet splitting problem and the suppression of the D=4 and D=5 baryon number violating operators.
Let us see how it happens in this specific case \cite{Witten:2001bf}.

The 3-sphere can be parametrized by two complex coordinates $z_1,z_2$ satisfying $|z_1|^2+|z_2|^2=1$. The $\Zbb_p$ acts on them as 
\bea\label{ProtDecZbbident}
 \Zbb_p : \: z_i \mapsto e^{2\pi i/p}z_i && i=1,2
\eea
It acts freely on $\Ss^3$ and so $Q=\Ss^3/\Zbb_p$ is a smooth manifold. Its fundamental group is $\Zbb_p$.
Moreover $Q$ admits a global symmetry $F\cong \Zbb_p$ that acts as:
\bea\label{ProtDecFaction}
 F : \: z_1 \mapsto z_1 & z_2\mapsto e^{2\pi i/p}z_2 
\eea
The fixed point set of $F$ consists of two circles: $S_1$ defined by $|z_1|=1,z_2=0$,  and $S_2$ defined by $z_1=0,|z_2|=1$. $S_1$ is left fixed by $F$, while $S_2$ is left fixed by $F$ once we consider the identification \eqref{ProtDecZbbident}. 
After having turned on a Wilson line around cycles given by $\Zbb_p$ identification, if we apply a $\Zbb_p$ transformation to a charged object, we have to apply to it also the corresponding Wilson line \cite{Witten:2001bf}. 

Since $F$ leaves fixed $S_1$ trivially, we can take $F$ to act trivially on fibers of the gauge bundle over $S_1$. On the other hand, since $F$ leaves fixed $S_2$ only modulo an element $\Gamma\in\Zbb_p$, the transformation \eqref{ProtDecFaction} must be accompanied by the $U_\Gamma$ transformation, on the charged objects over $S_2$.

Now we can place the matter and Higgs superfields on points of $S_1$ or $S_2$ in such a way to avoid the doublet-triplet splitting problem or dangerous operators. Only the terms invariant under $F$ survive in the Lagrangian. 

Let us begin with the doublet-triplet splitting problem. We place Higgs fields in the ${\bf 5}$ representation of $SU(5)$ on $S_1$ and those in the ${\bf \bar{5}}$ on $S_2$. So the $\Phi$ and $H_t$ in the ${\bf 5}$ transform in the same way under $F$, say ${\bf 5}_H\mapsto e^{i\alpha}{\bf 5}_H$. But the $\Phi$ and $H_t$ in the ${\bf \bar{5}}$ transform differently, say ${\bf(\bar{2}_H\oplus\bar{3}_H)\mapsto (e^{i\delta}\bar{2}_H\oplus e^{i\gamma}\bar{3}_H)}$. $e^{i\alpha},e^{i\delta}$ and $e^{i\gamma}$ are arbitrary $p^{th}$ roots of 1. The doublet-triplet splitting problem is solved by choosing the charges such that 
$e^{i(\alpha+\gamma)}=1$ but $e^{i(\alpha+\delta)}\not=1$. Then a ${\bf 3_H}{\bf\bar{3}_H}$ term in the superpotential is invariant, while a ${\bf 2_H}{\bf\bar{2}_H}$ is forbidden by the $F$ symmetry and must be generated at lower energies.

Let us now assign the $F$ charges to the matter fields. We assume that all  fields in the ${\bf 10}$ transform as 
${\bf 10}_i\mapsto e^{i\sigma} {\bf 10}_i$ $\forall i$, and all fields in ${\bf \bar{5}}$ as 
${\bf \bar{5}}_i\mapsto e^{i\tau} {\bf \bar{5}}_i$ $\forall i$.\footnote{They are assumed to be localized all on $S_1$.}
To give masses to the up quarks, we want ${\bf 5}_H {\bf 10}^2$ terms in the superpotential, that implies $e^{i(\alpha+2\sigma)}=1$. To give masses to the down quarks we need ${\bf \bar{5}}_H {\bf 10} {\bf \bar{5}}$ terms, and so $e^{i(\delta+\sigma+\tau)}=1$. To get mass terms for the neutrino, we need ${\bf 5}_H^2 {\bf \bar{5}}^2$ terms, implying $e^{2i(\alpha+\tau)}=1$. But we do not want terms such ${\bf 5}_H {\bf \bar{5}}$, and so $e^{i(\alpha+\tau)}=-1$. Solving these conditions one gets the constraints:
\bea
 \alpha = -2\sigma & \tau= 2\sigma +\pi & \delta = -3\sigma + \pi
\eea
Further constraining $\sigma$ one can prevent D=4 and D=5 baryon number violating interactions. In particular D=4 operators come from ${\bf 10} {\bf \bar{5}}^2$ terms, implying $5\sigma \not =0$, while D=5 operators come from ${\bf 10}^3 {\bf \bar{5}}$ terms, implying $5\sigma + \pi \not = 0$.

%$\clubsuit$ Yuk Unification pag16Wit.$\clubsuit$

\subsection{Proton Decay in Extradimensions}

% Following \cite{Friedmann:2002ty} we now describe how the Greens function on $Q$ appears in the calculation
% of the proton decay amplitude at dimension six in theories of this kind.
% 
% A matter current which can absorb or emit a massive gauge boson is of the form
% \be
% J_{\mu} = J_{\mu}^{\bf \bar{5}} + J_{\mu}^{\bf 10}
% \ee
% where the subscripts indicate the origin of the particles involved.
% 
% 
% In the case of four dimensional GUT theory, the gauge boson contribution to the proton
% decay amplitude is essentially
% \be
% g_{GUT}^2 \int d^4 x J_{\mu}(x) \tilde{J}^{\mu}(0) D(x,0)
% \ee
% where $J$ and $\tilde{J}$ are the two currents involved and $D$ is the propagator of the massive gauge
% boson. The latter transforms as $(\bf{3},\bf{2})^{-5/3}$ under the Standard Model gauge symmetry. Since
% the size of the proton is much bigger than the integration region which gives the dominant contribution, we can replace
% $J^\mu(x)$ by $J^\mu(0)$. In this case the integral gives the result
% \be\label{ampl4}
% {g_{GUT}^2 J_{\mu}(0) \tilde{J}^{\mu}(0) \over M^2}
% \ee
% with $M$ the boson mass. This is a consequence of the equation for the propagator
% \be\label{prop4}
% (\Delta_{4} + M^2 )D(x,0) = \delta(x,0)
% \ee
% where $\Delta_{4}$ is the  four dimensional Laplacian.

In this section we will see how the expression \eqref{ProtDecAmplOpApp} for the D=6 baryon number violating operators is modified in the higher dimensional theories under discussion here. In this case one must also include the contribution of all charged Kaluza-Klein modes in the $(\bf{3},\bf{2})^{-5/3}$ representation. The seven dimensional propagator $D(x, y;x', y')$ is a function of the coordinates $y$ on $Q$ as well as $x$ on $M^{3,1}$ and the currents are functions of $x$ but are labelled by the points $P_i$ which are the values of $y$ where the matter particles are located. So we get a term of the form
\bee
g_7^2 \int d^4 x J_{\mu}(x, {P_1}) \tilde{J}^{\mu}(0, {P_2}) D(x,P_1 ; 0, P_2 )
\ee
Again we can replace $J^\mu(x)$ by $J^\mu(0)$, so the previous expression is well approximated by
\bee\label{ampl7}
 g_{\rm GUT}^2 {\rm Vol}_Q J_{\mu}(0, {P_1}) \tilde{J}^{\mu}(0, {P_2})\int d^4 x D(x,P_1 ; 0, P_2 )
\ee
where we have substituted $g_{\rm GUT}^2 = \frac{g_7^2}{{\rm Vol}_Q}$.

This amplitude has to be compared with (\ref{ProtDecAmplOpApp}). Up to a factor of $M_X^2 {\rm Vol}_Q$, the difference is given by the $P_i$ dependent function:
\bee\label{GG}
   G(y_1,y_2)\equiv \int_{M^{3,1}} d^4x D(x,y_1;0,y_2).
\ee

The seven dimensional propagator satisfies
\bee\label{prop7}
(\Delta_{4} + \Delta_{Q})D(x,y_1;0, y_2) = \delta(x,0)\delta (y_1 , y_2)
\ee
where $\Delta_{Q}$ is the gauge covariant Laplacian on $Q$. From this we see that the eigenvalues of $\Delta_Q$ act as masses$^2$ from the four dimensional viewpoint.

$D(x,y_1;0,y_2)$ is the contraction of the Feynman propagator on $M^{3,1}\times Q$ of the seven dimensional gauge fields in the $(\bf{3},\bf{2})^{-5/3}+(\bf{\bar{3}},\bf{2})^{+5/3}$ representation of the Standard Model gauge group:
\bee
   D(x,y_1;0,y_2)=\frac{1}{(2\pi)^4}\sum_k\int d^4p \frac{e^{-i p\cdot x} \bar{\Psi}_k(y_1)\Psi_k(y_2)}{-p^2+\lambda_k}
\ee
where $\Psi_k$ are the eigenfunctions on $Q$ of $\Delta_Q$
\footnote{
When Wilson loops are turned on, the Laplacian on $Q$ is defined as $\Delta^{A}=g^{mn}\nabla_m^{A}\nabla_n^{A}$ where $\nabla^{A}$ includes both the spin connection on $Q$ and the gauge connection related to the Wilson loop.}
with eigenvalues $\lambda_k\leq 0$, and the integral over $p$ is considered after euclidean continuation.

When there are no zero modes of the Laplacian on $Q$, one can substitute this expression in (\ref{GG}) and get:
\bee\label{GG1}
   G(y_1,y_2)=\sum_k \frac{\bar{\Psi}_k(y_1)\Psi_k(y_2)}{\lambda_k}
\ee
{\it ie} the Green's function of the scalar Laplacian on $Q$ for scalar fields valued in  $(\bf{3},\bf{2})^{-5/3}$ 
representation.

When there is a non-zero background gauge field such that  the $SU(5)$ symmetry is broken to the Standard Model gauge group, the Laplacian typically has no zero modes in the space of functions with values in $(\bf{3},\bf{2})^{-5/3}+(\bf{\bar{3}},\bf{2})^{+5/3}$ and the expression (\ref{GG1}) is well defined. $G(y_1, y_2)$ is the Green's function of the Laplacian for scalar fields in this representation. 

As we have seen in the four dimensional case, the $JJ$ operator is decomposed as:
\bee
J_{\mu}J^{\mu} = J^{\bf 10}_{\mu} J^{\mu{\bf 10}} + J^{\bf 10}_{\mu} J^{\mu{\bf \bar{5}}}
+ J^{\bf \bar{5}}_{\mu}J^{\mu{\bf 10}} + J^{\bf \bar{5}}_{\mu}J^{\mu{\bf \bar{5}}}
\ee
Only the first term contributes to the cross-section for the decay of the proton into left-handed positrons. The second and third contribute to the decays into neutrinos whereas the last term does not contribute to the decay. So for the decays modes such as $p\rightarrow \pi^0e^+_L$  studied in \cite{Friedmann:2002ty}  both ${\bf 10}$ currents are localized at the same point on $Q$. The corresponding Greens function in (\ref{GG1}) is therefore evaluated at $P_1 = P_2$ for this decay channel and therefore the classical formula is {\it divergent}\footnote{Note that in the cases when $Q$ is one dimensional, the Green's function is 
not divergent when $P = P'$. This actually happens in some orbifold GUT models \cite{Hebecker:2002rc,Kawamura:2000ev,Altarelli:2001qj,Hall:2001pg,Hebecker:2001wq}}. This is presumably
regularized in M-theory \cite{Friedmann:2002ty}.

However, since generically the points supporting the ${\bf \bar{5}}$ and the ${\bf 10}$ are distinct (for example to generate reasonably small Yukawa couplings), for the decay channels involving neutrinos, the ${\bf 10}$ is at a point $P_1$ distinct from the point $P_2$ supporting the ${\bf \bar{5}}$ current. Therefore the current-current correlator depends explicitly on the Green's function on $Q$ evaluated at two different points $G(P_1 ; P_2)$. When $G(P_1 ; P_2)$ takes a small value the decay of the proton into neutrinos is suppressed accordingly. Generically, $Q$ is a curved, compact manifold and the Green's function will be a non-trivial function of the geodesic distance $d(P_1 , P_2)$ between the points. In order to investigate the behavior of such functions, in particular, whether or not they can take small values, we will present some explicit sample calculations in the M-theory context.

\

To compute the Green function, we need to know the scalars $\Psi_k$ in \eqref{GG1}. These are eigenfunctions of the Laplacian, that take values in the $(\bf{3},\bf{2})^{-5/3}$ representation. Moreover they must be well defined functions on $Q$. For simplicity we refer again to the example $Q=\Ss^3/\Zbb_p$. If we take a loop $\gamma_\Gamma$ in $Q$, a scalar $\Phi$ is well defined when $\Phi(y) = \Phi( \Gamma y )$. If $\Phi$ is charged under the gauge symmetry, then the Laplacian acting on $\Phi(y)$ depends explicitly on the background gauge field. This makes computing the spectrum difficult. However, since the background gauge field has zero field strength, $F=0$, we can locally eliminate the gauge field dependence by performing a non-single valued gauge transformation $g(y)$ (see %\cite{Hall:2001tn}
appendix \ref{AppPD1} for a simple example). The price we pay for this is  to change the periodicity condition on $\Phi(y)$ to 
\bea\label{ModBoundCond}
    \Phi(y)=U_\Gamma \Phi(\Gamma y) & {\rm where} & \Gamma\in\Zbb_p \mbox{   and   } y\in \mathcal{S}^3 .
\eea
where $U_\Gamma = g(\Gamma y)$ acts in the appropriate representation. 
Thus, in the presence of the Wilson line, a charged scalar field on $Q=\mathcal{S}^3/\Zbb_p$ is equivalent to a field on $\mathcal{S}^3$ satisfying the above invariance conditions. Since the spectrum of the ordinary Laplacian is known on the round $\mathcal{S}^3$ we can proceed.

In order to compute the Green's function $G(y_1,y_2)$, we will use the eigenmodes of the Laplacian on $Q$
which satisfy the boundary conditions (\ref{ModBoundCond}) and which take values in the $(\bf{3},\bf{2})^{-5/3}+(\bf{\bar{3}},\bf{2})^{+5/3}$ representation of $G_{SM}$.

\

We will now show the explicit computations of the Green's function done in \cite{Acharya:2005ks} in several examples when $Q$ has constant curvature. The details of most of these computations are given in appendix \ref{AppPD2}, but we will give some explicit
derivations below also.

\subsubsection{Constant Positive Curvature}

3-manifolds with constant positive curvature are all quotients of the round 3-sphere by a discrete group.
We will compute the relevant Green's function for quotients by $\Zbb_p$, beginning with the simplest
example.

\vskip 7mm
\noindent
{\sf The simplest case: $\mathbb{RP}^3=\mathcal{S}^3/\Zbb_2$}\\

% {\it The simplest case: $\Bbb{RP}^3=\mathcal{S}^3/\Bbb{Z}_2$}
% 
% \

This is a particular case of the example presented above, in which $p=2$, $q=1$ and
\bee\label{UG1} 
U_\Gamma=\left(
\begin{array}{ccccc}
  1&&&&\\& 1&&&\\&&1&&\\&&&-1&\\&&&&-1\\
\end{array}\right).
\ee
Under this transformation the generators of the $(\bf{3},\bf{2})^{-5/3}+(\bf{\bar{3}},\bf{2})^{+5/3}$ representation
are odd (because the adjoint of the Standard Model is the only invariant representation). Therefore
to get invariant eigenmodes on $\mathcal{S}^3/\Zbb_2$ we have to take the odd eigenfunctions on
$\mathcal{S}^3$ under the $\Zbb_2$ transformation.

The eigenvalues of the Laplacian on $\mathcal{S}^3$ are labelled by integers $k$ and
given by $\lambda_k=-k(k+2)$. The relative eigenspaces are
\bee
\mathcal{V}^k=\{T_{k; m_1,m_2} \,\, |-k/2\leq m_1,m_2\leq k/2\}
\ee
where
\bee
T_{k; m_1,m_2}(\chi,\theta,\varphi)=\sqrt{\frac{k+1}{2\pi^2}}\,\, D^{k/2}_{m_2,m_1}(\chi,\theta,\varphi)
\ee
where $D^{k/2}_{m_2,m_1}$ are the Wigner $D$-functions, written in terms of angular coordinates on $SU(2)$.
The $D$'s are just the matrix elements of the spin
$k/2$ representation of $SU(2)$.

Under a $\Zbb_2$ transformation, $T_{k; m_1,m_2}(y)\mapsto (-1)^k T_{k; m_1,m_2}(y)$. So the odd eigenfunctions are those 
relative to odd $k$. We have also to change the normalization of such functions, because the volume of $\mathcal{S}^3/\Zbb_2$ 
is half of the volume of the defining $\mathcal{S}^3$.

The sum (\ref{GG1}) becomes:
\bee\label{GGRP}
G(y_1,y_2)=\frac{1}{\pi^2}\sum_{k=1,3,...}^\infty\frac{k+1}{-k(k+2)}
\sum_{m_1,m_2}\bar{D}^{k/2}_{m_1,m_2}(g(y_1)) D^{k/2}_{m_1,m_2}(g(y_2))
\ee
From group theory we know that \cite{fink}:
\bee\label{DDgr}
\sum_{m_1,m_2}\bar{D}^{k/2}_{m_1,m_2}(g(y_1)) D^{k/2}_{m_1,m_2}(g(y_2))  = \frac{\sin[(k+1) d(y_1,y_2)]}{\sin[d(y_1,y_2)]}
\ee
where $d(y_1,y_2)$ is the geodesic distance on the 3-sphere between $y_1$ and $y_2$.

Inserting this relation in (\ref{GGRP}) one can do the sum explicitly:
\bea\label{GGRP1}
G(y_1,y_2)&=&\frac{1}{\pi^2}\sum_{k=1,3,...}^\infty\frac{k+1}{-k(k+2)} \frac{\sin[(k+1)d ]}{\sin[d ]} \nn\\
			&=& \frac{1}{\pi^2}\sum_{j=0}^\infty\frac{2j+2}{-(2j+1)(2j+3)} \frac{\sin[(2j+2)d ]}{\sin[d ]}\nn\\
			&=& -\frac{1}{2\pi^2 \sin d}\left(\sum_{h=1}^\infty\frac{h}{h^2-1/4} \sin[2hd ]\right)\nn\\
			&=& -\frac{1}{2\pi^2 \sin d}\left( \frac{\pi}{2}\frac{\sin(\pi/2-d )}{\sin (\pi/2)}\right)\nn
\eea
where we used \cite{integrals}
and doing the last step, one gets:
\bee\label{GGRP3}
G(y_1,y_2) = -\frac{1}{4\pi}\frac{1}{\tan d(y_1,y_2)}
\ee
where $d(y_1,y_2)\in[0,\pi/2]$ is restricted to the points representing
$\mathcal{S}^3/\Zbb_2$. We see that the absolute value of the Green's function takes all  values between $0$ 
and $\infty$. So in this example, if the ${\bf 10}$ multiplet and the ${\bf \bar{5}}$ multiplet are maximally
separated in $\mathbb{RP}^3$ the Green's function is zero and the cross-section vanishes. In this case the lifetime of
the decay channel into neutrinos receives no contribution at all from dimension six operators.

\vskip 7mm
\noindent
{\sf General Lens Space}\\

%\subsubsection*{General Lens Space}

The Lens space  $L(p,r)$ is the quotient of the
3-sphere by the cyclic group whose generator 
$\Gamma$ is the $SO(4)$ isometry given in $\Rbb^4$ by \cite{Weeks}:
\bee\label{Gamma} 
\Gamma=\left(
\begin{array}{cccc}
  \cos(2\pi/p)&-\sin(2\pi/p)&&\\ \sin(2\pi/p) &\cos(2\pi/p)&&\\
  &&\cos(2\pi r/p)&-\sin(2\pi r/p)\\&&\sin(2\pi r/p)&\cos(2\pi r/p)\\
\end{array}\right)
\ee
$U_\Gamma$ is given by (\ref{UG}). With the same procedure used for the previous case, one obtains 
the formula for the Green's function:
\bee\label{GGlens}
G(y_1,y_2) = \sum_{w=1}^p  u^w \frac{d(y_1,\Gamma^w y_2)-\pi}{4\pi^2 \tan d(y_1,\Gamma^w y_2)}
\ee
where $u\equiv e^{2\pi i 5wq/p}$, and $d\in[0,\pi]$ is again the geodesic distance on the sphere.

In order to study (\ref{GGlens}), we use the cartesian coordinates on $\Rbb^4$ 
where $\mathcal{S}^3$ is defined by $x^2+y^2+z^2+t^2=1$, and
choose, without loss of generality, $y_2=y_O\equiv(1,0,0,0)$.

At first, we note that it has a singularity only at $y_1 \rightarrow y_O$, at which 
$d\rightarrow 0$. In this limit $G\sim \frac{1}{4 \pi d}$, as one expects. One can check that
this is the only divergence.
%\footnote{
%By looking at the denominators, the other a priori possible singularities could 
%occur if  $d(y_1,\Gamma^w y_O)$ ($w\not=0{\rm mod}p$) 
%went to $0$ or to $\pi$. The first possibility is never realized, because 
%$\Gamma^w y_O$ has distance at least equal to $\pi/p$ from $y_2$ when $w\not=0{\rm mod}p$. 
%The second possibility does not indeed give singularities,
%because the numerator has the same zero as the denominator.}
Secondly, we note that the Green's function on a Lens space has always zeros. Actually,  
the points  $\tilde{y}_1 = (0,0,z,t)$ (with $z^2+t^2=1$)
%\footnote{
%Each domain of the sphere around $y_O$ that we choose to define $L(p,q)$, contains some such points.} 
have the same distance $d=\pi/2$ from each of the points  $\Gamma^w y_O=(x,y,0,0)$. This is because 
the distance on the sphere is given by $\cos d=1-\frac{d_E^2}{2}$ in terms of the euclidean 
distance on $\Rbb^4$, and the chosen points have always $d_E^2=2$. So. for this value of $d$
\bee
G=\frac{d-\pi}{4\pi^2 \tan d}\sum_w u^w=0
\ee 

%\newpage

\subsubsection{Constant Zero Curvature}

Any closed, compact zero curvature manifold is a quotient of the flat 3-torus by a discrete group. Here
we consider the case of the torus itself.

\vskip 7mm
\noindent
{\sf The 3-dimensional torus}\\

%\subsubsection*{The 3-dimensional torus}

We consider the square torus, with coordinates $\vec{x}$ and $-1/2\leq x_i<1/2$. 
It is a non-simply connected manifold, whose fundamental group 
has three generators. We choose a background gauge field such that the 
holonomy associated to each of three generators is
given by 
\bee
U_i=\left(
\begin{array}{ccccc}
  1&&&&\\& 1&&&\\&&1&&\\&&&-1&\\&&&&-1\\
\end{array}\right).
\ee
with $i=1,2,3$.
This choice breaks $SU(5)$ to the Standard Model gauge group. \\
The eigenfunctions on the torus with values in $(\bf{3},\bf{2})^{-5/3}+(\bf{\bar{3}},\bf{2})^{+5/3}$ 
are those which satisfy the boundary conditions:
\bee\label{BoundCondTor}
\Phi(\vec{x})=(-1)^{\sum_i k_i}\Phi(\vec{x}+\vec{k})
\ee
for arbitrary $\vec{k}$ with $k_i\in \Zbb$. This is because each lattice generator acts as $-1$
in the representation $(\bf{3},\bf{2})^{-5/3}+(\bf{\bar{3}},\bf{2})^{+5/3}$. 

Once we have found them, we can compute the Green's function, obtaining:
\bee\label{GreenTorus}
G(\vec{x},\vec{0})=- \sum_{\vec{m}} \frac{(-1)^{\sum_i m_i}}{4\pi|\vec{x}-\vec{m}|}
\ee
This is the same formula as the electrodynamic potential of a distribution of positive and
negative charges situated on nodes of the lattice given by $\vec{m}$, where the sign of the charge is given by $(-1)^{\sum_i m_i}$. It has the expected $\frac{1}{4\pi|\vec{x}|}$ singularity when $\vec{x}\sim\vec{0}$.
Moreover it has zeros when any of the $x_i$ is equal to $1/2$. 
Actually, the charges can be grouped
in pairs, one negative, one positive each of which has
the same distance from such points.
Summing all these contributions gives so zero since the contribution from each pair
is zero. One can check this more explicitly by evaluating the expression
(\ref{GreenTorus}) in the case $\vec{x}=(1/2,x_2,x_3)$.

%\newpage

\subsubsection{Constant Negative Curvature}

A constant negative curvature 3-manifold is a quotient of hyperbolic 3-space $\mathbb{H}^3$
by a discrete group. In the compact case such groups are very rich and complicated and a description
of the eigenfunctions of the Laplacian on charged scalars is difficult to give explicitly.
Instead of attempting an explicit computation, 
we will compute the Green's functions on $\mathbb{H}^3$ itself and we will give an argument for 
the large suppression of the Green's function on compact manifolds with negative curvature.

\vskip 7mm
\noindent
{\sf The Hyperbolic 3-space}\\

%\subsubsection*{The Hyperbolic 3-space}

In this case we get the Green's function, by computing the Heat Kernel $H(y_1,y_2;t)$ 
and then integrating on $t$. Actually
\bee\label{HeatK}
H(y_1,y_2;t)=\sum_k e^{-|\lambda_k| t}\bar{\Psi}_k(y_1)\Psi_k(y_2)
\ee
and, if the integral converges,
\bea
\int_0^\infty dt H(y_1,y_2;t)&=&  \int_0^\infty dt \sum_k e^{-|\lambda_k| t}\bar{\Psi}_k(y_1)\Psi_k(y_2) \nn\\
&=& -\sum_k \frac{\bar{\Psi}_k(y_1)\Psi_k(y_2)}{\lambda_k}\nn\\
&=& -G(y_1,y_2)
\eea
Following the explicit computation reported in the appendix \ref{AppPD3}, one gets:
\bee
G_{\mathbb{H}^3}(y_1,y_2)=-\frac{1}{4\pi}\frac{e^{-d(y_1,y_2)}}{\sinh d(y_1,y_2)}
\ee
In this case the Green's
function is suppressed already at distance of order $1$.

The Green's function on a quotient of $\mathbb{H}^3$ by a discrete group in the presence
of Wilson loops will be an infinite sum of the type:
\bee
G(y_1,y_2)=-\frac{1}{4\pi}\sum_\Gamma u(\Gamma)\frac{e^{-d(y_1,\Gamma y_2)}}{\sinh d(y_1,\Gamma y_2)}
\ee
For the Torus we have found a similar expression and we have seen that it has zeros. In this case
we have also the suppression of $G_{\mathbb{H}^3}$ at distance of order $L=V^{1/3}$, where $V$ is the volume of
the final compact manifold. 
So it is conceivable that the combined action of the
cancellation by the Wilson lines phases and the exponential suppression will bring $G(y_1,y_2)$, 
if not to have zeros, to be strongly suppressed for particular choices of the points $(y_1,y_2)$. This would allow us to 
make the same conclusions as for the previous cases.

\section{Summary and Discussions}

One of the main predictions of grand unified theories is the decay of the proton and the
experimental limits on the proton lifetime in various decay channels can
give strong constraints on GUT models. In the work \cite{Acharya:2005ks} we studied proton decay in theories with extra dimensions. In particular we discussed theories in which there are significantly different predictions for the proton lifetime
relative to four dimensional GUT's. In these theories GUT gauge fields propagate {\it in more than} four dimensions, but the chiral matter fields are {\it localized} in the extra dimensions. In these cases, the GUT gauge group can be broken to that of the Standard Model through compactification; for example it can be broken by a gauge field expectation value in the extra dimensions. We showed that in such models one can get an enhancement of the lifetime in some decay channels with respect to the four dimensional GUT prediction.

The mechanism for this is the following. Firstly the symmetries of the model are such that dimension five baryon number violating operators are suppressed. We showed that it is natural in M-theory context. The leading contribution at dimension six is through the mediation of colour triplet heavy gauge bosons. There is an infinite Kaluza-Klein tower of such massive lepto-quarks. These are analogous to the $X$ and $Y$ bosons of four dimensional
$GUT$'s the difference being in the number of such particles. Then, since generically (in the language of $SU(5)$)
the points where matter ${\bf 10}$'s are localized are distinct from the points supporting ${\bf \bar{5}}$'s,
there is a qualitative difference between the decay modes such as $p\rightarrow \pi^0e^+_L$ and those such as
$p\rightarrow \pi^0e^+_R$ or $p\rightarrow \pi^+ \bar{\nu}_R$. The reason is simple: the first decay mode
comes from a current-current correlator where  {\it both} fermion currents are of a single ${\bf 10}$
multiplet localized at the {\it same} point in the extra dimensions; on the other hand for the
other two channels the two currents involve a ${\bf {\bar 5}}$ and ${\bf 10}$ multiplet which are localized at {\it different} points in the extra dimensions.
The propagator for the Kaluza-Klein lepto-quarks in the extra dimensions can take a non-trivial form. The fact that the value of the propagator can become small, even zero, is what suppresses the latter two decay channels. In other words, cancellations to the amplitudes occur by including the contribution of all the relevant Kaluza-Klein modes.

Of course, the detailed prediction for the cross-section for the proton decay involving
currents in different multiplets is quite model dependent, since it depends both on the
particular metric on the extra dimensions and on the precise locations of the two currents involved in the decay.
To investigate this model dependence we calculated  the amplitude in a variety of different spaces.
In particular we took $Q$ to be a space with constant positive, zero or negative curvature and showed
that a significant effect can always occur.

\

For the channel $p\rightarrow \pi^0e^+_L$, the two currents are at the same point and the universal short distance behavior of the propagator leads to a divergence in the amplitude, which in the M-theory context studied in \cite{Friedmann:2002ty}
was argued to be regularized. On the contrary, for the channels $p\rightarrow \pi^0e^+_R$ and $p\rightarrow \pi^+ \bar{\nu}_R$, this divergence is absent in $SU(5)$
precisely because the two currents involved are separated in the extra dimensions.
Hence there can be a suppression of these two channels both with respect to the first channel and with respect to the four dimensional prediction.
For the case of $SO(10)$ where all the matter of one generation resides in a single ${\bf 16}$ multiplet,
all three channels suffer the {\it same} divergence, hence we do not expect any qualitative difference between the three amplitudes
in $SO(10)$. In particular we do not have suppression of the decay rate into neutrino. This gives a simple way to distinguish $SO(10)$ from $SU(5)$. 

Similarly, decays involving more than one generation e.g. $p \rightarrow K^0 \mu^+$ can also be suppressed by the
small value of the propagator. This happens because different families are generally put on different points in extradimensions. These families contains fermions in interaction basis. Going to the physical basis, mixing occurs. But it is still valid that the decay into $\mu^+_L$ can be suppressed with respect to the decay into $e^+_L$; this is because the leading contribution to the decay is suppressed in the first channel, but not in the second one.
The suppression of operators involving more than one generation can occur in $SO(10)$ as well. 

The fact that the decays of protons into pions and neutrinos or right handed
positrons {\it can} be highly suppressed could give a natural explanation if, for instance,
protons are observed to decay into positrons and the lifetime
for the decay channel into neutrinos is established to be significantly
longer than this decay time. Unfortunately super-Kamiokande is not sensitive
to the helicity of outgoing positrons. A measurement of the dominant
helicity would be a strong test of these  models with localized fermions
and should be considered when planning future proton decay experiments.

\

In the models described here {\it all} the fermions of the standard model
are {\it localized} in the extra dimensions. In this case, the {\it a priori} problem
that the $SU(5)$ mass relations for the first two generations are incorrect
can be solved by introducing additional vector-like localized matter (eg ${\bf 5 \oplus {\bar 5}}$) which
mix with these generations \cite{Witten:2001bf}. In  \cite{Witten:2001bf} it is also suggested that in M-theory models with Wilson lines, one can construct realistic Yukawa couplings by realizing them through many membrane instantons intersecting the three superfield positions. In many other models considered
in the literature, where proton decay has been considered in detail %\cite{hebmarch,Kawamura:2000ev,Altarelli:2001qj,Hall:2001pg,Hebecker:2001wq}, 
\cite{Hebecker:2002rc,Kawamura:2000ev,Altarelli:2001qj,Hall:2001pg,Hebecker:2001wq},
this problem can be solved by including fermions in the bulk of $Q$
which then mix with the localized fermions. In M-theory this option is not obviously available.
Furthermore, in the models of the sort considered in \cite{Hebecker:2002rc,Kawamura:2000ev,Altarelli:2001qj,Hall:2001pg,Hebecker:2001wq},
the extra dimensions have boundaries and $SU(5)$ is broken by boundary conditions.
These two considerations can then lead to models in which  decay channels involving
the {\it first generation only} are absent at dimension six. The dominant decays are then those
such as $p \rightarrow K^0 \mu^+$. By contrast, in the models under consideration in
this chapter, decays involving the first generation are allowed. Moreover, as we have explained, the same mechanism
which suppresses, say, $p \rightarrow \pi^+ {\bar \nu}_R$ can also suppress
$p \rightarrow K^0 \mu^+$. In principle therefore it is straightforward to distinguish between these
different types of models experimentally.

\chapter{Conclusions}

In this thesis we have illustrated various aspects of String Phenomenology. 

String Phenomenology is the branch of String Theory that studies how to relate the fundamental theory with the low energy theories that reproduce very well the experimental data. In fact, its primary goals are: a) to find models which reproduce all the experimentally observed physics, and from which to work out predictions for future observations; b) to discover new physical mechanisms that could have observable signatures, or could solve theoretical problems; c) to get an overall picture of the set of all phenomenologically interesting models arising in String Theory, and to find structures in this set that could help in making predictions or in understanding better the theory.
The works presented in this thesis tried to give contributions to all these points. 

\

The most serious obstacle to testing the theory is the problem of vacua multiplycity. As we have seen, including all the discrete data of compactification, such as fluxes and brane charges, generates a huge number of possible four dimensional models. The problem of searching throughout all of them is discouraging. Perhaps {\it a priori} selection principles or measure factors will help, but there is little agreement on what these might be. The most practical strategy is a statistical study of  this large space of vacua, trying to obtain as many informations as one can and to get a guiding line for the model builders. This approach is explained in {\it chapter 4}, where at the end we have given the detailed statistical study of one ensamble of String/M-theory vacua performed in \cite{Acharya:2005ez}. This work is strictly connected to the point c). To understand the full set of vacua, one has to investigate as many corners of the Landscape as one can. In \cite{Acharya:2005ez} we analyzed the statistics of a new set of four dimensional vacua, never studied before: the $G_2$ holonomy compactifications of M-theory. We got similar result as in Type IIB vacua statistics, such as the uniform distribution of vacua over the moduli space and large volume suppression. But we found also some differences, due to the limited discrete tunability of $G_2$ flux vacua, that makes the distributions of cosmological constant and supersymmetry breaking scale not uniform near zero. In \cite{Acharya:2005ez} we also studied the statistics of M-theory Freund-Rubin vacua. We have briefly reviewed the results and seen that their statistics is very different from the special holonomy vacua.

\

In {\it chapter 5} we have illustrated a more theoretical aspect of string compactifications. We have given the ten dimensional description of Type IIA flux vacua, previously described from a four dimensional point of view \cite{Acharya:2006ne}. It is an important step in understanding the connection between the ten dimensional theory and the four dimensional effective one. At first we have seen that the fluxes backreact to the geometry, giving compact manifold that are not Ricci-flat (and so not CY). Then we considered the so called "smeared approximation", in which the localized O6-plane is smeared through the compact manifold. We have found the same results obtained by using the four dimensional approach in the "CY with fluxes" approximation. 

\

In {\it chapter 6} we have presented a setup in which one could reproduce realistic models (a). In particular the phenomenological features we reproduced are the hierarchy between the weak scale and the Planck scale and the differences in the Yukawa couplings.
To do this we used an important effect of the backreaction of fluxes and D-branes on the geometry: the presence of a warp factor depending on the compact coordinates, in front of the four dimensional Minkowski metric. In fact, there is a class of Type IIB solutions of the ten dimensional supersymmetry equations in presence of fluxes, that have this form of the metric. 

The introduction of non-factorisable geometries was suggested by Randall-Sundrum \cite{Randall:1999ee} to give a solution to the problem of the large hierarchy between the weak scale and the Planck scale. It is realized by putting the Higgs on a four dimensional brane localized in regions of large warping. Five dimensional phenomenological models followed this seminal paper. They realized among other things the Yukawa hierarchy by having matter fields in the bulk. The fermion zero modes are localized in the extradimensions and have different overlaps with the Higgs, giving different and in some cases exponentially small Yukawa couplings. In \cite{Acharya:2006mx} we realized these properties in the setup of Type IIB warped compactifications, with matter on a D7-brane and localization generated by an instanton background on the D7. We found that we have more parameters that controll the Yukawa coupling than in the five dimensional models. In particular one can get small Yukawa coupling for fermions localized on the Higgs position, but that are zero modes of a fat instanton ({\it i.e.} with large size).

\

Finally, in chapter 7 we have described a mechanism to suppress the proton decay rate in some channels. It is peculiar of String/M-theory compactifications with localized fermions (such as M-theory on singular $G_2$ holonomy manifolds or Type IIA intersecting brane models). It gives predictions very different from four dimensional GUT theories or other phenomenological extradimensional GUT models (b). In summary, we saw that for $SU(5)$ GUT models, fermions in ${\bf \bar{5}}$ and ${\bf 10}$ representation are in general localized on different points of the compact space. When considering the decay amplitude of channels involving both the representations, we found that it is multiplied by the scalar Green function. This can take zero values when the distance between the two points have certain values. Actually this mechanism does not work for decay channels involving only the ${\bf 10}$ representation. This is peculiar of this kind of models. For $SO(10)$ GUT models this does not happen, and one does not find a suppression of the decay into neutrinos, as it happens for $SU(5)$ GUT models. However the mechanism works in suppressing decays into second generation fermions, both for 
$SO(10)$ and $SU(5)$ GUT models. This is because generally different generations reside on different points in the extradimensions. In conclusion, these models have very characteristic and attractive features, that could give signature in the future experiments.

\appendix

\chapter{String Frame and Einstein Frame}\label{StrEinFrame}

The ten dimensional effective actions that describe String Theory below the string scale $m_s\sim 1/\sqrt{\alpha'}$, are obtained studying the interactions of the massless modes of the string spectrum. One takes the string amplitude that describe such interactions and take the point-particle limit $\alpha'\rightarrow 0$ (one should also remember that we are working with perturbative string theories, so the limit $g_s\ll 1$ is implied). The dynamics of these states can be summarized in terms of a low energy field theory action. For any corner of String/M-theory there is a low energy ten dimensional action (plus an eleven dimensional one, for which, however, the original theory is unknown). These actions turn out to be the actions of supergravity theories in ten dimensions with $\NN=2$ and $\NN=1$ supersymmetry.

When one derive these actions from string amplitudes, the resulting action has not a canonically normalized Einstein term. There is a factor depending on the dilaton, multiplying the Riemann scalar $R$. To get rid of this factor, one has to do a field redefinition, by rescaling the metric. Let us see an example with Type IIA low energy effective action obtained from string amplitudes \cite{Polchinski:1998rq}:
\bea
 S_{IIA} &=& S_{NS}+S_{R}+S_{CS}, \mbox{ where } \nn\\
 S_{NS} &=& \frac{1}{2\kappa_{10}^2} \int d^{10}x \sqrt{-G}\,\, e^{-2\phi} \left( R+4\partial_\mu \phi \partial^\mu \phi - \frac{1}{2}|H|^2\right) \nn\\
 S_{R} &=& -\frac{1}{4\kappa_{10}^2} \int d^{10}x \sqrt{-G} \left( |F|^2 + |G|^2\right) \nn\\
 S_{CS} &=& -\frac{1}{4\kappa_{10}^2} \int B \wedge G \wedge G
\eea
The terms are grouped according to whether they belong to the NSNS or RR sector; the last term is the Chern-Simons term involving both sectors. The factor of the dilaton in the NSNS action comes from the fact that it is derived from tree-level string amplitude (sphere worldsheet). On the other hand, the RR tree-level action vanish and the first contribution is at one-loop (torus worldsheet), getting a further factor of $e^{2\phi}$ that cancels $e^{-2\phi}$. 

If one wants a canonically normalized gravity term, one has to rescale the metric:
\bee
 G_{\mu\nu} \mapsto e^{\phi/2}G_{\mu\nu}
\ee
The resulting action is then:
\bea
 S^{E}_{IIA} &=& S^{E}_{NS}+S^{E}_{R}+S_{CS}, \mbox{ where } \nn\\
 S^{E}_{NS} &=& \frac{1}{2\kappa_{10}^2} \int d^{10}x \sqrt{-G}  \left( R+4\partial_\mu \phi \partial^\mu \phi - e^{-\phi} \frac{1}{2}|H|^2\right) \nn\\
 S^{E}_{R} &=& -\frac{1}{4\kappa_{10}^2} \int d^{10}x \sqrt{-G}  \left( e^{3\phi/2}|F|^2 + e^{\phi/2}|G|^2\right) \nn\\
 S_{CS} &=& -\frac{1}{4\kappa_{10}^2} \int B \wedge G \wedge G
\eea

\chapter{M-theory Vacua}

In this appendix we will give the normalization we used in \cite{Acharya:2005ez} and the derivation of the volume distribution in the M-theory ensemble studied in that work and reviewed in chapter 4.

\section{Normalizations}\label{AppMth1}

We defined
\begin{equation}
 K = - 3 \ln \left( 4 \pi^{1/3} V_X \right); \qquad W_{flux}(z) = \frac{1}{\kappa_4^3} N_i
 z^i
\end{equation}
The expression for the supergravity potential is the standard one
for the case of dimensionless\footnote{The scalars can of course be
given standard dimensions by rescaling $\phi = z / \kappa_4$, which
absorbs the $\kappa_4^2$ factor for the first term in
(\ref{sugrapotential}).} scalars:
\begin{equation} \label{sugrapotential}
 V = \kappa_4^2 \, e^K \, (g^{i\bj} D_i W \bar{D}_{\bj} \bar{W} - 3 |W|^2)
\end{equation}
Note that when expressed in terms of the moduli $z$, none of these
effective four dimensional field theory quantities explicitly
contains the fundamental scale $\ell_M$, only the four dimensional
Planck scale $\kappa_4$, which is the directly measurable scale in
four dimensions. To check our normalizations, it is sufficient to
verify the tension of a domain wall corresponding to an M5 brane
wrapped around a supersymmetric 3-cycle $\Sigma$, which is
Poincar\'e dual to the jump in flux $\Delta G$ across the wall.
Let's take $C=0$ for simplicity. In the supergravity theory, we have
\begin{eqnarray*}
 T_{DW} &=& 2 e^{K/2} |\Delta W| = \frac{2}{V_X^{3/2} 8 \sqrt{\pi} \kappa_4^3} |\Delta N_i z^i| = \frac{2 \pi }{(4 \pi \kappa_4^2 V_X)^{3/2} \ell_M^3} |\int_{\Sigma} \varphi| \\ \\ &=& \frac{2 \pi }{\ell_M^6} \mbox{vol}(\Sigma) = T_5 \mbox{vol}(\Sigma),
\end{eqnarray*}
which is indeed the correct expression in M-theory.

\section{Volume Distribution}\label{AppMth2}

Here we give details on how to arrive at the distribution
(\ref{vdistr}) of volumes $v=V_X/V_X^{\rm max}$ when all $a_i = a =
7/3n$:
\begin{eqnarray*}
 d\NN[v]/dv &=& \sum_{\tilde{N_i}\geq 1} \delta(v - \prod_i \tilde{N}_i^{-a}) \approx \int_{\tilde{N}_i \geq 1} d^n \tilde{N} \, \delta(v - \prod_i \tilde{N}_i^{-a}) \\
 &=& \int_{U_i \geq 0} d^n U  \, e^{\sum_j U_j} \, \delta(v-e^{-a \sum_j
 U_j}) = \frac{1}{a}\, e^{-(1+a) \frac{\ln v}{a}} \int d^n U \, \delta(\sum_j U_j + \frac{\ln
 v}{a})\\
 &=& \frac{1}{a}\, v^{-(1+1/a)} \, \frac{(-\frac{\ln
 v}{a})^{n-1}}{(n-1)!} \, \Theta(1-v) = \frac{\left(\frac{3n}{7}\right)^n}{(n-1)!} (-\ln
 v)^{n-1} \, v^{-\frac{3n}{7}-1} \, \Theta(1-v)
\end{eqnarray*}

\chapter{Ten Dimensional Type IIA Vacua}

\section{$SU(3)$ Structure Conventions}

As said before, the existence of the spinor $\eta$ implies the existence of a globally defined 2-form $J$ and 3-form $\Omega$:
\begin{gather}\label{eq005bis}
J_{mn} \equiv i \eta_{-}^\dagger \gamma_{mn}\eta_{-} = -i \eta_{+}^\dagger \gamma_{mn}\eta_{+} \\
\label{eq005b}
\Omega_{mnp} \equiv \eta_{-}^\dagger \gamma_{mnp}\eta_{+} \qquad
\Omega^*_{mnp} = -\eta_{+}^\dagger \gamma_{mnp}\eta_{-} \:,
\end{gather}
with the normalization $\eta_+^\dag \eta_+ = \eta_-^\dag \eta_- = 1$. $J$ and $\Omega$ satisfy:
\begin{gather} 
 {J_m}^n {J_n}^p = - \delta_m^p \\
 {(\Pi^+)_m}^n \Omega_{npq}=\Omega_{mpq}   \qquad   {(\Pi^-)_m}^n\Omega_{npq}=0 \\ 
 {(\Pi^\pm)_m}^n \equiv \frac{1}{2}(\delta_m^n \mp i{J_m}^n) \:.
\end{gather}
So $J$ defines an almost complex structures with respect to which $\Omega$ is $(3,0)$. Moreover
\begin{equation} \label{eq025} 
\Omega \wedge J=0 \qquad \mbox{and} \qquad J^3 = \frac{3i}{4} \Omega\wedge\Omega^* =  6 d\mbox{vol}
\end{equation} 
and
\begin{equation}
\ast J = \frac{1}{2} J\wedge J  \qquad \ast (J \wedge J) = 2 J \qquad \ast \Omega = - i \Omega
\end{equation}
\begin{eqnarray}
\ast \tilde{F} = - \tilde{F}\wedge J && \ast (\tilde{F}\wedge J) = - \tilde{F}
\end{eqnarray}

\section{Check of the Equations of Motion} \label{AppTIIA1}

As we have said before, if the solution to the supersymmetry equations satisfies also the BI and the equations of motion for the forms, then it satisfies the Einstein and the dilaton equations as well \cite{Lust:2004ig}. 
Here we check that it is true for the dilaton and the 4-dimensional components of the Einstein equation. A complete proof in a more general $\NN=1$ supersymmetric context (type IIA/IIB $SU(3)\times SU(3)$ structure compactifications in presence of general supersymmetric sources) can be found in \cite{Koerber:2007hd}.

The dilaton eom \eqref{eq056} is the same as in \cite{Lust:2004ig}, but with the addition of the $O6$ term. 
Moreover, the fields take the same values on the solution as in 
\cite{Lust:2004ig}, except for $F$. The value of $F^2$ is the \cite{Lust:2004ig} one plus 
\begin{equation} \label{eq011}
\delta F^2=\frac{1}{4}\mu_6 \frac{\sqrt{-g_3}}{\sqrt{-g_6}} \delta^3(\Sigma)e^{-3\phi/4} \:.
\end{equation}
So if the \cite{Lust:2004ig} EOM are satisfied, all the terms in \eqref{eq056} 
sum up to zero, except for
\begin{equation}\label{eq012}
-\frac{3}{8} e^{3\phi/2} \delta |F|^2 + 
	\frac{3}{2}\mu_6 \frac{\sqrt{-g_3}}{\sqrt{-g_6}} \delta^3(\Sigma)e^{3\phi/4} \:.
\end{equation}
By substituting \eqref{eq011} into \eqref{eq012} one gets exactly zero and the dilaton 
EOM turns out to be correct.

Consider, now, the Einstein EOM in the $\mu,\nu=0,...,3$ directions. The piece of the equation 
which is not automatically zero if the \cite{Lust:2004ig} EOM are satisfied is:
\begin{equation}
\frac{1}{32}e^{3\phi/2}g_{\mu\nu}\delta |F|^2 -
	\frac{1}{8}\mu_6 \frac{\sqrt{-g_3}}{\sqrt{-g_6}} \delta^3(\Sigma) g_{\mu\nu}e^{3\phi/4} \:.
\end{equation}
Again the result is zero and the eom is satisfied.

\chapter{The ADHM Construction}\label{AppWMST1}

In this section we briefly review the ADHM formalism for instantons and how to use it to find bosonic and fermionic zero modes around their background \cite{Atiyah:1978ri} (to have a more complete review of the subject, see \cite{Dorey:2002ik} and references therein). We are interested in constructing finite action solutions of the four dimensional Euclidean Yang-Mills theory (instantons). The gauge potential satisfies a first order (anti-)self-duality equation
\begin{equation} \label{01}
F_{\mu\nu} = \pm (\ast F)_{\mu\nu} = \pm \frac{1}{2} \epsilon_{\mu\nu\rho\sigma} F_{\rho\sigma}
\end{equation}
% where the field strength $F_{\mu\nu}$ is given by
% \begin{gather}
% F_{\mu\nu} = \partial_\mu A_\nu - \partial_\nu A_\mu + [A_\mu,A_\nu] \\
% F_{\mu\nu}^a = \partial_\mu A_\nu^a - \partial_\nu A_\mu^a + \epsilon^{abc} A_\mu^b A_\nu^c \\
% \text{and} \qquad A_\mu = -\frac{i}{2} \tau^a A_\mu^a \qquad \qquad [\tau^a,\tau^b]=2i\epsilon^{abc}\tau^c \:.
% \end{gather}
% % and $\tau^a$ are the Pauli matrices.
% In the following, for reasons that will be clear, we will be interested in anti-self-dual solution, but it is clear that self-dual solution can be treated similarly. Moreover 
In the following we will restrict ourselves to $U(N)$ gauge groups.

In order to discuss the ADHM formalism, we introduce the quaternionic notation:
\begin{gather}
{\bf z} = z_\mu \sigma_\mu \qquad \qquad \bar{{\bf z}} = z_\mu \bar\sigma_\mu \\
z_\mu = \frac{1}{2} \mbox{tr} \:{\bf z} \:\bar\sigma_\mu
\end{gather}
where $\sigma_\mu=(i\tau^a,1)$ and $\bar\sigma_\mu=(-i\tau^a,1)$. 
% The following identities hold:
% \begin{gather}
% \sigma_\mu \bar\sigma_\nu = \delta_{\mu\nu} \mathbbm{1} + 2 \sigma_{\mu\nu} = \delta_{\mu\nu} \mathbbm{1} + i\eta_{a\mu\nu}\tau^a \:.
% \end{gather}

The ADHM formalism allows to obtain (anti-)self-dual field strength configurations by solving only algebraic equations.
The gauge field with instanton number $k$ for $U(N)$ gauge group is given by
\begin{equation}
A_\mu = v(z)^\dag \partial_\mu v(z) \:,
\end{equation}
where $v(z)$ is a $(N+2k)\times N$ matrix. It is defined by the equations
\begin{gather}
v(z)^\dag v(z) = 1 \label{02} \\
v(z)^\dag \Delta(z) = 0 \:. \label{03}
\end{gather}
Here $\Delta(z)$ is a $(N+2k)\times 2k$ matrix, linear in the position variable $z$, having the structure
\begin{equation} \label{04}
\Delta(z) = \begin{cases}
a - b \mathbf{z} & \text{self-dual instantons,} \\
a - b \bar{\mathbf{z}} & \text{anti-self-dual instantons,} \end{cases}
\end{equation} 
The matrices $a,b$ are constrained to satisfy the condition
\begin{equation}\label{eqR003}
\Delta(z)^\dag \Delta(z)=p^{-1}(z) \otimes {\bf 1}_2
\end{equation}
where $p^{-1}(z)$ is a $k\times k$ invertible matrix. This assures the (anti-)self-duality equation \eqref{01}. 

$a,b$ are $(N+2k)\times 2k$ matrices that contain the moduli of the instantonic configuration. 
Beacause of some symmetries of the equations above they can be brought to the form
\begin{equation}
a = \begin{pmatrix} \lambda \\ \xi \end{pmatrix} \qquad \qquad b = \begin{pmatrix} 0 \\ {\bf 1}_{2k} \end{pmatrix} \:,
\end{equation}
where $\lambda$ is an $N\times 2k$ matrix and $\xi$ is a $2k\times 2k$. There is no one-to-one correspondence between these two matrices and the moduli: some constraints and redundancies are left. The actual number of moduli is $4Nk$.

\section{Fermion Zero Modes}

We will be interested in the fermionic zero modes in the fundamental representation and with definite chirality, {\it i.e.} those solving:
\begin{equation}
 \sigma^\mu D_\mu \eta = \sigma^\mu (\partial_\mu+v^\dagger \partial_\mu v)\eta \:.
\end{equation}

One gets $k$ independent solutions for $\eta^T$ as an $N\times 2$ matrix:
\begin{equation}\label{eqR006}
 \eta^i_{u,\alpha}=(v^\dagger b p \sigma^2)_{u,i\alpha}
\end{equation}
where $u=1,...,N$, $i=1,...,k$ and $\alpha=1,2$. Thus we have found $k$ fermionic zero modes in the fundamental representation.

\section{$k=1$ $SU(2)$ Instanton}

We apply the machinery described above to the simplest case of one $SU(2)$ instanton. In this case, applying the further constraints on $a$ and $b$, one can put $\Delta$ in the form:
\begin{equation}
 \Delta = \left(\begin{array}{c} \rho \, \mathbf{1}_2 \\ \bar{\mathbf{Z}}-\bar{\mathbf{z}}\end{array}\right)
\end{equation}
with $\rho$ and $Z_\mu$ the $(4kN - N^2+1) = 5$ parameters of the solution in the case $k=1,N=2$. 

From here, using \eqref{eqR003}, we can get $f$:
\begin{equation}
 p(z) = \frac{1}{\rho^2+(z-Z)^2}
\end{equation}
Then solving for the normalized zero eigenvectors $v^\dagger \Delta=0$ and $v^\dagger v=1$, we have:
\begin{equation}
 v(z) = \left(\begin{array}{c} \left(\frac{(z-Z)^2}{\rho^2+(z-Z)^2}\right)^{1/2}\mathbf{1}_2   \\
             \left(\frac{\rho^2}{(z-Z)^2(\rho^2+(z-Z)^2}\right)^{1/2} (\mathbf{z-Z}) \end{array}\right)
\end{equation} 
And finally one gets the connection in singular gauge:
\begin{equation}
  A_\mu= \frac{\rho^2 (z-Z)_\nu}{(z-Z)^2(\rho^2+(z-Z)^2}\sigma_{\mu\nu}
\end{equation}

\subsection{Fermion Zero Modes}

We compute the fermion zero modes in this simple anti-instanton background, by using the formula \eqref{eqR006}:
\begin{equation}
  v^\dagger b p = \frac{\rho}{(\rho^2+(z-Z)^2)^{3/2}} \frac{\mathbf{z-Z}}{|z-Z|}
\end{equation} 
This is a $2\times 2$ matrix. One index is for the fundamental rep, while the other is a spinorial index.

\section{'t Hooft Solution}

Now we consider the case in which $k$ is general, the gauge group is $SU(2)$, 
and we will concentrate on a class of solutions described by $5k$ parameters (instead of $8k$): $\rho_i$ and $Z_i$, with $i=1,...,k$. It is called the {\it 't Hooft solution} \cite{tHooft:1976fv} and is characterized, in the ADHM construction, by:
\begin{equation}
 v(z)= \left(\begin{array}{c} \left[1+\sum_{i=1}^k\frac{\rho_i^2}{(z-Z_i)^2}\right]^{-1/2}\mathbf{1}_2   \\
             \left[1+\sum_{i=1}^k\frac{\rho_i^2}{(z-Z_i)^2}\right]^{-1/2}\frac{\rho_i^2(\mathbf{z-Z}_i)}{(z-Z_i)^2}\end{array}\right)
\end{equation}
It is obtained by taking
\begin{equation}
a = \begin{pmatrix} \rho_i\mathbf{1}_2 \\ \delta_{ji}\mathbf{Z}_i \end{pmatrix} \qquad \qquad 
b = \begin{pmatrix} 0 \\ {\bf 1}_{2k} \end{pmatrix} \:,
\end{equation}
From these, one can also get the expression for $p$. The diagonal entries are:
\begin{equation}
 p_{ii} = \left[1+\sum_{\ell=1}^k\frac{\rho_\ell^2}{(z-Z_\ell)^2}\right]^{-1}\frac{1}{(z-Z_i)^2}
	\left[1+\sum_{j\not=i}\frac{\rho_j^2}{(z-Z_j)^2}\right]\:,
\end{equation}
while the off-diagonal elements are:
\begin{equation}
 p_{ij}=-\left[1+\sum_{\ell=1}^k\frac{\rho_\ell^2}{(z-Z_\ell)^2}\right]^{-1} \frac{\rho_i\rho_j}{(z-Z_i)^2(z-Z_j)^2}\:.
\end{equation}

\

There are asymptotic regions of the parameters space where the multi-instanton configurations can be identified as being composed of well-separated single instantons. One can show that this limit is valid when
\begin{equation}\label{eqR012}
 (Z_i-Z_j)^2\gg \rho_i\rho_j \qquad \forall i\not = j
\end{equation}
In this limit the $Z_i$'s become the positions of the $k$ instantons, while the $\rho_i$'s are their sizes.

\subsection{Fermion Zero Modes}
As in the case $k=1$, we compute the fermion zero modes in the background described above, by using the formula \eqref{eqR006}:
\begin{eqnarray}\label{eqR014}
  (v^\dagger b p)_h &=& \left[1+\sum_{\ell=1}^k\frac{\rho_\ell^2}{(z-Z_\ell)^2}\right]^{-3/2}\frac{\rho_h}{(z-Z_h)^2}\times\\
	&&\times  \left\lbrace\left[1+\sum_{\ell=1}^k\frac{\rho_\ell^2}{(z-Z_\ell)^2}\right]\frac{\mathbf{z-Z}_h}{(z-Z_h)^2}-
		 \sum_{j=1}^k\frac{\rho_j^2}{(z-Z_j)^4}(\mathbf{z-Z}_j)\right\rbrace \nn
\end{eqnarray}
It is in the fundamental representation of $SU(2)$.

\

In the limit of well separated $k$ instantons, {\it i.e.} \eqref{eqR012}, the expression for the fermionic zero modes simplifies:
\begin{eqnarray}\label{eqR018}
  (v^\dagger b p)_h &\sim& \frac{\rho_h}{(\rho_h^2+(z-Z_h)^2)^{3/2}} \frac{\mathbf{z-Z}_h}{|z-Z_h|}\:.
\end{eqnarray}
It is the same expression for the fermion zero mode in the case of one instanton localized in $Z_h$. One can see that in regions around other instanton ($z\sim Z_j$, with $j\not =h$), the solution found above is of order $\frac{\rho_h \rho_j}{(Z_h-Z_j)^2}\ll 1$. So in this approximation there is one fermionic zero mode localized around each instanton. One has to note that the suppression of points distant from every instanton positions is larger than that obtained  around $Z_{j\not=k}$. On these points we have low peak, suppressed with respect to that on $Z_h$, but larger with respect to the value of the single instanton profile at that point.\label{page}

\subsection{Vector Zero Modes}

The vector zero modes in the adjoint representation of $SU(2)$ are those variations of $A_\mu$ that leave it a solution of the (anti-)selfdual equation (and that are not gauge transformations). They are associated to the parameter that describe the solution. 

In the ADHM construction it is given the expression of the zero modes:
\begin{equation}
 \delta A_\mu = -v^\dagger(\delta \! a p \sigma_\mu b^\dagger - b\bar{\sigma}_\mu p \delta\! a^\dagger)v
\end{equation}

Consider again the 't Hooft solution. There are $5k$ zero modes: $4k$ associated with changing positions of each instanton, and $k$ with changing their sizes. 

As an example, we give the expression for the zero mode relative to the translation of $Z_j$, by the vector $\Phi$:
\begin{eqnarray}\label{eqR015}
 \delta A^{\Phi}_\mu &=& \Phi_\nu\left[1+\sum_{\ell=1}^k\frac{\rho_\ell^2}{(z-Z_\ell)^2}\right]^{-2} \frac{\rho_j^2}{(z-Z_j)^4}
 	(\mathbf{z-Z}_j)^\dagger \sigma_{\mu\nu}\,\times \nn\\
		&&\times\,\left\lbrace \frac{(\mathbf{z-Z}_j)}{(z-Z_j)^2}\left[1+\sum_{i\not=j}\frac{\rho_i^2}{(z-Z_i)^2}\right]
	- \sum_{i\not=j}\frac{\rho_i^2}{(z-Z_i)^4}(\mathbf{z-Z}_i)\right\rbrace \nn\\
\end{eqnarray}
One can see that in the limit \eqref{eqR012} it becomes the zero mode of the single instanton solution localized on $Z_j$:
\begin{equation}
 \delta A^{\Phi}_\mu \sim \Phi_\nu\frac{\rho_j^2}{(z-Z_j)^2}
 	\frac{(\mathbf{z-Z}_j)^\dagger \sigma_{\mu\nu} (\mathbf{z-Z}_j)}{(\rho_j^2+(z-Z_j)^2)^2}
\end{equation}

\chapter{Warping Effects on the Dirac Operator}\label{AppWMST2}

We want to find the spin-connection relative to the metric:
\begin{eqnarray}
 ds^2 &=& f(r)^{-1/2}\tilde{g}_{(3,1)\mu\nu}\,dx^\mu dx^\nu + f(r)^{1/2} \tilde{g}_{(4)\alpha\beta}\,dz^\alpha dz^\beta \nn\\
      &=& f(r)^{-1/2}\eta_{mn} \, \tilde{e}^m\tilde{e}^n + f(r)^{1/2} \delta_{ab} \, \tilde{e}^a\tilde{e}^b\\
      &=& \eta_{mn}\, e^m e^n + \delta_{ab}\, e^a e^b \nn = \eta_{HK}\, e^H e^K
\end{eqnarray}
The corresponding 8-bein is given then by $e^m=f(r)^{-1/4}\tilde{e}^m$ and $e^a=f(r)^{1/4}\tilde{e}^a$. $r$ is the radial coordinate in the (4)-dimensional space spanned by the coordinates $z^\alpha$.

The spin connection is given by:
\begin{eqnarray}
 \omega_\Pi^{HK}&=&\frac{1}{2}e^{\Lambda H}(\partial_\Pi e^K_\Lambda-\partial_\Lambda e^K_\Pi) - [H\leftrightarrow K]\nn\\
 	&& -\frac{1}{2}e^{\Xi H}e^{\Upsilon K} (\partial_\Xi e_{\Upsilon Q}-\partial_\Upsilon e_{\Xi Q})e_\Pi^Q
\end{eqnarray}
Using this formula, one obtains:
\begin{eqnarray}
 \omega_\mu^{ab} &=& \tilde{\omega}_\mu^{ab} \nn\\
 \omega_\mu^{an} &=& \tilde{\omega}_\mu^{an}+\frac{f'}{4f^{3/2}}\tilde{e}_\mu^n \tilde{e}^{ra}\nn\\
 \omega_{\beta}^{\hat{a}\hat{b}} &=& \tilde{\omega}_{\beta}^{\hat{a}\hat{b}}\nn\\
 \omega_{\hat{\beta}}^{\hat{a}R} &=& \tilde{\omega}_{\hat{\beta}}^{\hat{a}R}
 		+\frac{f'}{4f} \tilde{e}_{\hat{\beta}}^{\hat{a}}\nn\\
 \omega_{r}^{\hat{a}R} &=& \tilde{\omega}_{r}^{\hat{a}R} \nn\\		
\end{eqnarray}
where $\tilde{\omega}$ is the spin connection associated to $\tilde{g}$ and the coordinates $x^\alpha$ are split in the radial coordinate $r$ and in the other three coordinates $x^{\hat{\alpha}}$ (and also $a=R,\hat{a}$).

\

The Dirac operator is given by
\begin{equation}
 \D_8 = e^\Pi_K \Gamma^K (\partial_\Pi + \omega_\Pi^{HQ} \frac{1}{4}\Gamma_H\Gamma_Q + A_\Pi)
\end{equation}
In the setup we are considering ($A_\mu=0$ and $\omega$ given above), it is equal to
\begin{eqnarray}
\D_8 &=& f^{1/4} \tilde{e}^\mu_m \Gamma^m (\partial_\mu + \omega_\mu^{HQ} \frac{1}{4}\Gamma_H\Gamma_Q)+
f^{-1/4} \tilde{e}^\alpha_a \Gamma^a (\partial_\alpha + \omega_\alpha^{HQ} \frac{1}{4}\Gamma_H\Gamma_Q + A_\alpha)\nn\\
&=& f^{1/4} \tilde{e}^\mu_m \Gamma^m ((\tilde{D}_{3,1})_\mu + \delta\omega_\mu^{HQ} \frac{1}{4}\Gamma_H\Gamma_Q)+
f^{-1/4} \tilde{e}^\alpha_a \Gamma^a ((\tilde{D}_4)_\alpha + \delta\omega_\alpha^{HQ} \frac{1}{4}\Gamma_H\Gamma_Q)\nn
\end{eqnarray}
Where $\delta\omega=\omega-\tilde{\omega}$ can be read off above. In particular 
\begin{eqnarray}
\tilde{e}^\mu_m \Gamma^m \, \delta\omega_\mu^{HQ} \frac{1}{4}\Gamma_H\Gamma_Q &=& 
-\frac{f'}{2\,f^{3/2}} \Gamma_r \\
\tilde{e}^\alpha_a \Gamma^a \,  \delta\omega_\alpha^{HQ} \frac{1}{4}\Gamma_H\Gamma_Q &=& 
\frac{3}{8}\frac{f'}{f} \Gamma_r
\end{eqnarray}
Putting all together one gets:
\begin{eqnarray}
\D_8 &=& f^{1/4} (\tilde{\D}_{3,1}\otimes {\bf 1}   +\frac{3}{8}\frac{f'}{f^{3/2}}\gamma^{(4)}\otimes\gamma_r) + f^{-1/4} (\gamma^{(4)}\otimes\D_4 -\frac{f'}{2f}\gamma^{(4)}\otimes\gamma_r)\nn\\
&=& f^{1/4} \tilde{\D}_{3,1}\otimes {\bf 1}+ f^{-1/4}\gamma^{(4)}\otimes\tilde{\D}_4 - \frac{1}{8f^{1/4}}\frac{f'}{f} \gamma^{(4)}\otimes\gamma_r
\end{eqnarray}

\

Splitting the eight dimensional spinor as $\Psi = \sum_k \chi_k(x) \otimes \psi_k(y) $, we see that the zero modes of $\tilde{\D}_{3,1}$ are associated to the zero modes of the operator $\hat{\D}_4=\tilde{\D}_4 - \frac{f'}{8f} \gamma_r$. If $\psi_0$ is a zero mode of $\tilde{\D}_4$, then $\psi=f^{1/8}\psi_0$ is a zero mode of $\hat{\D}_4$, since:
\begin{eqnarray}
\tilde{\D}_4 (f^{1/8}\psi_0) = \gamma_r (\partial_r f^{1/8}) \psi_0 = \frac{f'}{8f}\gamma_r(f^{1/8} \psi_0)\:.
\end{eqnarray}

\chapter{Modified Boundary Conditions Generated by Wilson Lines}\label{AppPD1}

We discuss a simple example in order to explain how a flat connection can be exchange with non-trivial boundary conditions \cite{Hall:2001tn}. We take a compactification on $S^1$, parametrized by 
$y\in[0,2\pi R]$ and $SU(2)$ gauge group. We also take 
\bee
A_{\rm bkg}(y)=A^y \frac{\bf{\tau^3}}{2} d y=A^y \frac{1}{2}\left(
	\begin{array}{cc}
		1&0\\0&-1\\
	\end{array}\right) dy
\ee
where $A^y$ does not depend on $y$. The corresponding holonomy is given by 
\bee
   T=e^{i\oint A^y \frac{\bf{\tau^3}}{2} d y}=\left(
	\begin{array}{cc}
		e^{\pi i A^y R}&\\&e^{\pi i A^y R }\\
	\end{array}\right) 
\ee
The KK modes of a scalar field $\Phi$ in the fundamental representation are given by the eigenmodes of the modified
Laplacian operator 
\bee
   \Delta^{(A)}=-\left(\partial_y-i A^y \frac{\bf{\tau^3}}{2}\right)^2 
\ee  
The eigenmodes, which must be periodic ($\Phi(y+2\pi R)=\Phi(y)$), are $\Phi_n^{+}=(e^{i n y/R},0)$ and $\Phi_n^{-}=(0,e^{i n y/R})$, 
while the relative eigenvalues are $m_n=|\frac{n}{R}\mp \frac{A^y}{2}|$.

If instead we gauge away the gauge field by $U(y)=e^{-i A^y  \frac{\bf{\tau^3}}{2}  y}$, the Laplacian is simply 
$-\partial_y^2$,  while the boundary condition is changed to 
\bee
   \Phi(2\pi R)=e^{-i \pi A^y {\bf \tau^3} R} \Phi(0) = T^{-1}\Phi(0)
\ee
The eigenmodes of the Laplacian are now given by $\Phi_n^{+}=(e^{i (\frac{n}{R} -A^y  \frac{y}{2})},0)$ 
and $\Phi_n^{-}=(0,e^{i (\frac{n}{R} +A^y  \frac{y}{2})})$, but the corresponding spectrum is identical to the 
previous one.

\chapter{Green's function on Lens spaces: details}\label{AppPD2}

In order to compute the Green's function on Lens spaces, one needs the eigenmodes on them.

\section{Eigenmodes of Laplacian on the 3-sphere}

In order to study the eigenmodes of the Laplacian on Lens spaces, we need to review the eigenmodes on the 3-sphere 
\cite{Weeks}.

At first, we introduce the toroidal coordinates on the 3-sphere $\mathcal{S}^3$. 
Let $x,y,z$ and $t$ be the usual coordinates in $\Rbb^4$, so $\mathcal{S}^3$ is defined by $x^2+y^2+z^2+t^2=1$, and can be 
parametrized by the coordinates $\chi,\theta$ and $\varphi$ as
\bea\label{TorCoord}
x&=&\cos\chi\,\,\cos\theta\\
y&=&\cos\chi\,\,\sin\theta\\
z&=&\sin\chi\,\,\cos\varphi\\
t&=&\sin\chi\,\,\sin\varphi
\eea
with $0\leq\chi\leq\pi/2$, $-\pi \leq \theta\leq \pi$ and $-\pi \leq\varphi\leq \pi$.

The eigenvalues of the Laplacian on $\mathcal{S}^3$ are given by $\lambda_k=-k(k+2)$. The relative eigenspaces are given by
\bee
\mathcal{V}^k=\{T_{k; m_1,m_2} \,\, |-k/2\leq m_1,m_2\leq k/2\}
\ee
where the $T$'s can be expressed in terms of the Wigner $D$-functions $D^{k/2}_{m_2,m_1}$:
\bee\label{TDapp}
T_{k; m_1,m_2}(\chi,\theta,\varphi)=\sqrt{\frac{k+1}{2\pi^2}}\,\, D^{k/2}_{m_2,m_1}(\chi,\theta,\varphi)
\ee

\section{Eigenmodes of Laplacian on Lens Spaces}

The Lens space $L(p,r)$ is the quotient of the 3-sphere by the cyclic group whose generator $\Gamma$ 
is the isometry \cite{Weeks}
\bea\label{TrZpqapp}
\chi \mapsto \chi; &   \theta \mapsto \theta + 2\pi/p; &\varphi \mapsto \varphi + 2\pi r/p
\eea
It can be described using toroidal coordinates, with limit 
$0\leq \chi\leq \pi/2$, $-\pi/p < \theta < \pi/p$ and $-\pi r/p < \varphi < \pi r/p$.
Obviously it cannot be covered only with  one such patch, but the set of non-covered points is of null measure.
Moreover it gives a good local description around 
the point $(\chi,\theta,\varphi)=(0,0,0)$.

Under the transformation (\ref{TrZpqapp}) the eigenfunctions found above transform as:
\bee\label{Ttransfapp}
T_{k; m_1,m_2}(\chi,\theta,\varphi)  \mapsto  e^{2\pi i(\ell+m r)/p} \,T_{k; m_1,m_2}(\chi,\theta,\varphi)
\ee
Moreover the $(\bf{3},\bf{2})^{-5/3}+(\bf{\bar{3}},\bf{2})^{+5/3}$ representation takes a factor $e^{2\pi i 5q/p}$
under the gauge transformation $U_\Gamma$. The condition (\ref{ModBoundCond}) then becomes
\bea\label{ModBoundCondLens}
   T_{k; m_1,m_2}(y) = e^{2\pi i(\ell+m r+5q)/p} \,T_{k; m_1,m_2}(y)
\eea
and the invariant eigenmodes are those satisfying the constraint $\ell+m r+5q=0 \,\,{\rm mod}\,\, p$.

If one wants the right normalization, in order to get an orthonormal base, the
$\sqrt{\frac{1}{2\pi^2}}$ factor has to be changed in the more general  
$\sqrt{\frac{1}{V}}$, where $V$ is the volume of $L(p,r)$. In what follows we will call $T$ 
the eigenmodes of Laplacian with appropriately modified normalization.

\section{Green's function}

Having the Laplacian eigenmodes on $L(p,r)$, we can compute the Green's function explicitly:
\bee 
G(y_1,y_2)=\sum_{{\begin{array}{c}  k;m_1,m_2\\constr\\ \end{array}}} \frac{1}{\lambda_k}\bar{T}_{k; m_1,m_2}(y_1) 
T_{k; m_1,m_2}(y_2)
\ee
where the sum over $\lambda_k=-k(k+2)$ and $\{ k,m_1,m_2  \}$ is constrained by 
$\ell+m r+5q=0 \,\,{\rm mod}\,\, p$ and $m_1$ and $m_2$ running from $-k/2$ to $k/2$ with integer step.
We implement these constraints by using the fact that
\bee
\frac{1}{p}\sum_{w=1}^p e^{2\pi i w(5q+\ell+mr)/p}
\ee
is equals to one if and only if $\ell+m r+5q=0 \,\,{\rm mod}\,\, p$ and is zero otherwise.

So we can write:
\bea\label{GG5}
G(y_1,y_2)&=&\sum_{{\begin{array}{c}  k\not=0;m_1,m_2\\unconstr\\ \end{array}}} \frac{1}{\lambda_k}\bar{T}_{k; m_1,m_2}(y_1)\left(
\frac{1}{p}\sum_{w=1}^p e^{2\pi i 5q w/p}  e^{2\pi i w(\ell+mr)/p}  T_{k; m_1,m_2}(y_2) \right) \nn\\
&=& \frac{1}{p}\sum_{w=1}^p  u^w \sum_{{\begin{array}{c}  k\not=0;m_1,m_2\\unconstr\\ \end{array}}} \frac{1}{\lambda_k}
\bar{T}_{k; m_1,m_2}(y_1)  T_{k; m_1,m_2}(\Gamma^wy_2)\nn\\
&=& \frac{1}{p}\sum_{w=1}^p  u^w \frac{2\pi^2}{V}G_{\mathcal{S}^3}(y_1,\Gamma^wy_2)
\eea
where $u\equiv e^{2\pi i 5wq/p}$ and
\bee
G_{\mathcal{S}^3}(y_1,y_2)\equiv \sum_{k\not=0;m_1,m_2} \frac{1}{\lambda_k}
\bar{T}_{k; m_1,m_2}^{\mathcal{S}^3}(y_1)  T_{k; m_1,m_2}^{\mathcal{S}^3}((y_2)
\ee
is the regulated Green's Function on the sphere ({\it e.i.} one neglects the zero mode in the sum and the modes have the appropriate normalization for
the sphere), which we will compute in a moment.

By using (\ref{DDgr}) and \cite{integrals} one gets:
\bea
G_{\mathcal{S}^3}(y_1,y_2) &=&  \frac{1}{2\pi^2}\sum_{k=1}^\infty\frac{k+1}{-k(k+2)} \frac{\sin[(k+1)d ]}{\sin[d ]} \nn\\
&=&-\frac{1}{2\pi^2}    \frac{1}{\sin[d]}\sum_{h=2}^\infty\frac{h}{h^2-1} \sin[h d ]       \nn \\
&=& -\frac{1}{4\pi\tan d}+\frac{1}{8\pi^2}+\frac{d}{4\pi^2 \tan d}
\eea
When we use it in order to compute (\ref{GG5}), we can neglect the constant $1/8\pi^2$ because it gives zero 
contribution: it factors out from the sum over $w$, which is so equal to zero since $5q \not = 0 {\rm mod}p$. So
\bee\label{GG6}
G(y_1,y_2) = \sum_{w=1}^p u^w \frac{d(y_1,\Gamma^w y_2)-\pi}{4\pi^2 \tan d(y_1,\Gamma^w y_2)}
\ee

We note that if we use the formula (\ref{GG6}) for the Green's function on $L(2,1)=\mathcal{S}^3/\Zbb_2$, 
we actually get the same result as (\ref{GGRP3}).

\chapter{Green's function on $\mathbb{H}^3$}\label{AppPD3}

In order to compute the fundamental solution to the the Heat equation (\ref{HeatK}) 
on $\mathbb{H}^3$, we will use the formula given 
at page $150$ of \cite{Chavel}:
\bee
H(y_1,y_2;t)= (4\pi t)^{-3/2}\,e^{-d^2(y_1,y_2)/4t} e^{-t}\frac{d(y_1,y_2)}{\sinh d(y_1,y_2)}  
\ee

We compute the following integral over $t$:
\bea
  \int_0^\infty \frac{dt}{t^{3/2}}e^{-d^2/4t} e^{-t} &=& 2^{3/2} d^{-1/2} K_{1/2}(d)\nn\\
  &=& 2^{3/2} d^{-1/2} \frac{e^{-d}(2\pi)^{1/2}}{2 d^{1/2}}\nn\\
  &=& 2 \pi^{1/2}\frac{e^{-d}}{d}
\eea
Where $K_\nu$ is the modified Bessel function.
So the Green's function is given by
\bea
  G_{\mathbb{H}^3}(y_1,y_2;t) &=& -\int_0^\infty dt \, H(y_1,y_2;t)\nn\\
  &=& -(4\pi)^{-3/2} \frac{d(y_1,y_2)}{\sinh d(y_1,y_2)} \,2 \pi^{1/2}\frac{e^{-d(y_1,y_2)}}{d(y_1,y_2)}\nn\\
  &=& -\frac{1}{4\pi}\frac{e^{-d(y_1,y_2)}}{\sinh d(y_1,y_2)}
\eea

%\bibliographystyle{utcaps}
%\bibliography{Thesis}

%NOTES

% Type IIA dual of G2 holonomy vacua ->proton decay in Type IIA
% Cosmological selection of vacua (Banks)
% Large volume is Type IIB

\end{document}